\documentclass[11pt]{article}
\usepackage{graphicx}
\usepackage[margin=1.25in]{geometry}
\usepackage[usenames,dvipsnames]{color}
\usepackage{url}
\usepackage[colorlinks = true,
            linkcolor = blue,
            urlcolor  = blue,
            citecolor = blue,
            anchorcolor = blue]{hyperref}
\usepackage{authblk}
\usepackage{arydshln}
\usepackage{amsmath,amssymb}
\usepackage{cite}
\usepackage{adjustbox}
\usepackage{multirow}

\makeatletter
\renewcommand\AB@affilsepx{, \protect\Affilfont}
\makeatother

\renewcommand*{\Affilfont}{\sl\small}

\def\beq{\begin{equation}}
\def\eeq#1{\label{#1}\end{equation}}
\def\eeqn{\end{equation}}

\newenvironment{Eqnarray}%
   {\arraycolsep 0.14em\begin{eqnarray}}{\end{eqnarray}}
\def\beqa{\begin{Eqnarray}}
\def\eeqa#1{\label{#1}\end{Eqnarray}}
\def\eeqan{\end{Eqnarray}}

\let\bar=\overbar

\def\lsim{\mathrel{\raise.3ex\hbox{$<$\kern-.75em\lower1ex\hbox{$\sim$}}}}
\def\gsim{\mathrel{\raise.3ex\hbox{$>$\kern-.75em\lower1ex\hbox{$\sim$}}}}

\def\del{\partial}
\def\Dslash{\not{\hbox{\kern-4pt $D$}}}
\def\dslash{\not{\hbox{\kern-2pt $\del$}}}
\def\pslash{\not{\hbox{\kern-2pt $p$}}}
\def\ETmiss{\not{\hbox{\kern-4pt $E$}}_T}

\def\Dlr{\mathrel{\raise1.5ex\hbox{$\leftrightarrow$\kern-1em\lower1.5ex\hbox{$D$}}}}

\def\MSB{{\bar{M \kern -2pt S}}}
\def\msb{{\bar{\scriptsize M \kern -1pt S}}}

\def\drb{{\bar{\scriptsize D \kern -1pt R}}}

\def\eV{{\rm eV}}
\def\keV{{\rm keV}}
\def\MeV{{\rm MeV}}
\def\GeV{{\rm GeV}}
\def\TeV{{\rm TeV}}

\newcommand\snowmass{\begin{center}\rule[-0.2in]{\hsize}{0.01in}\\\rule{\hsize}{0.01in}\\
\vskip 0.1in Submitted to the  Proceedings of the US Community Study\\ 
on the Future of Particle Physics (Snowmass 2021)\\ 
\rule{\hsize}{0.01in}\\\rule[+0.2in]{\hsize}{0.01in} \end{center}}

\definecolor{darkred}{rgb}{0.5,0,0}

\usepackage{subfiles}

\usepackage[utf8]{inputenc}
\usepackage[T1]{fontenc}
\usepackage[super]{nth}
\usepackage[caption=false]{subfig}
\usepackage{graphicx}
\usepackage{placeins}

\def\MET{\ensuremath{E_{\mathrm{T}}^{\mathrm{miss}}}} 

\def\TeV{\ifmmode {\mathrm{\ Te\kern -0.1em V}}\else
                   \textrm{Te\kern -0.1em V}\fi}%
\def\GeV{\ifmmode {\mathrm{\ Ge\kern -0.1em V}}\else
                   \textrm{Ge\kern -0.1em V}\fi}%
\def\MeV{\ifmmode {\mathrm{\ Me\kern -0.1em V}}\else
                   \textrm{Me\kern -0.1em V}\fi}%
\def\keV{\ifmmode {\mathrm{\ ke\kern -0.1em V}}\else
                   \textrm{ke\kern -0.1em V}\fi}%
\def\eV{\ifmmode  {\mathrm{\ e\kern -0.1em V}}\else
                   \textrm{e\kern -0.1em V}\fi}%

\def\TeVc{\ifmmode {\mathrm{\ Te\kern -0.1em V}/c}\else
                   {\textrm{Te\kern -0.1em V}/$c$}\fi}%
\def\GeVc{\ifmmode {\mathrm{\ Ge\kern -0.1em V}/c}\else
                   {\textrm{Ge\kern -0.1em V}/$c$}\fi}%
\def\MeVc{\ifmmode {\mathrm{\ Me\kern -0.1em V}/c}\else
                   {\textrm{Me\kern -0.1em V}/$c$}\fi}%
\def\keVc{\ifmmode {\mathrm{\ ke\kern -0.1em V}/c}\else
                   {\textrm{ke\kern -0.1em V}/$c$}\fi}%
\def\eVc{\ifmmode  {\mathrm{\ e\kern -0.1em V}/c}\else
                   {\textrm{e\kern -0.1em V}/$c$}\fi}%

\def\TeVcc{\ifmmode {\mathrm{\ Te\kern -0.1em V}/c^2}\else
                   {\textrm{Te\kern -0.1em V}/$c^2$}\fi}%
\def\GeVcc{\ifmmode {\mathrm{\ Ge\kern -0.1em V}/c^2}\else
                   {\textrm{Ge\kern -0.1em V}/$c^2$}\fi}%
\def\MeVcc{\ifmmode {\mathrm{\ Me\kern -0.1em V}/c^2}\else
                   {\textrm{Me\kern -0.1em V}/$c^2$}\fi}%
\def\keVcc{\ifmmode {\mathrm{\ ke\kern -0.1em V}/c^2}\else
                   {\textrm{ke\kern -0.1em V}/$c^2$}\fi}%
\def\eVcc{\ifmmode  {\mathrm{\ e\kern -0.1em V}/c^2}\else
                   {\textrm{e\kern -0.1em V}/$c^2$}\fi}%

\def\cm{\ifmmode  {\mathrm{\ cm}}\else
                   \textrm{~cm}\fi}%
\def\ggino{\ensuremath{\mathchoice%
      {\displaystyle\raise.4ex\hbox{$\displaystyle\tilde\chi$}}%
         {\textstyle\raise.4ex\hbox{$\textstyle\tilde\chi$}}%
       {\scriptstyle\raise.3ex\hbox{$\scriptstyle\tilde\chi$}}%
 {\scriptscriptstyle\raise.3ex\hbox{$\scriptscriptstyle\tilde\chi$}}}}

\def\chinop{\ensuremath{\mathchoice%
      {\displaystyle\raise.4ex\hbox{$\displaystyle\tilde\chi^+$}}%
         {\textstyle\raise.4ex\hbox{$\textstyle\tilde\chi^+$}}%
       {\scriptstyle\raise.3ex\hbox{$\scriptstyle\tilde\chi^+$}}%
 {\scriptscriptstyle\raise.3ex\hbox{$\scriptscriptstyle\tilde\chi^+$}}}}
\def\chinom{\ensuremath{\mathchoice%
      {\displaystyle\raise.4ex\hbox{$\displaystyle\tilde\chi^-$}}%
         {\textstyle\raise.4ex\hbox{$\textstyle\tilde\chi^-$}}%
       {\scriptstyle\raise.3ex\hbox{$\scriptstyle\tilde\chi^-$}}%
 {\scriptscriptstyle\raise.3ex\hbox{$\scriptscriptstyle\tilde\chi^-$}}}}
\def\chinopm{\ensuremath{\mathchoice%
      {\displaystyle\raise.4ex\hbox{$\displaystyle\tilde\chi^\pm$}}%
         {\textstyle\raise.4ex\hbox{$\textstyle\tilde\chi^\pm$}}%
       {\scriptstyle\raise.3ex\hbox{$\scriptstyle\tilde\chi^\pm$}}%
 {\scriptscriptstyle\raise.3ex\hbox{$\scriptscriptstyle\tilde\chi^\pm$}}}}
\def\chinomp{\ensuremath{\mathchoice%
      {\displaystyle\raise.4ex\hbox{$\displaystyle\tilde\chi^\mp$}}%
         {\textstyle\raise.4ex\hbox{$\textstyle\tilde\chi^\mp$}}%
       {\scriptstyle\raise.3ex\hbox{$\scriptstyle\tilde\chi^\mp$}}%
 {\scriptscriptstyle\raise.3ex\hbox{$\scriptscriptstyle\tilde\chi^\mp$}}}}

\def\chinoonep{\ensuremath{\mathchoice%
      {\displaystyle\raise.4ex\hbox{$\displaystyle\tilde\chi^+_1$}}%
         {\textstyle\raise.4ex\hbox{$\textstyle\tilde\chi^+_1$}}%
       {\scriptstyle\raise.3ex\hbox{$\scriptstyle\tilde\chi^+_1$}}%
 {\scriptscriptstyle\raise.3ex\hbox{$\scriptscriptstyle\tilde\chi^+_1$}}}}
\def\chinoonem{\ensuremath{\mathchoice%
      {\displaystyle\raise.4ex\hbox{$\displaystyle\tilde\chi^-_1$}}%
         {\textstyle\raise.4ex\hbox{$\textstyle\tilde\chi^-_1$}}%
       {\scriptstyle\raise.3ex\hbox{$\scriptstyle\tilde\chi^-_1$}}%
 {\scriptscriptstyle\raise.3ex\hbox{$\scriptscriptstyle\tilde\chi^-_1$}}}}
\def\chinoonepm{\ensuremath{\mathchoice%
      {\displaystyle\raise.4ex\hbox{$\displaystyle\tilde\chi^\pm_1$}}%
         {\textstyle\raise.4ex\hbox{$\textstyle\tilde\chi^\pm_1$}}%
       {\scriptstyle\raise.3ex\hbox{$\scriptstyle\tilde\chi^\pm_1$}}%
 {\scriptscriptstyle\raise.3ex\hbox{$\scriptscriptstyle\tilde\chi^\pm_1$}}}}

\def\chinotwop{\ensuremath{\mathchoice%
      {\displaystyle\raise.4ex\hbox{$\displaystyle\tilde\chi^+_2$}}%
         {\textstyle\raise.4ex\hbox{$\textstyle\tilde\chi^+_2$}}%
       {\scriptstyle\raise.3ex\hbox{$\scriptstyle\tilde\chi^+_2$}}%
 {\scriptscriptstyle\raise.3ex\hbox{$\scriptscriptstyle\tilde\chi^+_2$}}}}
\def\chinotwom{\ensuremath{\mathchoice%
      {\displaystyle\raise.4ex\hbox{$\displaystyle\tilde\chi^-_2$}}%
         {\textstyle\raise.4ex\hbox{$\textstyle\tilde\chi^-_2$}}%
       {\scriptstyle\raise.3ex\hbox{$\scriptstyle\tilde\chi^-_2$}}%
 {\scriptscriptstyle\raise.3ex\hbox{$\scriptscriptstyle\tilde\chi^-_2$}}}}
\def\chinotwopm{\ensuremath{\mathchoice%
      {\displaystyle\raise.4ex\hbox{$\displaystyle\tilde\chi^\pm_2$}}%
         {\textstyle\raise.4ex\hbox{$\textstyle\tilde\chi^\pm_2$}}%
       {\scriptstyle\raise.3ex\hbox{$\scriptstyle\tilde\chi^\pm_2$}}%
 {\scriptscriptstyle\raise.3ex\hbox{$\scriptscriptstyle\tilde\chi^\pm_2$}}}}

\def\nino{\ensuremath{\mathchoice%
      {\displaystyle\raise.4ex\hbox{$\displaystyle\tilde\chi^0$}}%
         {\textstyle\raise.4ex\hbox{$\textstyle\tilde\chi^0$}}%
       {\scriptstyle\raise.3ex\hbox{$\scriptstyle\tilde\chi^0$}}%
 {\scriptscriptstyle\raise.3ex\hbox{$\scriptscriptstyle\tilde\chi^0$}}}}

\def\ninoone{\ensuremath{\mathchoice%
      {\displaystyle\raise.4ex\hbox{$\displaystyle\tilde\chi^0_1$}}%
         {\textstyle\raise.4ex\hbox{$\textstyle\tilde\chi^0_1$}}%
       {\scriptstyle\raise.3ex\hbox{$\scriptstyle\tilde\chi^0_1$}}%
 {\scriptscriptstyle\raise.3ex\hbox{$\scriptscriptstyle\tilde\chi^0_1$}}}}
\def\ninotwo{\ensuremath{\mathchoice%
      {\displaystyle\raise.4ex\hbox{$\displaystyle\tilde\chi^0_2$}}%
         {\textstyle\raise.4ex\hbox{$\textstyle\tilde\chi^0_2$}}%
       {\scriptstyle\raise.3ex\hbox{$\scriptstyle\tilde\chi^0_2$}}%
 {\scriptscriptstyle\raise.3ex\hbox{$\scriptscriptstyle\tilde\chi^0_2$}}}}
\def\ninothree{\ensuremath{\mathchoice%
      {\displaystyle\raise.4ex\hbox{$\displaystyle\tilde\chi^0_3$}}%
         {\textstyle\raise.4ex\hbox{$\textstyle\tilde\chi^0_3$}}%
       {\scriptstyle\raise.3ex\hbox{$\scriptstyle\tilde\chi^0_3$}}%
 {\scriptscriptstyle\raise.3ex\hbox{$\scriptscriptstyle\tilde\chi^0_3$}}}}
\def\ninofour{\ensuremath{\mathchoice%
      {\displaystyle\raise.4ex\hbox{$\displaystyle\tilde\chi^0_4$}}%
         {\textstyle\raise.4ex\hbox{$\textstyle\tilde\chi^0_4$}}%
       {\scriptstyle\raise.3ex\hbox{$\scriptstyle\tilde\chi^0_4$}}%
 {\scriptscriptstyle\raise.3ex\hbox{$\scriptscriptstyle\tilde\chi^0_4$}}}}

\def\squark{\ensuremath{\tilde{q}}}
\def\gluino{\ensuremath{\tilde{g}}}
\def\stop{\ensuremath{\tilde{t}}}

\def\smu{\ensuremath{\tilde{\mu}}}
\def\stau{\ensuremath{\tilde{\tau}}}

\newcommand{\subsubsubsection}[1]{\vspace{0.5in}\noindent\textbf{#1}}

\makeatletter
\def\l@subsubsection#1#2{}
\makeatother

\begin{document}


\title{{\normalsize\snowmass}
Physics Beyond the Standard Model at Energy Frontier}

\newcommand{\wsu}{\affil{Department of Physics \& Astronomy, Wayne State University, Detroit, MI 48202, USA}}
\newcommand{\fnl}{\affil{Fermilab, P.O. Box 500, Batavia, IL 60510, USA}}
\newcommand{\Upenn}{\affil{Department of Physics \& Astronomy, University of Pennsylvania, Philadelphia, PA 19104, USA}}
\newcommand{\umd}{\affil{Maryland Center for Fundamental Physics, Department of Physics, University of Maryland, College Park, MD 20742, USA}}
\newcommand{\harvard}{\affil{Department of Physics, Harvard University, Cambridge, MA 02138, USA}}
\newcommand{\Argonne}{\affil{HEP Division, Argonne National Laboratory, Argonne, IL 60439}}
\newcommand{\UofC}{\affil{Physics Department, Enrico Fermi Institute and Kavli Institute for Cosmological Physics, University of Chicago, IL 60637}}
\newcommand{\stanford}{\affil{Stanford Institute for Theoretical Physics, Department of Physics, Stanford University, Stanford, CA 94305, USA}}
\newcommand{\Minnesota}{\affil{School of Physics and Astronomy, University of Minnesota, Minneapolis, MN 55455, USA}}
\newcommand{\CERN}{\affil{European Center for Nuclear Research (CERN) CH 1211, Geneva 23, Switzerland}}
\newcommand{\lbl}{\affil{Physics Division, Lawrence Berkeley National Laboratory, Berkeley, CA 94720, USA}}
\newcommand{\ucberkely}{\affil{Berkeley Center for Theoretical Physics, Department of Physics,
University of California, Berkeley, CA 94720, USA}}
\newcommand{\Ukansas}{\affil{Department of Physics and Astronomy, University of Kansas, Lawrence, Kansas 66045 U.S.A.}}
\newcommand{\mainz}{\affil{PRISMA+ Cluster of Excellence \& Mainz Institute for Theoretical Physics, Johannes
Gutenberg University, 55099 Mainz, Germany}}
\newcommand{\bethel}{\affil{Department of Physics \& Engineering, Bethel University, St. Paul MN, 55112, USA}}
\newcommand{\uwmadison}{\affil{Department of Physics, University of Wisconsin, 1150 University Avenue, Madison, WI 53706, USA}}
\newcommand{\uluman}{\affil{University of Manchester (United Kingdom) and Lund University (Sweden)}}
\newcommand{\ohiostate}{\affil{Department of Physics and Center for Cosmology and Astroparticle Physics, The Ohio State University, Columbus, OH 43210, USA}}
\newcommand{\lund}{\affil{Lund University (Sweden)}}
\newcommand{\triumf}{\affil{TRIUMF (Canada)}}
\newcommand{\warsaw}{\affil{University of Warsaw, Poland}}
\newcommand{\croatiacern}{\affil{Ruder Boskovic Institute, Croatia and CERN}}
\newcommand{\fsu}{\affil{8Florida State University, Tallahassee, FL, USA}}


 \newcommand{\instLAGLFA}{\affil{Munster U., ITP}}
 \newcommand{\instLEGAHG}{\affil{Northeastern U. (main)}}
 \newcommand{\instLALEIH}{\affil{American U., Sharjah}}
 \newcommand{\instLADFLL}{\affil{Oklahoma U.}}
 \newcommand{\instLADDEL}{\affil{Wisconsin U., Madison}}
 \newcommand{\instLADDDF}{\affil{Warsaw U.}}
 \newcommand{\instLACHHA}{\affil{DESY}}
 \newcommand{\instLAEGGB}{\affil{Taiwan, Natl. Taiwan U.}}
 \newcommand{\instLACIDI}{\affil{Hawaii U.}}
 \newcommand{\instLACLBG}{\affil{KEK, Tsukuba}}
 \newcommand{\instLBCDFB}{\affil{Sokendai, Tsukuba}}
 \newcommand{\instLBBCFE}{\affil{Tokyo U., IPMU}}
 \newcommand{\instLADCHD}{\affil{Tokyo U.}}
 \newcommand{\instLACHLA}{\affil{Eotvos U.}}
 \newcommand{\instBEHCDIG}{\affil{Nagoya U., IAR}}
 \newcommand{\instLBCAEF}{\affil{KMI, Nagoya}}
 \newcommand{\instLACLEG}{\affil{Kyushu U.}}
 \newcommand{\instLBADLC}{\affil{Sokendai, Kanagawa}}
 \newcommand{\instLADHAH}{\affil{Hokkaido U. of Education}}
 \newcommand{\instLACIFD}{\affil{Hokkaido U.}}
 \newcommand{\instLACGDC}{\affil{Alabama U.}}
 \newcommand{\instLACICF}{\affil{Chalmers U. Tech.}}
 \newcommand{\instLADCCL}{\affil{Stockholm U.}}
 \newcommand{\instBHEDIEI}{\affil{IP2I, Lyon}}
 \newcommand{\instLAHCIE}{\affil{Korea Inst. Advanced Study, Seoul}}
 \newcommand{\instLAIFID}{\affil{Paris, LPTHE}}
 \newcommand{\instLADDFF}{\affil{Wurzburg U.}}
 \newcommand{\instLADDBE}{\affil{Uppsala U.}}
 \newcommand{\instLADBDA}{\affil{Pittsburgh U.}}
 \newcommand{\instBCIDEBA}{\affil{SYSU, Guangzhou}}
 \newcommand{\instLADCFH}{\affil{Technion}}
 \newcommand{\instLADIIL}{\affil{Santa Barbara, KITP}}
 \newcommand{\instLACHFG}{\affil{Cracow, INP}}
 \newcommand{\instLACLCH}{\affil{King's Coll. London}}
 \newcommand{\instLACGCH}{\affil{Adelaide U.}}
 \newcommand{\instLACGHG}{\affil{Bonn U.}}
 \newcommand{\instLADBEI}{\affil{Queensland U.}}
 \newcommand{\instLACIGI}{\affil{Imperial Coll., London}}
 \newcommand{\instLADDAH}{\affil{UC, Santa Barbara}}
 \newcommand{\instLACGEH}{\affil{Arizona U.}}
 \newcommand{\instLADALE}{\affil{Oklahoma State U.}}
 \newcommand{\instLAFHBI}{\affil{Nanjing Normal U.}}
 \newcommand{\instLACHLG}{\affil{Fermilab}}
 \newcommand{\instLACIDC}{\affil{Hamburg U.}}
 \newcommand{\instLBAHGH}{\affil{Madrid, IFT}}
 \newcommand{\instLADABA}{\affil{Minnesota U.}}
 \newcommand{\instLACGEF}{\affil{Argonne}}
 \newcommand{\instLADBCG}{\affil{Pennsylvania U.}}
 \newcommand{\instLADIDI}{\affil{Taiwan, Natl. Central U.}}
 \newcommand{\instLACHIB}{\affil{Duke U.}}
 \newcommand{\instLACHDA}{\affil{Chicago U., EFI}}
 \newcommand{\instLAILFH}{\affil{Chicago U., KICP}}
 \newcommand{\instLBCAHF}{\affil{Unlisted, TR}}
 \newcommand{\instLACGIA}{\affil{Boston U.}}
 \newcommand{\instLACIHE}{\affil{Indiana U.}}
 \newcommand{\instBCLHFGL}{\affil{IBS, Daejeon}}
 \newcommand{\instLADCLA}{\affil{TRIUMF}}
 \newcommand{\instLAFHBD}{\affil{Jeonbuk Natl. U.}}
 \newcommand{\instLAGAIC}{\affil{Nankai U.}}
 \newcommand{\instLADDAF}{\affil{UC, San Diego}}
 \newcommand{\instLACHCF}{\affil{CERN}}
 \newcommand{\instLACIIE}{\affil{INFN, Padua}}
 \newcommand{\instLACHIC}{\affil{Durham U.}}
 \newcommand{\instLADBBL}{\affil{Paris U., VI-VII}}
 \newcommand{\instLACIBD}{\affil{Geneva U.}}
 \newcommand{\instLACLHA}{\affil{Louvain U.}}
 \newcommand{\instLACICH}{\affil{Graz U.}}
 \newcommand{\instLBBIFD}{\affil{Mainz U.}}
 \newcommand{\instLACLLL}{\affil{Melbourne U.}}
 \newcommand{\instLACGIG}{\affil{Brigham Young U.}}
 \newcommand{\instLACGGL}{\affil{Bhubaneswar, Inst. Phys.}}
 \newcommand{\instBCDCFEF}{\affil{HBNI, Mumbai}}
 \newcommand{\instLADDBB}{\affil{University Coll. London}}
 \newcommand{\instLCCIEL}{\affil{Kyushu U., Fukuoka (main)}}
 \newcommand{\instLADCGB}{\affil{Tennessee U.}}
 \newcommand{\instLACIGF}{\affil{IIT, Chicago}}
 \newcommand{\instLACGLC}{\affil{Brown U.}}
 \newcommand{\instLADDCC}{\affil{Victoria U.}}
 \newcommand{\instLACGEA}{\affil{Amsterdam U.}}
 \newcommand{\instLADIDC}{\affil{Nikhef, Amsterdam}}
 \newcommand{\instLACIGH}{\affil{Illinois U., Urbana}}
 \newcommand{\instBCBIAGI}{\affil{UC, Santa Cruz, Inst. Part. Phys.}}
 \newcommand{\instLADALH}{\affil{Oregon U.}}
 \newcommand{\instBGGEALC}{\affil{Tsung-Dao Lee Inst., Shanghai}}
 \newcommand{\instLAEHEA}{\affil{Shanghai Jiaotong U.}}
 \newcommand{\instLACLLC}{\affil{Massachusetts U., Amherst}}
 \newcommand{\instLAFCAF}{\affil{Caltech, Kellogg Lab}}
 \newcommand{\instLACHBB}{\affil{Caltech}}
 \newcommand{\instLADAFH}{\affil{Beihang U.}}
 \newcommand{\instLBAECL}{\affil{YITP, Stony Brook}}
 \newcommand{\instLACLFD}{\affil{LBL, Berkeley}}
 \newcommand{\instLADCLL}{\affil{UC, Berkeley}}
 \newcommand{\instLADCAG}{\affil{SLAC}}
 \newcommand{\instLACHGL}{\affil{Denver U.}}
 \newcommand{\instLACHFD}{\affil{Connecticut U.}}
 \newcommand{\instLADDAD}{\affil{UCLA}}
 \newcommand{\instLADADG}{\affil{Munich, Max Planck Inst.}}
 \newcommand{\instBCHCLGA}{\affil{UC, Irvine (main)}}
 \newcommand{\instLDACID}{\affil{Ec. Polytech., Palaiseau (main)}}
 \newcommand{\instLAHLAH}{\affil{Valencia U., IFIC}}
 \newcommand{\instLACLLA}{\affil{Maryland U.}}
 \newcommand{\instBEHBADF}{\affil{EPFL, Lausanne, LPTP}}
 \newcommand{\instLACHCL}{\affil{Chicago U.}}
 \newcommand{\instLADCGC}{\affil{Texas A-M}}
 \newcommand{\instLACIDF}{\affil{Harvard U.}}
 \newcommand{\instLAEEHF}{\affil{Rome U., Tor Vergata}}
 \newcommand{\instLAHGLB}{\affil{INFN, Rome2}}
 \newcommand{\instLALECB}{\affil{Enrico Fermi Ctr., Rome}}
 \newcommand{\instBFLBFIL}{\affil{U. Montpellier, IMAG}}
 \newcommand{\instLADCBC}{\affil{Southampton U.}}
 \newcommand{\instLADBHE}{\affil{Rutherford}}
 \newcommand{\instLACGDE}{\affil{Alberta U.}}
 \newcommand{\instLAEBGA}{\affil{Rabat U.}}
 \newcommand{\instLBBABD}{\affil{Johannesburg U.}}
 \newcommand{\instLBABIH}{\affil{BNL, NSLS}}
 \newcommand{\instLACGGD}{\affil{Ben Gurion U. of Negev}}
 \newcommand{\instLALLAA}{\affil{NCTS, Taipei}}
 \newcommand{\instLACIAH}{\affil{Frascati}}
 \newcommand{\instLADDDD}{\affil{Warsaw, Inst. Nucl. Studies}}
 \newcommand{\instLADABG}{\affil{MIT, LNS}}
 \newcommand{\instLACHDE}{\affil{Cincinnati U.}}
 \newcommand{\instLAGDHC}{\affil{Harish-Chandra Res. Inst.}}
 \newcommand{\instLAGALG}{\affil{North Florida U.}}
 \newcommand{\instLADDBF}{\affil{Utah U.}}
 \newcommand{\instBHILLDF}{\affil{IFPU, Trieste}}
 \newcommand{\instLACIII}{\affil{INFN, Trieste}}
 \newcommand{\instLACGHI}{\affil{Boskovic Inst., Zagreb}}
 \newcommand{\instBCGICFI}{\affil{Brookhaven Natl. Lab.}}
 \newcommand{\instLADBCD}{\affil{Beijing, Inst. High Energy Phys.}}
 \newcommand{\instLADGAD}{\affil{Peking U.}}
 \newcommand{\instLBBGHB}{\affil{Peking U., CHEP}}
 \newcommand{\instLAIFCE}{\affil{Beihang U.}}
 \newcommand{\instLBBFBI}{\affil{Ohio State U., CCAPP}}
 \newcommand{\instLACHBC}{\affil{Cambridge U.}}
 \newcommand{\instLACLHD}{\affil{Lund U.}}
 \newcommand{\instLADDEC}{\affil{Weizmann Inst.}}
 \newcommand{\instLAHEFF}{\affil{MIT}}
 \newcommand{\instLAIBGI}{\affil{Antonio Narino U.}}
 \newcommand{\instLACGGG}{\affil{DESY, Zeuthen}}
 \newcommand{\instLACIIC}{\affil{INFN, Milan}}
 \newcommand{\instLADDAC}{\affil{UC, Irvine}}
 \newcommand{\instLACLIC}{\affil{Mainz U., Inst. Phys.}}
 \newcommand{\instLAEFAF}{\affil{Lafayette Coll.}}
 \newcommand{\instCB}{\affil{University of Michigan}}
 \newcommand{\instCC}{\affil{Perimeter Institute for Theoretical Physics}}
 \newcommand{\instCD}{\affil{INFN-TIFPA Trento}}
 \newcommand{\instCE}{\affil{Columbia University}}
 \newcommand{\instCF}{\affil{University of California Davis}}
 \newcommand{\instCG}{\affil{University of Zurich}}
 \newcommand{\instETH}{\affil{Zurich, ETH}}
 \newcommand{\instCH}{\affil{University of Notre Dame}}
 \newcommand{\instCI}{\affil{Stony Brook University}}
 \newcommand{\instCL}{\affil{Illinois Mathematics and Science Academy}}
 \newcommand{\instCBA}{\affil{University of California}}
 \newcommand{\instCBB}{\affil{Universit\`a degli Studi and INFN Roma Tre}}
 \newcommand{\instCBC}{\affil{University of Rochester}}
 \newcommand{\instCBD}{\affil{CNRS}}
 \newcommand{\instCBE}{\affil{Wayne State University}}
 \newcommand{\instCBF}{\affil{Instituto Superior Técnico (IST)}}
 \newcommand{\instCBG}{\affil{The Ohio State University}}
 \newcommand{\instCBH}{\affil{University of Washington}}
 \newcommand{\instCBI}{\affil{University of Kansas}}
 \newcommand{\instCBL}{\affil{Paul Scherrer Institute}}
 \newcommand{\instCCA}{\affil{Shanghai Jiao Tong University}}
 \newcommand{\instCCB}{\affil{University of Illinois}}
 \newcommand{\instCCC}{\affil{Northern Illinois University}}
 \newcommand{\instCCD}{\affil{Universitat de Val\`encia}}
 \newcommand{\instCCE}{\affil{Kyungpook National University}}
 \newcommand{\instCCF}{\affil{National Research Nuclear University MEPhI}}
 \newcommand{\instCCG}{\affil{University of Iowa}}
 \newcommand{\instCCH}{\affil{Kennesaw State University}}
 \newcommand{\instCCI}{\affil{Sam Houston State University}}
 \newcommand{\instCCL}{\affil{UC Santa Cruz}}
 \newcommand{\instCDA}{\affil{Shenzhen Campus of Sun Yat-sen University}}
 \newcommand{\instCDB}{\affil{University of Massachusetts}}
 \newcommand{\instCDC}{\affil{California Institute of Technology}}
 \newcommand{\instCDD}{\affil{University of Toronto}}
 \newcommand{\instCDE}{\affil{McGill University}}
 \newcommand{\instCDF}{\affil{TIFR}}
 \newcommand{\instCDG}{\affil{Karlsruhe Institute of Technology}}
 \newcommand{\instCDH}{\affil{Carleton University}}
 \newcommand{\instCSICUV}{\affil{Instituto de F\'isica Corpuscular (CSIC-UV), Valencia, Spain}}

\author{{\bf Convenors:} Tulika Bose}
\uwmadison
\author{ Antonio Boveia}
\ohiostate
\author{ Caterina Doglioni}
\uluman
\author{ Simone Pagan Griso}
\lbl
\author{ James Hirschauer}
\fnl
\author{ Elliot Lipeles}
\Upenn
\author{ Zhen Liu}
\Minnesota
\author{ Nausheen~R.~Shah}
\wsu
\author{ Lian-Tao Wang}
\UofC


\author{\newline\newline Kaustubh Agashe}\umd
% ...


\maketitle

\clearpage
\setcounter{tocdepth}{1}
\tableofcontents

\captionsetup{justification=justified,singlelinecheck=off}

\newpage

\section{Executive Summary}\label{ExecSum}


Searching for BSM physics directly through new particle production or indirectly through precision measurements is a vibrant and evolving field. It is a primary science
driver in high-energy physics, and the energy frontier provides many
unique approaches and discovery opportunities. The energy frontier
casts a wide net across thousands of searches and observables.
These can be interpreted in the context of complete top-down models
and more model-independent ad-hoc simplified models. Furthermore,
the frontier provides an opportunity to directly constrain or discover possible
dark matter candidates and associated sectors.

The energy frontier has laid out a vision for the immediate, intermediate, and
long-term future. For BSM physics, each phase would provide significant new
insights into fundamental physics and discovery potential.



{\vspace{\baselineskip} \noindent\bf Immediate future: High Luminosity LHC}

The immediate future of the energy frontier is the
the HL-LHC program, with accelerator and detector upgrades of the LHC
already under construction. The HL-LHC will provide a significant
extension of the already enormously-successful LHC program. A huge BSM
parameter space has been explored since the LHC startup, dramatically
changing theoretical perspectives on the most pressing questions of high energy physics. Representative accomplishments
include
\begin{itemize}
  \item major advances in sensitivity to additional gauge bosons and fermions, such as a $Z^\prime\rightarrow \ell\ell$ reach more than five times
    what it was 12 years ago;
  \item a tightening grip on minimal supersymmetry, with gluino and light-squark limits over two TeV and stop-squark limits above one TeV;
  \item the development of novel techniques to constrain challenging model signatures, such as phenomenologically-interesting compressed spectra;
  \item innovative searches for long-lived particles and other exotic topologies;
  \item and a rich program of searches for particle dark matter and the physics connecting it to the Standard Model.
\end{itemize}

The HL-LHC will expand this exciting program through
\begin{itemize}
\item a factor of 20 enhancement of the current Run-2 luminosity;
\item new detector capabilities, including the extension of the tracking
  coverage to the forward regions, pico-second-precision timing detectors, and extended
  trigger systems;
\item new analysis techniques such as jet-substructure, machine learning, data scouting, trigger-level analysis, and automated anomaly detection;
\item and, finally, new auxiliary detectors that make further use of the
  collisions produced by the HL-LHC for a variety of dedicated
  searches.
\end{itemize}
For example, the new sensitivity provided by the HL-LHC will increase SUSY particle reaches by a factor of 1.5 to 2. This increase has a significant impact in the model's context. For RPC stop-squarks, a large portion of phenomenologically interesting parameter space will be covered, and for RPC compressed Higgsinos, the sensitivity in the ``natural SUSY'' parameter space will be greatly increased. Other examples include an order
of magnitude increase in sensitivity to dark photon coupling and a
factor of 4 improvement in the Higgs to invisible branching ratio.
A dedicated program of auxiliary experiments can also extend the sensitivities to hidden sectors. 


{\vspace{\baselineskip} \noindent\bf Intermediate future: a Higgs factory}

The $e^+e^-$ Higgs factory envisioned in the Energy Frontier report will
provide several potential avenues for discovery. Precision measurements of Higgs properties are sensitive to BSM physics regardless of the
direct production signature. In some cases, they are sensitive to BSM
physics that may be beyond the reach of the HL-LHC or even
higher-energy colliders. The clean, large sample of Higgses will
provide unique access to exotic Higgs decays, which a hadron collider 
may find challenging to identify. Such a machine can see
most new direct production processes below $\sqrt{s}/2$ and have
indirect sensitivity to higher energy scales. Very large
luminosities at lower energies, such as the $Z$-pole, or near the top-quark pair
and $WW$ thresholds can also provide high-precision measurements which
indirectly constrain new physics (e.g., the $W$-mass). Such samples
would also significantly increase constraints on moderate-energy, weakly-coupled
phenomena such as dark photons.



{\vspace{\baselineskip} \noindent\bf Long-term future: A multi-TeV discovery machine (hadron, muon or $e^+e^-$/$\gamma\gamma$ collider) }

In the longer term, a multi-TeV machine will provide the largest and most comprehensive
increase in the BSM discovery reach of the energy frontier. The reach would dramatically increase in almost all categories for both
model-dependent and model-independent searches. For several machines under discussion,
the thermal WIMP target could be reached for minimal WIMP candidates
(e.g., Higgsino- and Wino-like models with no other relevant
particles), which is representative of the power of the experiments. A discovery machine would push constraints on the Higgs-mass scale naturalness by the square of the increase in energy reach, which is potentially two
orders of magnitude beyond the HL-LHC. It could open up access to new
hidden sectors by producing a substantially higher mediator
mass or probing even smaller couplings. The reach of each of the
proposed colliders differs in non-trivial ways, as has
been summarized in this document for various representative
cases. To realize such a machine, a strong R\&D
program for both accelerators and detectors is critical, with continued
tight interaction between the Accelerator, Instrumentation, and Energy
Frontiers.

{\vspace{\baselineskip} \noindent\bf Final words}

The Energy Frontier BSM physics program is tightly integrated
with the rest of the high-energy physics landscape. There are clear
connections to the enabling technologies on the Accelerator, Instrumentation, and
Computing Frontiers and to the underlying theory. There is a wide
variety of strong motivations to continue searching for BSM physics at the Energy Frontier,
including the existence of dark matter
and unsolved theoretical puzzles such as the origin and naturalness of the
Higgs boson mass. The Energy Frontier is also poised to help address
anomalies and discoveries in other frontiers, such as dark matter
signals observed at other frontiers, and $B$-physics anomalies and the
g-2 measurements from the Rare and Precision Process
Frontier. Finally, these proposed experiments are a step into the
unknown with potential discovery possibilities that may be beyond
our imaginations.


\section{Introduction}

There are several reasons why physics beyond the Standard Model (BSM) of particle physics is likely and, in some cases, unavoidable.  
Such reasons are connected to fundamental questions,  answering which is among the highest priorities of particle physics.
Current and future experiments at the energy frontier offer unique capabilities to explore many of these questions.

A subset of the most relevant questions to energy frontier approaches can be grouped in three broad categories:


\begin{enumerate}

\item 
Phenomena that have been observed but where a fundamental explanation is still lacking.
These include
\begin{itemize}
 \item What is the fundamental composition of Dark Matter? An overview of this motivation is given in Section \ref{Sec:DM}
 \item What is the additional source of CP violation needed to explain the matter-antimatter asymmetry observed in the universe? 
 \item 
 Possible observations of BSM physics referred to broadly as {\it Anomalies}, which are discussed in Sec.~\ref{Sec:Anm} and underly the strong connection of the energy frontier to the rare and intensity frontiers.
 \item Cosmological tensions, such as the Hubble tension, which might indicate new particles.
 \item How is particle physics unified with gravity?
\end{itemize}

\item 
Guiding principles forming the basis of the successful stories behind the current Standard Model~(SM) and, more generally, of modern theoretical physics. These may offer us insight on where the theoretical framework is ``hinting'' for a more complete description of Nature:
\begin{itemize}
 \item Naturalness has many faces, and is a hotly debated topic. Sec.~\ref{Sec:Nat} discusses both the Naturalness considerations underlying our hunt for BSM physics and how future colliders address these puzzles.
 \item The flavor structure of the SM is complex and begs for an underlying explanation. This may be related to questions regarding the scale/origin of electroweak symmetry breaking and the Higgs mechanism. We discuss some of the motivations related to these questions in Sec.~\ref{Sec:HEWSB}, however these topics are addressed in detail in the Snowmass Higgs Report~\cite{sm21:higgs-report}.
 \item Why are neutrinos so light? What is the mechanism of their mass generation?
 \item Why is there no CP violation in the strong force?
 \item Is there a simpler "grand unified" theory unifying gauge couplings and/or quarks and leptons?
\end{itemize}

\item 
As history has shown many times that particle physics should maintain a wide open view for possible new phenomena that might not fit in the simplest theoretical extensions of the SM. Sec.~\ref{Sec:Gen-intro} discusses approaches to broad questions such as:
\begin{itemize}
\item Are there new interactions or new particles around or above the electroweak scale? 
\item Is lepton universality violated? 
\item Are there long-lived or feebly-interacting particles which have evaded traditional BSM searches?
\item Finally, there is a broad question of how  to reduce biases in our searches 
and conduct them in a more model-independent way.
\end{itemize}

\end{enumerate}

Two main theoretical approaches in exploring BSM physics can be commonly identified. 
The first consists in seeking self-consistent theories that aim to address the questions above and can significantly boost our understanding of the fundamental laws of Nature. 
These well-motivated models of BSM physics which are self-consistent to high-energy scales are excellent test cases for exploring possible experimental signatures and their interrelation. 
Section \ref{Sec:CHED_intro} discusses models with composite particles and extra dimensions, which 
can have similar phenomenology if the energy scale of the extra dimensions is high. Sec.~ \ref{Sec:susy_intro} discusses Supersymetry (SUSY) which as well as being 
a well-motivated and broadly studied model, is also a good example of a model with partner particles with a broad range of properties. The projections for SUSY models are at the same time representative of other such models, for example Kaluza-Klein excitations in extra-dimensions.  Section  \ref{Sec:Leptoquark} discusses leptoquarks which can arise in R-parity violating SUSY models as well as in  grand-unified theories and are motivated by some of the flavor physics anomalies.

Looking beyond these prominent models, the landscape of possible experimental and theoretically-motivated models and signatures is very large. In this approach,  well-defined but incomplete theories extend specific areas without the expectation of full self-consistency. These {\it simplified} models and {\it portal} models are in some cases simplifications of complete theories. It is not practical nor useful 
to try to be exhaustive in projecting the scientific output of projects targeting all such models.
Instead, we focus on a representative set of models and signatures that are deeply connected with the fundamental questions above and
represent a wide range of physics that can be explored at the energy frontier.
Such an approach has the advantage of providing a manageable framework where different experimental results can be easily compared and, eventually, mapped into the parameter space of complete theories. However, the drawback is that those have intrinsically a larger degree of arbitrariness and should be viewed as simpler guiding frameworks for the more general exploration of BSM physics.

Section~\ref{Sec:NB} looks into the experimental reach for new bosons; such new bosons, if discovered, can provide clues to the existence of new forces beyond the ones currently known. The important case of heavy new scalars, that are Higgs-like, is addressed in the Snowmass Higgs Report~\cite{sm21:higgs-report}.
Section~\ref{Sec:NF} seeks to assess the potential to discover new fermions at the high-energy frontier.
A set of signatures that have been drawing increased attention in the recent years and during this Snowmass process comprise of experimental collider signatures where new particles are created that possess a macroscopic lifetime before, if at all, decaying. The observation of such new particles and their theoretical motivation is distinct enough that merits its own section and will be discussed in Section~\ref{Sec:LLP}.

Dark matter is well established in astrophysical observations and is a pressing issue for particle physics and cosmology. Section \ref{Sec:DMFP} gives the future prospects of energy frontier experimental programs targeting the nature of Dark Matter and its interrelation to cosmology.

Finally, given the exploratory nature of particle physics, it is equally important to maximize the ability to identify phenomena that are either not expected or corners of parameter-space of specific theories that tend to be harder to probe extensively. Section~\ref{Sec:Gen} discuss such scenarios and how different techniques and experimental setups can be used to ensure a coverage as wide as possible and limit the theoretical and experimental bias in looking for new phenomena at the energy frontier.

This report builds on much of the effort and material collected and discussed during the update of the European Strategy, summarized in \cite{Strategy:2019vxc}. 
For this reason, throughout this document we emphasize updates and work that has been done since the release of the Briefing Book and specifically submitted as whitepapers to the Snowmass process.


\section{Experimental guidance \& motivation}\label{Sec:Mot-exp}
\subsection{Dark Matter}\label{Sec:DM}


The existence of dark matter (DM) in our universe is one of the most concrete pieces of evidence for physics beyond the Standard Model. 
However, very little has been observed so far about dark matter beyond its gravitational effects. 
Any signal pointing to dark matter interactions beyond gravity would bring us closer to answering one of the central questions of particle and astroparticle physics: \textit{What is the nature of dark matter and how does it interact with ordinary matter?}. 

Given our experimental ignorance so far, there are a myriad of theoretical possibilities in terms of dark matter candidates and interaction strengths. 
Therefore, the search for dark matter must be conducted in synergy between different Frontiers, using multiple probes and assumptions. 
The landscape of complementary theoretical and experimental efforts needed in the search for dark matter will be discussed in an upcoming cross-Frontier report. 

When searching for particle DM, many experiments target theoretical hypotheses that foresee some kind of interaction between the DM and the SM. 
Such interactions are the key to directly produce massive dark matter particles at the highest possible energies, via SM particle collisions in a terrestrial lab.
This gives collider experiments the unique possibility to probe the dark interaction and study the properties of DM in detail. 

Since DM-SM interactions are generally feeble (as a direct consequence to the dark matter's \textit{darkness}), DM particles escape detection at collider experiments and can be discovered in the products of collisions where some of the total transverse momentum required by conservation laws is missing (missing transverse momentum). 
A discovery of an invisible particle at colliders must be complemented by an observation of DM in astrophysics experiments, to verify the new particle's cosmological connection.
On the other hand, if particle dark matter signals are discovered elsewhere, it would be imperative to produce and study these particles and their associated DM-SM interactions in the controlled conditions of high-energy accelerator-based experiments.
The interplay of searches for dark matter across experimental frontiers is discussed further in the Snowmass 2022 Dark Matter Complementarity report \cite{DMComplementarityPlaceholder}.
The document, developed by the combined effort of the experimental and cross-cutting frontiers, sketches opportunities for cooperation, lays out a road map for a coherent strategy and presents case studies for discovery.

Among the possible theoretical scenarios for particle DM, collider experiment are well-suited to test whether DM is a Weakly Interacting Massive Particle (WIMP). 
In this scenario, the mass of the DM particle is within roughly 1 GeV and 100 TeV, and it has sizable couplings with the SM so that they are in thermal equilibrium in the early universe. 
As a result, its relic abundance is determined by particle physics parameters, such as mass and coupling, and is largely independent of the initial condition and evolutionary history of the Universe relevant for DM production. 
WIMP candidates feature in a number of theories with connections with other electroweak scale new physics, including Supersymmetry. 
In the past decade, there has been significant progress made in the search for WIMPs with experiments at colliders as well as at underground facilities, and with astrophysical observations. 
The null results obtained by these experiments do not yet cover the full scenario for the parameter space of such benchmarks, and WIMPs remain a compelling target for DM searches at colliders and beyond. 

Broadly speaking, WIMP scenarios can be classified according to the way in which the DM particle couples the SM particles. 
A particularly simple and predictive class of models is the so called \textit{minimal dark matter}, where no new force is needed since the interaction between the DM particle and the SM particles are mediated by the SM weak interaction.  
In this scenario, the dark matter particle is the lightest member of an electroweak multiplet, with candidates known in the context of supersymmetry as Higgsino and Wino, respectively. 
In order to give the correct thermal relic abundance, the Higgsino and Wino masses are predicted to be in the TeV(s) range. 
Testing this class of models is one of the main physics drivers for future high energy colliders, and DM candidates with these properties are also a target for complementary discoveries in next-generation direct detection and indirect detection experiments. 
The prospects for the discovery of these benchmark models are discussed in Section \ref{sec:dm:wimp}.  

The DM can also couple to the SM via so-called \textit{portals}, which include a direct coupling between SM and DM particles via gauge-invariant operators. 
The Higgs boson provides a prime example of such a portal: as a spin-0 particle, it gives us the possibility to write down a renormalizable coupling with the DM, that can have a sizable effect on SM Higgs decay and properties.  
At high energy colliders, such a coupling gives rise to dark matter production, mediated by the Higgs boson. 
If the DM has a mass that is less than half the Higgs mass, then experiments at colliders can  directly detect decays of the Higgs to invisible particles, and interpret excesses in terms of the Higgs portal. 
Precision measurements of the Higgs couplings, which are one of the objectives of future collider, can also contribute to discover or constrain the Higgs portal scenario, complementing possible future direct and indirect detection observations of DM particles. 
A precise understanding of the Higgs boson can also probe DM models with extended Higgs sectors. 
Future prospects on searches for dark matter with the Higgs boson are discussed in Section \ref{Sub:HiggsPortal}. 

A next-to-minimal scenario in terms of particles and interactions with the SM is that the DM particle couples to the SM via a new force mediator. 
A set of so-called \textit{simplified mediator} or \textit{portal} models have been put forward to provide search targets to explore this possibility. 
These classes of models have several free parameters, including the interaction strength between the mediator and the SM and the DM, as well as the mass of the mediator and the DM mass. 
A feature that makes colliders very well suited to probe these scenarios is that the mediator itself is a new physics particle which has been created by SM particle interactions, and must therefore also decay into the same SM particles, leading to visible signals in the detector. 
The combination of the searches for signals of both the DM signal and the mediator can give the most comprehensive coverage of the model space. 
In these scenarios, the mass of the DM particle spans a large spectrum of values, from DM that is lighter than 1 GeV to the canonical WIMP mass range  -- in this case searches at colliders can also provide complementary information to experiments searching for light dark matter in the Snowmass Rare Processes and Precision Frontier~\cite{sm21:rf-report}. 
Future discoveries and constraints of mediator and portal models of dark matter are discussed in Sections \ref{Sec:DMst} and \ref{Sec:OtherDM}.

Just like the SM has many particles and a variety of interactions, the DM can belong to a dark sector with rich dynamics. 
Many such extensions cover a much broader range of masses and couplings in comparison with the WIMP scenario. 
A number of these models feature extremely small coupling with the SM, which means that the DM or dark sector particles cannot be produced in sizable amounts at high-energy colliders. 
At the same time, colliders are well suited to discover scenarios with heavier dark sector states connected to the DM portals, or dark sector models where the dark matter particles appear together with other dark sector particles that lead to unusual, visible signals such as in the case of confining dark sectors subject to interactions that mirror QCD. 
The model-building and connections with cosmological dark matter in these latter cases are still in active development, with ongoing collaborations between theory and experiment. 
Considerations on these DM models at present and future colliders can be found in Section \ref{Sec:OtherDM}.


\subsection{Anomalies in Indirect Measurements (g-2, $m_W$, etc)}\label{Sec:Anm}


As discussed in the introduction, there are strong arguments for BSM physics.
Important clues about what physics lies beyond the SM may first appear in precision measurements of SM observables giving us targets for direct BSM searches. A flexible strategy to target present or future anomalies with direct searches is therefore of paramount importance.
As examples, in this section we focus on three types of existing measurements
that are in tension with their corresponding SM prediction and whose possible implications have received strong interest in the high-energy community: 

\subsubsection{Status of Anomalies}
{\bf $B$-physics anomalies:} Since 2012, a number of measurements of (semi-)leptonic $B$-meson decays have shown deviations from SM predictions~\cite{BaBar:2012obs,BaBar:2013mob,LHCb:2013ghj,LHCb:2014vgu,CMS:2014xfa,LHCb:2015gmp,Belle:2015qfa,LHCb:2015svh,Belle:2016dyj,LHCb:2017rmj,LHCb:2017avl,LHCb:2017smo,LHCb:2017rln,ATLAS:2018cur,CMS:2019bbr,LHCb:2019hip,LHCb:2020lmf,LHCb:2020gog,LHCb:2021trn,LHCb:2021awg,LHCb:2021vsc}. The significance of the deviation of the measurement of any particular $B$-physics observable and its SM prediction remains $\lesssim 3\,\sigma$. Combinations of the individual measurements follow an intriguing pattern and a combined preference for BSM effects of order $5\,\sigma$ (see, for example, Refs.~\cite{Altmannshofer:2021qrr,Cornella:2021sby,Isidori:2021vtc} for recent work). More data and better theory calculations are needed to settle if (and which of) the $B$-physics observables differ significantly from their SM predictions.  

{\bf Muon g-2:} In 2021, the Fermilab Muon (g-2) collaboration published its first measurement of the muon's anomalous magnetic dipole moment~\cite{Muong-2:2021ojo}, which is in excellent agreement with the previous measurement by the E821 experiment at BNL~\cite{Muong-2:2006rrc}, yielding a combined value of $a_\mu^{\rm exp} = 116\,592\,061(41) \times 10^{-11}$ [$a_\mu \equiv (g_\mu-2)/2$, with $g_\mu$ the muon's ``$g$-factor'']. Leading up to the Fermilab measurement, a significant effort was made to produce an up-to-date consensus value of the SM prediction for $a_\mu$, see Ref.~\cite{Aoyama:2020ynm} and references therein, $a_\mu^{\rm SM} = 116\,591\,810(43) \times 10^{-11}$. Taking $a_\mu^{\rm exp}$ and $a_\mu^{\rm SM}$ at face value, the measurement is in $4.2\,\sigma$ tension with the SM prediction. However, the hadronic vacuum polarization (HVP) contribution to $a_\mu^{\rm SM}$, a particular class of hadronic corrections, cannot be computed perturbatively and drives the uncertainty of $a_\mu^{\rm SM}$. The recommended value for $a_\mu^{\rm SM}$ from Ref.~\cite{Aoyama:2020ynm} uses a semi-analytic determination of the HVP contribution, extracting its value from low-energy hadron production cross sections via dispersion relations. Ref.~\cite{Borsanyi:2020mff} presented an {\it ab initio} (lattice QCD) calculation of the HVP contribution with precision comparable to the dispersive determination. However, the central value of the HVP contribution inferred by Ref.~\cite{Borsanyi:2020mff} is $\Delta a_\mu^{\rm HVP} \sim 2.3 \times 10^{-9}$ {\it larger} than the value recommended in Ref.~\cite{Aoyama:2020ynm} -- taken at face value, this shift brings the measured value of $a_\mu$ into good agreement with the SM prediction. Both the data-driven extraction of the HVP contribution via the dispersion relations and the lattice QCD calculations are currently undergoing scrutiny, see, e.g., Refs.~\cite{Aubin:2022hgm,Ce:2022kxy}. Until the HVP question is settled, the significance of the muon g-2 anomaly remains unclear. Note that, if the lattice QCD calculation of the HVP contribution is correct, it is diffcult to explain the low-energy hadron production cross sections~\cite{Lehner:2020crt,Crivellin:2020zul,Keshavarzi:2020bfy,deRafael:2020uif}.

{\bf $W$-boson mass:} Recently the CDF collaboration published a new direct measurement of the $W$-boson mass, $m_W = 80\,433 \pm 9\,$MeV~\cite{CDF:2022hxs} while the inferred value from electroweak precision fits is $m_W = 80\,359 \pm 5\,$MeV~\cite{deBlas:2021wap}, corresponding to a $7\,\sigma$ deviation. However, the CDF measurement of the $W$-boson mass is also in tension with direct $m_W$ measurements at the LHC: the ATLAS collaboration reported $m_W = 80\,370 \pm 19\,$MeV~\cite{ATLAS:2017rzl}, and LHCb recently measured $80,354 \pm 36\,$MeV~\cite{LHCb:2021bjt}, values which agree well with each other and with $m_W$ inferred from electroweak precision fits. Further scrutiny of the measurements of each experimental collaboration are required before the discrepancy between the direct measurement of the $W$-boson mass and the electroweak precision test can be assessed, see, e.g., Refs.~\cite{Isaacson:2022rts,Gao:2022wxk} for first steps in this direction. Furthermore, the published ATLAS and LHCb results use only a small fraction of the data recorded at the LHC, and as yet the CMS collaboration has not published a $W$-boson mass measurement.

\subsubsection{Implications of Anomalies}
Despite the questions about the significance of these anomalies, it is prudent to ask which hints about the nature of new physics we can collect from these anomalies. A vast literature of BSM explanations for each anomaly exists, and attempts have been made to explain multiple anomalies at the same time. Here, we will name only a few examples that are particularly relevant in the context of this report. We first comment on possible implications of the $B$-physics anomalies. Given the comparatively small energy scale of $B$-meson decays, a standard tool to analyze the $B$ anomalies are effective field theories (EFTs), see Refs.~\cite{Altmannshofer:2021qrr,Geng:2021nhg,Hurth:2021nsi,Ciuchini:2021smi} for recent work. The most relevant realizations of the EFT operators addressing the $B$ anomalies are perhaps leptoquark models, see, e.g., Ref.~\cite{Altmannshofer:2022aml} for a brief summary. Leptoquark as light as a few TeV that can address these anomalies, just above current LHC constraints~\cite{Angelescu:2018tyl,Cornella:2019hct,Popov:2019tyc}. Realizations of such leptoquarks in compositness models can be found in Refs.~\cite{Gripaios:2014tna,Barbieri:2016las,Marzocca:2018wcf}, and R-parity violating SUSY constructions can, e.g., be found in Refs.~\cite{Deshpande:2016yrv,Das:2017kfo,Earl:2018snx,Trifinopoulos:2018rna}. A particularly interesting (and ambitious) realization of the leptoquarks was proposed in Refs.~\cite{Bordone:2017bld,Bordone:2018nbg,Fuentes-Martin:2022xnb}, constructing a UV-complete embedding of the leptoquarks into a framework that explains the flavor structure of the SM via a generation-dependent Pati-Salam theory. Collider searches for extra gauge bosons necessarily accompanying the leptoquark have been presented in Ref.~\cite{Baker:2019sli}, while connections to DM have been explored in Refs.~\cite{Guadagnoli:2020tlx,Baker:2021llj}.

For the muon g-2 anomaly, hundreds of publications have discussed phenomenological BSM models that may explain the value of $a_\mu^{\rm exp}$. A class of models that are interesting in light of combined explanations of the muon g-2 and the $B$-physics anomalies are leptoquarks, see, e.g., Refs.~\cite{Bauer:2015knc,Crivellin:2020tsz,Crivellin:2020mjs,Hiller:2021pul,Nomura:2021oeu,Du:2021zkq,Ban:2021tos,FileviezPerez:2021lkq}. The leptoquarks explaining these anomalies may have masses in the few-TeV region, just beyond current LHC limits, and these leptoquarks can be embedded in UV complete models. The muon g-2 anomaly is also straightforward to explain in conventional (R-parity conserving) SUSY model such as the MSSM, see, e.g., Refs.~\cite{Barbieri:1982aj,Ellis:1982by,Kosower:1983yw,Moroi:1995yh,Carena:1996qa,Feng:2001tr,Martin:2001st,Marchetti:2008hw,Athron:2015rva,Chakraborti:2020vjp,Chakraborti:2021kkr,Chakraborti:2021dli,Baum:2021qzx,Baer:2021aax}. The most straightforward explanation for the muon g-2 anomaly in the MSSM is via one-loop contributions from light charginos and muon sneutrinos (the ``chargino-sneutrino'' contribution) or from light binos and left-right mixed smuons (the ``neutralino-smuon'' contribution). The contribution to $a_\mu$ from these loops is approximately $\Delta a_\mu \sim 10^{-9} \tan\beta \left(100\,{\rm GeV}/\widetilde{m}\right)^2$, where $\tan\beta$ is the usual ratio of the down- to up-type Higgs vacuum expectation values, and $\widetilde{m}$ is the mass scale of the SUSY particles participating in the one-loop diagram. Given the observed discrepancy of $a_\mu^{\rm exp} - a_\mu^{\rm SM} \sim 2.5 \times 10^{-9}$, we can see that for moderately large values $10 \lesssim \tan\beta \lesssim 50$, the electroweakly interacting sparticles contributing to $a_\mu$ must have masses in the range $200\,{\rm GeV} \lesssim \widetilde{m} \lesssim 500\,$GeV. While null-results from SUSY searches at the LHC have long excluded strongly charged sparticles (such as gluinos and squarks) in this mass range, LHC searches only recently began putting strong constraints on this range of $\widetilde{m}$ for sparticles that interact only electroweakly such as the electroweakinos and smuons that could explain the g-2 anomaly~\cite{ATL-PHYS-PUB-2022-013,CMS:SUSY}, and these bounds should improve with further analyses of existing LHC Run2 data as well as with the Run3 data to be recorded in the next years. Future lepton colliders also play an important role for probing electroweak sparticles, especially for (semi-)compressed spectra~\cite{Freitas:2003yp,Berggren:2013vna,Baer:2019gvu,Habermehl:2020njb,Baum:2020gjj,Agashe:2022uih,NunezPardodeVera:2022izz,Buttazzo:2020ibd,Yin:2020afe,Arakawa:2022mkr,Paradisi:2022vqp,Capdevilla:2021kcf}. It is worth noting that explanations of the muon g-2 anomalies in the MSSM suggest electroweakinos in the few-hundred GeV mass range, a parameter region of the MSSM yielding viable dark matter candidates~\cite{Chakraborti:2020vjp,Chakraborti:2021kkr,Chakraborti:2021dli,Baum:2021qzx}.

The $W$-boson mass anomaly is the most recent of the anomalies discussed here, and thus its possible BSM explanations are, as yet, less well explored. Most works appearing after the publication of Ref.~\cite{CDF:2022hxs}  have considered phenomenologically inspired models and it remains to be seen how possible explanations of the $m_W$-anomaly can be embedded in more complete models. Refs.~\cite{Bagnaschi:2022qhb,Yang:2022gvz,Athron:2022isz} have considered simultaneous explanations of the muon g-2 and the $m_W$ anomalies in the MSSM and, while not impossible, addressing both anomalies at the same time seems challenging. 

\subsubsection{Summary}

New experimental data and improved theoretical calculations could settle the status of the $B$-physics, the muon g-2, and the $W$-boson mass anomalies in the near future. If these observables are indeed discrepant from their SM predictions, they can offer us important hints for where to focus direct BSM searches. For example, both the muon g-2 anomaly and the $B$-physics anomalies could be explained with leptoquarks in the few TeV mass region, which can partially be probed in future runs of the (HL-)LHC, but may require high-energy colliders to reach. The muon g-2 anomaly could also point towards realizations of SUSY models with electroweak sparticles in the few-hundred GeV mass region, which can be probed with searches for electroweakinos and sleptons with existing and future LHC data, but may require a lepton-collider or higher-energy collider to fully explore. These examples of anomalies act as example scenarios for the interplay of energy frontier and rare and precision measuments. The energy frontier machines play a critical role in constraining and potentially uncovering the sources of any measured anomalies.

\subsection{General Exploration}\label{Sec:Gen-intro}

\begin{center} “{\it Experiment is the source of imagination. All the philosophers in the
world thinking for thousands of years couldn't come up with quantum mechanics}", Sidney Coleman.
\end{center}
The properties and interaction of the Standard Model particles would have been just as opaque to theoretically understand without experiment. The progress in discovering, measuring, and constructing a theoretical edifice to codify our understanding of the world of elementary particles has made remarkable progress over the past 75 years. Experiments at high energy accelerators have driven these discoveries and remain the only way to directly examine higher energy processes that have been so critical to the understanding of the building blocks of our universe. Although there are many specific motivations that are laid out in this document, the exploration of the energy frontier will allow us to observe any exotic particle and/or interaction whether they have been predicted from our current theoretical understanding or not.  The initial motivation for considering a TeV scale collider~\cite{Edwards:1985fap} (which would later be realized by the Tevatron at FNAL) in the early 1960s and the LHC was to be able to explore higher energy and the questions of the future. To quote Ellis, Gelmini and Kowalski in Ref.~\cite{Jacob:1984vnf}, “The most exciting experiments with a new accelerator are those which discover new particles. One of the main motivations for a Large Hadron Collider in the LEP tunnel is the opportunity it offers for exploring a new energy range, and perhaps discovering new particles with masses up to 0(1) TeV”. As the time scales for planning and construction are long, it was only later that the detailed motivation of probing the TeV scale was advanced enough to make precise theoretical predictions. Imagine if our scientific predecessors had decided to stop the exploration of the natural world because the model with a proton, a neutron, and an electron was ‘enough’ to explain the chemical elements? A direct frontal assault collider frontier has been more productive and remains intrinsically more likely to yield major breakthrough discoveries than less direct methods and remains our most likely path to success in the future.

\section{Theoretical guidance \& motivation}\label{Sec:Mot}

\subsection{Naturalness}\label{Sec:Nat}

The Standard Model, while fully full consistent, contains many unexplained small parameters, such as the electroweak hierarchy (the ratio of electroweak and Planck energies), small Yukawa couplings, and the strong CP angle $\bar \theta$. Finding a more fundamental explanation of these parameters is one of the major goals of high-energy physics. The QCD scale provides an example of what such an explanation could look like: it is generated dynamically, through dimensional transmutation, as the scale $\propto \exp[-8\pi^2/(bg_s^2)]$ at which a mildly small coupling $g_s$ at high energies runs to become strong, triggering confinement. The electroweak hierarchy has been a particular focus of efforts to build similarly compelling dynamical models and to experimentally test them. There are two especially prominent classes of examples: supersymmetric (SUSY) extensions of the Standard Model (reviewed in~\cite{Martin:1997ns}), in which SUSY can be dynamically broken at a low scale explained by dimensional transmutation and SUSY breaking can, in turn, trigger electroweak symmetry breaking~\cite{Ibanez:1982fr, Alvarez-Gaume:1983drc}; and models in which the Higgs boson is a composite particle, especially those in which it is a pseudo-Nambu-Goldstone boson~\cite{Kaplan:1983fs, Kaplan:1983sm, Dugan:1984hq} (see~\cite{Panico:2015jxa} for a modern review). These models provide {\em natural} explanations of the electroweak hierarchy, and in particular they are technically natural in the sense that radiative corrections to the Higgs boson mass are at most logarithmically divergent in such theories.

Natural models for the electroweak scale typically replace the quadratic {\em divergence} of the Higgs mass with a quadratic {\em sensitivity} to the mass scale of new particles. Because the Higgs field interacts much more strongly with third generation fermions than with the light generations, many models correlate the flavor puzzle with the electroweak hierarchy problem. For example, in supersymmetry, the electroweak hierarchy is obviously sensitive to the higgsino mass~\cite{Barbieri:1987fn}, but it is also very sensitive to the stop masses, motivating models in which the third-generation squarks are the lightest~\cite{Dimopoulos:1995mi, Pomarol:1995xc, Dvali:1996rj, Cohen:1996vb} and prompting a number of experimental searches targeting such models. In the context of composite Higgs models, the leading paradigm involves ``partially composite'' fermions~\cite{Kaplan:1991dc, Contino:2004vy}, with the top quark mixing strongly with composite states and fermionic top partners providing some of the first expected signals in collider experiments. So far, the LHC has not discovered any of these predicted signals. This has spurred investigation of variant models in which the collider signals are more difficult to observe, such as $R$-parity violating or Stealth SUSY models (see, e.g.,~\cite{Fan:2011yu,Evans:2013jna,Fan:2015mxp,Buckley:2016kvr} for natural simplified models the LHC can target) or ``neutral naturalness'' models where quadratic divergences are canceled by particles without strong Standard Model interactions, e.g., the Twin Higgs scenario~\cite{Chacko:2005pe} (see the Snowmass white paper~\cite{Batell:2022pzc} for more on neutral naturalness). However, given the continued lack of new physics signals at the LHC, including in measurements of Higgs couplings, many of these models must be at least mildly tuned to fit the data.

The lack of experimental evidence so far for any natural explanation of the electroweak hierarchy has resulted in the concept of naturalness itself becoming a subject of much debate. Various claims that naturalness predicted particles below a particular mass scale have been falsified by data, leading some to conclude that the concept of naturalness itself should be abandoned. It is important to keep in mind, however, that naturalness was never the source of a quantitative no-lose theorem. Rather, it is a heuristic, and one based on simple logic. A natural theory is one in which a large part of parameter space leads to qualitatively similar physics to the world around us, e.g., to an exponentially large hierarchy between the electroweak and Planck scales. An unnatural theory, like the Standard Model, is one in which ultraviolet parameters must be chosen in an exponentially tiny sliver of parameter space to lead to physics like the world around us. The {\em naturalness heuristic} is the simple statement that, all else being equal, a theory that describes a universe like the one we live in over a substantial portion of its UV parameter space is more likely to be correct than one that describes a universe like the one we live in over only a tiny sliver of parameter space.  Naturalness, then, is an argument about what new physics is most {\em plausible}. This heuristic has had successes in the past. One prominent example of a successful prediction of the naturalness heuristic in particle physics was the prediction that the charm quark should exist with $m_c \lesssim 1.5\,\mathrm{GeV}$, based on the quadratically divergent mass difference between the $K_L$ and $K_S$ mesons in the theory without charm~\cite{Vainshtein:1973md, Gaillard:1974hs} (see also~\cite{Ma:1974fg}).

The naturalness heuristic continues to be an important guide suggesting physics beyond the Standard Model might exist near the TeV scale. Naturalness does not give a sharp quantitative prediction, so the absence of superpartners or other signs of natural new physics at the LHC cannot invalidate the naturalness heuristic. Conversely, the naturalness heuristic cannot produce a precise upper bound on the mass scale of new physics. It is a guideline, and its failure in experiments so far suggests that we may live in a universe that is somewhat fine-tuned. That said, the heuristic would still indicate that supersymmetry with 10 TeV superpartners, despite its mild fine tuning, is a more plausible theory of nature than a theory where the Standard Model is valid up to the Planck scale. Attempts at more formal quantification of the limits in various scenarios, including anthropically tuned stringy landscapes cases, can indicate which parameters might be more strongly constrained; e.g. \cite{Baer:2020kwz,Baer:2021tta,Baer:2022naw}. These give predictions of superpartner masses significantly below the 10 TeV scale, and in particular constrain the Higgsinos to be less than ${\cal O}(500 GeV)$, but there is not a general consensus of such quantifications.

In the context of minimal supersymmetry, further support for a mildly fine-tuned spectrum arises from the observation that the Higgs mass of 125 GeV points to somewhat heavier superpartners, with stops at the 10 TeV scale or above (though a large left-right stop mixing could allow lighter squarks to be compatible with the measured Higgs mass; see~\cite{Draper:2016pys} for a review of Higgs mass calculations in SUSY). Such moderately tuned spectra could simply be an accident of the universe in which we live, or could be argued for from the perspective of the landscape of a more complete theory (see, e.g., the Snowmass white paper~\cite{Baer:2021tta}). The connection between the Higgs mass and heavy scalar superpartners has led to an increased interest in ``split'' supersymmetric spectra~\cite{Arkani-Hamed:2004ymt,Wells:2004di} in which the gaugino (and possibly higgsino) masses could remain at the TeV scale while the scalar superpartners are at the 10 TeV to 1 PeV scale~\cite{Arkani-Hamed:2012fhg, Arvanitaki:2012ps}. Such spectra are predicted by many simple models of SUSY breaking. A mildly long-lived gluino could provide a key collider signal of such a scenario.

Although moderately tuned, but mostly natural, scenarios remain compelling search targets that are compatible with all existing data, theorists have also begun to investigate more radical deviations from older paradigms. A key example is the cosmological relaxation scenario~\cite{Graham:2015cka}, in which dynamics in the early universe drives a light scalar field to become trapped at a value at which the effective Higgs potential appears to be fine-tuned from the viewpoint of traditional effective field theory. See the Snowmass White Paper~\cite{Craig:2022uua} for a more thorough discussion of novel approaches to naturalness, many of which have interesting experimental consequences.

\subsection{Higgs and Electroweak Symmetry Breaking}\label{Sec:HEWSB}

The Higgs boson is the newest and least understood member of the SM. The Higgs mechanism gives us an origin for the mass of the SM particles, but itself lacks an explanation. We don't have any understanding of what sets the scale for electroweak symmetry breaking, nor what drives the dynamics for it. As discussed in Sec.~\ref{Sec:Nat}, a search for an explanation and origin of the electroweak scale underlies a significant focus for our BSM explorations. 

Many of our BSM models have a SM-like Higgs embedded in "extended" Higgs sectors, which incorporate a combination of additional $SU(2)$ singlets, doublets and/or triplets. There is vast literature on such extensions, as well as very strong experimental programs that have been carried out at the LHC and prior. The exploration of such additional scalars is a huge thrust of the future energy program, and as such is addressed in detail in the Snowmass Higgs Report~\cite{sm21:higgs-report}. Here we will briefly summarize the main points. 

In a generic extended Higgs sector, the quartic couplings between the Higgs fields and their resulting vacuum structure define the inter-Higgs and Higgs-gauge couplings. Of particular relevance is the impact of mixing between the different scalars due to these. Current experimental observations very strongly constrain the couplings and mass of the 125 GeV state observed at the LHC in 2012. In particular, its couplings must be very "SM-like"~\cite{Sopczak:2020vrs}. 
Hence, precision studies of the Higgs couplings are a major avenue of exploration for the presence of additional scalars at the weak scale, and provide  constraints on the possible quartic couplings in a given model. 

The prospects for direct searches of these additional Higgs scalars depend strongly on their couplings to the SM fermions, which govern both the production and decay of these particles at colliders. These fermion couplings to the Higgs fields must be defined such that tree-level flavor changing neutral currents are forbidden to be consistent with stringent experimental constraints. As an example, in one of the most commonly studies extension, the two Higgs doublet model~(2HDM), this is accomplished by assigning $Z_2$ charges to the Higgs doublets and the SM fermions.

Such additional Higgs fields may be able to provide a source of CP-violation in addition to the SM, and also perturb the SM Higgs potential in the early Universe such that electroweak Baryogenesis gives rise to the required matter-antimatter asymmetry observed in nature. Additionally, the Higgs may serve as a portal to Dark Matter, as discussed later in Sections ~\ref{Sub:HiggsPortal} and \ref{Sec:OtherDM}. 

In more complete models, such as SUSY, which require at least two Higgs doublets, the quartic couplings are not completely free parameters, as is the case in generic extensions. In the Minimal Supersymmetric Standard Model~(MSSM) for example, requirements of the mass and SM-like nature of the observed Higgs boson provide stringent constraints on the additional Higgs states at the weak scale, as well as requiring additional top superpartners, as discussed in Sec.~\ref{Sec:SUSY}. Less minimal models of SUSY such as the Next to minimal supersymmetric SM~(NMSSM), have an additional SM singlet superfield, which alleviates some of these constraints, but are still a subset of the more generic 2HDM + singlet models.   

In contrast to models with  explicit additional particle content, the SM Higgs may arise as emergent phenomena in models with extra dimensions or in Composite Higgs models. Such models will generically have additional states at the weak scale which are also a target of searches at colliders, both directly and via their impact on precision physics as discussed in more detail in Secs.~\ref{Sec:CHED},~\ref{sec:nb:composite-higgs} and \ref{sec:nb:composite-kk}.

\subsection{Composite Higgs and Extra Dimensions}\label{Sec:CHED_intro}
\subsubsection{Extra Dimensions}



Extra dimensions (see review in \cite{ParticleDataGroup:2020ssz}, for example) and SUSY are the only possible extensions of relativistic space-time in field theory.
%
%
Both 
%
%
concepts
%
%
%
feature in super-string theory, which is perhaps the only well-developed, consistent theory
of
quantum gravity.
In light of the above facts, there is a strong possibility that extra dimensions exist in Nature.
The crucial question then is are there any observable effects of extra dimensions?
The answer 
depends on the size of the extra dimensions: 
first of all, extra dimensions must be compact, otherwise
the 
inverse
square
law of gravitational
and 
electrostatic
forces would be modified at long distances, in contradiction with observations.
A SM field (as well as gravitational)  propagating
in such a finite-size extra dimension manifests as a tower of Kaluza-Klein (KK) modes from the four-dimensional (4d)
viewpoint, with masses quantized increasingly in units of inverse size of extra dimension (called the compactification or KK scale).
The zero-mode (typically massless) of this KK decomposition is identified with the observed SM field/particle.
In this framework, the other/heavier modes are new, beyond SM, particles whose effects are potentially observable.
In addition, there is a particle called the radion, which corresponds to the modulus associated with the fluctuations of the size of the extra dimension.
%

%
%
One can then ask which KK scales are motivated
%
%
from theory and experimental perspectives?
In particular, the lightest KK scale being $\sim {\cal O}$ (TeV) would be relevant for the solving
the Planck-weak hierarchy problem of the SM: the lightest KK top quark could cancel the quadratic divergence in the SM Higgs mass$^2$ from the SM top quark loop.
It could also 
provide
a dark matter candidate 
in the form of the lightest KK particle.
On the other hand,
%
%
if the KK scale is much heavier than $\sim {\cal O}$ (TeV),
then such an extra dimension could 
still address, for example, the flavor hierarchy of the SM via varying profiles for the SM fermions in the extra dimension.
Moving onto 
signals from these KK particles, 
in case of the lightest KK scale being $\sim {\cal O}$ (TeV), these KK particles might of course be {\em directly} accessible to the LHC/future colliders.
In addition (and this point is valid even if KK particles are beyond
LHC/future collider reach), there are
indirect/virtual effects of these KK particles on flavor/CP violation and electroweak (EW) fits of the SM.
Indeed, $\sim {\cal O}$ (TeV) KK scale could be strongly constrained by such precision tests {\em in general}; remarkably, 
this tension might be ameliorated by implementing suitable symmetries, which could be simple extensions of well-known protection mechanisms of the
SM itself, thus 
%
%
rendering $\sim {\cal O}$ (TeV) KK scale  viable.
Overall,
%
%
the above discussion strongly motivates direct search at the LHC/future colliders for such KK particles.

In fact, current extensive direct searches at the LHC are already probing $\sim$ TeV mass scale, say, for KK gauge bosons/gravitons, 
produced in $q \bar{q}$ or $gg$ annihilation, followed by 
%
%
decay into 2 SM particles.
Note that, unlike universal
$Z^{ \prime } / W^{ \prime}$, 
these KK particles often decay preferentially into heavier SM particles, such as top quark/Higgs particles (including longitudinal $W/Z$).
Given the hierarchy between KK scale and masses of these SM particles, the top quarks/Higgs bosons are produced boosted, so their
decay products (in turn) merge. This direction has resulted in development of sophisticated jet-substructure techniques
to identify such novel
%
%
objects (from the detector angle), which could be useful 
in other contexts as well.
More recently, non-standard decays of such KK particles have also been studied,
%
%
for example, decay into other new particles, such as heavier into lighter KK particles or
KK particles into radion (see, for example, \cite{Agashe:2022gbb}), which subsequently
%
%
decay into SM particles.

In summary, extra dimensions have a a variety
of 
motivations
and lead to a rich set of signals, especially from direct production at the LHC/future colliders, where there remains 
a 
reasonable chance of their discovery (possibly via non-standard modes), in spite of the current bounds (direct and indirect).
Thus, continuing
and 
diversifying such 
searches at the LHC and future colliders is
justified.

\subsubsection{Composite Higgs}


The question of whether the Higgs boson is an elementary or composite particle remains a fundamental mystery.  A composite Higgs boson (see \cite{Panico:2015jxa}, for a review) requires a new strong gauge interaction whose coupling becomes strong (via dimensional transmutation) above the TeV scale and binds together new elementary constituents (possible UV completions identifying the underlying new gauge group and involving elementary fermion constituents have been proposed in ~\cite{Barnard:2013zea,Ferretti:2013kya}). These constituents not only form a Higgs boson, but inevitably the new strong force binds these constituents into many other bound states (or resonances), much like the structure seen with QCD dynamics. Importantly, to separate the Higgs boson resonance from possible heavy resonances, composite Higgs models assume that the new strong dynamics does not break the electroweak symmetry and instead the symmetry is explicitly broken by the gauge and Yukawa couplings of the Standard Model. This means that the Higgs boson becomes a pseudo-Nambu Goldstone boson of the underlying strong dynamics (similar to the pions in QCD). In this way the Higgs boson is identified as the lightest bound state, protected by an approximate global symmetry, much below the compositeness scale. Therefore, observing the remaining heavier resonances above the Higgs boson mass at the compositeness scale is a tell-tale sign of Higgs compositeness.

The resonance mass approximately corresponds to the scale of compositeness, and current searches generically constrain the lowest lying resonances to be heavier than the TeV scale. For instance (see \cite{ParticleDataGroup:2020ssz}), searches for spin 1 resonances (similar to the $\rho$ meson in QCD) place a lower limit of $\sim 3$ TeV, while limits on spin 2 resonances are constrained to be heavier than $\sim 2$ TeV. An interesting feature of composite Higgs models is that they can also explain the fermion mass hierarchy via partial compositeness~\cite{Kaplan:1983fs,Contino:2004vy} and in particular, contain a spectrum of composite (spin 1/2) top quark states. Furthermore, in order to be compatible with the observed Higgs boson mass, these top partner resonances 
must be lighter than the spin 1 and 2 resonances and lead to 
distinctive signals due to the fact that they can have exotic electromagnetic charges such as $Q=-1/3,2/3$ and $5/3$.  For instance, the decay of a $Q=5/3$ fermionic resonance provides a pair of same-sign leptons in the final state, with limits reaching 1.3 TeV. It is also possible that in non-minimal models the new strong dynamics could produce a neutral, heavy spin 0 resonance that plays the role of dark matter~\cite{Frigerio:2012uc} and therefore dark matter searches provide complementary ways to look for resonances.  Clearly, there is strong motivation to continue searching for a variety of resonances and will be a primary goal at the LHC and future colliders.

Another possible sign of compositeness would be to search for deviations in the couplings of the Higgs boson to gauge bosons and the (composite) top quark. These deviations are inversely proportional to the scale of compositeness and therefore requires precision measurements. Current limits constrain these deviations at the 10\% level~\cite{Panico:2015jxa}. Another possible signature of compositeness is the large decay rate of the Higgs boson into a Z boson and photon, which can be much more enhanced than in the Standard Model~\cite{Giudice:2007fh}. Finally, if the scale of compositeness is much higher than the TeV scale then in some models there is also the possibility of observing long-lived decays of colored spin-0 resonances in displaced vertex searches~\cite{Barnard:2014tla}. Current limits constrain these resonances to be above the TeV scale~\cite{Barnard:2015rba} and indirectly probe compositeness scales $\gtrsim 10$ TeV.

In summary, understanding whether the Higgs boson is composite or not is an important and  fundamental question. Current limits on resonance searches suggest that the compositeness scale is beyond the TeV scale. Future experiments searching for deviations in the Higgs couplings could also reveal composite structure, as would future colliders that search for resonances up to the 10 TeV scale and should remain a top priority for the ongoing experimental effort.

\subsection{Supersymmetry (SUSY)}\label{Sec:susy_intro}


Supersymmetry~\cite{Nilles:1983ge,Haber:1984rc,Martin:1997ns,Baer:2006rs} provides extensions of the Standard Model (SM) in which the particles related by this symmetry share the same gauge quantum numbers, but differ 
in the value of the spin. Supersymmetry relates also the value of
their masses and hence the effective Yukawa couplings to the Higgs fields of the theory.  The
non-observation of supersymmetric partners of the SM particles, with the same masses, demands the introduction of so-called soft supersymmetry
breaking terms, which are governed by dimensionful and gauge invariant interactions. The breaking terms preserve the
important feature of supersymmetry that is its lack of sensitivity on physics at the ultraviolet scale, that at the
loop-level leads to the cancellation of a quadratic dependence on the mass of very heavy particles~\cite{Dimopoulos:1981zb}. The motivation for a supersymmetric extension of the Standard Model comes from the unification of gauge and Yukawa couplings at high energies~\cite{Dimopoulos:1981yj,Langacker:1992rq,Dimopoulos:1991yz,Carena:1993ag}, the radiative breakdown of the electroweak symmetry~\cite{Ibanez:1982fr,Alvarez-Gaume:1983drc,Ibanez:1982fr,Carena:1994bv,Baer:2012up}, the natural existence of a Dark Matter candidate~\cite{Drees:1992am} and the possibility of obtaining these theories as a low energy realization of grand unified theories and superstrings, that allow for a consistent quantization of gravity. Experimental searches
for supersymmetric particles have  set bounds on the soft supersymmetry breaking terms, due to
the absence of any direct observable signature of these particles. 

In the so-called Minimal
Supersymmetric extension of the SM, MSSM,  each chiral fermion of the SM has a complex scalar superpartner
and each gauge-boson degree of freedom has a Weyl-fermion superpartner.  The Higgs sector needs to
be extended to two Higgs doublets (and four Weyl fermions) in order to cancel anomalies and be able to provide masses to up and
down quarks, due to the restrictions that supersymmetry impose on the possible Yukawa couplings. 
Both scalar top partners couple with a Yukawa coupling, related to the SM top Yukawa coupling. Such a strong
coupling is important, since it leads to radiative corrections to the Higgs mass which grows logarithmically
with the scalar-top mass scale~\cite{Haber:1990aw,Ellis:1990nz},. These corrections are very important since in the MSSM supersymmetry
relates the quartic Higgs couplings to the gauge couplings leading to an
upper bound on the Higgs mass equal to $M_Z$, before radiative corrections. Raising the Higgs mass via radiative corrections demands
scalar top masses  to be above approximately the TeV scale~\cite{Carena:1995bx,Heinemeyer:1998np,Bagnaschi:2014rsa,Draper:2013oza,Lee:2015uza,Vega:2015fna,Bahl:2017aev,Slavich:2020zjv}. 
The Large Hadron Collider (LHC) is currently looking for scalar tops and is setting limits close to the TeV scale,
that as emphasized before is the lowest scale possible to lead to compatibility with 
the observed Higgs mass in the MSSM~\cite{ATLAS:2020xzu,CMS:2021eha}. The LHC also set bounds of similar order on the masses of the superpartner of the gluon 
and of lighter generation quarks, with the gluon superpartner mass constraint being the strongest
one, of the order of 2~TeV~\cite{ATLAS:2021twp}.  If strongly interacting supersymmetric particles get  masses of the order
of a few to several TeV, as they should in order to avoid large radiative corrections on the Higgs mass parameters,
they may be probed at the high luminosity LHC or at higher energy lepton or hadron colliders. 

Let us stress that most of the collider bounds on strongly coupled supersymmetric particles have been obtained by assuming the existence of a discrete R-Parity symmetry, that assigns an even parity to all SM particles and an odd parity to their supersymmetric partners. This parity symmetry ensures the stability of the lightest supersymmetric particle, which tends to be neutral and constitutes therefore an excellent Dark Matter candidate, and eliminates all renormalizable interactions that can lead to proton decay. In the presence of R-Parity, the production of supersymmetric particles lead to large missing energy signatures. R-Parity violation~\cite{Dreiner:1997uz,Barbier:2004ez}, however, is possible and can lead to an explanation of observed anomalies in flavor physica~\cite{Deshpande:2016yrv,Altmannshofer:2020axr}. At colliders, R-Parity violation may lead to copious hadronic or leptonic decays of the heavy, colored supersymmetric particles, as well as to long lifetimes of these particles. The LHC bounds on the supersymmetric colored particle masses in such a case tend to be of similar strength as in the case of R-Parity conservation~\cite{CMS:2016vuw,ATLAS:2018umm,ATLAS:2021fbt}.

Regarding the weakly interacting particles, their masses are less constrained experimentally, with bounds
of the order of a few hundred GeV up to 1~TeV, depending on their decay properties~\cite{ATLAS:2021moa,ATLAS:2021yqv,CMS:2022sfi,Liu:2020ctf,Liu:2020muv}.  The future sentivity to SUSY weak states at HL-LHC and possible new experiment are discussed in Sections \ref{sec:SUSY_ewk} and \ref{sec:SUSY_slepton}.
Possible information about their masses may also be obtained under the assumption that this sector may explain the observed Dark Matter
in the Universe~\cite{Drees:1992am} or  the recent measurement of the anomalous magnetic moment of the
muon, $(g-2)_\mu$~\cite{Muong-2:2021ojo}, which shows  a significant deviation with respect to its expected value within the SM. 
If the mass difference of the scalar leptons and the superpartners of the gauge bosons is not very large, and the characteristic
mass scale is of the order of a few hundred GeV, one can easily explain the Dark Matter relic density, with the 
Dark Matter particle identified with the superpartner of the hypercharge gauge boson. This can be done avoiding 
the current experimental bounds from the LHC and at the same time obtain $(g-2)_\mu$ values consistent with experiment~\cite{Moroi:1995yh}.
The mass parameter  associated with the mass of the supersymmetric partners of the Higgs bosons,
can be constrained by demanding, in addition to $(g-2)_\mu$ and the relic density,  compatibility with direct 
and indirect Dark Matter detection, and vacuum stability constraints, which sets a correlation between the
value of the SUSY $|\mu|$ parameter and the relative sign of the electroweak gauge superpartner mass terms~\cite{Chakraborti:2021dli,Baer:2021aax,Baum:2021qzx}.  In general, values
of $|\mu|$ lower than about 1~TeV are preferred by these considerations~\cite{Chigusa:2022xpq}. 

The NMSSM~\cite{Ellwanger:2009dp} is an extension of the MSSM in which a singlet under the SM gauge group is added to the spectrum.
The main difference is hence in the Higgs sector and its superpartners. In particular, the coupling between the
singlet and the two Higgs doublets affects the Higgs quartic couplings and allows to obtain the proper Higgs
mass for lighter scalar tops. These corrections, however, are small unless the two Higgs vacuum expectation
values are of the same order. If they are, however, one can not only obtain the proper Higgs mass but also
suppress its mixing with the heavier Higgs sector coming from the second Higgs doublet, hence leading
to compatibility with Higgs data even if the heavy, non-standard Higgs particles, are relatively light~\cite{Carena:2015moc}.  Finally, in the NMSSM
one can also get $(g-2)_\mu$ and the Dark Matter density~\cite{Baum:2017enm,Abdughani:2021pdc,Cao:2021tuh}, with properties similar to the case of the MSSM.


\section{Methods}\label{Sec:Mot}

\subsection{Benchmark collider scenarios}

While all colliders offer a multifaceted approach to the search for new physics, different types offer different strengths and features.  For example some colliders focus  on extending the energy reach, some on reaching the highest precision possible. For the Snowmass 2021 exercise, we will focus on two main classes of colliders identified as \textit{Higgs factories} and \textit{multi-TeV colliders} respectively. We define \textit{Higgs factories} as lepton colliders with center-of-mass energy up to 1~TeV that will substantially improve the Higgs-boson precision physics program beyond the HL-LHC reach. On the other hand lepton and hadron colliders with center-of-mass energies beyond 1~TeV will be labeled as \textit{multi-TeV colliders} and will primarily be identified by the potential of allowing for the direct exploration of energy scales beyond the reach of the HL-LHC. Of course, any such separation is intrinsically arbitrary. Higgs factories can also complement the discovery reach of the HL-LHC in the low-mass region, and will provide a wealth of precision measurements beyond Higgs-physics alone. At the same time, multi-TeV colliders will produce huge numbers of Higgs-bosons and continue to indirectly test new physics via SM precision measurements. The labeling of \textit{Higgs factories} versus \textit{multi-TeV colliders} is only meant to organize the possible benchmark scenarios considered in the report, as illustrated in Tables~\ref{tabHiggsFactory} and \ref{tabBSMColliders}. 
\begin{table}[h!]
\begin{center}
 \caption{Benchmark scenarios for Snowmass 2021 Higgs factory studies.}
\begin{tabular}[c]{||l l|c|c|c||}
\hline
 \hline
Collider	&	Type	&	$\sqrt{s}$	&	$\mathcal{P} [\%]$	&	$\mathcal{L}_{\rm int}$	\\
	&		&		&	$e^-/e^+$	&	${\rm ab}^{-1}$	\\ \hline\hline
HL-LHC	&	pp	&	14 TeV	&		&	6	\\ 
\hline
ILC and C$^3$	&	ee	&	250 GeV	&	$\pm80/\pm30$	&	2	\\
c.o.m almost &		&	350 GeV	&	$\pm80/\pm30$	&	0.2	\\
similar	&		&	500 GeV	&	$\pm80/\pm30$	&	4	\\
	&		&	1 TeV	&	$\pm80/\pm20$	&	8	\\
	\hline
CLIC	&	ee	&	380 GeV	&	$\pm80/0$	&	1	\\
\hline
CEPC	&	ee	&	$M_Z$	&		&	60	\\
	&		&	2$M_W$	&		&	3.6	\\
	&		&	240 GeV	&		&	20	\\ 
	&		&	360 GeV	&		&	1 \\
	\hline
FCC-ee	&	ee	&	$M_Z$	&		&	150	\\
	&		&	2$M_W$	&		&	10	\\
	&		&	240 GeV	&		&	5	\\
	&		&	2 $M_{top}$	&		&	1.5	\\
\hline
muon-collider (higgs)	&	$\mu\mu$	&	125 GeV &		&	0.02\\		
\hline \hline
\end{tabular}
\label{tabHiggsFactory}
\end{center}
\end{table}
\begin{table}[h!]
\begin{center}
 \caption{Benchmark scenarios for Snowmass 2021 energy frontier multi-TeV collider studies.}
\begin{tabular}[c]{||l l|c|c|c||}
\hline
 \hline
Collider	&	Type	&	$\sqrt{s}$	&	$\mathcal{P} [\%]$	& $\mathcal{L}_{\rm int}$	\\
	&		&		&. $e^-/e^+$ &  ${\rm ab}^{-1}$	\\ \hline\hline
HE-LHC & 	pp	&	27 TeV	&	& 15	\\ 
\hline
FCC-hh	&	pp	&	100 TeV	&	&  30	\\
    	\hline\hline
LHeC	&	ep	&	1.3 TeV	&   &  1	\\
FCC-eh	&		&	3.5 TeV	&	&	2	\\
\hline\hline
CLIC   &  ee  &	  1.5 TeV	&	$\pm 80/0$	&	2.5	\\
	   &	  &   3.0 TeV	&	$\pm 80/0$	&	5	\\
\hline \hline
	    High energy muon-collider	&	$\mu\mu$	&	3 TeV	&	&	1	\\
	&		&	10 TeV	&	&	10	\\
\hline\hline
\end{tabular}
\label{tabBSMColliders}
\end{center}
\end{table}

\subsection{Discovery versus Exclusion Limits}\label{Sec:DiscVsLimit}

\begin{figure}[h!]
    \subfloat[ATLAS Wino-Bino HL-LHC sensitivity ]{\includegraphics[width=0.55\textwidth]{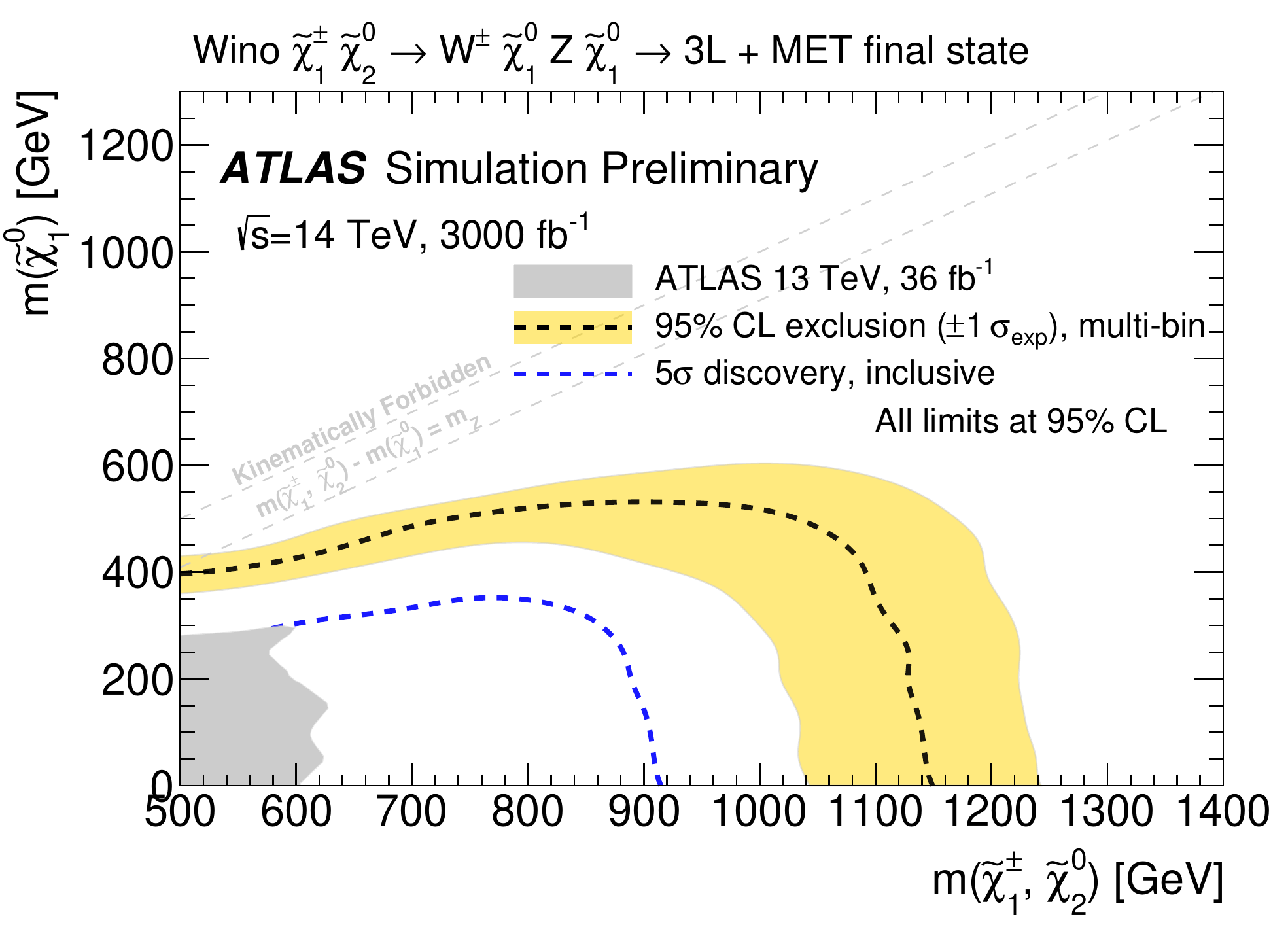}}
    \subfloat[CMS Higgsino HL-LHC sensitivity]{\includegraphics[width=0.45\textwidth]{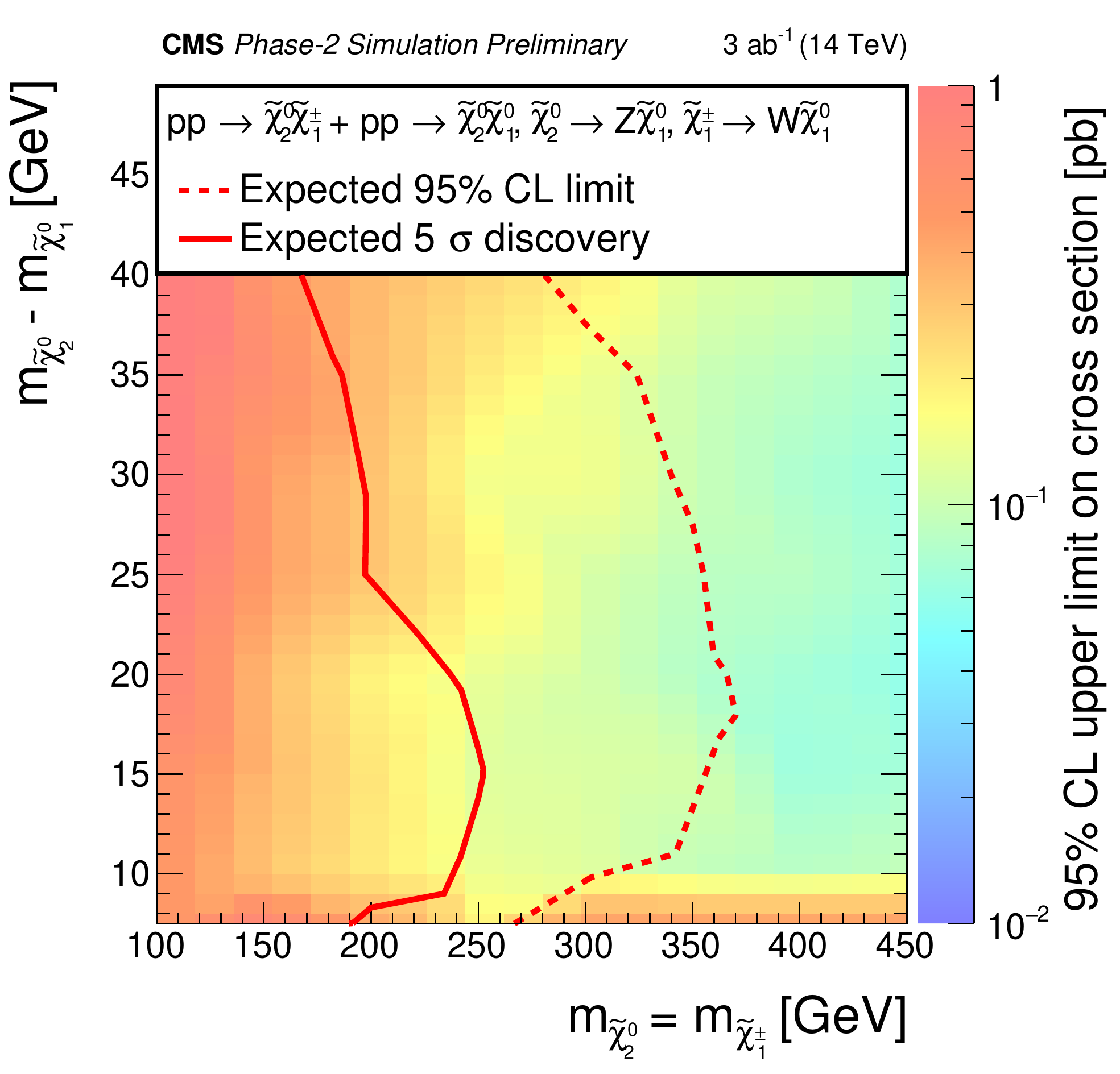}}
    \caption{Example comparisons of 5$\sigma$ discovery to 95\% CL exclusion regions.}
    \label{fig:ef08_ExclusionVsDiscovery}
\end{figure}

The BSM physics sensitivity presented in this document includes
95\% confidence-level (CL) exclusion limits and in some cases,
also the 5$\sigma$ discovery reach. A 5$\sigma$ discovery requires more events than a 95\% CL exclusion. For lepton colliders at lower energies ($\lesssim 1 \TeV$), the 5$\sigma$ discovery region is close to the 95\% CL exclusion region, because the events are produced primarily at the center-of-mass
energy. As the lepton-collider energy is increased, more of the interesting cross-section is due to vector-boson fusion, where the collision energy of the vector bosons is a distribution with a maximum at the center-of-mass energy. In this case the 5$\sigma$ discovery reach will be significantly below the a 95\% CL exclusion. Similarly, beamstrahlung effects reduce the fraction of energy at the nominal center-of-mass (CM) energy. For example for CLIC, 60\% of the luminosity is over 99\% of the CM energy for a 380 \GeV collider, but this decreases to 33\% at 3 \TeV \cite{Robson:2744946}.  For hadron colliders, the energy reach and the luminosity are strongly coupled because the proton structure. The result is that the 5$\sigma$ discovery region is smaller than the 95\% CL exclusion region for hadron colliders. Figure \ref{fig:ef08_ExclusionVsDiscovery} show an example of this, where discovery sensitivity is 30-40\% lower than the 95\% CL exclusion reach. An additional example is shown in Table \ref{tab:ZpTable} in the New-Bosons Section.

\subsection{Estimation Methods}\label{Sec:ColliderReach}

There are a variety of procedures for estimating the future sensitivity of colliders. Sensitivity can be estimated directly by simulating the collider beams, the collision physics, and the detector response, and then analyzing the simulated data. Alternatively, results from existing experiments, and existing direct estimates can be extrapolated to different scenarios. For example in lepton colliders, the limits are very frequently limited by the center-of-mass energy, so the exclusion and discovery reach for many particles are approximately $\sqrt(s)/2$. For hadron colliders, scaling can be performed using parton luminosity assuming reconstruction efficiencies, background rejection, and signal-to-background cross-section ratios remain roughly constant \cite{ColliderReach}. While the direct method is in principle more robust, such analyses are often simplified compared to existing analyses of real data. Extrapolation, on the other hand, is subject to the assumptions listed above but is based on comparatively optimized analyses. Where both methods are available, both or the range of results is indicated in plots; this is particularly relevant for the complex binned analyses, see for example Figure \ref{fig:ef08_higgsinoplot}.



\section{Composite Higgs and Extra Dimensions}\label{Sec:CHED}


\textit{Note: This section is not yet complete and more content is expected to be added during the next iteration. Comments on specific missing content are still appreciated.}


\subsection{Kaluza-Klein Excitations}
\label{sec:nb:composite-kk}

In models with one or more extra spatial dimensions, as mentioned above, the extra dimensions are generally compactified with a small radius to satisfy table-top experimental constraints measuring the gravitational interaction at the sub-mm length scale~\cite{Adelberger:2009zz, Murata:2014nra}.  As a consequence, any SM fields that propagate in the extra dimensions give rise to a Kaluza-Klein tower of excitations in the 4D theory, as a result of the equation of motion in the compactified dimensions.  At colliders, these Kaluza-Klein modes can give rise to resonance signatures since they share the same spin and gauge quantum numbers as their familiar SM counterparts, which are simply the zero modes of the KK tower.

Sensitivity to the Randall-Sundrum model of the KK gluon decaying to $t\bar{t}$ was discussed at Snowmass. This KK model is one of the many models that are constrained by general $t\bar{t}$ resonance searches, which are regularly performed at the LHC~\cite{CMS:2018rkg,ATLAS:2020lks} and are planned for future $pp$ colliders~\cite{Helsens:2019bfw}. The mass reach, and hence the inverse scale of the extra dimensions, for the KK gluon is 5.7 (6.6) TeV for 5$\sigma$ (95\% CL) for the HL-LHC at $\sqrt{s}=14$ TeV with 3 ab$^{-1}$, and 9.4 (10.7) TeV for the HE-LHC at $\sqrt{s}=27$ TeV with 15 ab$^{-1}$~\cite{CMS-PAS-FTR-18-009}. 
A model of a KK gluon decaying to a SM boson and a radion was also discussed~\cite{Agashe:2022gbb}, in particular the three-gluon final state which has been searched for at the LHC~\cite{CMS:2022tqn}. 

Weakly produced KK states, such as the the KK graviton, are often  searched for in di-jet resonances at pp colliders, $pp \to X \to 2$ jets, discussed further in Sec.~\ref{sec:dijet-resonances}.  The $5-\sigma$ discovery mass at future colliders is shown as a function of the integrated luminosity in Fig.~\ref{fig:strong-weak}, and the corresponding summary of sensitivities in presented in Table \ref{tab:dijet-sensitivity-table}.

\subsection{Composite Higgs}
\label{sec:nb:composite-higgs}

As discussed in detail previously, in composite Higgs models, the minimal Higgs mechanism for electroweak symmetry breaking is extended by new confining degrees of freedom that ameliorate the gauge hierarchy problem in the Standard Model.  The corresponding 125~GeV Higgs scalar observed is then generally a mixture of a fundamental scalar boson and a composite meson of the confining group.  The combination of these ingredients then cures the perturbative unitarity problem of longitudinal EW gauge boson scattering amplitudes when the SM Higgs is neglected.  Moreover, since the fundamental scalar boson is only partially responsible for unitarity restoration, the composite degrees of freedom are matched to EW diboson resonances necessary for the remaining restoration of unitarity.  These EW diboson resonances are thus a central prediction in composite Higgs models and emblematic of their rich phenomenology.

The phenomenology of a Composite Higgs model is mainly governed by two parameters: the mass~(compositeness) scale $m_*$, and the coupling $g_*$~(which sets the scale of the couplings in the EFT Lagrangian). In comparison with the SM couplings, we expect a strongly interacting sector to have $g_*>1$ couplings, while unitarity requires $g_* < 4\pi$. The Wilson Coefficients, defined in Ref.~\cite{de_Blas_2020}, can be all parameterized in terms of this mass scale and coupling, modulo order 1 factors. Different colliders have complementary sensitivities to the various operators, the most relevant ones being~\cite{Strategy:2019vxc, de_Blas_2020} 

\begin{eqnarray}
\frac{c_\phi}{\Lambda^2} \frac{1}{2} \partial_\mu \left(\phi^\dagger \phi\right) \partial^\mu \left(\phi^\dagger \phi\right) &\qquad;\qquad& C_\phi = \frac{c_\phi}{\Lambda^2} \sim \frac{g_*^2}{m_*^2}\; ,  \\
\frac{c_W}{\Lambda^2}\frac{i g}{2}\left( \phi^\dagger \overleftrightarrow{D}\phi \right) D_\nu W^{a\;\mu\nu}  &\qquad;\qquad& C_W = \frac{c_W}{\Lambda^2} \sim \frac{1}{m_*^2} \; , \\
\frac{c_{2W}}{\Lambda^2}\frac{g^2}{2}\left( D_\mu W^{a\; \mu \nu}\right) \left( D_\rho W^{a\; \rho \nu}\right) &\qquad;\qquad& C_{2W} = \frac{c_{2W}}{\Lambda^2} \sim \frac{1}{g_*^2 m_*^2} \; .
\end{eqnarray}

\begin{figure}[h!]
  \begin{center}
    \includegraphics[width=1.0\textwidth]{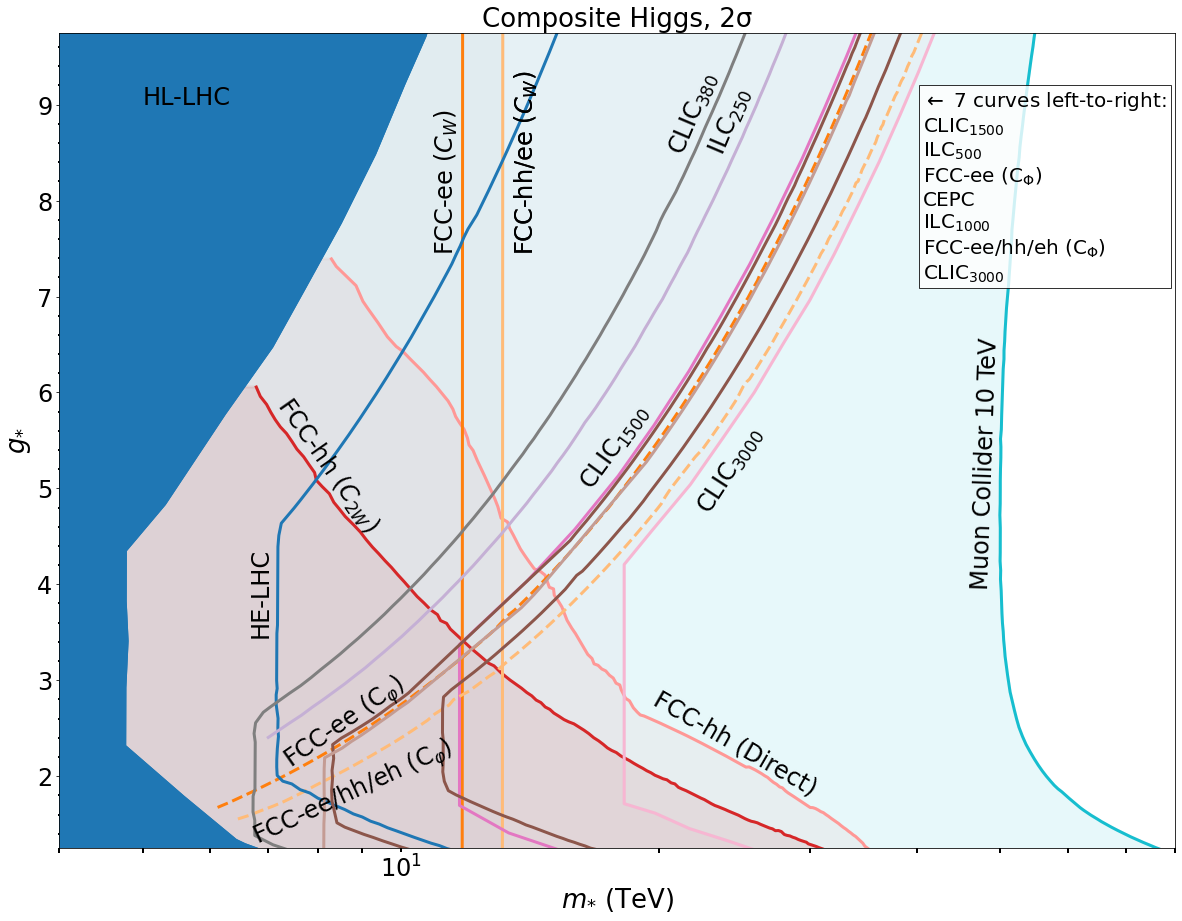} 
  \end{center}
  \caption{Exclusion (2-$\sigma$) sensitivity projections for future colliders~(as labeled) in the mass scale vs. coupling strength plane for Composite Higgs models. 
  Plot based on Refs.~\cite{Strategy:2019vxc, Chen:2022msz}. See text for details.}
    \label{fig:ef08_compHiggs}
\end{figure}

As can be seen in Fig.~\ref{fig:ef08_compHiggs}, while the FCC (both ee and hh/ee) sensitivity to $C_W$ is almost independent of $g_*$, $C_\phi$ can probe larger couplings for higher mass scales, and would also impact Higgs coupling measurements. $C_{2W}$ on the other hand has lower sensitivity for larger couplings as the mass scale increases. Also shown is the direct search sensitivity for a triplet vector $\rho$ resonance. The muon collider is also expected to probe such models, and we show projections for the 10 TeV Muon collider~\cite{Chen:2022msz} considering the tree level process $\mu^+ \mu^- \rightarrow hh\nu\nu$, which is minimally sensitive to $C_\phi$. Probes of $C_\phi$ from Higgs coupling measurements~\cite{Buttazzo:2020uzc} may be competitive or stronger at lower energy muon colliders, but they are not considered here. A further degree of freedom is associated with the compositeness of the fermionic sector, and can have significant effects on $t\bar{t}$ and $b\bar{b}$ production. For  example, the sensitivity of a 10 TeV muon collider to a scenario where both the left and right handed tops are assumed to have equal amounts of compositeness $\epsilon_t = \epsilon_q \sqrt{y_t}/g_*$~(not shown) can significantly impact the possible reach. For details, see Ref.~\cite{Chen:2022msz} and references within. For further discussions of the sensitivity of future colliders to composite Higgs models see Refs.~\cite{Thamm:2015zwa,FCC:2018byv, Baker:2022zxv}.

One particular example of a composite Higgs model that was discussed at Snowmass~\cite{Lane:2022ybv}, and its phenomenology includes a new Higgs doublet $(H^\prime,H^\pm, A)$ with mass less than about $0.5$ TeV, and new vector mesons $\rho^{\pm,0}_H$ and $a^{\pm,0}_H$ with mass greater than about 1 TeV that decay to pairs of EW bosons $WW$, $WZ$, $WH$ and $ZH$. Additional exotic states such as Vector-like quarks may also be found in Composite Higgs models, and are discussed in Sec.~\ref{Sec:HeavyQuarks}.

\section{Supersymmetry (SUSY)}\label{Sec:SUSY}

As discussed in Sec.~\ref{Sec:susy_intro}, MSSM SUSY is a complex model with partners for every SM particle plus additional Higgs states which mix with the SM Higgs. This wide variety of phenomena, which provide benchmarks  for a panoply of possible experimental signatures, and the fact that it is widely studied make it a good context to make comparison plots between different collider scenarios. The relevance of these plots goes beyond the SUSY context. The relative sensitivity to weak and strong, large and small mass-splitting scenarios are representative of what sensitivity might be observed in other models with new states. The following studies focus on $R$-parity conserving SUSY where there is a stable lightest-supersymmetric state that is only weakly interacting. This is a challenging scenario, particularly for hadron colliders which have pile-up effects, a range of parton collision energies, and reduced resolution and information about the momentum conservation in the beam direction. It should be noted that models such as some RPV scenarios has significatly less constrained and the LEP limits have not yet been exceeded by the LHC.

\subsubsubsection{ \bf Squarks and Gluinos}

\begin{figure}[h!]
  \begin{center}
    \includegraphics[width=1\textwidth]{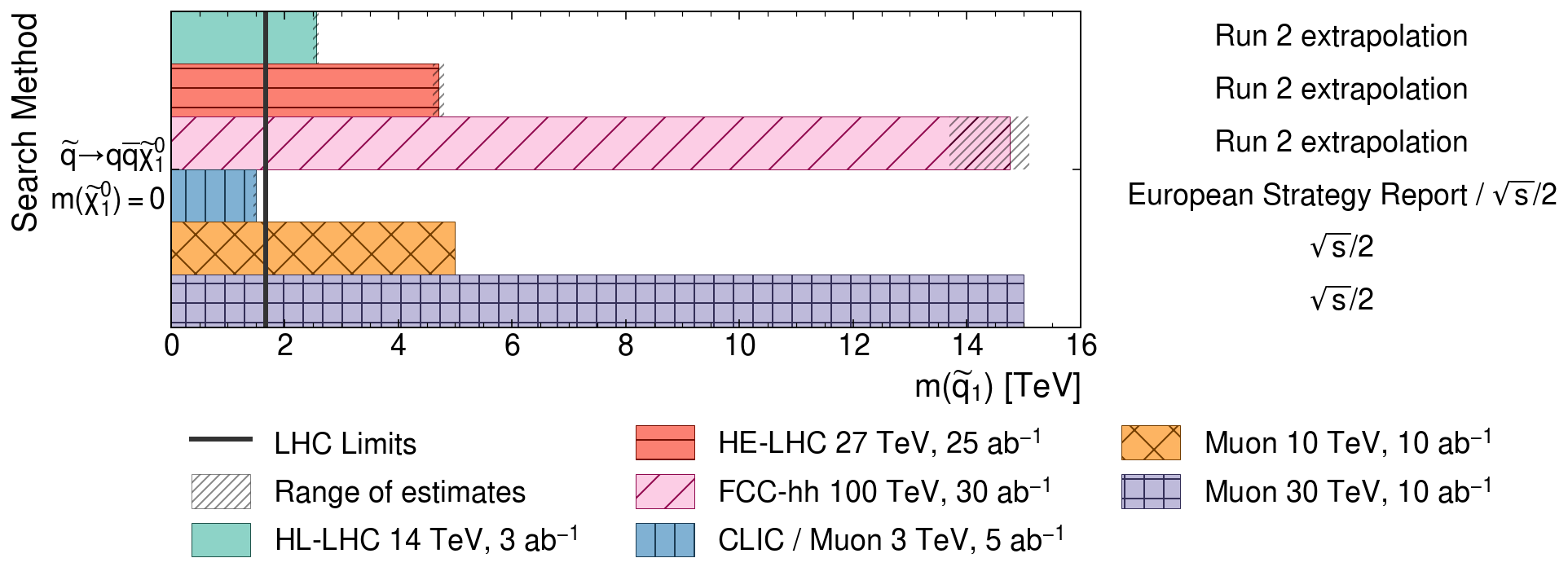} 
  \end{center}
  \caption{Estimated first- and second-generation squark exclusion reaches for various colliders.  Lepton ($e$ or $\mu$) collider limits are estimated to be $\sqrt{s}/2$. Current expected limits from the LHC \cite{ATLAS:2020syg,CMS:2019zmd} are shown as vertical lines. A table detailing the origin of each line is given in Table \ref{tab:ef08_lightsquark}}. The hashed gray band indicates the range of estimates in the case where both a dedicated study and Run-2 extrapolation are available.
    \label{fig:ef08_lightsquark}
\end{figure}

\begin{figure}[h!]
    {\centering
    \includegraphics[width=1\textwidth]{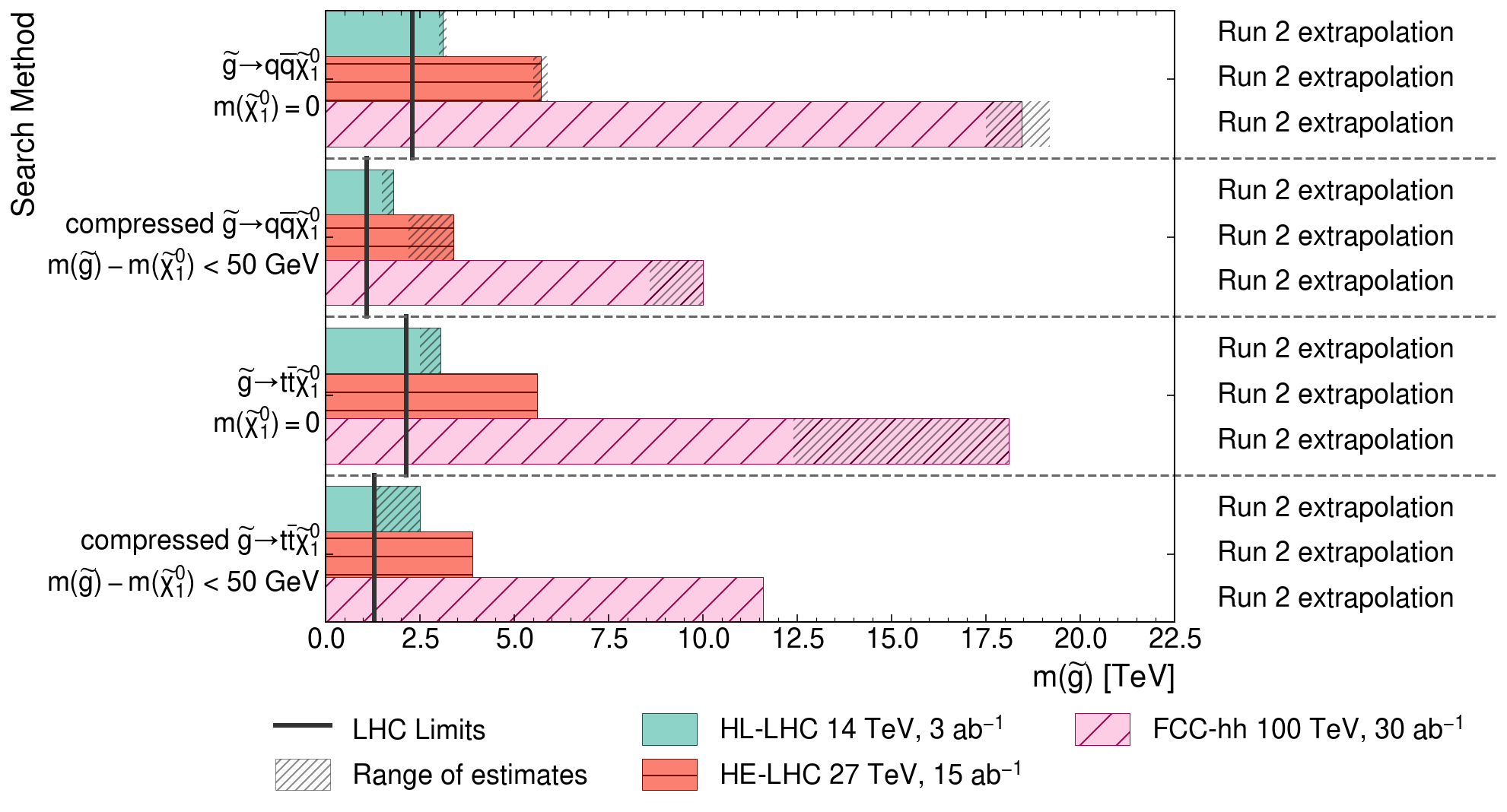}}
    \caption{Estimated gluino exclusion reaches for various colliders and search methods. Current expected limits from the LHC \cite{ATLAS:2020syg,CMS:2019zmd} are shown as vertical lines. A table detailing the origin of each line is given in Table \ref{tab:ef08_gluino}. The results are given for a set of scenarios including large squark-neutralino mass splitting and compress, and for decays to top quarks and direct decays to light quarks. The hashed gray band indicates the range of estimates in the case where both a dedicated study and Run-2 extrapolation are available.}
    \label{fig:ef08_gluino}
\end{figure}

Squarks (\squark) and gluinos (\gluino) are strongly interacting states which can decay weakly (at some point in their decay chain). Because of their strong coupling, they can be produced through strong interactions in hadron colliders resulting in a large cross-section. In lepton colliders, the squarks can be produced through their electroweak couplings, gluinos are more challenging to produce. There are a large variety of possible decay chains. For the purposes of the squarks and gluinos sensitivity studies, we focus on direct decays to the neutralino LSP and decays via an intermediate chargino, e.g.  $\gluino\gluino \rightarrow qq\prime \chi^{+} qq\prime \chi^{-} \rightarrow   qq\prime  qq\prime \chi^{0}_1 \chi^{0}_1 W^+W ^-$ will result in an up to 8 jet plus \MET signature. The various decay chains are lumped into a single multijet plus \MET\ signature. More details can be on the decay chains and analysis methods can be found in the Run-2 ATLAS and CMS papers \cite{ATLAS:2020syg,CMS:2019zmd}.  Figures \ref{fig:ef08_lightsquark} and \ref{fig:ef08_gluino} show the 95\% CL expected exclusion limits for squarks  and gluinos, respectively. Hadron colliders have a significantly greater reach for these strongly produced states than electron colliders and lower energy muon colliders, while a 30 \TeV\ muon collider would be comparable to FCC-hh.  
For the lepton colliders, we assume that the luminosities are sufficient to saturate their energy reach leading a limits of $\sqrt{s}/2$. This has been observed to be a good approximation in the cases for which there are studies available.
The gluino results in Figure \ref{fig:ef08_gluino} show that the sensitivity is reduced for compressed scenarios by a factor of order 1.5 to 2.

%

Top partners are particularly relevant to naturalness motivations as
the top quark is one of the main drivers of the quadratic divergence
in the Higgs boson mass in the SM.

\begin{figure}[h!]
    {\centering
    \includegraphics[width=\textwidth]{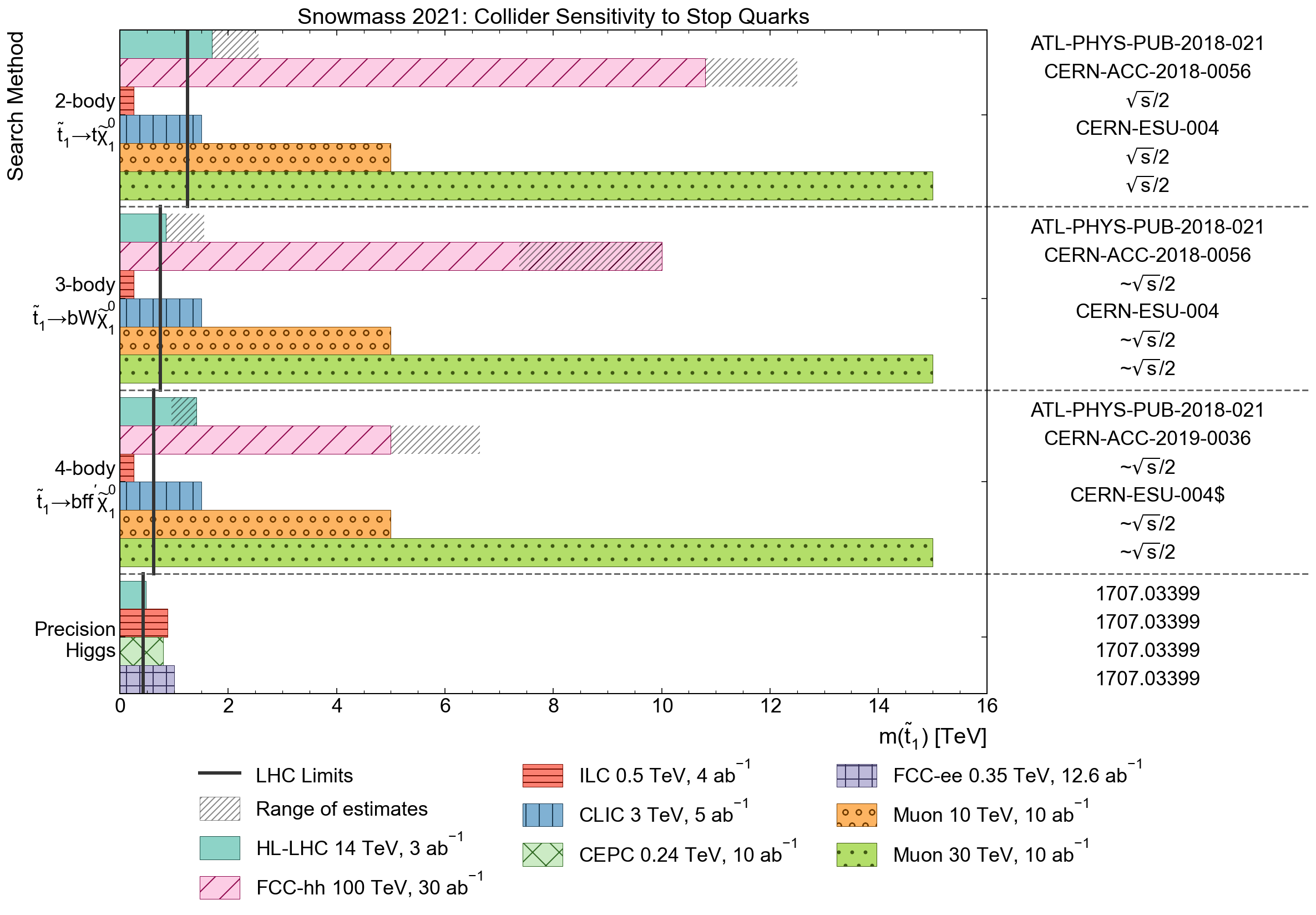}}
    \caption{Estimated stop exclusion reaches for various colliders and search methods. 
    The two, three, and four-body decay searches target the regions $\Delta m(\widetilde{t}_1, \widetilde{\chi}_1^0) \in (m_t, \infty)$, $(m_b + m_W, m_t)$, and $(0, m_b + m_W)$ respectively. 
    The bars show the largest limit on $m(\widetilde{t}_1)$ in the $m(\widetilde{t}_1)-m(\widetilde{\chi}_1^0)$ phase-space for each region. 
    The Precision Higgs constraints are based on measuring production rates of the Higgs boson assuming the only BSM contributions are from stops. 
    ILC, CLIC, and Muon Collider limits are estimated to be $\sqrt{s}/2$, with slight inefficiencies in the three and four-body decay searches due to soft decay products. Current expected limits from the LHC \cite{ATL-PHYS-PUB-2022-013} are shown as vertical lines. A table detailing the origin of each line is given in Table \ref{tab:ef08_stop_sources}. The hashed gray band indicates the range of estimates in the case where both a dedicated study and Run-2 extrapolation are available.}
    \label{fig:ef08_stop}
\end{figure}

Stop squarks (\stop) are another hadronically produced state at hadron colliders. The production
cross-sections for stop are somewhat lower than light squarks and
gluinos because there are fewer stop squark states, and because the
gluino has larger strong interaction. The decay chains are also more
complex because the top quarks decay. Three kinematic regions are
considered, all with the stop decaying to a top quark and a neutralino,
but in compressed scenarios, the top quark maybe be off-shell
resulting in a 3-body ($bW\ninoone$). In even more compressed
scenarios, the $W$ may also be off-shell resulting in a 4-body
($bff^\prime\ninoone$) final-state. Figure \ref{fig:ef08_stop}, shows
the corresponding sensitivities, which are somewhat lower than the
light squarks in hadron colliders, while the lepton collider
sensitivity remains at approximately $\sqrt(s)/2$. Again sensitivity
in compressed scenarios is lower by approximately a factor of $2$. 
  Precision
measurements can also be indirectly sensitive to stop squarks through
a loop in the Higgs boson to $\gamma\gamma$ and $gg$ coupling
\cite{Essig_2017}. Figure \ref{fig:ef08_stop} also shows the estimated
sensitivity for future precision measurements, but it does not exceed
the HL-LHC sensitivity in the scenarios considered.

%

\subsubsubsection{{\bf Charginos and Neutralinos}}
\label{sec:SUSY_ewk}

Charginos ($\chinopm$) and neutralinos ($\nino$) are produced through
electroweak couplings in both lepton and hadron colliders. For hadron
colliders, this means the production cross-sections are lower than for
the strongly interacting states, lowering the energy reach compared
to the strongly produced states. The charginos and neutralinos are mixtures
of the SUSY partners of the Higgs boson, $W$, and $B$ fields. The masses, production
cross-sections, and decays branching fraction depend on this mixing. For
simplicity, two extreme cases are considered. 

\begin{figure}[h!]
    \centering
    \includegraphics[width=1\textwidth]{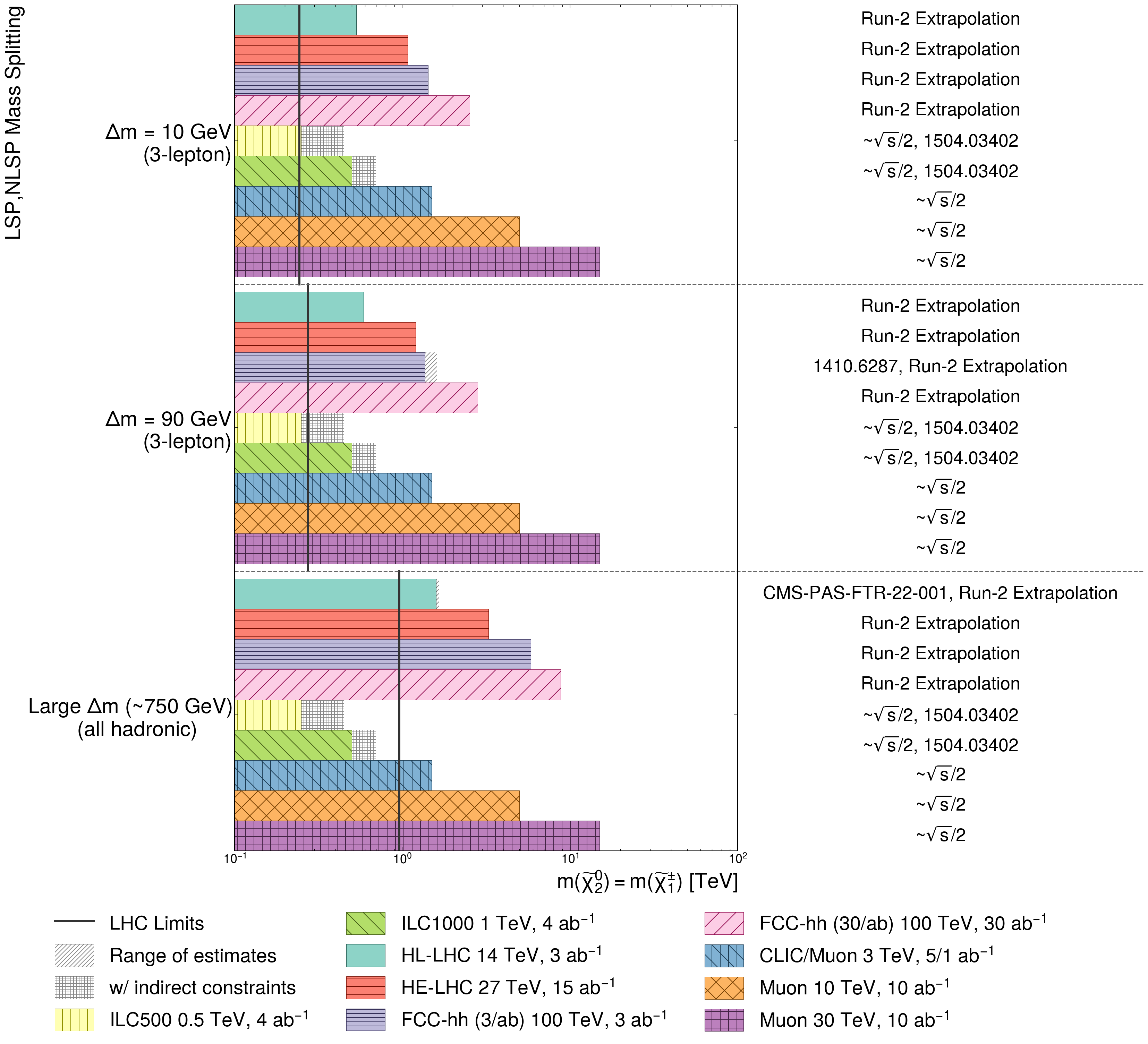}
    \caption{Wino NLSP bino LSP sensitivity comparison for various collider scenarios.  A table detailing the origin of each line is given in Table \ref{tab:ef08winochargino}. Most of the results are based on Run-2 using Collider Reach, but where a dedicated study is also available the results are consistent. Lepton colliders are assumed to be energy limited with a limit of $\sqrt(s)/2$. }
    \label{fig:ef08winochargino}
\end{figure}

The first set of studies, shown in Figure \ref{fig:ef08winochargino},
targets larger mass splittings with Wino-Bino coupling for
production. In this model, the lightest chargino ($\chinoonepm$) and
next-to-lightest neutralino ($\ninotwo$) are produced as a pair.
These can then decay as $\chinoonepm \rightarrow W^{\pm}\ninoone$ and
$\ninotwo \rightarrow Z/h \ninoone$ resulting a $WZ$ or $Wh$ plus \MET
final state. For the lower chargino-neutralino mass-splitting range
($\Delta M \approx 10\ \GeV$ and $\Delta M \approx 90\ \GeV$), the
results are based on the leptonic decays of the $W$ and $Z$ bosons
($WZ \rightarrow 3\ell$).  For larger mass-splitting scenarios
($\Delta M \gtrsim 750\ \GeV$), the hadronic decays are more sensitive.
These limits assume that the $\chinoonepm\ninotwo$ decays 100\% to $WZ$,
which is unrealistic, but combinations with $Wh$ searches should
mitigate this assumption.  The results are primarily based on the
collider reach extrapolation of the current LHC Run-2 results
\cite{ATLAS:2021yqv,CMS:2018xqw}, with one dedicated FCC-hh study
relevant for the $\Delta M \approx 90\ \GeV$ region
\cite{Golling:2016gvc}. These results are consistent where
comparable. The dedicated FCC-hh result is not considered for the higher
mass-splitting region where the sensitivity is dominated by the all
hadronic final state. For the lepton colliders, the energy reach
is expected to $\sqrt(s)/2$.

\begin{figure}[h!]
    \centering
    \includegraphics[width=1.0\textwidth]{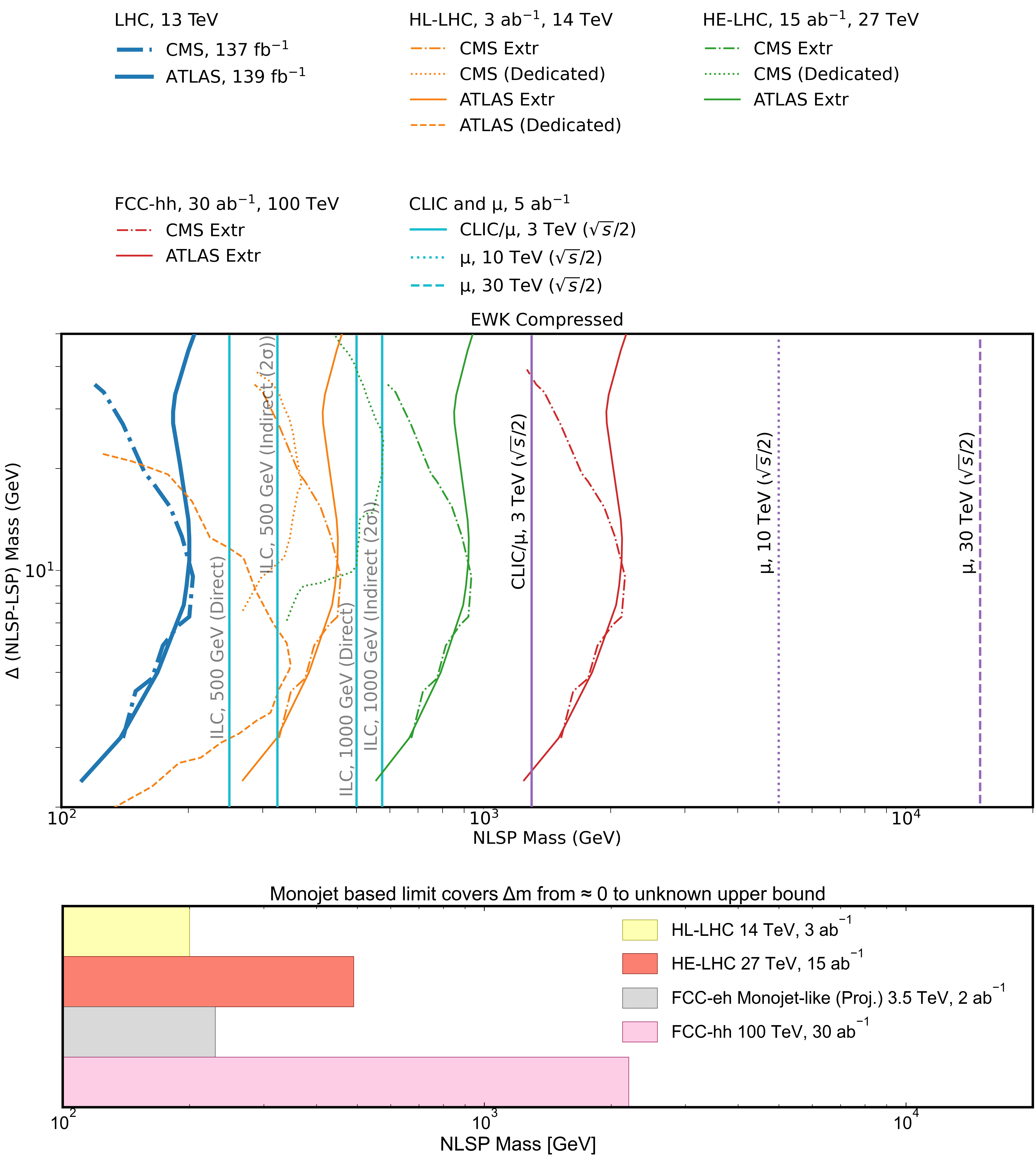}
    \caption{Compressed EWK reaches for CMS (dotted and dot-dashed) and ATLAS (full line and dashed). Plot references are included in Table \ref{tab:ef08higgsino}.}
    
    \label{fig:ef08_higgsinoplot}
\end{figure}

A second model, shown in Figure \ref{fig:ef08_higgsinoplot}, is considered where the LSP is primarily the Higgsino
with small mixings with the other states. Such a scenario is of particular interest in naturalness-motivated scenarios, because the fine-tunning
of the Higgs mass is particularly sensitive to the Higgsino mass parameter which contributes to the Higgs mass at tree-level. This leads to small mass
splitting between the neutralino LSP and the lightest chargino and
the next-to-lightest neutralino. CMS and ATLAS have performed dedicated
searches for this compressed region using a combination of the two and
three lepton and \MET final-states using Run-2 LHC data\cite{CMS:2021edw,Aad:2019qnd}.
They also have dedicated studies for the HL-LHC and HE-LHC
sensitivity. These provide a good opportunity to compare dedicated
studies with the collider reach extrapolations. For the current Run-2,
we see that the reach in CMS and ATLAS are comparable for some mass
range, but because of different optimizations as a function of $\Delta
M$, they are not identical. When these results are extrapolated to
HL-LHC, and compared to the dedicated ATLAS HL-LHC result, there is
good agreement at one $\Delta M$, but not at others. This is because
the ATLAS dedicated result only considers one bin as opposed to the
complex many-bin analysis in the Run-2 result. So while the dedicated
analysis is based on a full calculation of the signal and background
cross-sections and a realistic simulation of the detector performance,
it does not have analysis optimization Run-2 result. It should also
be noted that there is potential for improvement using more sophisticated
selections \cite{Baer:2022qrw}, which gives a comparable result to the Run-2
extrapolation.
Again, for the
lepton colliders, the energy reach is expected to $\sqrt(s)/2$, with
a possible small reduction at the lowest $\Delta M$ region.

For the Higgsino model, there are two other important search
strategies. If the $\Delta M$ is small enough, then there are essentially
no visible signs of the Higgsinos, but they can be limited by the
\MET distribution in mono-jets \cite{Low:2014cba}. The upper $\Delta M$ reach
of this method has not been studied but is likely to be large enough
to close the $\Delta M$-gap with the dilepton searches. For that reason,
those limits are shown as a bar chart beneath the other results.
For pure Higgsinos, the mass splitting is so small, that
particle becomes long-lived (see Section~\ref{sub:disappearingTracks}). This case is relevant for dark matter
and is addressed in Section \ref{sec:dm:wimp}.

%

%
%

\subsubsubsection{{\bf Sleptons}}
\label{sec:SUSY_slepton}

Sleptons aren't motivated to be light by naturalness considerations,
but they may still be light. There are two other strong considerations
that could point to light smuons ($\smu$) and staus ($\stau$). One
region of parameter space that can potentially give the right dark
matter relic abundance is the neutralino-stau coanhilation region
\cite{Ellis:1998kh}.  The g-2 anomaly points toward a light smuon
and/or chargino \cite{Aboubrahim:2021ily}. If the lightest charginos, neutralinos, staus, and smuons are measurable at a linear collider, the logic can be inverted and it can be confirmed that new particles account for the g-2 discrepancy \cite{Endo:2022qnm}.

Sleptons have even lower production cross-sections than charginos and
neutralinos, coupled with difficult signatures, this makes them
challenging for hadron colliders.  The smuon limits are only estimated
from Run-2 LHC using collider reach studies \footnote{We exclude here
  the special case of long-lived sleptons.}. There are no dedicated
studies of smuon limits for lepton colliders, however, it is reasonable
to expect they'd be close to the $\sqrt{s}/2$ limit. For stau limits,
there is a dedicated from ATLAS for the HL-LHC. For FCC-hh, we use
collider reach extrapolations as representative of the expected
reach. For ILC, a dedicated study \cite{NunezPardodeVera:2022izz} shows that for ILC500 and
ILC1000, the exclusion reach is close to $\sqrt{s}/2$ of most of the
range of $\Delta M(\stau,\ninoone)$, but degrades slightly at the
lowest $\Delta M \lesssim 5 \GeV$.  We assume that the
$\approx\sqrt{s}/2$ limit also applies to CLIC3000 and muon colliders.

\begin{figure}[h!]
    \centering
    \includegraphics[width=1\textwidth]{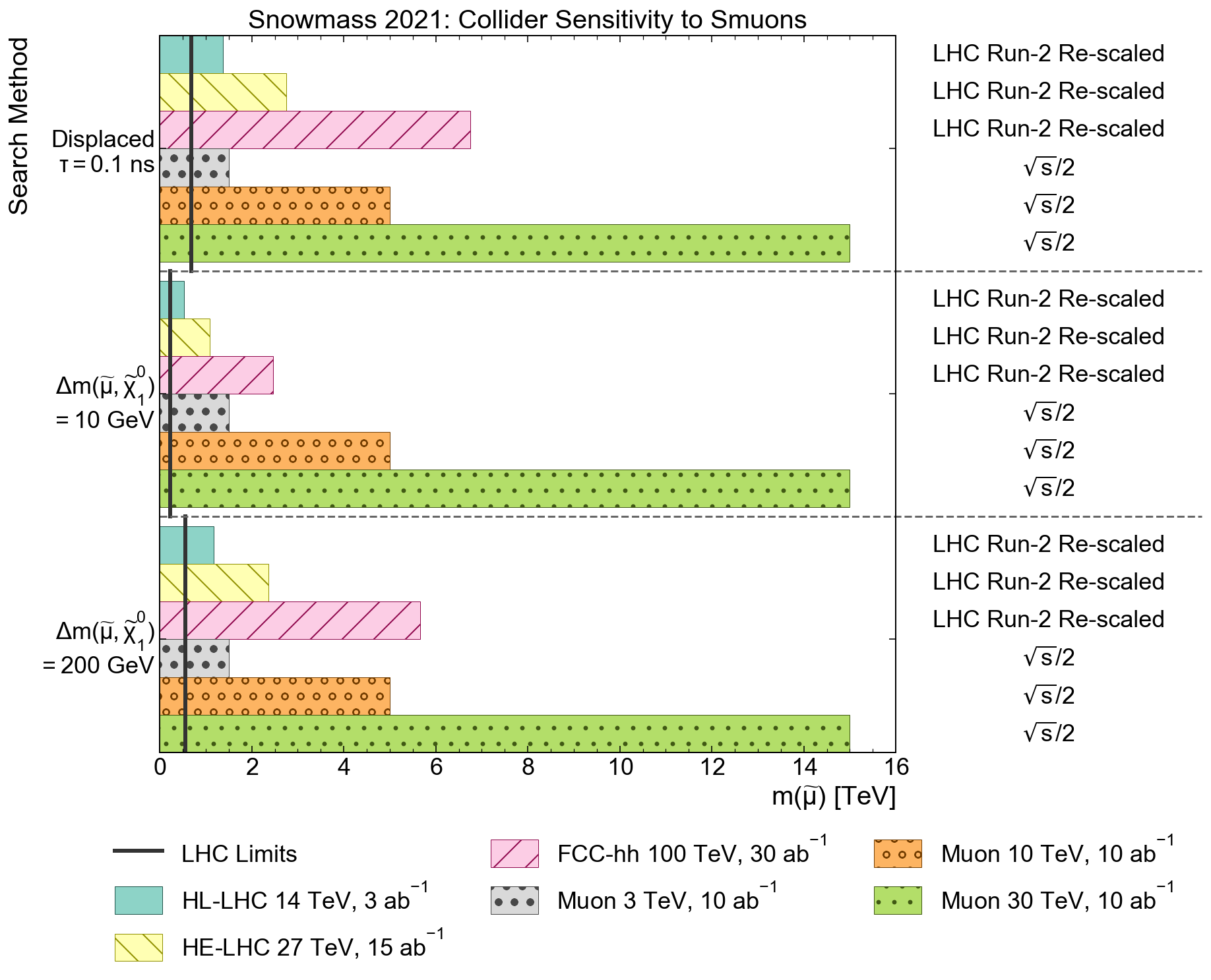}
    \caption{Smuon sensitivity comparison for various collider scenarios. A table detailing the origin of each line is given in Table \ref{tab:ef08_smuon}}.
    \label{fig:ef08_smuon}
\end{figure}

%

\begin{figure}[h!]
    \centering
    \includegraphics[width=1\textwidth]{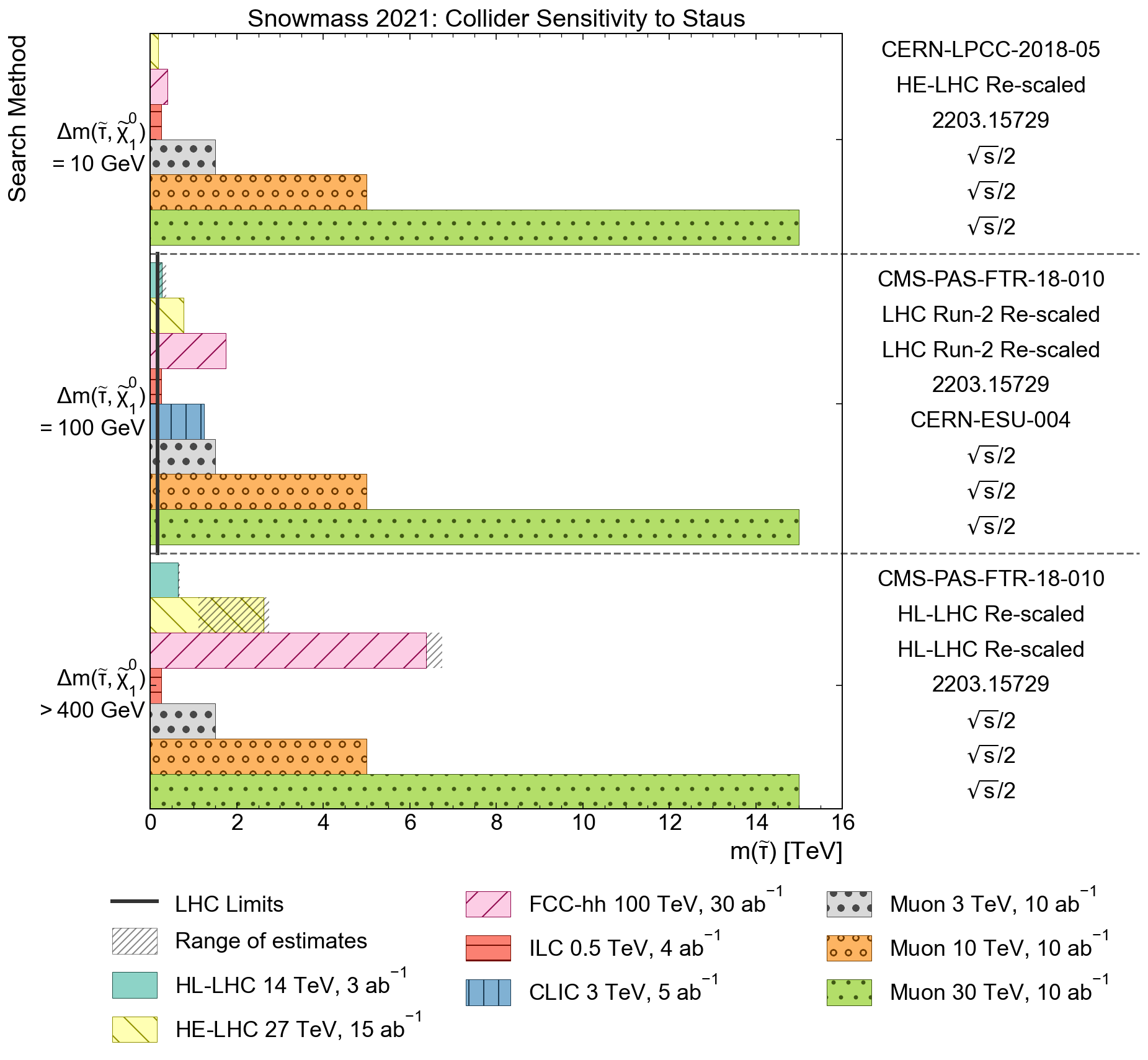}
    \caption{Stau sensitivity comparison for various collider scenarios. A table detailing the origin of each line is given in Table \ref{tab:ef08_stau}}. Results include extrapolations of Run-2, and dedicated studies. The HE-LHC high-mass-splitting result is one of the few cases where the extrapolation is not very consistent with the dedicated study.
    \label{fig:ef08_stau}
\end{figure}

%
\FloatBarrier
\pagebreak

\subsection{pMSSM Scans}

Interpretation of experimental results within the MSSM framework traditionally relies on two primary methods to reduce the large number of free parameters in the theory. By assuming a specific mechanism for SUSY breaking at a high unification scale, boundary conditions can be imposed on the soft SUSY breaking parameters in order to reduce the number of free parameters of the model. The most prominent example is the constrained MSSM (cMSSM)~\cite{Chamseddine:1982jx,Barbieri:1982eh,Ibanez:1982ee,Hall:1983iz} with five free parameters. The other interpretation method, known as "simplified model spectra" (SMS)~\cite{sms2009,sms2012,cms2013}, analyzes the production of only one set of SUSY particles at the electroweak (EW) scale with a fixed decay chain. While these frameworks allow for efficient interpretation of results, they do so at the expense of sampling only a very small part of the phase space of the MSSM and potentially focusing on signatures that may not be realized in nature.  Recently, the ATLAS and CMS collaborations and theory community have attempted to ameliorate the limitations of interpretations based on the cMSSM and SMS by use of the phenomenological MSSM (pMSSM) \cite{atlaspmssm2015,cmspmssm2016,pmssm2009,pmssm2012,laa2017coverage,AbdusSalam:2011fc}, which reduces the 120-parameter MSSM space to 19 free parameters, specified at the  EW scale, based on assumptions related to current experimental constraints (including those from flavor, CP violation, and EW symmetry breaking) rather than details of the SUSY breaking mechanism.

An updated scan of the pMSSM, based on Markov chain Monte Carlo procedure, was performed specifically for the 2021 Snowmass process as described in detail in Ref.~\cite{snowmassPMSSM}. The scan was designed to cover the accessible energy ranges of many future collider scenarios, including electron, muon, and hadron colliders with center-of-mass energies from a few hundred GeV to 100 TeV.  By including calculated values of interesting physical observables for each scan point, including the muon anomalous magnetic moment, the dark matter relic density, and properties of the Higgs boson, the scan also provides a means to determine how improvements in sensitivity to those observables from potential future experiments  exclude or favor regions of the pMSSM parameter space.

Of particular interest is a comparison of how precision measurements and direct searches from future colliders impact the allowed pMSSM parameter space.  The next generation of colliders are expected to improve precision on Higgs boson couplings from the 2-10\% level at HL-LHC to the 0.1-1\% level~\cite{deBlas:2019rxi}.  The analysis of the Snowmass 2021 pMSSM scan confirms previous studies~\cite{Bahl:2020kwe} showing that MSSM effects cause deviations of Higgs couplings from the SM expectation primarily via tree-level impacts of the MSSM Higgs sector rather than from loop  contributions of new SUSY particles.  The MSSM-related Higgs coupling deviations are largest (few percent scale) for couplings to the b quark and $\tau$, which scale as $\kappa_{b,\tau} \propto \sin(\beta - \alpha) - \cos(\beta - \alpha) \tan\beta$ where $\alpha$ is the angle that diagonalizes the CP-even Higgs sector. Deviations from the SM for other Higgs couplings are of order 0.1\% or less, which make them essentially inaccessible for precision measurements from any future collider scenario considered in this Snowmass process.  

\begin{figure}[htb]
\begin{center}
\includegraphics[width=0.49\hsize]{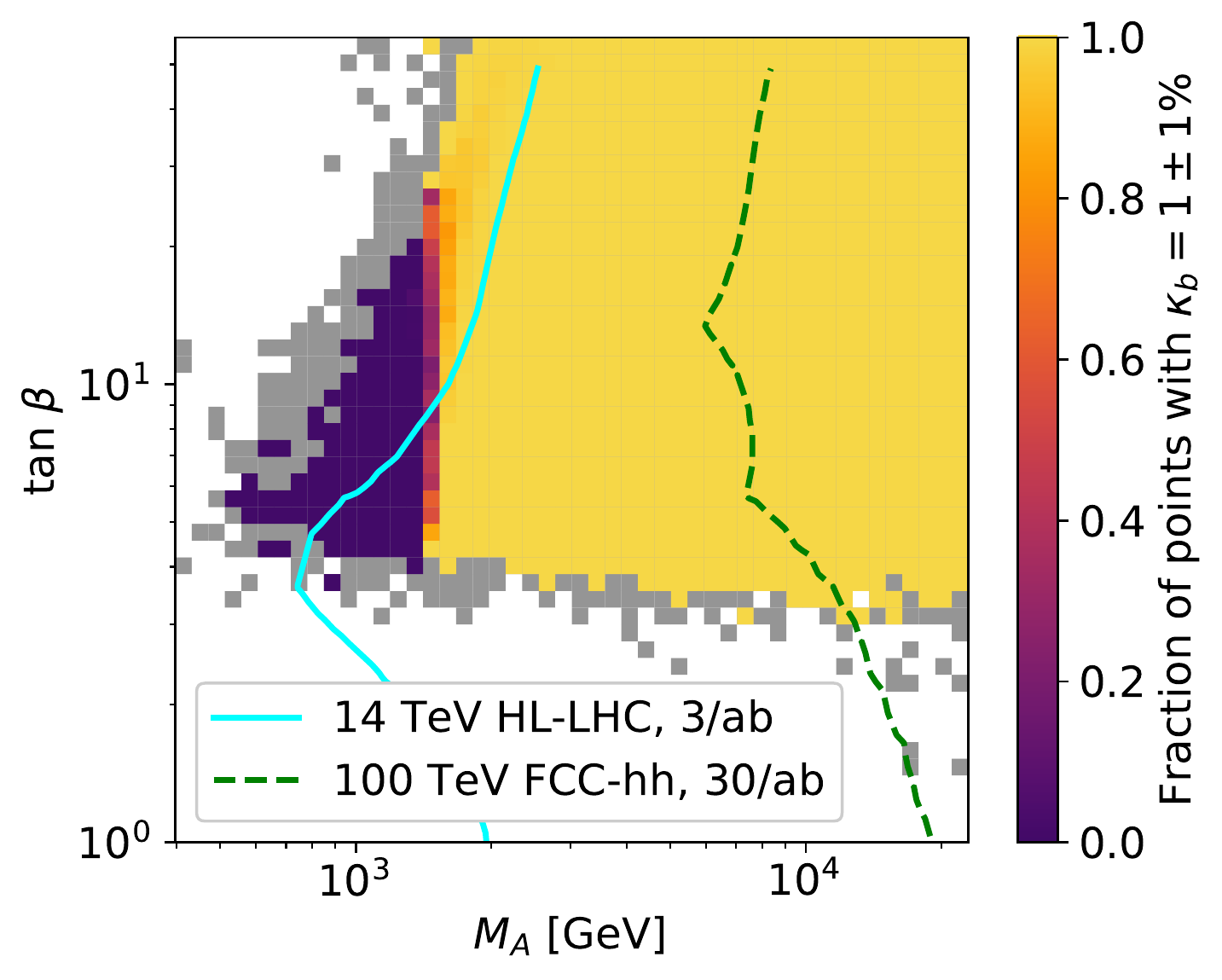}
\end{center}
\caption{The fraction of pMSSM scan points with $\kappa_{b}$ within $1\%$ of the SM expectation of unity as a function of $\tan\beta$ and $M_A$. The range of $1\%$ is chosen to approximately reflect the 95\% CL corresponding to the 0.48\% precision on $\kappa_b$ expected from the FCC-ee/eh/hh combination~\cite{deBlas:2019rxi}.  Expected 95\% CL exclusions from HL-LHC~\cite{Craig_2017, ATL-PHYS-PUB-2022-018} and FCC-hh~\cite{Craig_2017} are overlaid for reference. White bins include no scan points generated by the Markov chain Monte Carlo (McMC) procedure. Gray bins include scan points generated by the McMC, but rejected at a later step because of lack of consistency with current precision measurements and direct searches.}
\label{fig:pmssm}
\end{figure}

The impact of expected improved precision of the measurement the Higgs boson coupling to $b$ quarks ($\kappa_{b}$ in the $\kappa$-framework~\cite{kappa, kappa2}) as a function of $\tan\beta$ and the mass of the pseudoscalar Higgs boson of the MSSM ($M_A$) is shown in Fig.~\ref{fig:pmssm}.  The plot shows the fraction of pMSSM scan points with $\kappa_{b}$ within $1\%$ of the SM expectation of unity, where the range of $1\%$ is chosen to approximately reflect the 95\% CL corresponding to the 0.48\% precision on $\kappa_b$ expected from a  combination of precision measurements at FCC-ee, FCC-eh, and FCC-hh~\cite{deBlas:2019rxi}.  Expected 95\% CL exclusions from direct searches for pseudoscalar Higgs boson ($A$) at the  HL-LHC and FCC-hh are overlaid for reference; points to the left of the lines are excluded.  Exclusions at low $\tan\beta$ are obtained from studies of $A\rightarrow bb/tt$~\cite{Craig_2017}, and those at high $\tan\beta$ come from projections for $A\rightarrow \tau^{+}\tau^{-}$~\cite{Craig_2017, ATL-PHYS-PUB-2022-018}.  As is evident in the plot, direct searches for $A$ at the HL-LHC are expected to provide better sensitivity to the MSSM than the highest precision measurements of $\kappa_b$, which shows the strongest MSSM-related deviation of any Higgs coupling parameter.

\section{Leptoquarks} \label{Sec:Leptoquark}


Leptoquarks~(LQ) are particles that mediate conversions between quarks and leptons. As such they need to carry color charge as well as baryon and lepton number. As discussed in Sec.~\ref{Sec:Anm}, LQs arise in many BSM models, such as GUT-inspired scenarios~\cite{Bordone:2017bld,Bordone:2018nbg,Fuentes-Martin:2022xnb}, $R$-parity violating SUSY~\cite{Deshpande:2016yrv,Das:2017kfo,Earl:2018snx,Trifinopoulos:2018rna} or composite Higgs models~\cite{Gripaios:2014tna,Barbieri:2016las,Marzocca:2018wcf}, and may resolve some of the  observed $B$-physics anomalies~\cite{Altmannshofer:2022aml}. However, care must be taken since LQs which allow diquark interaction may allow the proton to decay. 

Leptoquarks have been searched for extensively at the LHC, and current mass exclusion bounds extend $\gtrsim$ 1 TeV~\cite{Sopczak:2021vlw, Angelescu:2018tyl,Cornella:2019hct,Popov:2019tyc, CMS:2018yke}. While leptoquark production at hadron colliders are mainly governed by their strong interaction, lepton colliders are sensitive to their electroweak couplings and flavor/gauge structure, and thus provide a complimentary probe.  

LQs may be either scalar or vector, however, the signatures of different LQs are closely related. Most of the current work focuses on the LQ $U_1$ in the notation of Ref.~\cite{Dorsner:2016wpm}, and is a vector in the (3, 1, 2/3) representation of the SM gauge group. However, phenomenology of different spins or representations is expected to be qualitatively similar. Parameterizing the couplings of a LQ to muons and second and third generation squarks, a high energy muon collider may be able to probe LQ mass scales an order of magnitude higher than $\sqrt{s}$ and with perturbative couplings, considering pair production, single production and DY production for $b$-quarks~\cite{MuonCollider:2022xlm, Asadi:2021gah}.


\section{New Bosons and Heavy Resonances}\label{Sec:NB}

In this section, we present an overview of the prospects for new heavy vector bosons that are probed by resonance searches at colliders.  We begin with the canonical example of a $Z'$ boson, which is a neutral vector particle coupling to a SM fermion and antifermion. This example demonstrates the discovery capability of lepton colliders, hadron colliders, and the necessity of an HEP program that includes both machines. There was a diversity of ideas for new bosons and heavy resonances presented at Snowmass, so we conclude this section with a discussion of those other models and channels.

\subsection{$Z^{\prime}$ Bosons: the Standard Candle of BSM Physics}

In the Standard Model, the $Z$ and $\gamma$ bosons mediate the flavor-conserving weak and electromagnetic neutral currents between SM fermions, respectively.  In general, we can model the interactions of a new vector boson via the current coupling to SM fields, where the overall proportionality constant is the new gauge coupling.  The remaining model dependence is dictated by the pattern of charge assignments of the SM fields in the current coupling, which could include both vector and axial-vector interactions as long as the ultraviolet theory fulfills gauge anomaly cancellation requirements.  Hence, the $Z'$ boson benchmark paradigm is characterized by the new physics parameters of the $Z'$ mass, the new gauge coupling, and a fixed pattern of SM fermion charge assignments.  

From the phenomenological perspective, resonance searches are generally characterized by the production coupling, the decay coupling, and the resonance mass, where the decay coupling is typically traded for the branching fraction to the desired final state.  Correspondingly, any given $Z'$ model will have a prediction for the final state branching fraction dictated by the pattern of charge assignments of the SM fermions, which reduces the parameter space to the $Z'$ coupling and $Z'$ mass.  This is also true even when the $Z'$ is too heavy to be probed directly as a resonance at the given collider energy and instead is tested via angular correlations or effective operators of SM fermions.  In this way, different $Z'$ models are essentially distinguished by their patterns of charge assignments and otherwise can be characterized by their common phenomenological parameters of coupling and mass.

This coupling vs.~mass framework for $Z'$ searches~\cite{Dobrescu:2013cmh, Dobrescu:2021vak} thus fulfills a twofold purpose especially suited for the Snowmass process.  First, the framework helps distill the $Z'$ resonance signal from disparate ultraviolet models into the minimal new physics parameter space relevant for resonance searches at colliders.  Second, the framework also affords the direct comparison of experimental reach across different collider proposals, including a comparison of $e^+ e^-$, $pp$, and $\mu^+ \mu^-$ colliders as well as other collider options.  This will be illustrated and discussed in our summary table~\ref{tab:ZpTable}, presented at the end of this subsection.  We first discuss the specific $Z'$ models studied in different Snowmass contributions.

\subsubsection{Universal $Z^{\prime}$}

\begin{figure}[htb]
\begin{center}
\includegraphics[width=0.49\hsize]{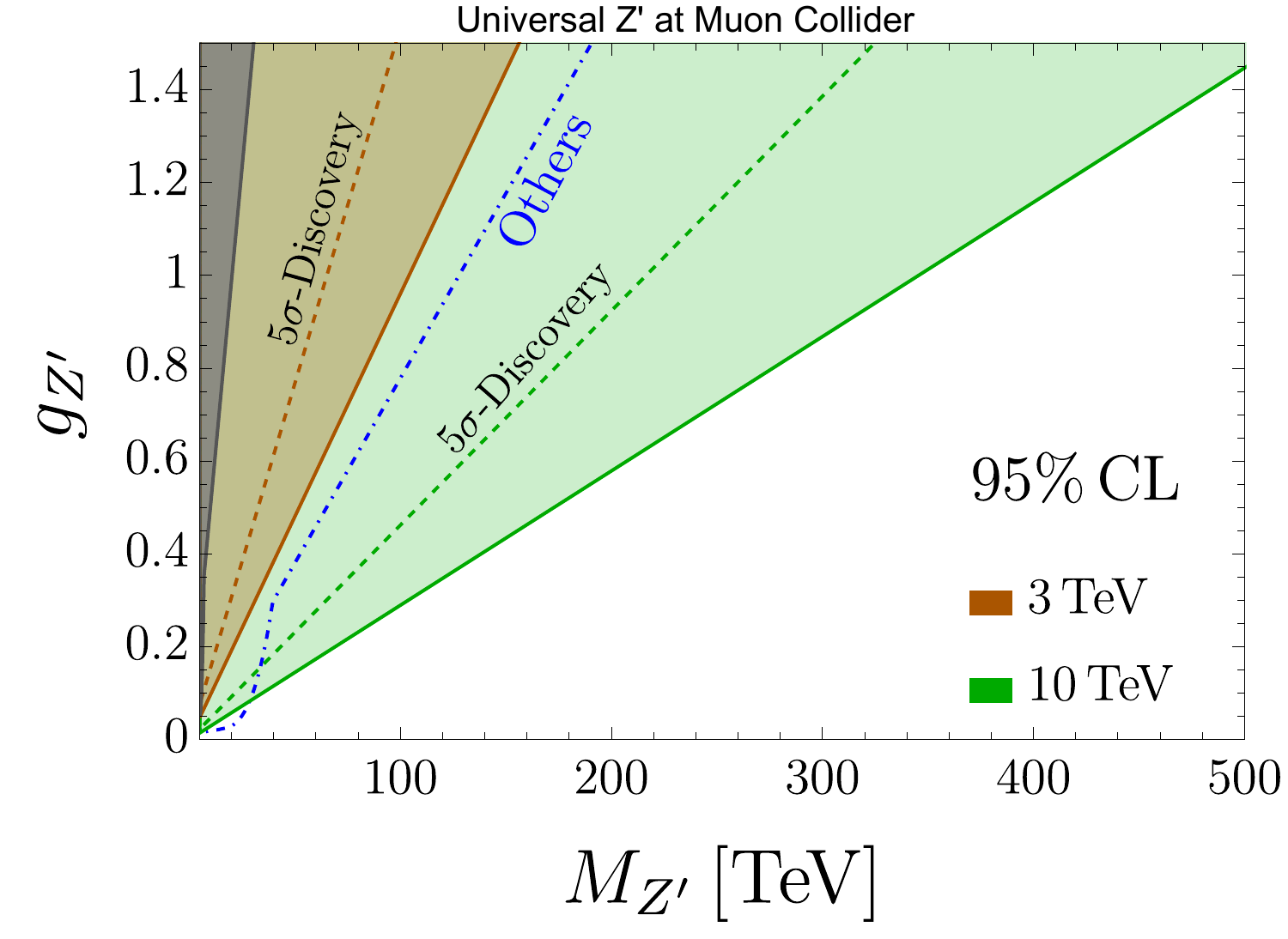}
\includegraphics[width=0.5\hsize]{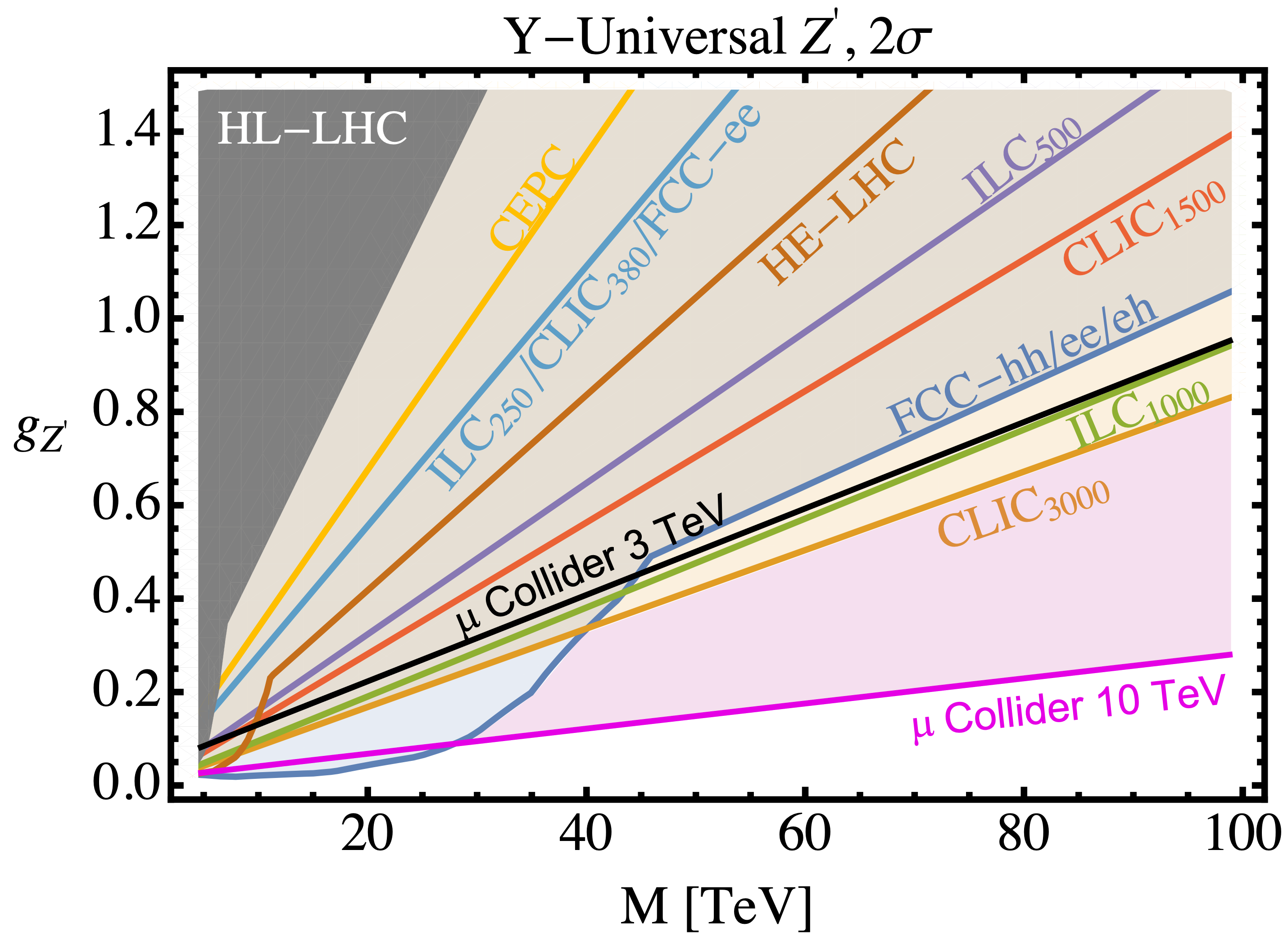}
\end{center}
\caption{ (left) The coupling versus mass reach for a universal $Z^{\prime}$ at the muon collider~\cite{Aime:2022flm}, for 95\% CL exclusion (solid) and 5$\sigma$ discovery (dashed), and the envelope of other colliders~\cite{Strategy:2019vxc} for 95\% CL exclusion (dashed blue). (right) Coupling versus mass reach at 95\% CL for electron-positron colliders (CEPC, ILC, CLIC and FCC-ee) and proton-proton colliders (HL-LHC, HE-LHC and FCC-hh) and an electron-proton collider (FCC-eh) from Ref.~\cite{Strategy:2019vxc} and the muon collider~\cite{Aime:2022flm}.}
\label{fig:UnivZp}
\end{figure}
The universal $Z^{\prime}$ model features a $Z^{\prime}$ boson with unit charges for all SM fermions, hence its universal designation.  Figure~\ref{fig:UnivZp} compares a Snowmass result on the sensitivity to a universal $Z^{\prime}$ at the muon collider~\cite{Aime:2022flm} with other colliders~\cite{Strategy:2019vxc}. A muon collider at $\sqrt{s}=3$ TeV is competitive with other colliders, with sensitivity nearly identical to ILC at $\sqrt{s}=1$ TeV. A muon collider at $\sqrt{s}=10$ TeV has the highest mass reach for a universal $Z^{\prime}$ with large couplings $g_{Z^{\prime}}$, uniquely probing masses $M_{Z^{\prime}}>100$ TeV. A muon collider at $\sqrt{s}=10$ TeV is sensitive to smaller couplings than the other colliders, with the exception of FCC-hh, which has the highest sensitivity from direct searches within the mass region $M_{Z^{\prime}}<28$ TeV.   Lepton colliders have an edge in sensitivity when the boson is so heavy that only indirect effects can be measured, arising from the fact that in the signal kinematic distributions, the lepton collider experiments benefit from relatively smaller systematic uncertainties.

\subsubsection{Lepton-Specific $Z^{\prime}$ Boson}

Motivated by the $B$-physics anomalies and indications for violation of lepton universality, Ref.~\cite{MuonCollider:2022xlm} considered the benchmark of a $Z^{\prime}$ mediating gauged $L_\mu - L_\tau$ number, which can also address the anomalous magnetic moment of the muon from the Muon g-2 experiment.  Here, the special nature of the specific lepton charges highlights the capability of a 3~TeV muon collider since it produces the $Z^{\prime}$ boson directly, as shown in Fig.~\ref{fig:LFV-SSM-Zp} from Ref.~\cite{MuonCollider:2022xlm}.

\subsubsection{Sequential Standard Model $Z^{\prime}$}

\begin{figure}[htb!]
\begin{center}
\includegraphics[width=0.47\hsize]{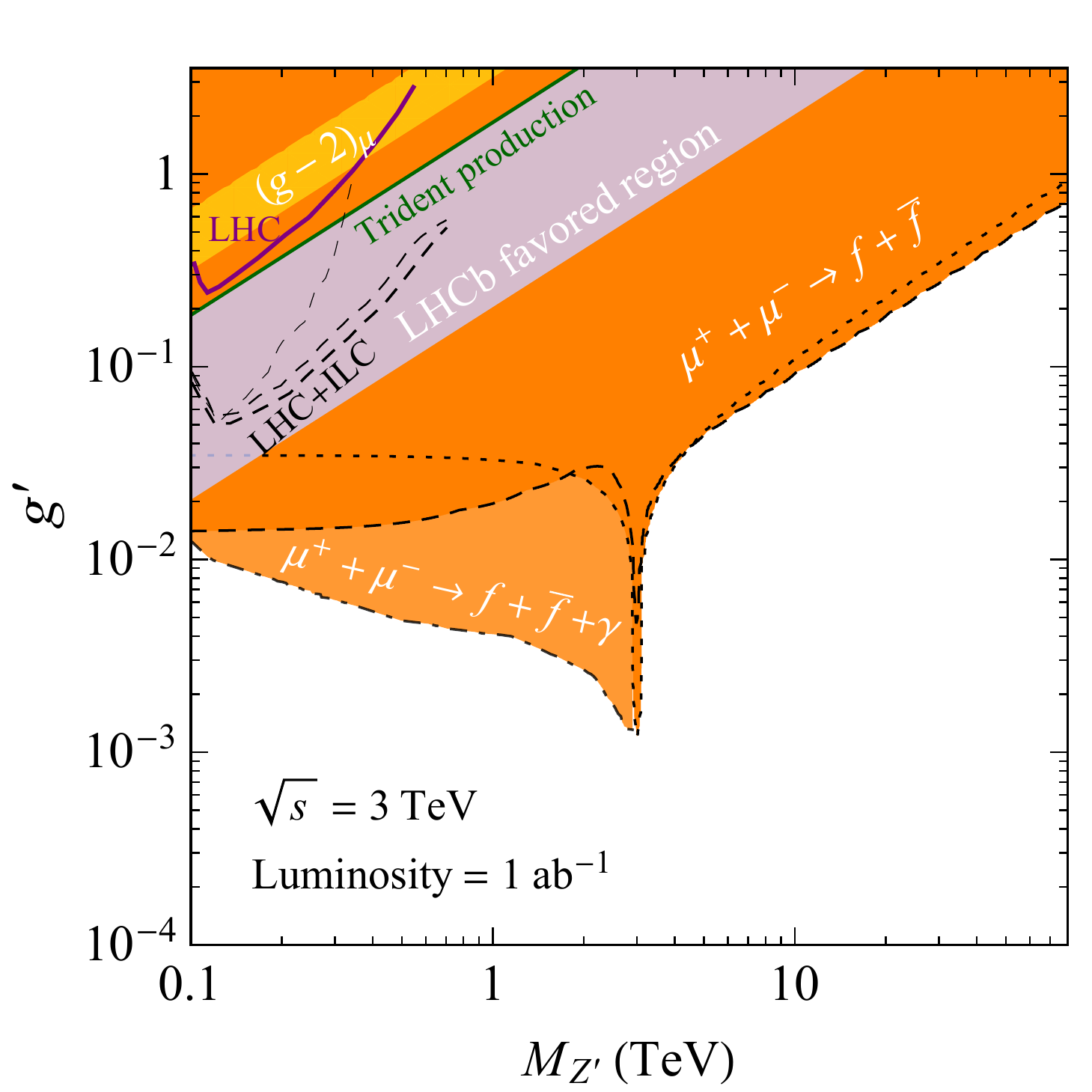}
\includegraphics[width=0.52\hsize]{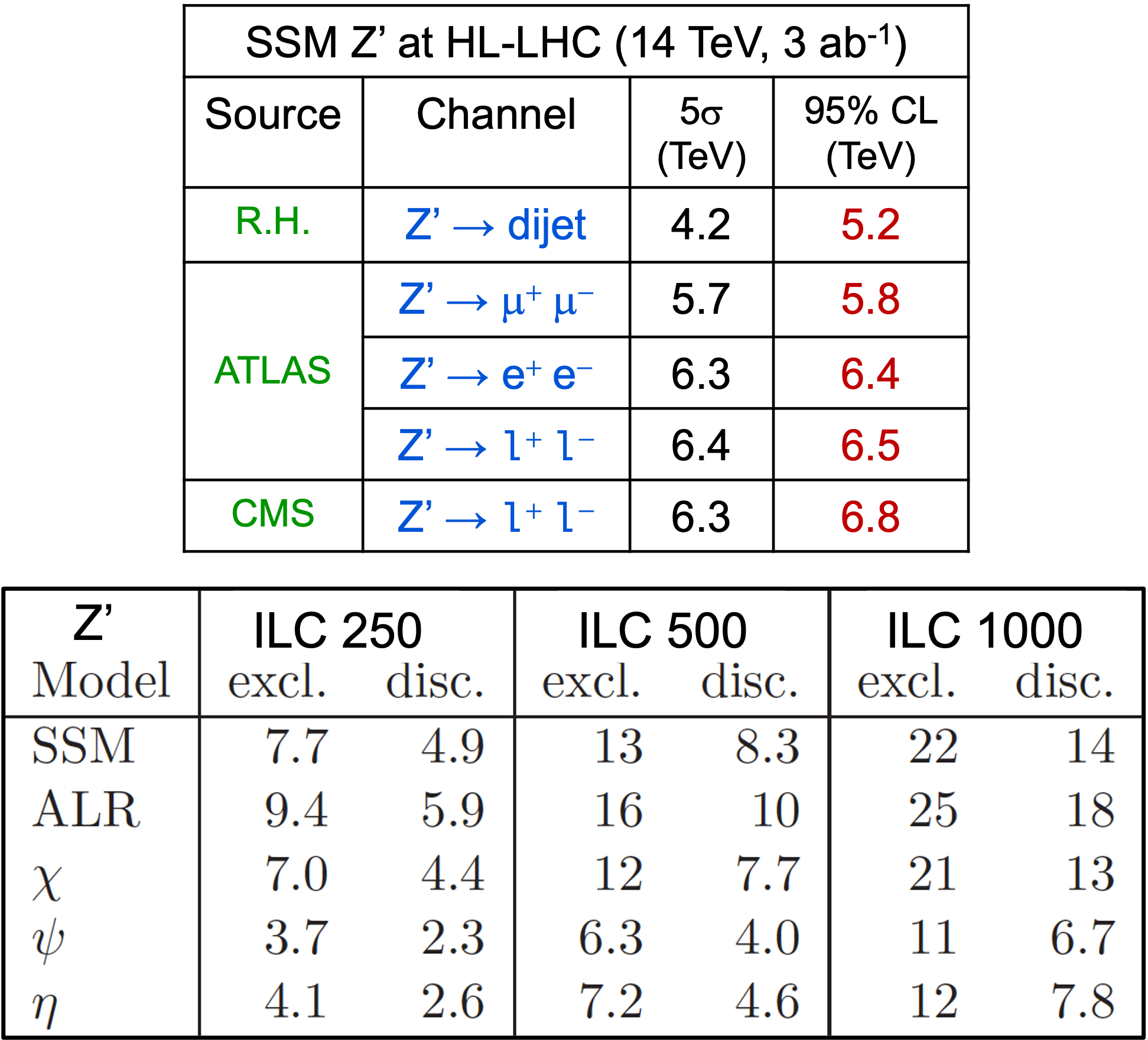}
\end{center}
\caption{ (left) The coupling versus mass limits for a lepton universality violating $Z^{\prime}$ at the muon collider~\cite{MuonCollider:2022xlm} (orange), compared to the mass reach of LHC (black curve) and LHC+ILC (black dashed curves), and the favored regions of parameter space from measurements at Muon g-2 (yellow) and LHCb (violet). (right) SSM $Z^{\prime}$ mass reach for 5$\sigma$ discovery and 95\% CL exclusion from (top) HL-LHC in the dijet channel~\cite{harris2022sensitivity}, dimuon, dielectron and combined dilepton channel in ATLAS~\cite{ATL-PHYS-PUB-2018-044} and in CMS~\cite{CMS-PAS-FTR-21-005} and (bottom) from ILC at three operating energies for SSM and four other $Z^{\prime}$ models decaying to pairs of fermions.~\cite{Suehara:2022pwv}.}
\label{fig:LFV-SSM-Zp}
\end{figure}

The sequential standard model (SSM) $Z^{\prime}$ boson follows the same coupling pattern of the SM $Z$ boson, and is the benchmark model most commonly used by experimental searches. The recently begun running period of LHC, Run 3 at $\sqrt{s}=13.6$ TeV with 160 fb$^{-1}$, when combined with the previous running period, Run 2 at $\sqrt{s}=13$ TeV with 140 fb$^{-1}$, is expected to result in a mass reach of 5.4 TeV for SSM $Z^{\prime}$ at 95\% CL in the dilepton channel~\cite{CMS-PAS-FTR-21-005}. HL-LHC will significantly extend LHC sensitivity to the SSM $Z^{\prime}$ model, as shown in the upper right panel of Fig.~\ref{fig:LFV-SSM-Zp} for several channels and experiments that were studied for Snowmass ~\cite{harris2022sensitivity, ATL-PHYS-PUB-2018-044, CMS-PAS-FTR-21-005}.  ILC would extend the sensitivity beyond HL-LHC, and the lower right panel of Fig.~\ref{fig:LFV-SSM-Zp} shows a set of $Z^{\prime}$ benchmarks including the SSM $Z^{\prime}$ for different ILC machine operating conditions~\cite{Suehara:2022pwv}. Further increases in sensitivity to the SSM $Z^{\prime}$ would come from future proton-proton (pp) colliders, and table~\ref{tab:dijet-sensitivity-table} lists that sensitivity for decays to dijets, and the dependence on integrated luminosity is shown in Fig.~\ref{fig:strong-weak}.

\subsubsection{Machines Ordered by $Z^{\prime}$ Sensitivity}

In this subsection, we organize the Snowmass contributions into a summary table to enable an illustrative comparison between the various $Z^{\prime}$ models and current and possible collider scenarios.  To enable the comparison and focus on the mass reach of the different colliders, we adopted the $g_Z^{\prime} = 0.2$ coupling parameter for the universal $Z^\prime$ model, since it roughly aligns with the mass reach for the SSM $Z^{\prime}$ model in the resonance channels studied.  As we move down the table shown in table~\ref{tab:ZpTable}, the $Z^{\prime}$ mass reach steadily increases.

\begin{table}[htb!]
\begin{center}
\includegraphics[width=\hsize]{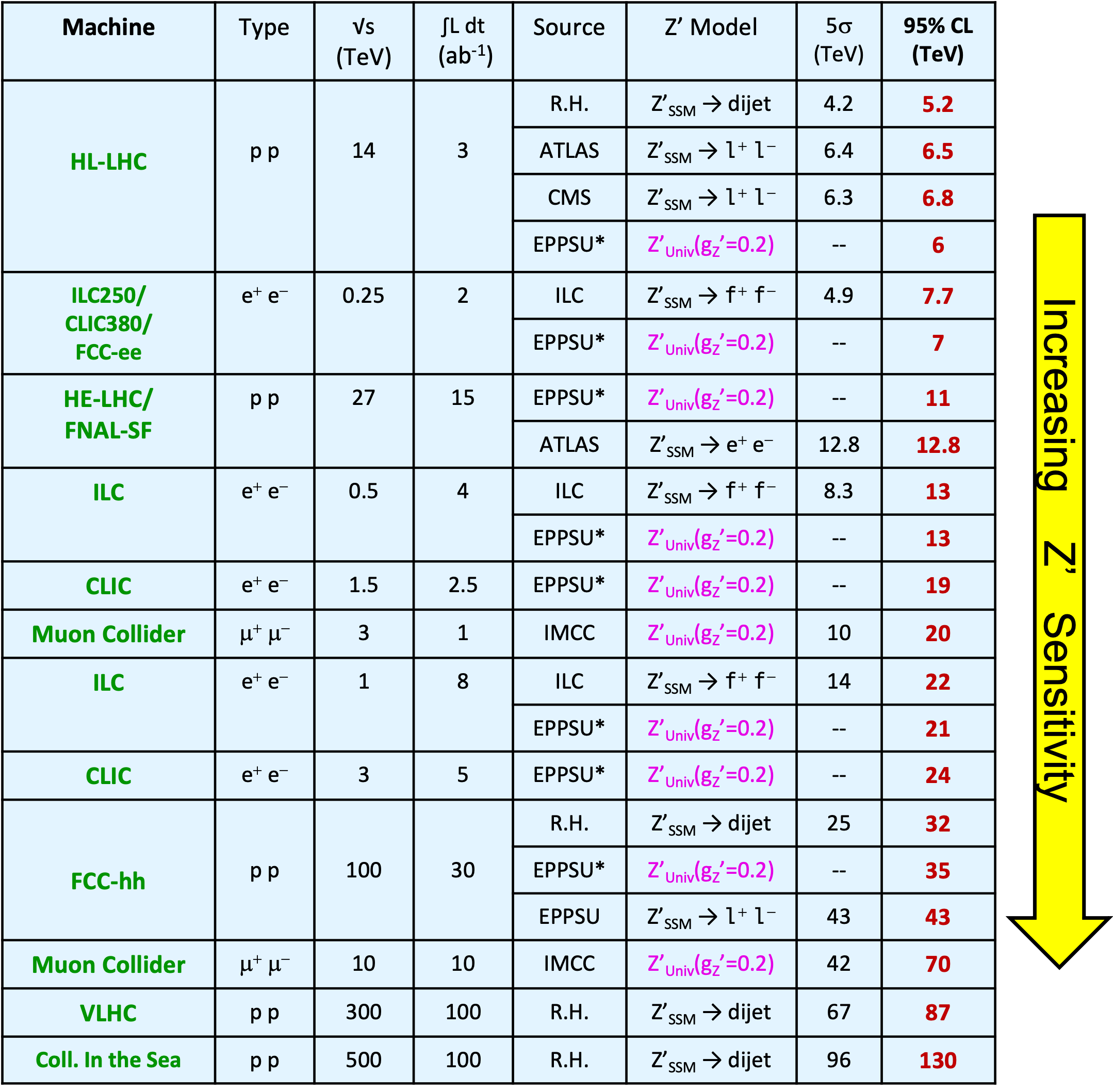}
\end{center}
\caption{ For each collider we list the operating point and mass reach, for 5$\sigma$ discovery and 95\% CL exclusion, of the SSM $Z^{\prime}$ model taken from Refs.~\cite{harris2022sensitivity,Helsens:2019bfw,ATL-PHYS-PUB-2018-044,CMS-PAS-FTR-21-005,Suehara:2022pwv}, and the mass reach of the universal $Z^{\prime}$ model with a coupling $g_{Z^{\prime}}=0.2$ from Refs.~\cite{Aime:2022flm,Strategy:2019vxc} that we determined from Fig.~\ref{fig:UnivZp}. }
\label{tab:ZpTable}
\end{table}

At first glance, this table shows the obvious correlation that higher center of mass collider energy affords higher reach in $Z^\prime$ mass, where the orders of magnitude spanned in collider energy pay off in orders of magnitude in $Z^{\prime}$ mass reach.  This is justified since the resonance signal is assured when the $Z^{\prime}$ boson is within the kinematic reach of the collider.  Moreover, for a given operating point of a collider, we see that the two $Z^{\prime}$ model benchmarks have very comparable results, which reflects the fact that the underlying charge assignments of SM fermions to the $Z^{\prime}$ currents only differ by $\mathcal{O}(1)$ factors, and so these results would be broadly applicable in other models where $Z^{\prime}$ bosons couple to all SM fermions, such as in gauged $B - L$ models.  For more fermion-specific models, such as $L_{\mu} - L_{\tau}$ or gauged baryon number, which are equally relevant to the model benchmarks shown in table~\ref{tab:ZpTable}, the distinction between the different colliders becomes dramatically more important since the $Z^{\prime}$ resonance would be produced via a tree-level coupling in some colliders while only produced via a kinetic mixing coupling or a loop-induced coupling in others.  As a first estimate, the corresponding reach for a point of comparison to table~\ref{tab:ZpTable} would then adopt a coupling suppressed by a loop factor when the model does not couple to the initial partons at tree-level.

In table~\ref{tab:ZpTable} the relationship between the $Z^\prime$ mass reach at 95\% CL and the mass reach at 5$\sigma$ depends on the machine type and final state. The two sensitivities are roughly equal for dilepton final states at pp colliders, because the $Z^\prime$ peak is beyond the highest masses of the dilepton continuum background from electroweak production via Drell-Yan, a convincing and background-free exclusion or discovery. For dijet final states at pp colliders, the direct searches for a $Z^\prime$ dijet mass bump has a 95\% CL mass reach that is roughly 20-30\% larger than the $5\sigma$ mass reach, because here the continuum background is larger from strong production of dijets via QCD. Finally, lepton colliders search within the kinematic distributions of fermion pairs for the indirect effects of a $Z^{\prime}$, with huge backgrounds at di-fermion masses significantly lower than the $Z^{\prime}$ pole mass, resulting in a 95\% CL mass reach that is roughly 60-100\% larger than the $5\sigma$ mass reach. Therefore, table~\ref{tab:ZpTable} illustrates both the power of lepton colliders for indirect discovery of new physics, and the subsequent necessity of a hadron collider to directly produce and confirm that new physics, within a complete program for the future of HEP like that discussed at Snowmass for FCC~\cite{Bernardi:2022hny}.

\subsection{$W^{\prime}$ Bosons}

Models that feature $W^{\prime}$ bosons differ from $Z^{\prime}$ models since $W^{\prime}$ bosons mediate a charged current interaction and hence necessarily extend the SM electroweak gauge symmetry either via a product gauge group or embedding the SM electroweak group in a larger symmetry.  We show the sensitivity of HL-LHC to the di-fermion decays of a right-handed $W_R^\prime$ gauge boson or an SSM $W^{\prime}$ boson in the table on the left side of Fig.~\ref{fig:Wprime}. The sensitivity for SSM $W^{\prime}$ decaying to dijets at future pp colliders is also shown in table~\ref{tab:dijet-sensitivity-table} and Fig.~\ref{fig:strong-weak}. For all di-fermion decays, the $W^{\prime}$ mass reach is larger than $Z^{\prime}$, due to the larger production cross section. Di-boson decays are more challenging, and no mass reach projections were done for snowmass but were studied in Refs.~\cite{Thamm:2015zwa,Baker:2022zxv}. 
The multi-body mass distributions of ${W^{\prime}_{\mbox{SSM}}}\rightarrow W {Z^{\prime}_{\mbox{SSM}}}$ at HL-LHC were studied~\cite{Chekanov:2021huv} and are shown on the right side of Fig.~\ref{fig:Wprime}.


\begin{table}[htb]
\begin{minipage}{0.5\linewidth}
\begin{tabular}{|c|c|c|c|} \hline
\multicolumn{4}{|c|}{$\mathbf{W^{\prime}}$\textbf{ at HL-LHC (14 TeV, 3 ab$^{-1}$)}} \\ \hline
\textbf{Source} & \textbf{Model \&} & $\mathbf{5\sigma}$ & \textbf{95\% CL} \\
 & \textbf{Channel} & \textbf{(TeV)} & \textbf{(TeV)} \\ \hline
 ATLAS~\cite{ATL-PHYS-PUB-2018-044} &  $W^\prime_{R}\rightarrow$ tb & 4.3 & 4.9 \\ \hline
 RH~\cite{harris2022sensitivity} & $W^\prime_{SSM}\rightarrow$ dijet & 4.8 & 5.6 \\ \hline
 CMS~\cite{CMS-PAS-FTR-18-030} & $W^\prime_{SSM}\rightarrow \tau \nu$ & -- & 7.2 \\ \hline
                               & $W^\prime_{SSM}\rightarrow \mu \nu$ & 7.1 & 7.3 \\ \cline{2-4}
ATLAS~\cite{ATL-PHYS-PUB-2018-044} & $W^\prime_{SSM}\rightarrow e \nu$ & 7.5 & 7.6 \\ \cline{2-4}
                               & $W^\prime_{SSM}\rightarrow l \nu$ & 7.7 & 7.9 \\ \hline
\end{tabular}
\end{minipage}
\begin{minipage}{0.48\linewidth}
    \centering
    \includegraphics[width=\linewidth]{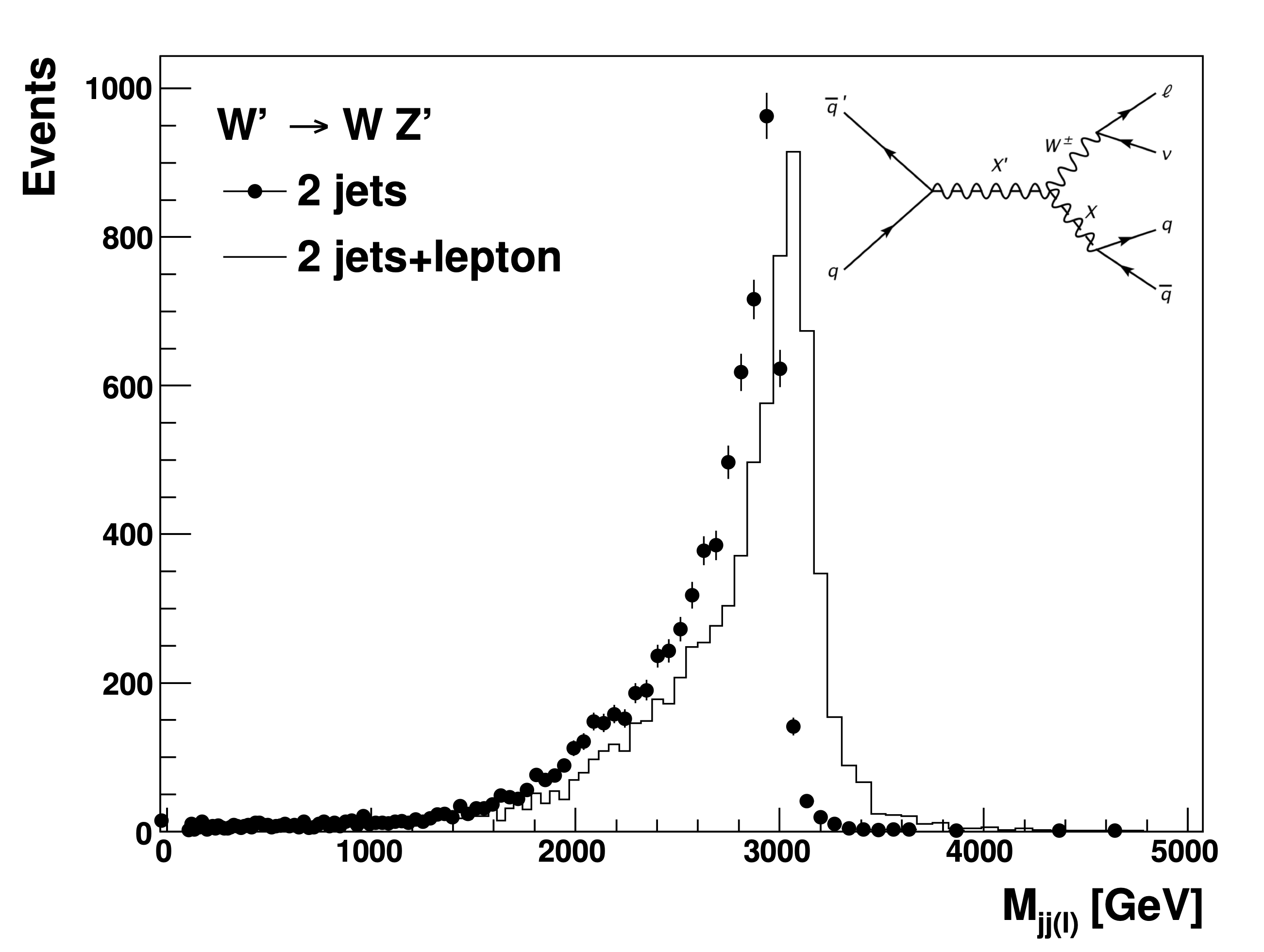}
\end{minipage}
\caption{ (left) $W^{\prime}$ mass reach for 5$\sigma$ discovery and 95\% CL exclusion from HL-LHC, for a right-handed model decaying to a top and bottom quark in ATLAS~\cite{ATL-PHYS-PUB-2018-044}, and for the SSM model in the dijet channel~\cite{harris2022sensitivity}, the tau channel in CMS~\cite{CMS-PAS-FTR-18-030}, and the muon, electron, and combined channel in ATLAS~\cite{ATL-PHYS-PUB-2018-044}. (right) The distribution of dijet mass (points) and dijet + lepton mass (histogram) of the simulated decay 
${W^{\prime}_{\mbox{SSM}}}\rightarrow W {Z^{\prime}_{\mbox{SSM}}} \rightarrow (l\nu)(q\bar{q})$ for HL-LHC~\cite{Chekanov:2021huv}.}
\label{fig:Wprime}
\end{table}

\subsection{Axion-Like Particles}

In contrast to $Z'$ and $W'$ bosons, axion-like particles (ALPs) are new pseudoscalar particles whose Lagrangian interactions are generally governed by a discrete shift symmetry.  These pseudoscalars arise as pseudo-Nambu Goldstone bosons from a spontaneously broken global $U(1)$ symmetry in the ultraviolet theory.  In analogy with the QCD axion arising from the Peccei-Quinn mechanism or the pion from the QCD chiral Lagrangian, ALP interactions are characterized by a decay constant $f_a$ associated with their PNGB nature and, unlike traditional QCD axions, ALP masses are free parameters and provide the leading explicit shift symmetry breaking.  While ALP Lagrangians have a rich phenomenology, the main phenomenological target at experiments is the ALP coupling to two photons, allowing a smooth transition between traditional QCD axion and ALP parameters.

\begin{figure}[htb]
\begin{center}
\includegraphics[width=\hsize]{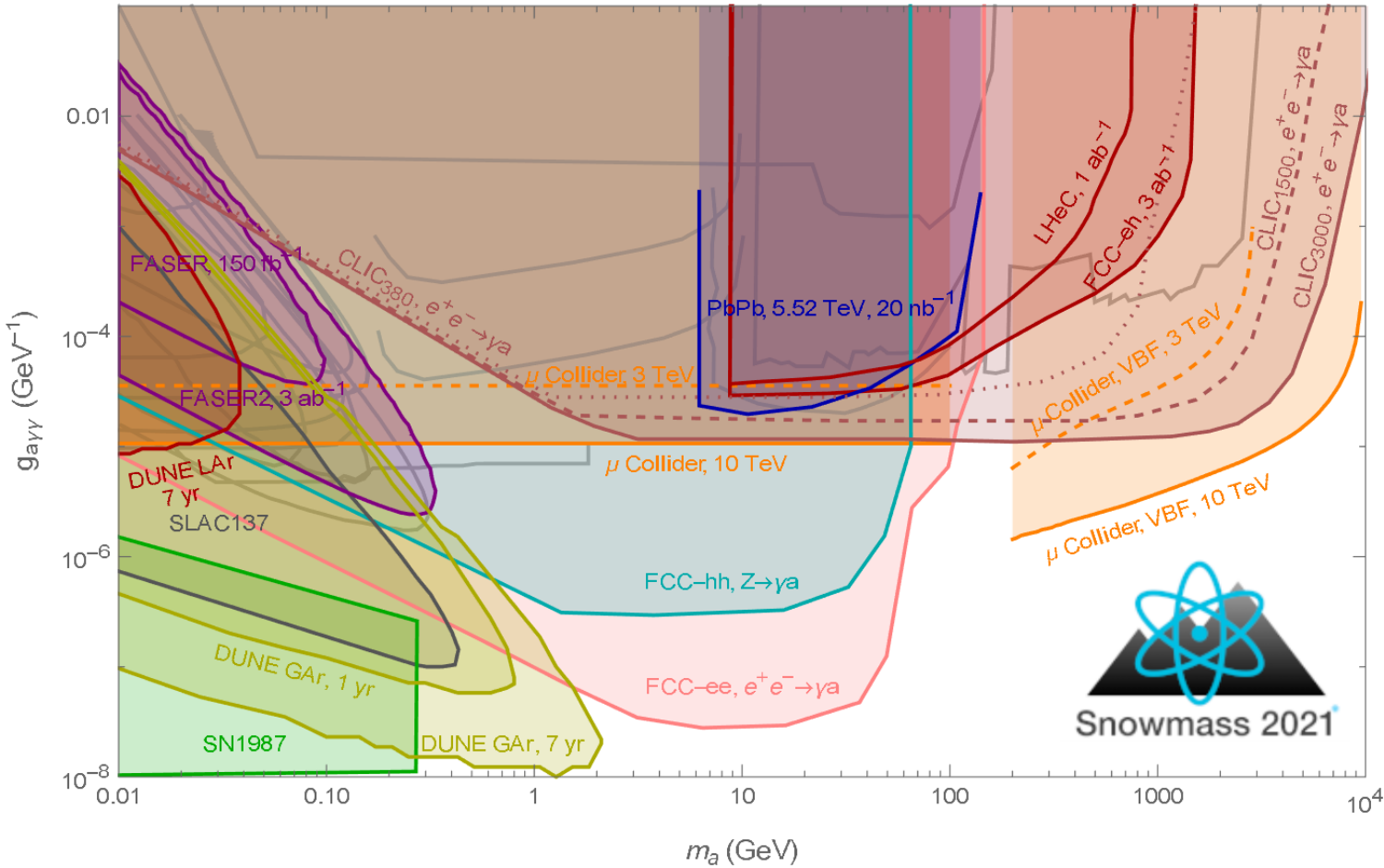}
\end{center}
\caption{ The axion-like particle (ALP) coupling in the diphoton channel $g_{a\gamma\gamma}$ versus 95\% CL mass reach is shown for multiple  colliders~\cite{Bernardi:2022hny,Strategy:2019vxc}, including Snowmass studies on the muon collider~\cite{MuonCollider:2022xlm,Han:2022mzp} (orange) at $\sqrt{s}=3$ TeV (dashed) and 10 TeV (solid) as well as from the DUNE near detector~\cite{Brdar:2020dpr} with liquid Argon technology (dark red) and gaseous Argon technology (dark yellow).}
\label{fig:ALPs}
\end{figure}

Figure~\ref{fig:ALPs} shows the results of Snowmass studies on the sensitivity of the muon collider to an ALP~\cite{MuonCollider:2022xlm, Han:2022mzp} compared with other colliders~\cite{Bernardi:2022hny, Strategy:2019vxc,Bauer:2018uxu} in the $m_a$ vs.~$g_{a \gamma \gamma}$ plane. For ALP decays to diphotons, a 10 TeV muon collider is the most sensitive to high ALP masses $m_a>200$ GeV, from vector boson fusion production processes ($VV\rightarrow a\rightarrow \gamma \gamma$). FCC-ee has the best sensitivity in the medium mass range $1 \lesssim m_a < 100$~GeV, from the associated production process ($Z\rightarrow \gamma a$), thanks to the potentially very large integrated luminosity expected at the $Z$ pole in circular $e^+e^-$ colliders. In the more near term, the best collider limits on ALPs coupling to photons over the range $m_a \approx 0.1-100$~GeV will be set by exploiting photon-photon collisions in ultraperipheral interactions of heavy-ions during the HL-LHC phase \cite{Bruce:2018yzs,dEnterria:2022sut}.\footnote{There are other interesting physics considerations for ultraperipheral interactions of heavy-ions, e.g., Refs.~\cite{dEnterria:2018uly,Klein:2020fmr,dEnterria:2013zqi,Buhler:2022knp}.}
By searching for a diphoton resonance above the light-by-light continuum, ALP-photon couplings as low as $g_{a\gamma\gamma}\approx 10^{-2}$~GeV$^{-5}$ can be probed over this broad $m_a$ range.  Importantly, the DUNE near detector can provide complementary sensitivity to the collider probes acting as an improvement over the existing SLAC137 constraints, with the gaseous Argon technology providing leading sensitivity for $m_a \lesssim 1$ GeV~\cite{Brdar:2020dpr}.

It is worth noting that the ALP is typically expected to have non-suppressed coupling to gluons, in particular in its connection to the Strong CP puzzle of QCD~\cite{Hook:2019qoh,DiLuzio:2020wdo}. Having gluonic couplings changes the considerations for the search channels and the performance at different facilities appreciably (see recent phenomenological studies~\cite{Kelly:2020dda,Agrawal:2021dbo,Feng:2022inv,ArgoNeuT:2022mrm}).

\subsection{Dijet Resonances}

\label{sec:dijet-resonances}

\begin{figure}[bht]
\begin{center}
\includegraphics[width=0.49\hsize]{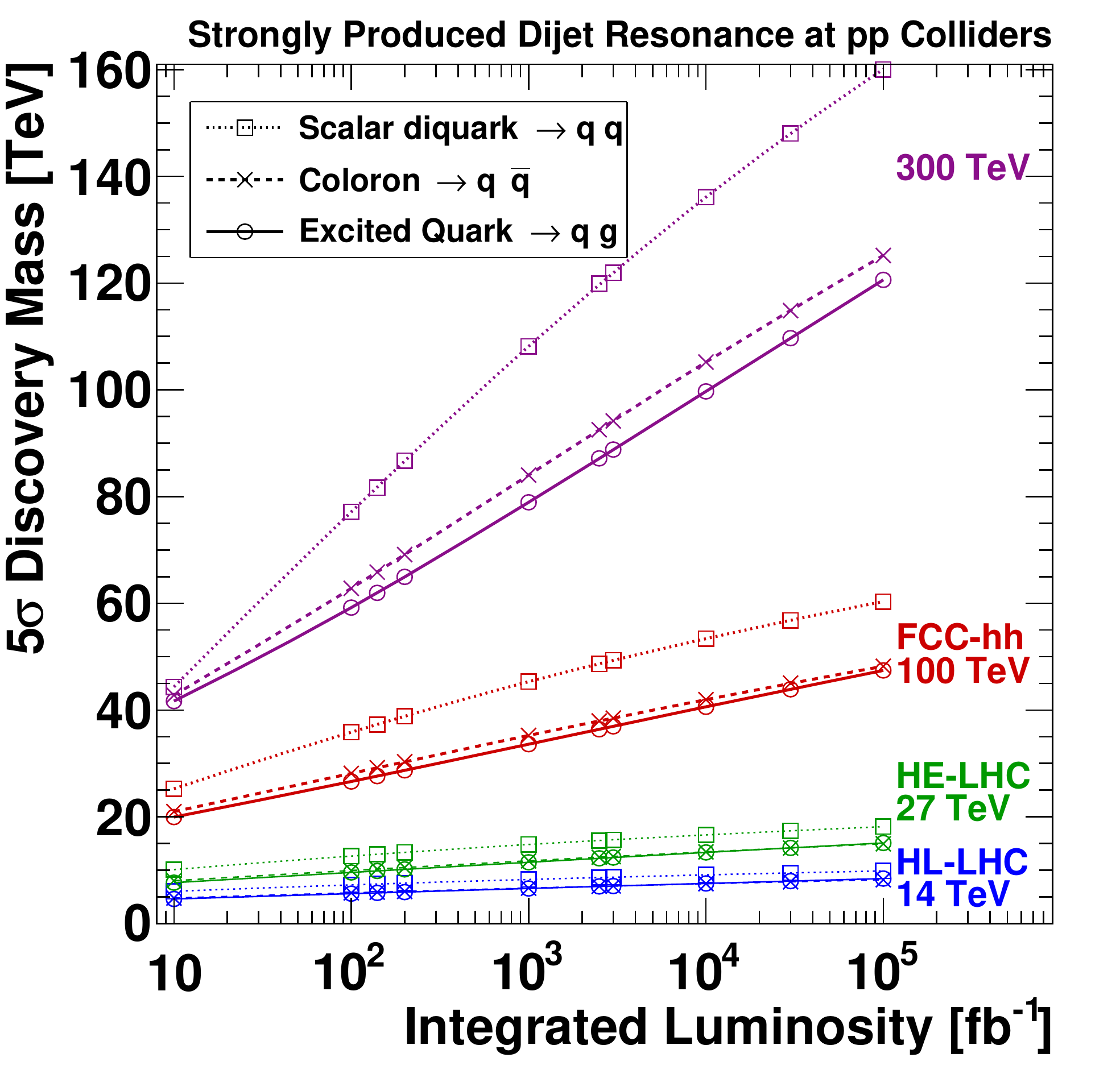}
\includegraphics[width=0.49\hsize]{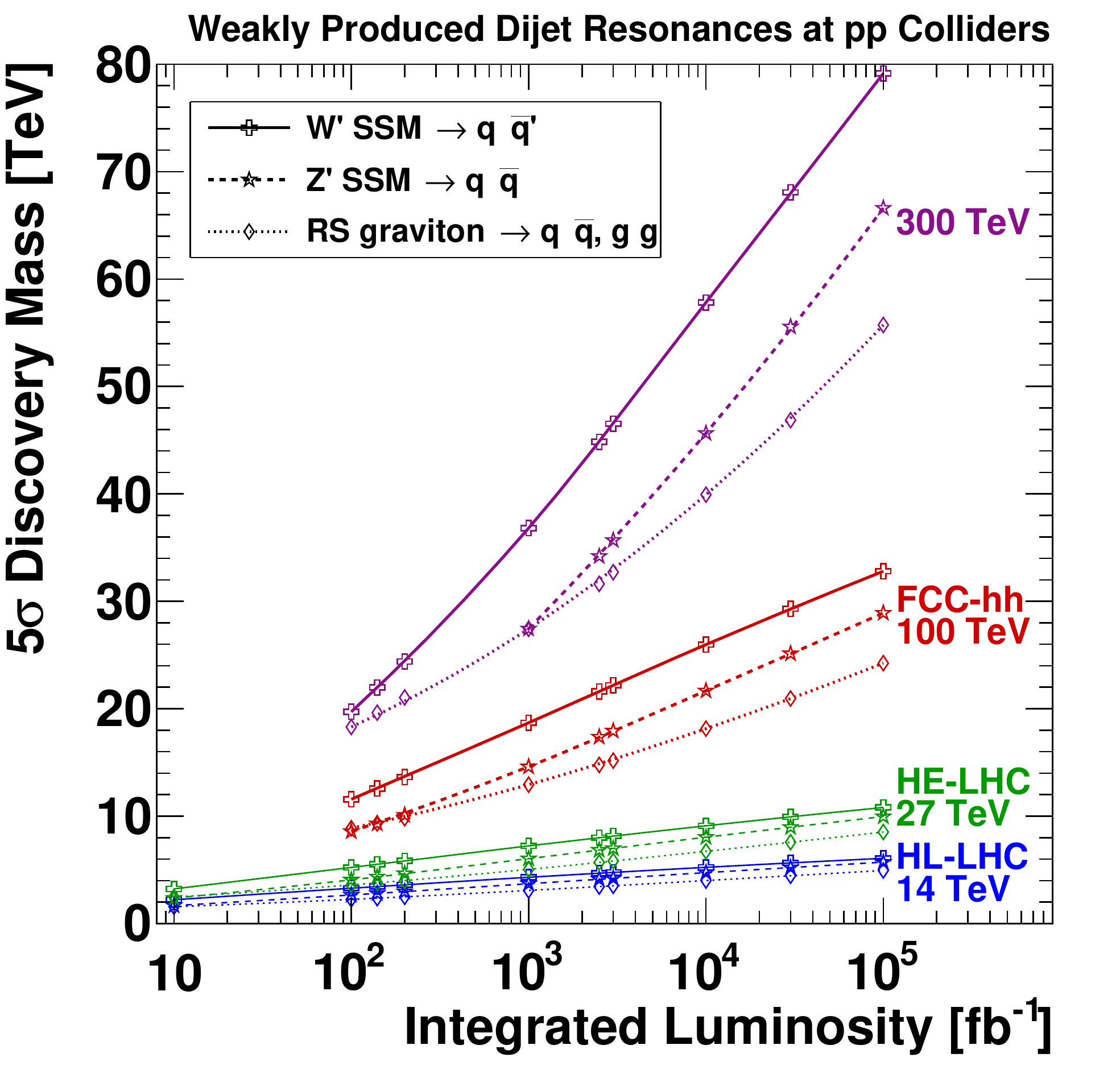}
\end{center}
\caption{ Sensitivity to strongly and weakly produced dijet resonance models. The $5\sigma$ discovery mass for four values of pp collider $\sqrt{s}$ (colors) as a function of integrated luminosity for dijet resonances from (left) the large cross section models of diquarks (boxes), colorons (Xs), and excited quarks (circles) and from (right) the smaller cross section models of W$^{\prime}$ SSM bosons (crosses)  $Z^{\prime}$ SSM bosons (stars) and Randall-Sundrum gravitons (diamonds). From Ref.~\cite{harris2022sensitivity}. }
\label{fig:strong-weak}
\end{figure}

The sensitivity to dijet resonances at pp colliders was explored during Snowmass 2021 as discussed in Refs.~\cite{harris2022sensitivity,Bernardi:2022hny,Guler:2022czi}. The process, $pp \rightarrow X \rightarrow 2\mbox{ jets}$, is an essential benchmark of discovery capability of pp colliders and is sensitive to a variety of models of new physics at the highest mass scales.  The sensitivity to a dijet resonance is mainly determined by its cross section. The study considered strongly produced models, those with large production cross sections, that include scalar diquarks, colorons and excited quarks. At the highest resonance masses these strongly produced models can only be observed at a pp collider, as lepton colliders can only produce diquarks and excited quarks in pairs at significantly lower masses. Also considered are weakly produced models, with production cross sections that are roughly two orders of magnitude smaller, that include $W^{\prime}$s, $Z^{\prime}$s and Randall-Sundrum gravitons, which can also be observed at lepton colliders as previously discussed. The $5\sigma$ discovery mass is shown as a function of integrated luminosity in Fig.~\ref{fig:strong-weak} 
for these two sets of models organized by production strength. Note two scaling behaviors of the sensitivity: the mass sensitivity is roughly proportional to the collision energy, and for any fixed value of $\sqrt{s}$, the sensitivity is roughly proportional to the logarithm of the integrated luminosity. In table~\ref{tab:dijet-sensitivity-table} the study summarizes the sensitivity of five major options for future colliders, a selection of the results contained in the full study~\cite{harris2022sensitivity}, conducted for eight collision energies and ten integrated luminosities. Again, the production strength is useful for organizing the results, and very roughly speaking, the discovery mass reach of a proton-proton collider is about half it's collision energy for strongly produced dijet resonances, and about one quarter of it's collision energy for weakly produced dijet resonances. These masses are all very large compared to the mass scales directly accessible by other colliders, and demonstrate a pp collider's unique ability to deeply explore the energy frontier.

\begin{table}[thb]
\begin{center}
\includegraphics[width=\hsize]{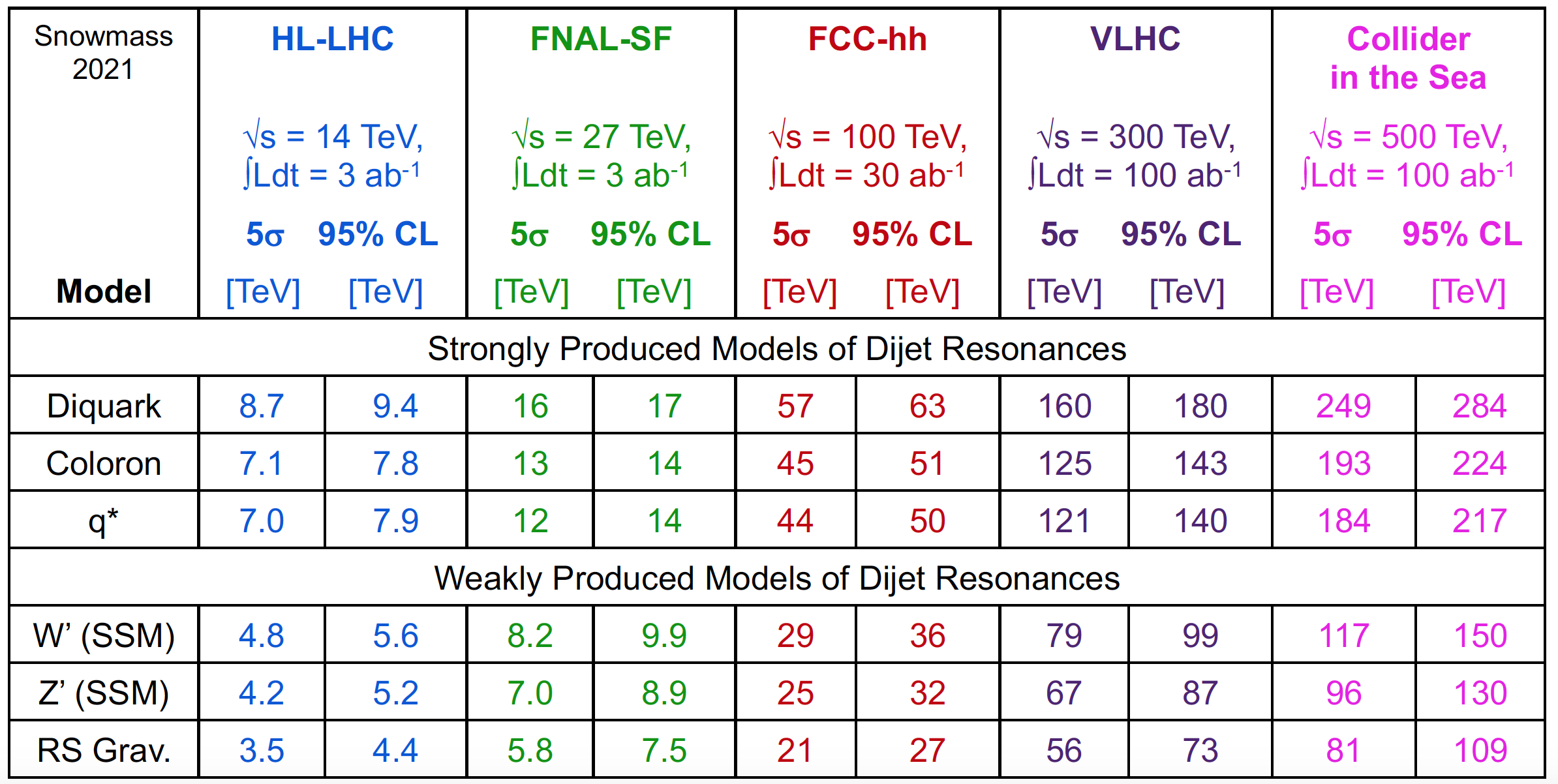}
\end{center}
\caption{Comparison of the sensitivity to dijet resonances of five pp colliders at their approximate baseline integrated luminosities. The mass for discovery at $5\sigma$, or exclusion at 95\% confidence level, is listed  for six models in descending order of model cross section. From Ref.~\cite{harris2022sensitivity}.}
\label{tab:dijet-sensitivity-table}
\end{table}

\section{New Fermions}\label{Sec:NF}

The fermion sector revealed the first laboratory evidence for physics beyond the SM in the form of neutrino oscillations and the consequent requirement that SM neutrinos have mass. New fermions with masses from MeV -- TeV now appear in many possible extensions to the SM. In this section we will discuss models that propose new neutral leptons, charged leptons, and heavy quarks whose masses may lie within the reach of future collider physics programs.

\subsection{Neutral Leptons}

Extensions to the SM that account for neutrinos masses typically incorporate right-handed neutrinos with mass terms that either conserve lepton flavor (``Dirac'') or do not (``Majorana''). In most models, the additional right-handed neutrinos are ``sterile'' with very small mixings to SM neutrinos, and they have masses much larger than the eV scale. Neutral leptons with masses on the MeV scale or higher are referred to as \emph{heavy} neutral leptons (HNLs).

A common benchmark model for HNLs is the Seesaw Mechanism. In type-1 Seesaw models, singlet HNL is added to the SM, and in type-3 Seesaw models, a triplet of heavy leptons is added, one of which is neutral. These neutral HNLs could mix with the known ``active'' neutrinos, making detection possible if their mass range is accessible to colliders. However, the couplings to SM particles are predicted to be small in this mass range. Hence, high luminosity future colliders will be critical for any potential discovery.

\subsubsection{Type-1 Seesaw}

An HNL denoted $N$ can be produced through a variety of processes at $e^+e^-$, $pp$, and $ep$ colliders. Phenomenological studies typically assume a single $N$ that decays to SM particles, mixing with SM neutrinos via a matrix $U$. The HNL could decay via $N \rightarrow eW$ or $N \rightarrow \nu Z$, making multi-lepton signatures a popular search route with small SM backgrounds. As the couplings proportional to $|U|$ decrease, $N$ becomes long-lived and may decay with significant displacement from the original collision.

\begin{figure}[htb]
    \centering
    \includegraphics[width=\textwidth]{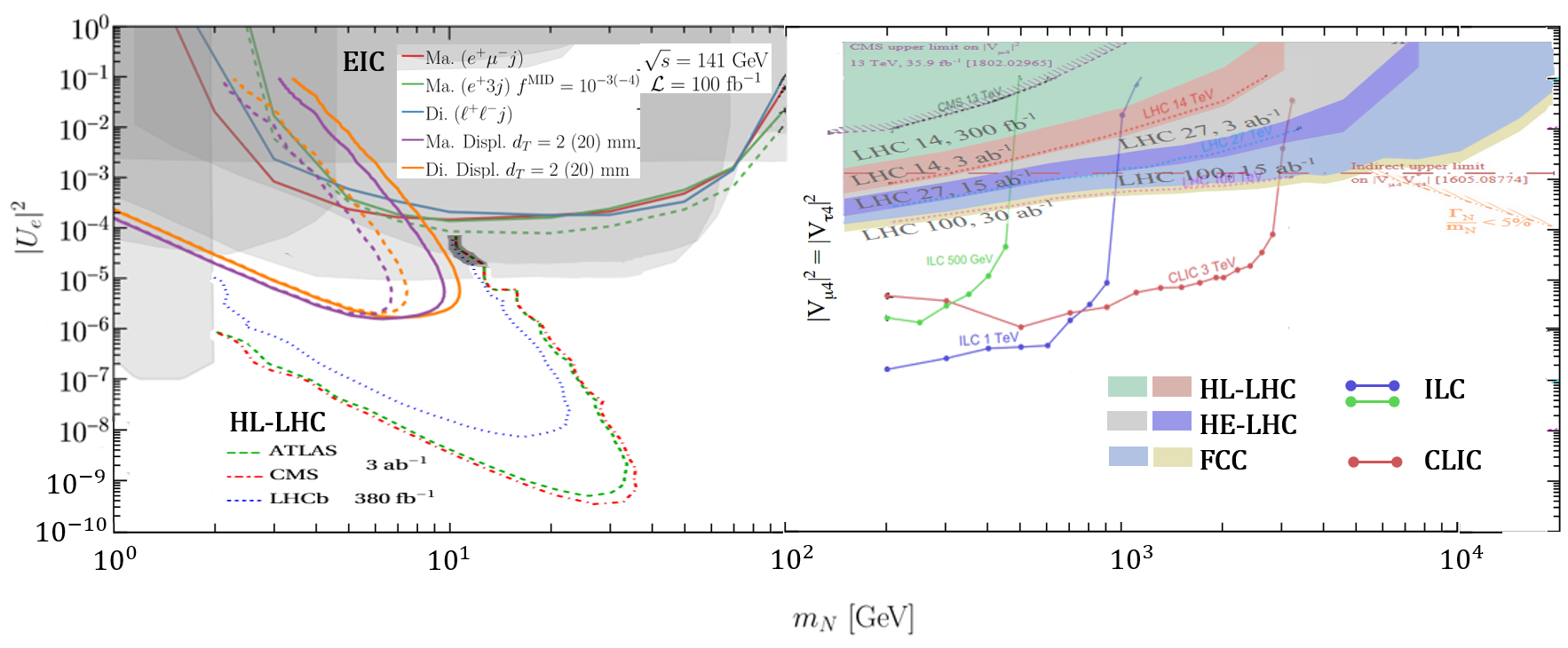}
    \caption{Expected bounds for HNLs in the type-1 Seesaw model. Low mass constraints on the coupling of the HNL to electrons are set for the EIC and HL-LHC~\cite{HNL_EIC,Abdullahi:2022jlv}. High mass constraints from the HL-LHC, HE-LHC, FCC-hh, ILC, and CLIC assume either two (hadron colliders) or three (lepton colliders) non-zero couplings between the HNL and SM leptons~\cite{Abdullahi:2022jlv,Mekala:2022cmm}.}
    \label{fig:HNLtype1}
\end{figure}

At the EIC, the lepton number violating signatures $e^-p \rightarrow j(N \rightarrow e^+W^{-*})$, with $W^{-*} \rightarrow e^+\mu^-\bar{\nu_\mu})$ or $W^{-*} \rightarrow e^+jj$ would strongly suggest a Majorana HNL, and the $e^+\mu^-$ final state would have very small SM background. These searches could constrain $|U_e|^2$, the mixing of $N$ with the electron neutrino, down to $\sim{10^{-4}}$ for $m_N$ on the order of 10 GeV, as shown in Fig.~\ref{fig:HNLtype1}~\cite{HNL_EIC}. The EIC could also constrain this coupling at lower $m_N$ via similar searches that require the $N$ decay vertex to be displaced by 2 -- 20 mm transverse to the beam direction. In both types of search, a Dirac HNL with $e^-$ in the final state allows a similar reach as found in the Majorana case. In the same $m_N$ region, ATLAS, CMS, and LHCb may be able to constrain single-flavor couplings to much lower values using the muon chambers as displaced vertex detectors, subject to the detector conditions described in~\cite{Abdullahi:2022jlv}. This potential reach for $|U_e|^2$ is also shown in Fig.~\ref{fig:HNLtype1}.

The HL-LHC experiments and other future high energy colliders can also search for high-mass HNL signatures, in which the $W/Z$ bosons produced in the HNL decay are on-shell. When considering only one flavor of mixing, a fast-simulation study by CMS in a high-momentum three-lepton final state predicts exclusions up to 2.0 TeV for $|U_e|=1$ and 2.4 TeV for $|U_{\mu}=1|$ at 14 TeV with the full HL-LHC dataset of 3 ab$^{-1}$~\cite{HNL_CMS}. Figure~\ref{fig:HNLtype1} shows the potential reach of high-mass searches at $pp$ colliders of various energies and integrated luminosities, in the scenario with $|U_{\mu}|^2 = |U_{\tau}|^2$ and $|U_e|^2=0$ (though bounds are similar when replacing muon coupling with electron)~\cite{Abdullahi:2022jlv}. These projections are compared in Fig.~\ref{fig:HNLtype1} to bounds expected from the ILC or CLIC at $\sqrt{s}=500$ GeV (1.6 ab$^{-1}$), 1 TeV (3.2 ab$^{-1}$), and 3 TeV (4 ab$^{-1}$)~\cite{Mekala:2022cmm}. The ILC/CLIC fast-simulation projection for the $\ell\nu jj$ final state utilized a boosted decision tree to discriminate signal from the background and assumes equal coupling of the $N$ to all three SM flavors. For prompt HNL searches, Fig.~\ref{fig:HNLtype1} shows that $ep$ collisions allow access to the smallest couplings for low $m_N$, $e^+e^-$ collisions for $m_N \sim 100$--$1000$ GeV, and $pp$ collisions for even higher masses.

\begin{figure}[htb]
    \centering
    \includegraphics[width=0.99\textwidth]{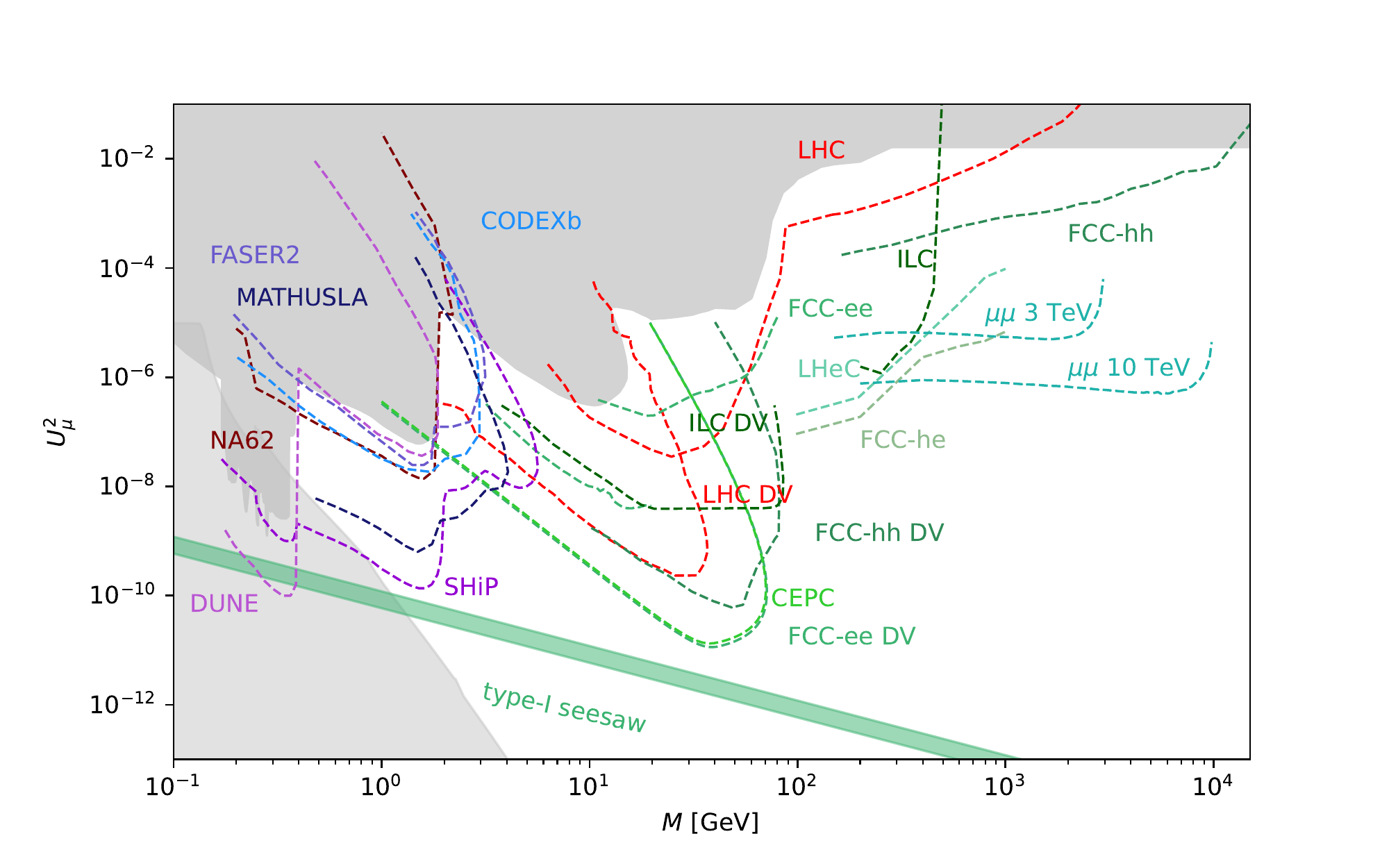}
    \caption{
    Constraints and future sensitivities for HNLs with mass $M$ and mixing $U_\mu^2$ with muon neutrinos (summed over three HNL flavours). \emph{Medium gray:} Constraints on the mixing of HNLs from past experiments~\cite{CHARM:1985nku,Abela:1981nf,Yamazaki:1984sj,E949:2014gsn,Bernardi:1987ek,NuTeV:1999kej,Vaitaitis:2000vc,CMS:2018iaf,DELPHI:1996qcc,ATLAS:2019kpx,CMS:2022fut}.
 \emph{Colourful lines:}
Estimated sensitivities of the main HL-LHC detectors (adapted from \cite{Izaguirre:2015pga,Drewes:2019fou,Pascoli:2018heg}) and NA62 \cite{Drewes:2018gkc},
with the sensitivities of selected planned or proposed experiments (DUNE~\cite{Ballett:2019bgd}, FASER2~\cite{FASER:2018eoc}, SHiP~\cite{SHiP:2018xqw,Gorbunov:2020rjx}, MATHUSLA~\cite{Curtin:2018mvb}, CODEX-b~\cite{Aielli:2019ivi}, cf.~\cite{Agrawal:2021dbo} for a more complete list) as well as selected proposed future colliders 
(FCC-ee or CEPC \cite{Alimena:2022hfr,Shen:2022ffi,BayNielsen:2017yws}, 
FCC-hh~\cite{Antusch:2016ejd,Pascoli:2018heg},
ILC \cite{Mekala:2022cmm,Antusch:2016vyf}
LHeC and FCC-he \cite{Antusch:2019eiz},
and muon colliders \cite{ZhenLiuTeamHNLMuC,TaoLiuTeamHNLMuC}, with DV indicating displaced vertex searches).
\emph{Green band:} Indicative lower bound on the total HNL mixing $U_e^2+U_\mu^2+U_\tau^2$ from the requirement to explain the light neutrino oscillation data \cite{Esteban:2020cvm} when varying the lightest neutrino mass and marginalizing over light neutrino mass orderings.
 The matter-antimatter asymmetry of the universe \cite{Canetti:2012zc} can be explained via low scale leptogenesis  \cite{Akhmedov:1998qx,Asaka:2005pn,Pilaftsis:2003gt} along with the light neutrino masses in most of the white region above this band \cite{Drewes:2021nqr}.
\emph{Light gray:} Lower bound on $U_\mu^2$
 from BBN~\cite{Sabti:2020yrt,Boyarsky:2020dzc}.
Plot adapted from \cite{Abdullahi:2022jlv}.
    }
    \label{fig:HNLtype1disp}
\end{figure}

Many other proposed future experiments could offer discovery potential for Type-1 Seesaw HNLs, particularly in the low-mass / small-coupling region where long-lived searches will be required. Figure~\ref{fig:HNLtype1disp}, adapted from~\cite{Abdullahi:2022jlv,Alimena:2022hfr}, shows the expected reach of experiments such as FASER2, MATHUSLA, CODEXb, and the FCC (see Section~\ref{sec:dedicateddetectors}.) Many of these experiments are proposed to be realized within the HL-LHC timescale. In a longer timescale, the FCC-ee could probe the deepest into small couplings for GeV-scale HNLs. The ``type-I seesaw'' line indicates the approximate parametric scaling that would be satisfied in a simplified model with just a single neutrino flavor. Realistic three-generation models can have specific Yukawa textures and/or additional symmetries which can modify this scaling and make the regions above or below this line physically relevant. Concretely, one expects a specific Yukawa texture or approximate symmetry in the region above the line to keep the masses of the SM neutrinos small enough, or one can realize it through the inverse seesaw mechanism. Below the dashed line, additional contributions to the neutrino masses (e.g.~a set of additional, heavier neutral leptons) are needed to explain the neutrino oscillation data. 
The heavy neutrino mass range displayed in figure 16/2-35 allows for successful leptogenesis during heavy neutrino freeze-in \cite{Akhmedov:1998qx,Asaka:2005pn} or freeze-out \cite{Pilaftsis:2003gt}. Updated parameter space scans \cite{Klaric:2021cpi,Drewes:2021nqr,Hernandez:2022ivz} show that the LHC, FCC-ee and other future colliders can access the parameter region where heavy neutrinos can simultaneously explain the light neutrino masses and the baryon asymmetry of the universe.
We can view the extra Higgs doublet as a new portal to the VLLs, similarly to SUSY scenarios heavy Higgs portal to electroweakino and scalar lepton sectors~\cite{Gori:2018pmk}.

\clearpage

\subsubsection{Type-3 Seesaw}

\begin{figure}[htb]
    \centering
    \includegraphics[width=0.48\textwidth]{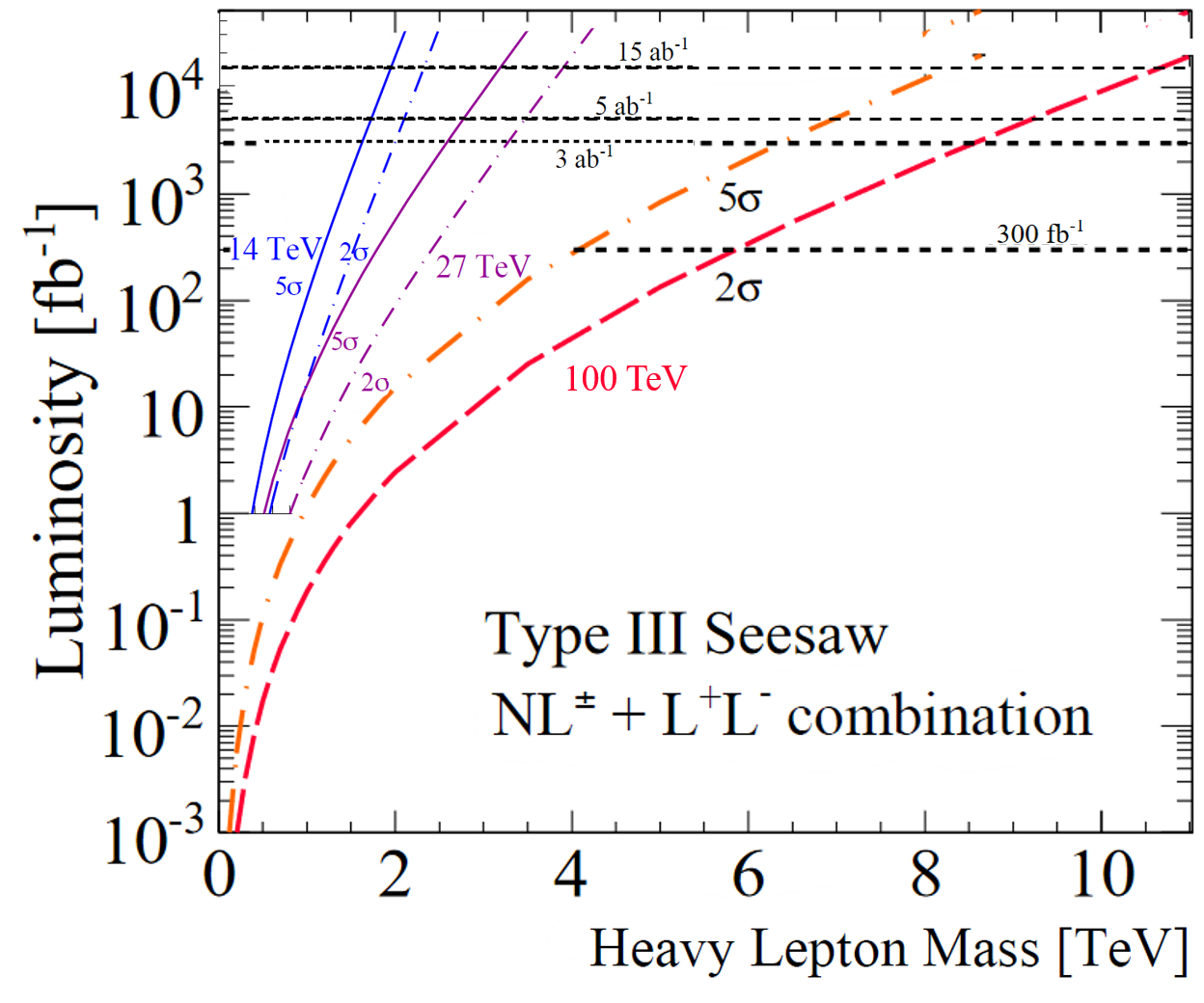}
    \includegraphics[width=0.48\textwidth]{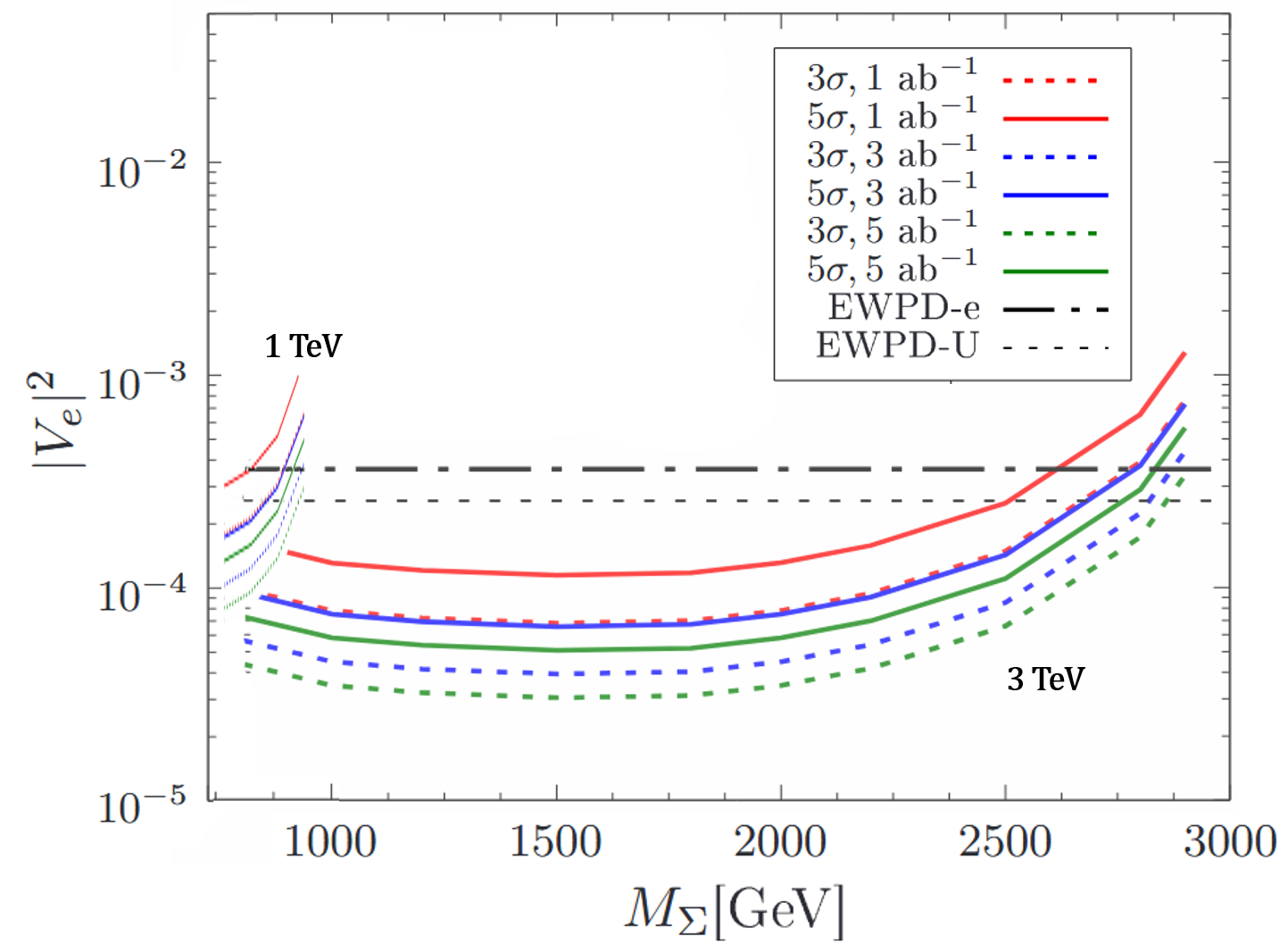}
    \caption{Expected bounds for HNLs in the Type-3 Seesaw model at $pp$ colliders (left, from~\cite{Abdullahi:2022jlv,HNL_LNV}) and lepton colliders (right, from~\cite{ILCInternationalDevelopmentTeam:2022izu}).}
    \label{fig:HNLtype3}
\end{figure}

In the Type-3 Seesaw model, the HNL is a member of a triplet of new fermions added to the SM, labeled $N$,$E^\pm$, or $\Sigma^0$,$\Sigma^\pm$, or similar. In this model, the gauge charges of the triplet fermions allow for pair production without suppressing mixing between active and sterile neutrinos~\cite{Abdullahi:2022jlv}. Production of $NE^\pm$ followed by $N \rightarrow W\ell$ can yield final states with many leptons, or lepton and high-momentum jets. ATLAS and CMS currently exclude $\Sigma$ masses below about 1 TeV, depending on the final state, and could extend that reach to 2 TeV with the full HL-LHC  dataset~\cite{Abdullahi:2022jlv}. Figure~\ref{fig:HNLtype3} (left) shows the projected luminosities required for exclusion and discovery of a heavy-fermion triplet at $pp$ colliders with HL-LHC, HE-LHC, and FCC-hh energies, assuming a combination of searches for $NE^{\pm}$ and $E^{+}E^{-}$ production~\cite{Abdullahi:2022jlv,HNL_LNV}.

At $e^+e^-$ colliders, single production of a triplet HNL is possible, based on the coupling of the triplet to the electron. This coupling is fixed at 0.00035 (EWPD-e) and 0.00025 (EWPD-U) by electroweak precision observables, regardless of $\Sigma$ mass. Bound from future colliders can be considerably stronger, as summarized in Fig.~\ref{fig:HNLtype3} (right), where the $e^+e^- \rightarrow (N \rightarrow W\ell)\nu$ process was analyzed in a final state with a lepton, $\slash{E}_T$, and large-radius jet with 2-prong substructure~\cite{ILCInternationalDevelopmentTeam:2022izu}.


\subsection{Charged Leptons}




Vectorlike leptons (VLL) are weakly interacting fermions whose left-handed and right-handed
components transform the same way under the Standard Model gauge group. Unlike chiral fermions,
they decouple from low-energy observables as they get heavier, since the mass arises from a bare electroweak-singlet term.
Following ref.\cite{Kumar:2015tna}, consider two minimal models: a charged $SU(2)_L$-singlet pair (``Singlet VLL") and a $SU(2)_L$-doublet pair (``Doublet VLL").
In both models, VLLs promptly decay to SM states via a mass mixing
with the SM $\tau$ leptons.
The mass eigenstates include a new charged Dirac pair $\tau^{\prime +} \tau^{\prime -}$
with mass $M_{\tau^\prime}$.
For large $M_{\tau^\prime}$, the branching ratios
$\left\{{\rm BR}(\tau^\prime \rightarrow W \nu), {\rm BR}(\tau^\prime \rightarrow Z \tau), {\rm BR}(\tau^\prime \rightarrow h \tau)\right\}$
asymptotically approach
\{0.5, 0.25, 0.25\} in the Singlet VLL model,
and 
\{0, 0.5, 0.5\} in the Doublet VLL model,
illustrating the Goldstone equivalence principle.
The Doublet VLL model also contains a new neutral Dirac pair $\nu^\prime \overline{\nu}^\prime$
with mass $M_{\nu^\prime} \approx M_{\tau^\prime}$ and 
${\rm BR}(\nu^\prime \rightarrow W \tau) = 1$,
reflecting the assumption that $\nu^\prime$ does not mix with the SM neutrinos.

\begin{figure}[h]
\includegraphics[width=0.49\textwidth]{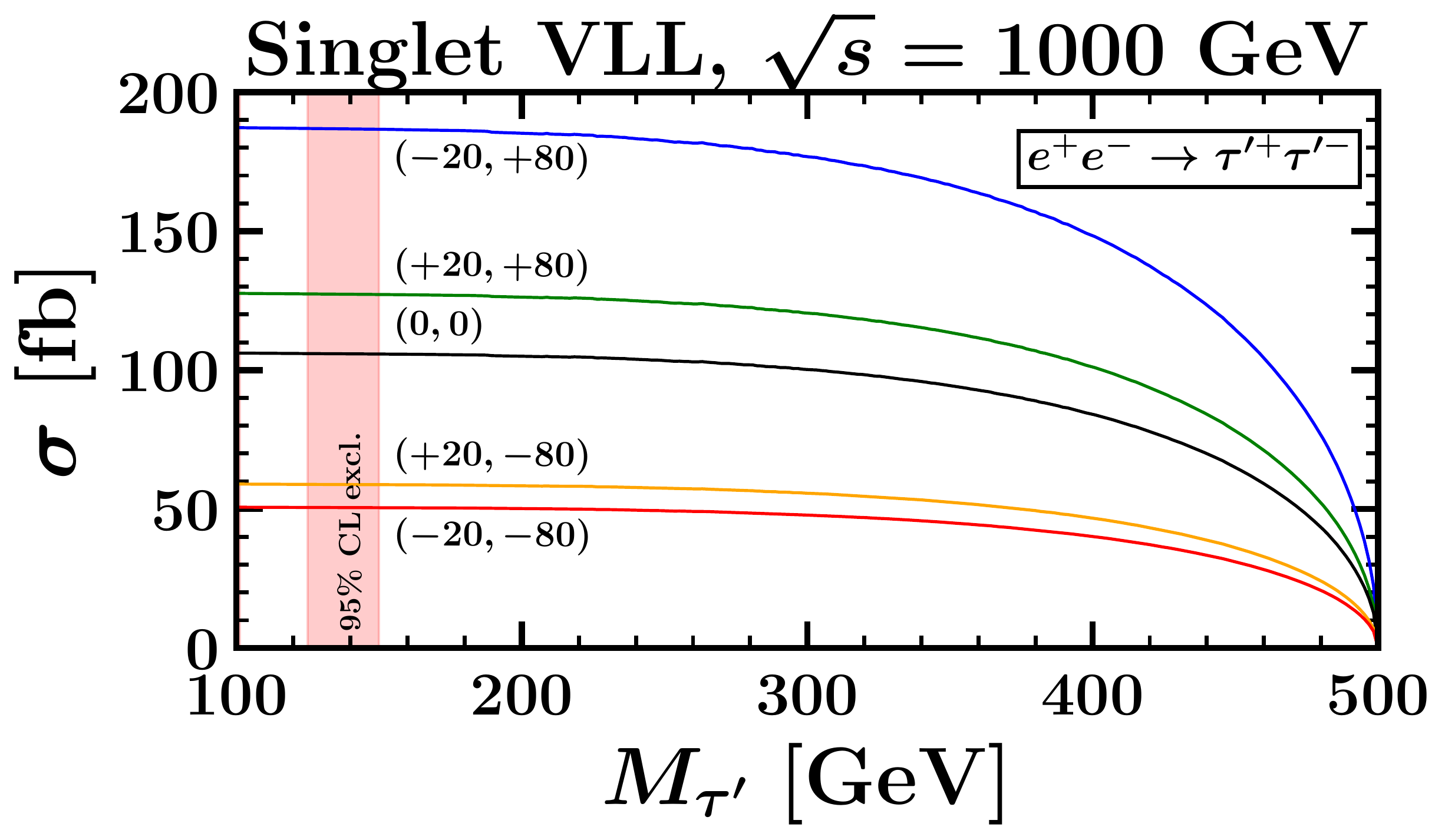}
\includegraphics[width=0.49\textwidth]{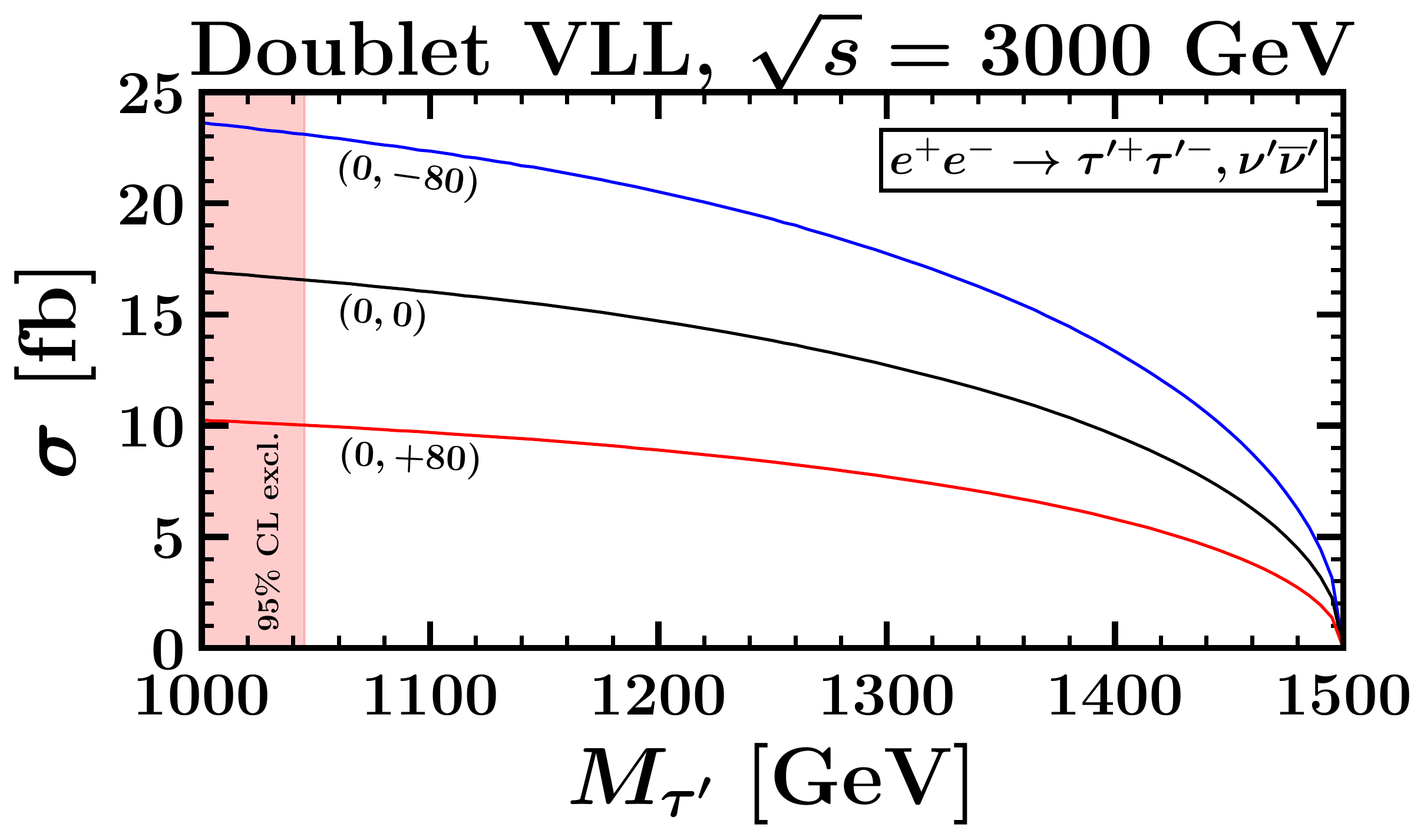}
\caption{The pair-production cross-sections for the Singlet (left) and Doublet (right)  VLL for $e^+ e^-$ collisions at $\sqrt{s} = $ 1000 (left) and 3000 (right) GeV as a function of $M_{\tau^\prime}$ for various choices of the beam polarizations $(\mathcal{P}_{e^+}, \mathcal{P}_{e^-})$.
The shaded regions correspond to the current 95\% CL exclusion on $M_{\tau^\prime}$.\label{fig:sigmavsM}}
\end{figure}

\begin{table}[ht]
\caption{Mass reaches at future $pp$ colliders for discovering (at $5\sigma$) or excluding (at 95\% CL) doublet VLL, using multi-lepton signatures, from~\cite{Bhattiprolu:2019vdu}. The relative uncertainty in the background is taken to be 10\%, and reaches are reported for the best individual signal region.}
\label{tab:futureproj_VLLD}
\begin{center}
\begin{tabular}{| c | c | c | c |}
\hline
~$pp$~$\sqrt{s}$ [TeV]~ & ~$\mathcal{L}$ [ab$^{-1}$]~ & ~$5\sigma$ discovery reach [GeV]~ & ~95\% CL exclusion reach [GeV]~\\
\hline
\hline
    ~14~ & ~3~ & ~900~ & ~1250~
    \\
    ~27~ & ~15~ & ~1700~ & ~2300~
    \\
    ~70~ & ~30~ & ~3400~ & ~4700~
    \\
    ~100~ & ~30~ & ~4000~ & ~5750~
    \\
\hline

\end{tabular}
\end{center}
\end{table}
Based on the latest LHC exclusions at 95\% CL by the CMS collaboration \cite{CMS:2022nty},
the masses are constrained to be $101.2 \leq M_{\tau^\prime} \leq 125$ GeV
or $M_{\tau^\prime} \ge 150$ GeV in the Singlet VLL model,
and $M_{\tau^\prime} \ge 1045$ GeV in the Doublet VLL model.
Ref.\cite{Bhattiprolu:2019vdu} estimated the mass reaches of the Doublet VLL, summarized in Table\ref{tab:futureproj_VLLD}, for future $pp$ colliders with $\sqrt{s} = (14, 27, 70, 100)$ TeV and
$\mathcal{L} = (3, 15, 30, 30)$ ab$^{-1}$, respectively, using multi-lepton signatures.
On the other hand, ref.\cite{Bhattiprolu:2019vdu} did not find significant exclusion or discovery potential
for charged weak-isosinglets at future $pp$ colliders, due to smaller cross-sections and large ${\rm BR}(\tau^\prime \rightarrow W \nu)$.
The Singlet VLL, which seem to pose a much more difficult challenge at $pp$ colliders, thus makes an excellent case for future lepton colliders, even at low masses.
Figure\ref{fig:sigmavsM} shows the VLL pair-production cross-sections as a function of $M_{\tau^\prime}$
for $e^+ e^-$ collisions at $\sqrt{s} = 1000$ GeV in the Singlet VLL model
and $\sqrt{s} = 3000$ GeV in the Doublet VLL model for various choices of the beam polarizations. One sensitivity study for the ILC and CLIC, performed in a fully leptonic final state, suggests that almost the entire accessible range of $M_{\tau^\prime}$ for each collider energy could be excluded with 15 fb$^{-1}$ (ILC 500 GeV), 23 fb$^{-1}$ (ILC 1.0 GeV), and 200 fb$^{-1}$ (CLIC 1.5 TeV)~\cite{Shang:2021mgn}. In fully hadronic final states the required luminosities decrease considerably, offering strong promise for discovery of a Singlet VLL.

\subsection{Heavy Quarks}\label{Sec:HeavyQuarks}

Heavy vector-like quarks (VLQs) appear in many beyond the SM models.  In particular, they are ubiquitous in compositie Higgs models.  At hadron colliders, the typical heavy VLQ  production mode is pair production through the strong force:
\begin{eqnarray}
pp\rightarrow \Psi\bar{\Psi}
\end{eqnarray},
where $\Psi$ is a vector-like quark (VLQ).  Up-type VLQs can also be singly produced with a $W$ or additonal jet through EW interactions:
\begin{eqnarray}
pp\rightarrow TW/T+{\rm jet},
\end{eqnarray}
were $T$ is a VLQ with charge $+2/3$.  Charge $+5/3$ VLQs can also be produced in association with a jet via interactions with EW gauge bosons and third generation quarks.  Once produced, common search strategies are for the decays of VLQs into SM gauge and Higgs bosons together with a top or bottom quark.

In the $pp\rightarrow T\bar{T}\rightarrow tZtZ$ assuming 100\% branching ratio of $T\rightarrow tZ$, the HL-LHC can discover heavy vector like quarks up to masses~\cite{Ilten:2015hya}
\begin{eqnarray}
1250-1450~{\rm GeV}.
\end{eqnarray}
The range of masses depends on the precise background systematics. 


Prior to Snowmass 2021, there have been past studies of singly produced charge $2/3$ and charge $5/3$ VLQs at the FCC-ee~\cite{Mangano:2270978}.  Unlike pair production which only depends on the strong force interactions, and hence the quantum numbers of of the VLQs, single production depends on the model independent coupling of a VLQ with EW gauge or Higgs bosons and third generation SM quarks  These model independent couplings can be the same as those that are responsible for VLQ decays.  Hence, assuming that the width to mass ratio is less than 30\% can limit the coupling responsible for single production. With this assumptions, the FCC-ee can exclude single production of charge $+2/3$ ($T$) and charge $+5/3$ ($X$) VLQs with masses of ~\cite{Mangano:2270978}
\begin{eqnarray}
M_T\lesssim 12~{\rm TeV~ and~} M_X\lesssim 14~{\rm TeV},
\end{eqnarray}
respectively, with 1 ab$^{-1}$.  With 10 $ab^{-1}$ these exclusion regions increase to~\cite{Mangano:2270978}
\begin{eqnarray}
M_T\lesssim 16.5~{\rm TeV~ and~} M_X\lesssim 19~{\rm TeV}.
\end{eqnarray}

\subsection{Exotic Signals}  

While traditional production and decay modes provide useful benchmarks to gauge the capabilities of colliders.  However, extending beyond the the simplest simplified model can greatly change the phenomenology.  Indeed, in composite Higgs type models, it may be expected that VLQ prefer to decay into non-SM final states~\cite{Bizot:2018tds}.

Some examples of altered phenomenology:
\begin{itemize}
\item 2HDM: Adding an extra Higgs doublet to a VLF model can change both the production and decay modes.  It is possible for VLQs to preferentially decay into the additional Higgs bosons and SM fermions~\cite{Dermisek:2021zjd,Dermisek:2020gbr,Dermisek:2022kyh}:
\begin{eqnarray}
T\rightarrow tH,\,T\rightarrow bH^\pm\nonumber\\
B\rightarrow bH,\,B\rightarrow tH^\pm,
\end{eqnarray}
where $H$ is a heavy CP-even neutral Higgs, and $H^\pm$ are heavy charged Higgs.  With 3 ab$^{-1}$ a 14 TeV LHC will be sensitive to these decays up to VLQ masses of 2.25-2.4~TeV~\cite{Dermisek:2022kyh}. The heavy Higgs cascade decay is also possible. The reach of the HL-LHC with 3~ab$^{-1}$ is expected to cover the heavy Higgs boson mass up to 2-2.5 TeV, heavy vectorlike top quark mass up to 1.4 TeV, and heavy vectorlike bottom quark mass up to 1.8 TeV~\cite{Dermisek:2019heo}.

Vectorlike leptons in 2HDMs introduce new production resonant production modes of VLLs with multi-lepton signals ~\cite{Dermisek:2022kyh}
\begin{eqnarray}
gg\rightarrow H\rightarrow L^\pm\mu^\mp\rightarrow h\mu^\pm\mu^\mp,\,Z\mu^\pm\mu^\mp
\end{eqnarray}
Since these are resonant, the production rates can be favorable to production through off-shell SM gauge bosons allowing for signal of TeV scale VLLs~\cite{Dermisek:2022kyh}.
\item Composite Higgs models often have additional pseudoscalars and scalars in addition to VLQs.  If kinematically allowed, the VLQ can preferentially decay to these new scalar states~\cite{Dolan:2016eki,Bizot:2018tds,Kim:2018mks,Banerjee:2022xmu,Benbrik:2019zdp}.  Pair production and decay into scalars is a promising search at the HL-LHC~\cite{Banerjee:2022xmu}:
\begin{eqnarray}
pp\rightarrow \bar{\Psi}\Psi \rightarrow t\bar{t}SS,\,b\bar{b}SS,
\end{eqnarray}
where $S$ is an additional neutral or chaged scalar or pseudoscalar, and $\Psi$ can be an up-type, down-type, or charge $5/3$ VLQ.  Depending on how $S'$ decays, this can lead to top or bottom production in association with additional jets, SM gauge boson, or tau leptons.  At the HL-LHC, many of this searches will be sensitive to VLQ quark masses of $1-1.5$~TeV and scalar masses about 500 GeV~\cite{Banerjee:2022xmu}.
\item Beyond specific model realizations, VLQs in effective field theories can lead to novel signatures.  In particular, an up-type VLQ can couple to a top quark via a magnetic or chromo-magnetic effective interaction, and a down-type VLQ with the bottom quark.  These couplings can lead to novel single VLQ production at hadron colliders via intermediate gluon~\cite{Kim:2018mks,Criado:2019mvu,Belyaev:2021zgq}:
\begin{eqnarray}
pp\rightarrow Tt/\,Bb,
\end{eqnarray}
and novel decays into gluons and photons~\cite{Kim:2018mks,Alhazmi:2018whk,Criado:2019mvu}:
\begin{eqnarray}
T\rightarrow tg/t\gamma,\,\,B\rightarrow bg/b\gamma.
\end{eqnarray}
The HL-LHC could be sensitive to these singley produced VLQs up to 2-3 TeV for combined $Tt$ and $Bb$ searches.~\cite{Belyaev:2021zgq}.  A study previous to Snomwass 2021 has also shown that the HL-LHC could discover pair produced $T\bar{T}$ with subsequent decays into gluon and photons up to masses of 1.4 TeV~\cite{Alhazmi:2018whk}.
\end{itemize}


\section{Long Lived Particles}\label{Sec:LLP}

\begin{figure}
    \centering
    \includegraphics[width=0.6\textwidth]{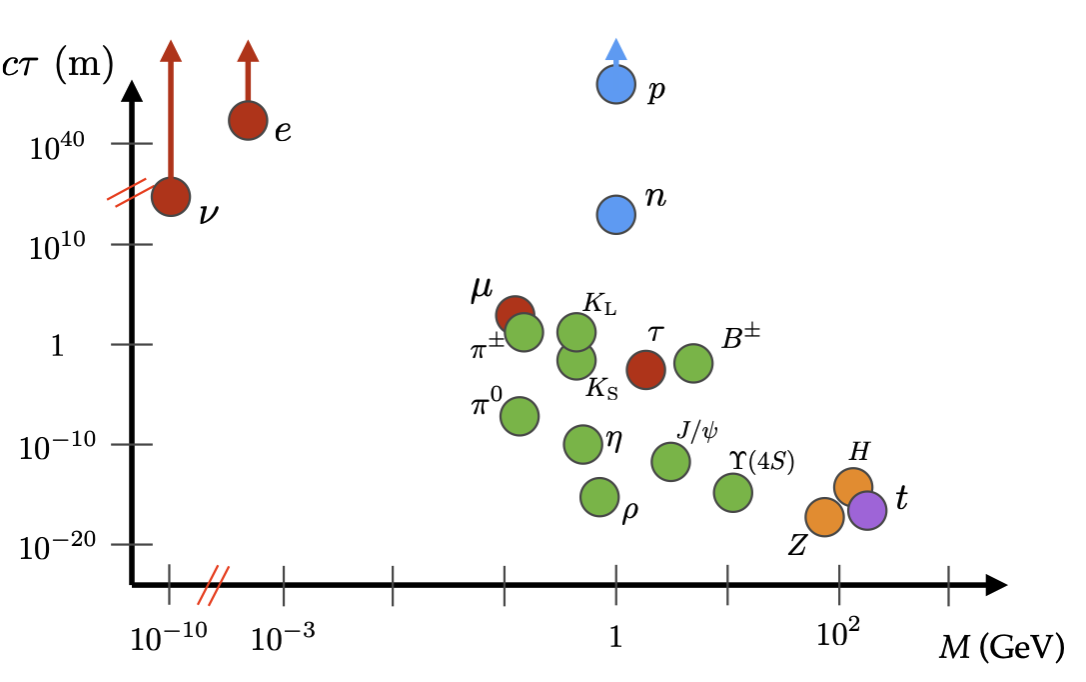}
    \caption{The $c\tau$ of SM particles versus their mass. Plot by Brian Shuve and reproduced in~\cite{Alimena:2019zri}.}
    \label{fig:LLP_plot}
\end{figure}

Particles with long lifetimes are an important opportunity in the search for new phenomena. The Standard Model gave us many examples, spanning a wide range of lifetimes (See Fig.~\ref{fig:LLP_plot})\footnote{Results are often presented as a function of $c\tau$, where $c$ is the speed of light and $\tau$ is the proper lifetime.} Long-lived particles exist because the SM is equipped with an approximate symmetry (flavor) and a hierarchy of scales (QCD vs electroweak), which results in some particles being vastly more long-lived than others, even if their masses are similar. Approximate symmetries and multiple scales are also ingredients in most BSM models, including supersymmetry (both R-parity conserving and violating), extended neutrino sectors and models of dark matter and/or baryogenesis~\cite{Alimena:2019zri}. As such, new long-lived particles (LLPs) are a \emph{generic} signature of BSM physics and appear in many models. Alternatively, one may also make a bottom-up argument in some cases: if light ($\lesssim$ 1 GeV) new particles exist, they must necessarily be very weakly coupled for phenomenological reasons. As a result, they are often long-lived and excellent candidates for auxiliary LLP detectors, as explained in Section~\ref{sec:dedicateddetectors}.

When produced at the LHC, LLPs have distinct, unconventional experimental signatures: they can either decay far from the primary proton-proton interaction but within a detector, such as ATLAS or CMS, or even completely pass through the detector before decaying. Some specific examples of LLP signatures include displaced and delayed leptons, photons, and jets; disappearing tracks; and nonstandard tracks produced by monopoles, quirks or heavy stable charged particles (HSCPs). Standard triggers, object reconstruction, and background estimation are usually inadequate for LLP searches because they are designed for promptly decaying particles, and custom techniques are often needed to analyze the data. The challenging background environment near the LHC collision points moreover has motivated a number of (proposals for) dedicated LLP detectors. Such auxiliary detectors can involve a large amount of shielding and therefore are particularly well motivated for light LLPs. Some designs can also access the ultra-high rapidity regime, which is powerful for models where the LLPs are primarily produced within the beamline.  

The content of this section has strong synergies with other sections of this report, including the New Bosons discussion in Section~\ref{Sec:NB} and the New Fermions discussion in Section~\ref{Sec:NF}, as, for example, heavy neutral leptons and axion-like particles could be long-lived. LLPs also have synergies with the Rare Processes Frontier, especially RF06~\cite{Gori:2022vri}, which also has close ties to heavy neutral leptons, dark photons and light hidden sectors more generally, as well as the opportunity to explore external detectors at the LHC.

In Section \ref{sub:stategiesRandD}, we discuss strategies and detector R\&D that are particularly relevant for LLPs, followed by a brief overview of the (proposed) dedicated LLP detectors in Section~\ref{sec:dedicateddetectors}. In Section \ref{sub:signaturesAndModels}, we discuss the projected results for different LLP signatures and models. Throughout these sections, we mention only a few examples and encourage the reader to refer to the white papers and other citations for more complete information.

\subsection{Strategies and detector R\&D} \label{sub:stategiesRandD}

As the high-energy particle physics community and particularly the Snowmass community begins to design future detectors, it is important to keep the many, varied LLP signatures in mind, lest we design new detectors that are biased against them. For example, overly-aggressive filtering can introduce biases that limit the acceptance for displaced tracks~\cite{Jindariani:2022gxj}. At the same time, we can develop technologies, such as dedicated trigger algorithms~\cite{Alimena:2021mdu}, displaced tracking algorithms~\cite{Kotwal:2019zia}, and timing detectors~\cite{CMS:2667167,Chekanov:2020xco}, to explicitly reconstruct and identify LLPs. We are also designing detectors dedicated to searching for LLPs, which are meant to fill in the gaps in sensitivity of ATLAS, CMS and LHCb  (see Section~\ref{sec:dedicateddetectors}).

In addition, the collider environments themselves can play a large role in the LLP sensitivity. Of course, factors like the achievable integrated luminosity will help determine the LLP reach. But in addition, careful studies of beam-induced backgrounds will be necessary in order to reduce and/or quantify these background contributions without removing possible LLP signals. Other important factors to consider for LLPs include the time between collisions and how that interplays with the detector readout, as well as the size of the beamspot, the amount of pileup and the material budget of the detector areas closest to the interaction point. Many of the detector topics discussed below need to be studied explicitly in the foreseen collider environment.

\subsubsection{General detector requirements}

When designing particle detectors capable of finding LLPs, there are a few general requirements to keep in mind. First, the signal geometric acceptance for LLPs is maximized for detectors which  \emph{i)} are hermetic, \emph{ii)} have a large active volume and \emph{iii)} are constructed on or near the interaction point. Second, different geometry choices that provide similar hermeticity for prompt particles can differ drastically in their ability to reconstruct particles that do not originate from the interaction point. In particular, high granularity at large radii enables better reconstruction efficiency of displaced tracks and vertices, and helps to distinguish them from beam-induced and noncollision backgrounds. The choice of magnetic field and the material budget of the magnet also have important implications for reconstruction capabilities at large radii. In addition, the particle reconstruction techniques, both at the trigger and the offline reconstruction level, must not prevent LLP searches and must be capable of reconstructing and identifying displaced particles.

In a realistic experiment, background mitigation is as important as signal efficiency, especially at a high intensity hadron machine, where trigger capabilities and even data-rate reduction have to consider the needs of LLPs. In particular, the muon chambers at ATLAS and CMS do a prioi not satisfy all the criteria in the previous paragraph, but are nevertheless some of our best subdetectors for LLPs thanks to the shielding provided by the hadronic calorimeter. A high volume, (partially) shielded subdetector system like the current muon systems would therefore play an important role in searches for LLPs at future hadron colliders. For a future $e^+e^-$ collider, on the other hand, the background yields are expected to be much lower and it is not a priori obvious whether a large radius, partially shielded detector is a net gain from the point of view of searching for LLPs: It may instead be beneficial to invest the equivalent amount of space into a larger inner detector, and restrict the muon system to the minimum required for prompt muon identification. Finally, muon colliders~\cite{MuonCollider:2022xlm} come with a new set of challenges for LLP searches, as their detectors are bombarded from both sides with ultrahigh energy electrons/positrons from the in-flight decay of the muon beam \cite{Jindariani:2022gxj,Ally:2022rgk}. It is difficult to shield the detectors from this qualitatively new beam background, but over $99\%$ background rejection can be achieved by making use of timing and angular measurements from paired layers~\cite{Ally:2022rgk}. Whether simultaneously a good signal efficiency for LLPs can be maintained needs to be studied further.
 
Detector limitations notwithstanding, it is also worth noting that dedicated reconstructions algorithms are needed to reconstruct displaced phenomena. It is already difficult to access low-level detector information and to run non-standard reconstruction on large subsets of events. This is expected to become more challenging at the HL-LHC.

\subsubsection{Tracking detectors}
\label{sec:LLP:tracking-detectors}

In the inner tracker, we need to maintain the ability to measure ionization energy loss ($dE/dx$) and time of flight. The specifications needed for these techniques will depend on e.g. the collider environment, but in general they need to be maintained because they are very important for identifying LLP signatures. For example, $dE/dx$ is essential for several searches from ATLAS, namely, a search for heavy, long-lived, charged particles with large $dE/dx$ as measured in the pixel detector~\cite{ATLAS:2022pib} and also for a search for multi-charged particles, which makes use of $dE/dx$ as measured in several inner tracker subsystems~\cite{ATLAS-CONF-2022-034}. Furthermore, $dE/dx$ in the tracker and time of flight from the muon system are the primary discriminating variables in searches for HSCPs at CMS~\cite{CMS-PAS-EXO-16-036}.
The reduced resolution in $dE/dx$ measurement of the Phase II ATLAS and CMS inner trackers won't fundamentally limit these searches, as shown in Ref.~\cite{Chen:2017yzr}.

Furthermore, we need to maintain the ability to perform precision tracking at large radii of the tracker, for example by maintaining good granularity. The minimum radius of the first layers of the tracker and the number of small radius layers are other important factors to consider, in order to eliminate any gaps between ``large'' and ``small'' displacements and to maximize the sensitivity for missing energy and disappearing track signatures.

We should also consider the tracking capabilities needed for LLPs that live long enough to decay within the muon system. Here, the capability for nonpointing and displaced particle reconstruction is paramount \cite{CERN-LHCC-2017-012}, as well as a high granularity for vertexing. For example, displaced and nonpointing muons reconstructed only in the muon system were used to improve the sensitivity at large lifetimes in searches for displaced muons at CMS and ATLAS~\cite{CMS:2015pca,CMS:2022qej,CERN-LHCC-2017-012,ATLAS:2018rjc} and in a search for stopped particles that decay to muons at CMS~\cite{CMS:2017kku}. We should also make careful assessment of the detector volume needed, and consider the ability to distinguish muons from cosmic rays with displaced muons coming from collider-produced LLPs.

Light LLPs in particular can be produced strongly in the forward direction, as is the case at LHCb or the various beam dump facilities covered in the the Rare Processes Frontier. In such cases, a disc-shaped tracker, such as the  LHCb Vertex Locator (VELO), is highly advantageous. The VELO moreover has no magnetic field, which further contributes to its excellent vertex resolution.  By combining this capability with LHCb's new, triggerless readout, LHCb's already impressive sensitivity for light LLP's will be further enhanced going forward \cite{Craik:2022riw}. See \cite{Borsato:2021aum} for comprehensive summary of LHCb's physics potential for LLPs and BSM physics more broadly during Run 3 and beyond.

Finally, at future lepton colliders, gas-based trackers such as wire and Time Projection Chambers are being considered, which could have excellent capabilities for reconstructing displaced and kinked tracks.

\subsubsection{Calorimeters}

For the calorimeters, several factors are important to keep in mind for LLPs. First of all, the segmentation and geometry choices of the calorimeters have a large impact on LLP sensitivity~\cite{ATLAS:2022zhj,CMS:2020iwv,LHCb:2017xxn}. In addition, we should also consider the stopping power of the calorimeters and other highly dense detector material, such as the return yoke of the CMS muon system. This stopping power is important for searches that consider an LLP that could be sufficiently heavy and ionizing that it comes to a stop in the detector material~\cite{CMS:2017kku,ATLAS:2021mdj}. Then, the readout and powering during non-colliding bunches needs to be considered to maximize the sensitivity to stopped particle decays. We should also consider the ability to look for converted photons from LLP decays in the calorimeters, as well as the need for pileup and beam-induced background suppression.

\subsubsection{Timing detectors}
Timing information in the calorimeters is already been used by ATLAS and CMS in searches such as delayed photons \cite{CMS:2019zxa,ATLAS-CONF-2022-051} and delayed jets \cite{CMS:2019qjk}.
Dedicated timing layers with a few tens of picosecond resolution will however greatly improve the potential for LLPs during the HL-LHC phase, for example with 4-dimensional vertexing, as well as for pileup and background suppression and particle identification~\cite{Chekanov:2020xco}. CMS's planned MIP Timing Detector~\cite{CMS:2667167} and High-Granularity Calorimeter~\cite{CERN-LHCC-2017-023} will both have precision timing capabilities and have been shown to increase the coverage and sensitivity to HSCPs, displaced and delayed photons, as well as displaced and delayed jets~\cite{Liu:2018wte,CMS:2019qjk,ATLAS:2019mfr,Liu:2020vur}. For example, assuming a 30 ps timing resolution will be available from the MIP Timing Detector, the reach of displaced photons in a GMSB scenario will be significantly extended for $0.1 < c\tau < 1000$ cm~\cite{CMS:2667167}. In addition, ATLAS plans to focus on the challenging forward region with their High-Granularity Timing Detector~\cite{CERN-LHCC-2020-007}, which may also have physics potential for LLPs. Generally, the usage of precision timing information and correlating them with different subdetectors could enhanced sensitivities to a broad class of long-lived particles signature~\cite{Liu:2018wte,Liu:2020vur,Chiu:2021sgs}.

\subsubsection{Triggers}

Being able to trigger on the unusual signatures of LLPs is especially crucial, because there is no search without the data collected by the trigger. At the trigger level, in ATLAS and CMS, tracking has only been available so far at the high-level trigger (HLT). The HLT is latency limited, and as such only simplified tracking algorithms are available, which sometimes only cover small regions of the detector, require an associated object in the calorimeter or muon system, or come with more stringent kinematic requirements than offline tracking. Displaced tracking at the HLT has been exploited, for example with displaced jet triggers in CMS~\cite{CMS:2020iwv}, but this has clear limitations. 
Further improvement to track reconstruction will allow the experiments to take advantage of triggers at the HLT based on the inner tracking systems and sensitive to LLPs already during Run~3~\cite{ATLAS:2017zsd,url:atlas_run3_lrt}.

CMS is also investing heavily in tracking in the Level 1 (L1) trigger for the Phase II Upgrades, that could allow to qualitatively expand the reach for LLPs. To take full advantage of these new capabilities, it is important that these developments take note of the specific requirements for such signatures, such as the ability to reconstruct tracks with non-zero impact parameters.
CMS plans to address this challenge in Phase II by implementing double tracking layers, which can independently from each other reconstruct track ``stubs''. A rudimentary $p_T$ measurement of these stubs will allow them to reduce the combinatorics sufficiently to render track finding feasible in the L1 trigger \cite{collaboration:2283192}. Though this method is aimed at prompt tracks, it was shown that it has the potential to find displaced tracks as well, if certain software and latency requirements are met \cite{Gershtein:2017tsv,CERN-LHCC-2020-004}. The full potential of this technique can be unlocked if also displaced vertices can be reconstructed \cite{Gershtein:2019dhy,Hook:2019qoh,Gershtein:2020mwi}, though the feasibility of this is more speculative at the moment.

A general, truth-level study of track-based triggers at the HL-LHC~\cite{DiPetrillo:2022nsz} was performed for 3 unconventional signatures: soft unclustered energy patterns (SUEPs) that produce high multiplicity, low transverse momenta tracks; GMSB long-lived staus that produce displaced leptons or anomalous prompt tracks; and Higgs portal long-lived scalars that decay to displaced hadrons. The trigger efficiency was measured as a function of the baseline parameters of a track trigger, including transverse momentum and impact parameters. This study concludes that, in order to address all of the unconventional signatures simultaneously, the best track-trigger design would keep the transverse momentum threshold as low as possible for prompt tracks, and for increasing transverse momentum values, the allowed transverse impact parameter range can be increased. In addition, it was shown that unsupervised machine learning algorithms implemented on FPGAs could be beneficial for triggering on disappearing tracks \cite{Kotwal:2019zia}.

In addition, dedicated LLP triggers have been developed for calorimeter and muon-system signatures. For example, displaced jets were identified in the ATLAS calorimeters with a trigger that featured a lack of tracks associated to the jet and a small fraction of the energy deposited in the electromagnetic calorimeter~\cite{ATLAS:2019qrr,ATLAS:2021tnq}. In addition, calorimeter timing can also be used to identify nonprompt jets~\cite{CMS:2019qjk}. Decays of neutral LLPs in the muon system can be triggered on in ATLAS with the L1 trigger~\cite{ATLAS:2013bsk}, and dedicated displaced muon L1 triggers are being developed in CMS as well. Dedicated triggers were also developed in ATLAS and CMS for calorimeter and muon signatures that are out-of-time with respect to collisions, in order to trigger on the decays of stopped LLPs~\cite{CMS:2017kku,ATLAS:2021mdj}. Attention to specific requirements on jet-based and calorimeter-based triggers need to remain a central point of attention for any future development, where even small changes that are implemented to reduce trigger rates with negligible loss of efficiency for promptly-produced objects can have disastrous consequences in the case of LLPs.

As of Run 3, LHCb will perform its full event reconstruction online \cite{CERN-LHCC-2014-016}, bypassing most of the trigger related challenges laid out above. This is a major contributor to its excellent prospects for discovering or constraining light LLPs, in particular dark photons \cite{Craik:2022riw}.

With future colliders, particularly future lepton colliders, triggers may not be needed at all, and instead, all of the data may be able to be saved. This option is very advantageous for LLPs, which are rare signatures that could otherwise be missed.

\subsubsection{Alternative data taking strategies}

\label{sec:scouting}
Since the trigger systems are typically limited in bandwidth and data storage rather than in computing power, a very intriguing option is to perform a large piece of the analysis online and only write a very reduced data format to tape. This idea is known as the turbo stream at LHCb~\cite{Aaij:2019uij}, data scouting at CMS~\cite{CMS:2016ltu} and trigger-level analysis at ATLAS~\cite{ATLAS:2018qto}. LHCb has successfully deployed this concept for charm physics in particular and plans to further build on this in Run 3, by performing full online event reconstruction at 40 MHz. This is expected to provide a big gain in sensitivity to displaced signatures of dark photons and dark scalars in particular~\cite{Ilten:2015hya,Ilten:2016tkc}. Furthermore, CMS plans to extend the scouting program to the L1 trigger for the HL-LHC~\cite{Badaro:2020kkb}.

Until recently, ATLAS and CMS had utilized the scouting technique only for prompt signatures, most notably low mass dijet resonances. A recent CMS data scouting search that used Run 2 data showed excellent sensitivity for displaced dimuon pairs, which can have transverse momenta as soft as a few GeV~\cite{CMS:2021sch}. As a result, it is an excellent probe for dark scalar models as well as certain exotic Higgs decays. The power of these (projected) searches provides strong motivation to preserve and expand the data scouting program into Run 3 and the HL-LHC.

Data parking~\cite{Mukherjee:2019anz} is a complementary strategy that makes use of triggers with lower thresholds and saves the full event content, yet reconstructs physics objects only during shutdowns when extra computing resources become available. Thus, data parking is another potential avenue to access low transverse momentum, displaced objects. For example, during Run 2, CMS was able to park nearly 10 billion b-hadron pairs by triggering on events with slightly displaced muons. These data could be used to search for LLPs in unbiased b-hadron decays. CMS plans to extend data parking to other scenarios as well.

Finally, given that pileup is a major source of background for LLP searches, a dedicated low pileup run may be beneficial for some searches for light LLPs. Such a low pileup run may also be well motivated for precision Standard Model measurements, in particular the W boson mass.

\subsection{Dedicated detectors for LLPs\label{sec:dedicateddetectors}}

General-purpose particle detectors like ATLAS and CMS are advantageous when searching for LLPs because these hermetic detectors are close to the collision point, providing a large acceptance. In addition, they are relatively large, with radii of several meters, which provides sensitivity to a wide range of lifetimes, and since general-purpose detectors have multiple detector subsystems, they are sensitive to a wide range of experimental signatures. However, these general-purpose detectors are limited by the large amount of prompt and long-lived SM backgrounds and by the challenge of triggering on and reconstructing these unusual signatures. They are thus largely sensitive to relatively heavy LLPs, with some important exceptions (see Section~\ref{sub:lightNeutralLLPs}). 

\begin{figure}
    \centering
    \includegraphics[width=0.48\textwidth]{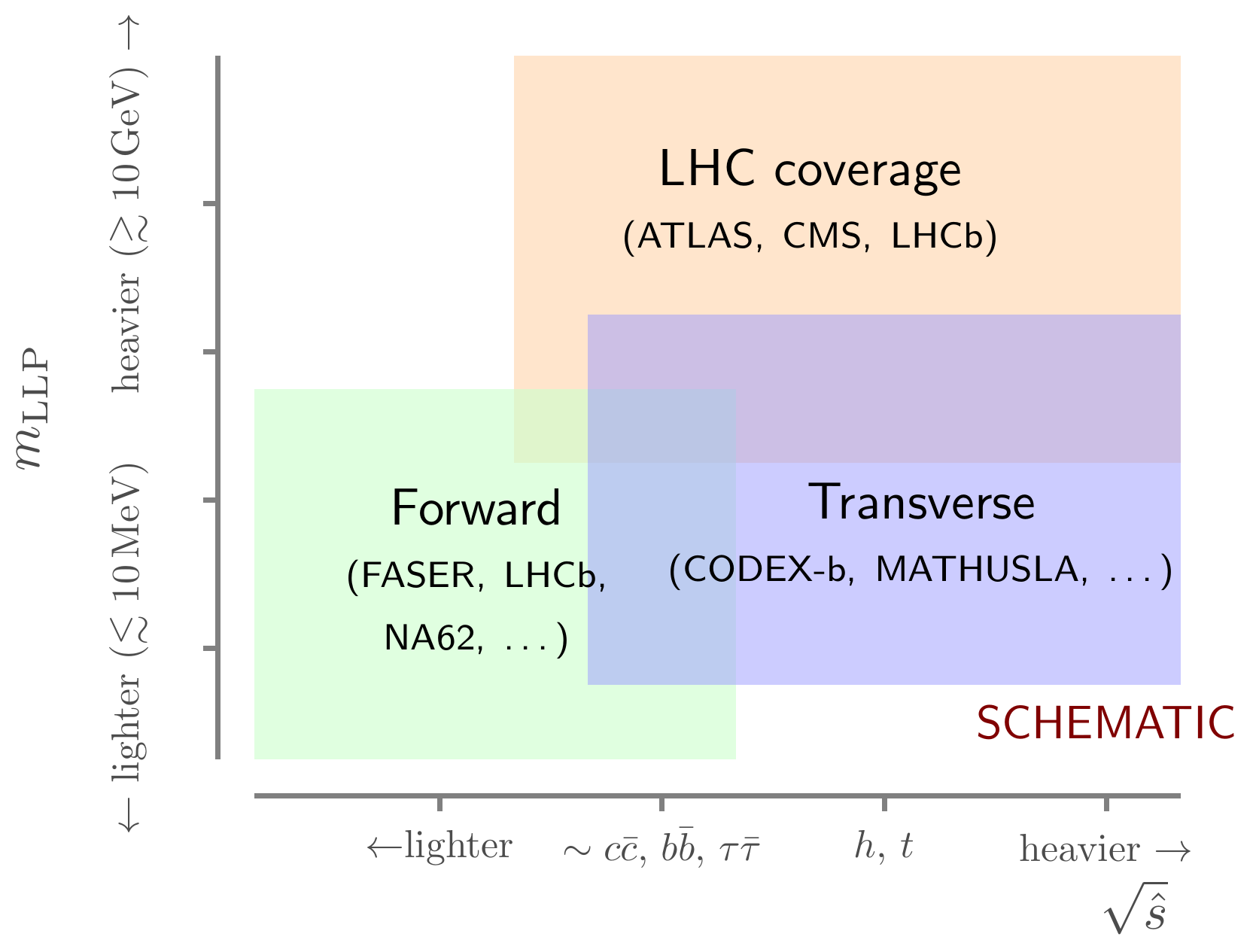} \hfill
    \includegraphics[width=0.49\textwidth]{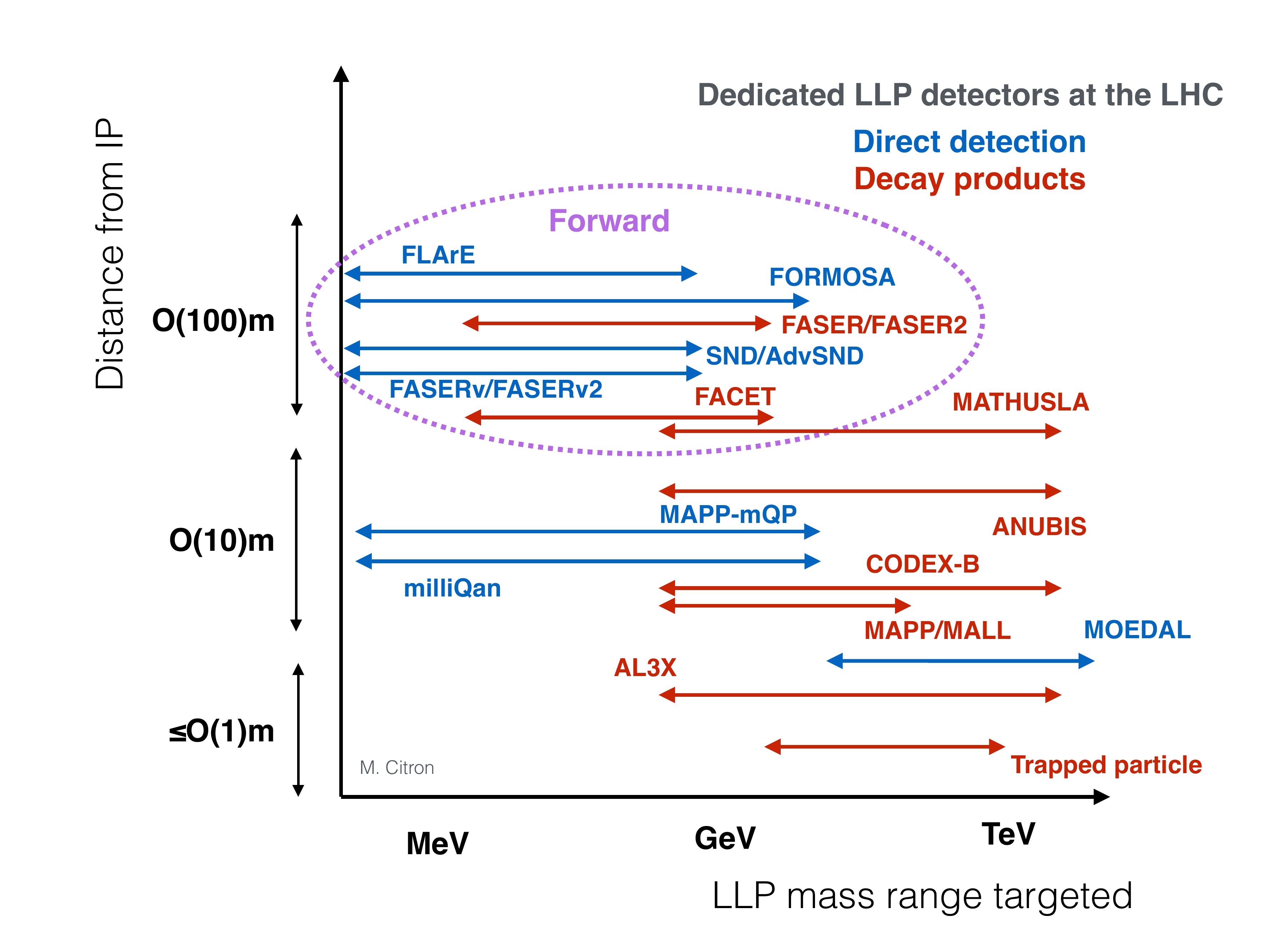}
    \caption{\textbf{Left:} Complementarity of different experiments searching for LLPs~\cite{Aielli:2019ivi}. \textbf{Right:} Overview of (proposed) dedicated LLP detectors at the LHC. Plot by Matthew Citron.}
    \label{fig:dedicatedLLPdetectors}
\end{figure}

Detectors that are, on the other hand, designed specifically to search for (light) LLPs can circumvent many of these limitations. For example, SM backgrounds coming from particle collisions can be mitigated by rock or dedicated shielding. In addition, the trigger can usually be relatively simple or in some cases, not needed at all, and the reconstruction is designed for specific LLP signatures. In short, these dedicated detectors are designed and positioned optimally for the targeted LLP signature. The rapidity range of the detector is particularly important: Forward detectors such as FASER excel at searching for LLP's produced at low $\sqrt{\hat s}$, for example in $\pi^0$ decays. Central detectors on the other hand are needed to probe particles produced at relatively high invariant mass, for instance in exotic Higgs decays. The interplay between both types of detectors and their complementarity with ATLAS, CMS and LHCb is illustrated schematically in the left hand panel of Figure~\ref{fig:dedicatedLLPdetectors}.
As a result, several different dedicated detectors are also needed, many of which have benefited from close engagement with the theory community \cite{Essig:2022yzw}. The right hand panel of Figure \ref{fig:dedicatedLLPdetectors} gives an overview of the proposed dedicated LLP detectors, which we will summarize below.

\subsubsection{Forward detectors}
\label{Sub:ForwardDetectors}

The FASER experiment~\cite{FASER:2018bac} is located 480 m downstream of ATLAS, covering the approximate rapidity range of $\eta \gtrsim 9$. This extreme forward location allows FASER to take advantage of the enormous flux of forward pions from the ATLAS interaction point. If a fraction of these $\pi^0$ decay to an exotic particle, such a particle is almost guaranteed to have a macroscopic proper lifetime and its enormous boost can let it survive long enough to decay within the FASER acceptance.  FASER is ready to take data in the upcoming LHC run, and is expected to have sensitivity to hidden sector particles such as dark photons.

To upgrade the FASER detector and allow for other forward experiments, the construction of a dedicated forward physics facility (FPF) has been proposed \cite{Feng:2022inv,Anchordoqui:2021ghd}. The corresponding studies have zoomed in on two preferred options: \emph{1)} a new purpose-built facility, approximately 617–682 m west of the ATLAS IP, and \emph{2)} constructing alcoves extending the existing UJ12 cavern, which is 480–521 m east of the ATLAS IP. (See \cite{Feng:2022inv} for comparison of the required civil engineering between both locations.) The FPF would house a number of experiments, probing a range of SM and BSM physics. In particular, it would house FASER2, which would have an acceptance to LLPs produced in exotic $\pi^0$ decays that is about 20 times larger than that of FASER. The larger rapidity range of FASER2 would also enable it to probe LLPs originating from exotic decays of bottom and charm mesons. As a result, FASER2 would have extensive acceptance for dark photon and dark scalar models, light axion-like particles, and heavy neutral leptons. Its reach for heavy neutral leptons is shown in  Fig.~\ref{fig:HNLtype1}; we refer to the report of the \emph{``Dark Sector Studies at High Intensities''} working group in the rare processes frontier for a quantitative comparison of the physics reach of FASER and FASER2 with other intensity frontier experiments for the remaining models. The FPF would also house a number of experiments targeting SM measurements of neutrino cross sections and/or parton distribution functions at very small $x$ (FASER$\nu$, advSND and FLArE), which are covered in the report of the \emph{rare processes frontier}. In contrast to FASER2, these detectors are high density detectors, to increase the chances of a neutrino scatter within the detector volume. As a result, these detectors can also be sensitive to certain dark matter candidates that are part of a strongly coupled dark sector. Finally, the FPF would host FORMOSA, a dedicated experiment for milicharged particles, as discussed in Section~\ref{sec:chargedexternal}. 

There are two proposed forward detectors which do not make use of the proposed FPF. The first is the Forward-Aperture CMS ExTension (FACET)~\cite{Cerci:2021nlb}, a proposed extension of the CMS detector that is designed to search for very forward decays of LLPs such as dark photons, heavy neutral leptons, axion-like particles, and dark Higgs bosons. It would be constructed around the LHC beam pipe and would be sensitive to any particle that can penetrate at least 50~m of magnetized iron and decay in an 18~m long, 1~m diameter vacuum pipe. Due to its proximity to the beamline, its expected backgrounds are highly non-trivial and deserve further study. The MoEDAL-MAPP detector \cite{Mitsou:2022vka} would consist out of a nested boxes of scintillator hodoscope detectors, located in the forward region at the LHCb interaction point, targeting similar parameter space as FACET. These options are further explored in the Snowmass Rare Processes and Precision Frontier Report~\cite{sm21:rf06report}.

\subsubsection{Central detectors}

The Massive Timing Hodoscope for Ultra Stable neutraL pArticles (MATHUSLA)~\cite{MATHUSLA:2022sze} is a large-scale (100x100 $\mathrm{m}^2$) surface detector above the CMS interaction point, which would have excellent acceptance to the decay of neutral LLPs. The main challenge for this experiment is to instrument a very large volume cost effectively, which the collaboration aims to address with extruded scintillator bars. MATHUSLA has the potential supply an L1 trigger signal to CMS, enabling a suite of joint searches between both experiments.

The COmpact Detector for EXotics at LHCb (CODEX-b)~\cite{Aielli:2022awh} follows a similar concept as MATHUSLA but would be installed in the DELPHI/UXA cavern next to LHCb’s interaction point. This allows for a much more compact design and higher granularity detectors, at the expense of a reduction in exposure. CODEX-b requires a few meters of additional shielding, equipped with an active muon veto. CODEX-b has the potential supply trigger signal to LHCb, enabling a suite of joint searches between both experiments. A 2 m $\times$ 2 m $\times$ 2 m demonstrator detector (CODEX-$\beta$) will be installed in the course of 2023.

The AL3X proposal \cite{Gligorov:2018vkc} aimed to reuse much of the ALICE infrastructure, should their operations conclude after Run 4 of the LHC. In its current form, the proposal is however not compatible with the ALICE upgrade plans.  The ANUBIS proposal \cite{Bauer:2019vqk} aimed to suspend a large area RPC detector in one of the service shaft above ATLAS or CMS. In its current form it faces non-trivial engineering and background suppression challenges.

For the performance of these detector for LLP produced in the decay of heavy particle, e.g. the SM Higgs, we refer to Section~\ref{sub:lightNeutralLLPs}. For their performance for LLPs produced in exotic meson decays, we refer to the report of the \emph{``Dark Sector Studies at High Intensities''} working group in the rare processes frontier~\cite{sm21:rf06report}.

\subsubsection{Detectors for charged LLPs\label{sec:chargedexternal}}

The MoEDAL experiment \cite{MoEDAL:2014ttp,Pinfold:2022quc} is a largely passive detector, designed to search for magnetic monopoles and highly ionizing charged particles. It has been operation at the LHCb interaction point since Run 1; see \cite{Mitsou:2022vka} for a recent overview of results. In addition to Drell-Yan production in pp collisions, magnetic monopoles may also be produced through the Schwinger process in heavy ion collisions \cite{dEnterria:2022sut,Gould:2017zwi}.  The MoEDAL collaboration is planning for an extension, MAPP-mQP, which targets millicharged particles.

The milliQan experiment~\cite{milliQan:2021lne} will search for particles that have a small fractional electron charge (millicharge) in the shielded environment of the PX56 drainage and observation gallery located located above the CMS interaction point. The experiment consists of a detector with four layers each composed of 16 $5\times5\times60$~cm scintillator "bars" read out with PMTs, as well as a another four layer detector with 12 larger ($40 \times 60 \times 5$~cm) "slabs" in each layer that adds sensitivity to higher charges. The collaboration has already installed and operated a small fraction of the full detector (“milliQan demonstrator”), which enabled them to exclude some previously unconstrained parameter space for millicharged particles \cite{PhysRevD.102.032002}. The funding for the full experiment has been secured, and installation will be complete by June 2022. Should the FPF be constructed, it could house the FORMOSA experiment, which would use the same detector concept as the miliQan experiment \cite{Foroughi-Abari:2020qar}. 

\subsubsection{Detectors at future colliders}

There are a few general considerations to take into account when designing dedicated LLP detectors at future colliders. New facilities could have pre-excavated, shielded experimental halls in forward and central regions, already anticipating dedicated detectors. Another point to consider is whether a new LLP experiment should have a trigger and data acquisition system integrated with the nearby general-purpose detector, which would facilitate combined analyses and tracing the entire flight path of the LLP, or whether such systems should be independent from those of the general-purpose detector, which would allow for continuous data-taking and maximum acceptance.

The possibility of auxiliary LLP detectors at future facilities has already been investigated for future lepton colliders \cite{Wang:2019xvx,Tian:2022rsi,Chrzaszcz:2020emg} as well as future hadron colliders \cite{Bhattacherjee:2021rml}.

\subsection{Signatures \& models} \label{sub:signaturesAndModels}

\begin{figure}
    \centering
    \includegraphics[width=0.6\textwidth]{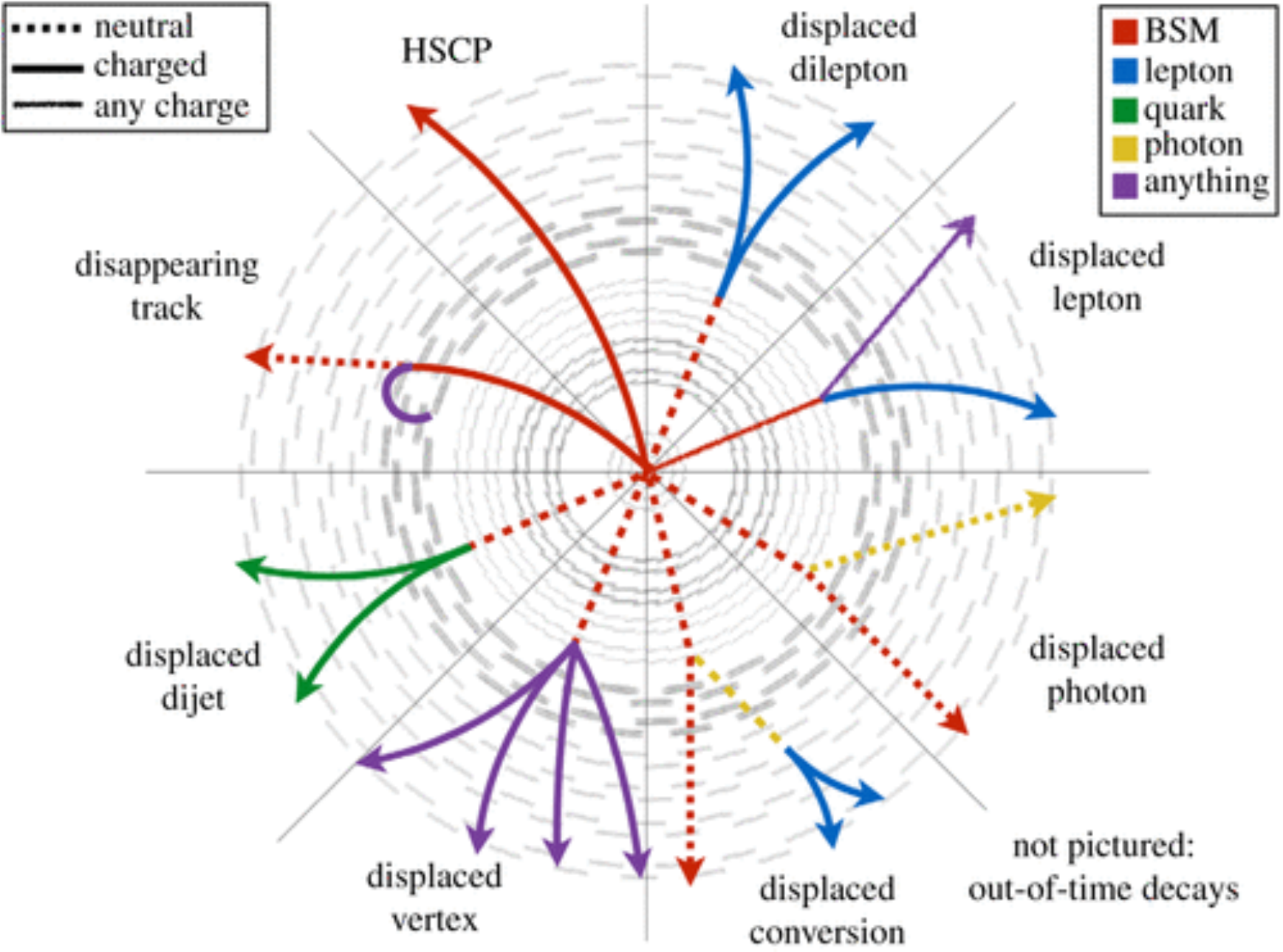}
    \caption{Overview of representative LLP signatures at the LHC. Plot by Jamie Antonelli.}
    \label{fig:LLP_overview}
\end{figure}

The space of LLP signatures is very rich and complicated, ranging from exotic-looking tracks to heavy stable charged particles to various types of displaced vertices, a subset of which is illustrated Fig.~\ref{fig:LLP_overview}. A comprehensive overview is beyond the scope of this summary, and we instead highlight an incomplete set of examples, which are chosen because of their strong theory motivation, their usefulness to benchmark experiments and colliders, or their unique nature.

Concretely, we discuss the disappearing track signature in Section~\ref{sub:disappearingTracks}, light neutral LLPs in Section~\ref{sub:lightNeutralLLPs}, and heavy neutral LLPs in Section~\ref{sub:heavyNeutralLLPs}. Here, ``light'' refers to LLPs with masses such that they are difficult to trigger on at the LHC, whereas ``heavy'' LLPs are relatively easy to trigger on at the LHC, due to their high momenta.

\subsubsection{Charged LLPs} \label{sub:disappearingTracks}

LLPs that are electrically charged can be produced by many different models, and they produce several signatures that can be directly detected at current and future experiments.
For example, if the charged LLP decays within the detector, the LLP could produce a disappearing track signature if it decays to neutral and/or very soft particles that cannot be reconstructed.

\begin{figure}[htb]
\begin{center}
\includegraphics[width=0.9\hsize]{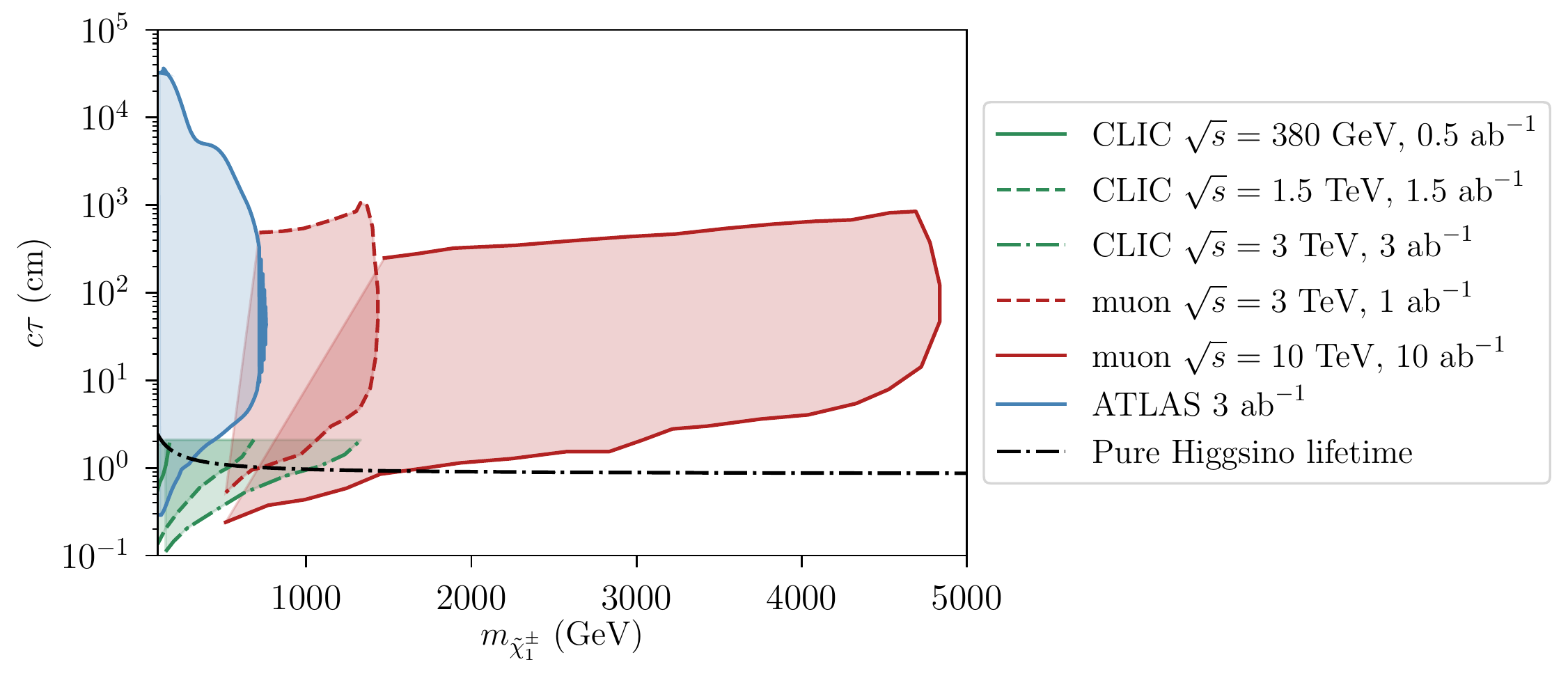}
\end{center}
\caption{Projected reach of disappearing track signatures in the chargino mass--$c\tau$ plane at 95\% CL exclusion from CLIC (green curves)~\cite{Klamka:2790402}, a muon collider~\cite{Capdevilla:2021fmj}, and ATLAS at the HL-LHC (blue curve)~\cite{ATL-PHYS-PUB-2022-018}.
}
\label{fig:disappearingTracks}
\end{figure}

 The latter scenario is very natural for electroweak multiplets, either in the context of supersymmetry or minimal WIMP dark matter models. An electroweak doublet (``pure Higgsino'') and electroweak triplet (``pure wino'') each contain an electrically charged state and respectively two and one neutral state(s)\footnote{Strictly speaking the pure Higgsino as a dark matter candidate has long been excluded by direct detection experiments. However, even a very small correction from a higher dimensional operator suffices to make the dominant $Z$ exchange kinematically inaccessible for direct detection experiments, with no impact on the collider searches.}. In the absence of higher dimensional operators, the (lightest) neutral state will always be slightly lighter than the charged state due to a one-loop electroweak corrections. For the pure Higgsino and pure wino cases, this natural splitting is respectively 344~MeV and 166~MeV. This implies that the $\chi^\pm_1\to \chi^0_1 \pi^\pm$ decay is extremely phase space suppressed. Taking the pure Higgsino case as an example, its predicted proper lifetime \cite{Thomas:1998wy} is shown by the dashed black curve in Fig.~\ref{fig:disappearingTracks}, along with the projected 95\% exclusion limits for the HL-LHC~\cite{ATL-PHYS-PUB-2022-018}, CLIC~\cite{Klamka:2790402} and two benchmark muon colliders~\cite{Capdevilla:2021fmj}. (The assumptions for these studies vary somewhat; we refer to the references cited above for details.) The presence of other particles and/or higher dimensional operators can modify the splitting between the charged and neutral components and thus alter the decay width significantly. Their generic effect is to increase the splitting and therefore shorten the lifetime. To obtain longer lifetimes, an increasingly delicate cancellation against the SM contribution is needed, which can be viewed as a fine tuning.  

\begin{figure}[htb]
\begin{center}
\includegraphics[width=0.9\hsize]{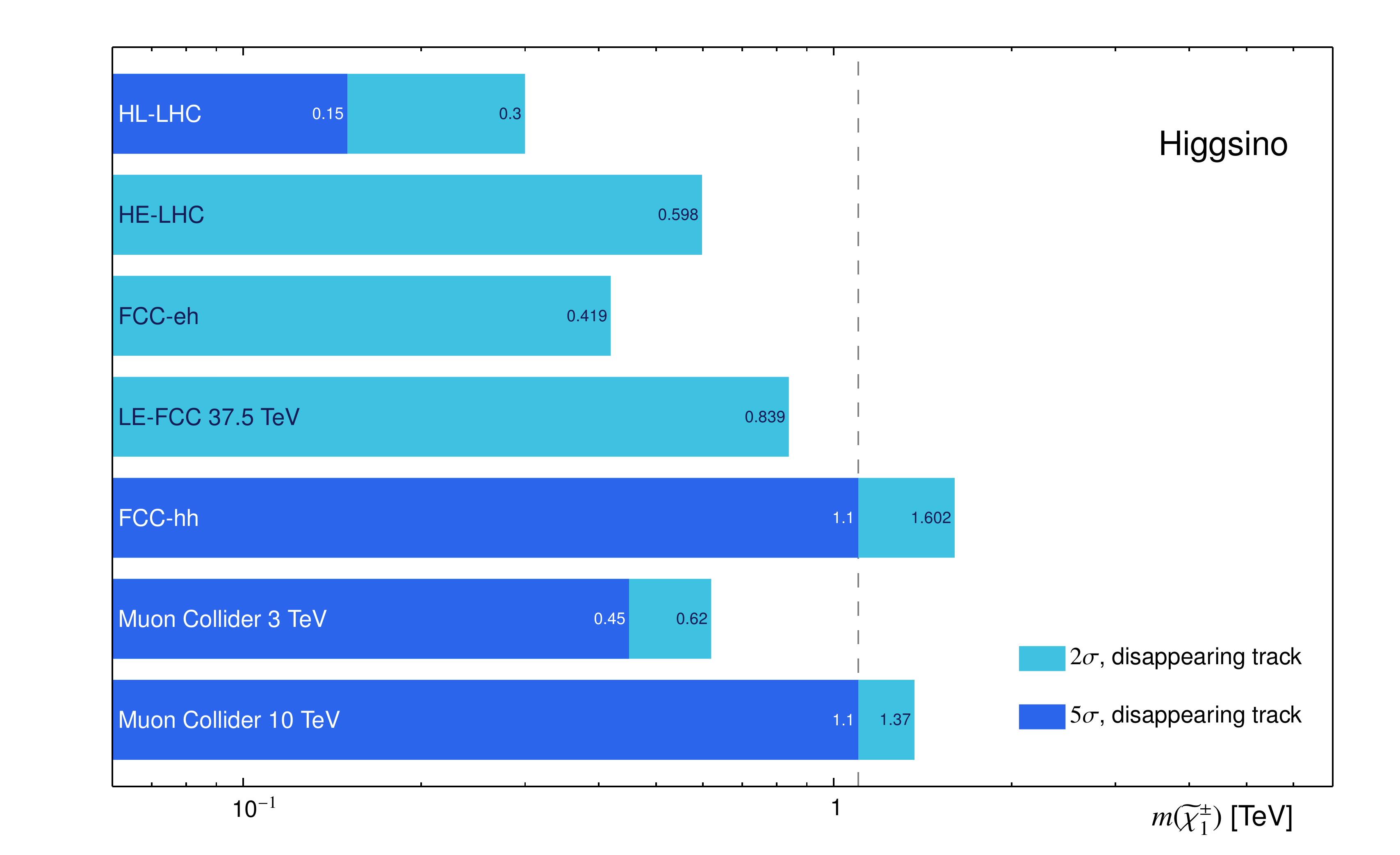}
\end{center}
\caption{Overview plot for the sensitivity of the HL-LHC~\cite{ATL-PHYS-PUB-2022-018}, HE-LHC~\cite{Han:2018wus}, LE-FCC~\cite{Strategy:2019vxc}, FCC-eh \cite{Curtin:2017bxr}, FCC-hh~\cite{Saito:2019rtg} and several high energy muon colliders~\cite{Capdevilla:2021fmj} to the pure Higgsino, assuming its natural mass splitting. Figure adapted from ~\cite{Capdevilla:2021fmj}.}
\label{fig:disappearingTracks2}
\end{figure}

Figure \ref{fig:disappearingTracks2} shows the projected reach of disappearing track signatures at the HL-LHC~\cite{ATL-PHYS-PUB-2022-018}, HE-LHC~\cite{Han:2018wus}, LE-FCC~\cite{Strategy:2019vxc}, FCC-eh \cite{Curtin:2017bxr}, FCC-hh~\cite{Saito:2019rtg}, and several high energy muon colliders~\cite{Capdevilla:2021fmj}, assuming a pure Higgsino with its natural mass splitting. Further discussion on these constraints and their implications for dark matter can be found in Section~\ref{sec:dm:wimp}.

If the charged LLP does not decay within the detector, but instead is heavy and long-lived enough, it could travel all the way through the detector before decaying, thus exhibiting an HSCP signature. HSCPs typically travel at significantly slower than the speed of light, could have significant $dE/dx$, and may have an electrical charge not equal to $\pm1e$~\cite{Fairbairn:2006gg,Kusenko:1997si,Koch:2007um}. As a result of these features, standard particle trigger and reconstruction algorithms may be unable to identify HSCPs. HSCPs are predicted for instance in split SUSY or gauge mediation scenarios, in which the (next-to-)lightest supersymmetric particle could be stable on collider lengths scales. If the (N)LSP in a squark or gluino, it will hadronize with SM quarks to form ``R-hadrons''~\cite{Kraan:2004tz,Mackeprang:2006gx}, which can carry electric charge and will interact strongly with the detector material. A slepton or chargino (N)LSP will behave like a very heavy lepton. There is a great potential for improvements in HSCP searches in the future, for example by making use of the MIP timing detector at CMS, which has been shown to greatly improve the velocity measurements of LLPs, and thus the analysis sensitivity~\cite{CMS:2667167}. Care will also be needed in future experimental designs to ensure particle identification capabilities are matched for these new physics scenarios, as for instance briefly mentioned earlier in Section~\ref{sec:LLP:tracking-detectors}. 

\subsubsection{Low mass displaced vertices} \label{sub:lightNeutralLLPs}

LLPs that are electrically neutral can produce many different signatures, as they can show up in different final states and with varying multiplicities. It is moreover useful to make a distinction between ``light''($\lesssim 100$ GeV) and ``heavy'' ($\gtrsim 100$ GeV) LLPs: light LLPs must be neutral under the SM gauge interactions, given the robust bounds from LEP II in particular. This means that they must be produced in the decay of a heavier particle, which itself must have an appreciable production cross section. Obvious candidates for the LLP's mother particle are the SM Higgs, $Z$ or $W$ bosons, as well as new, heavy particles. Light LLPs moreover often require innovative trigger strategies and/or auxiliary LLP detectors to optimally cover their parameter space. 

Canonical examples of light LLPs are heavy neutral leptons and axion-like particles, which can be produced easily in exotic $W/Z$ and exotic $Z$ decays respectively. They were covered in sections~\ref{Sec:NB} and \ref{Sec:NF}. Here we instead highlight a simplified topology where the SM Higgs decays to a pair of light LLPs ($h\to SS$), for which we consider the $S\to \mu\mu$ and $S\to$ hadrons decay modes. The former is a typical feature of models where $S$ is identified with a kinetically mixed dark photon, while the latter is generic if $S$ is a scalar field which mixes with the SM Higgs.

\begin{figure}
    \centering
    \includegraphics[width=0.9\textwidth]{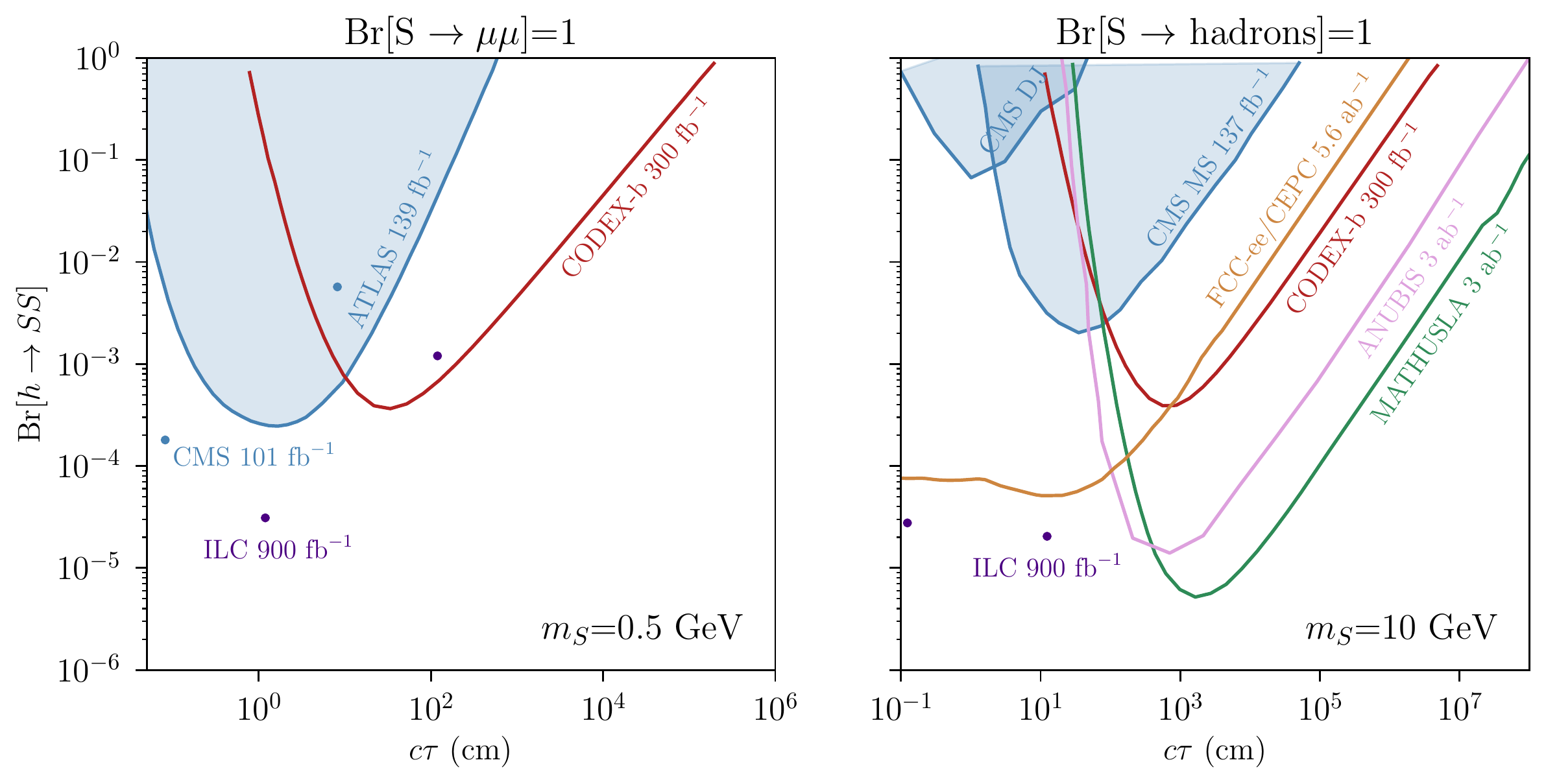}
    \caption{Sample sensitivity plots for the Higgs boson branching ratio to a pair of LLPs.
    \textbf{Left:} Assuming $S$ decays exclusively to muons, we show the existing limits from the CMS scouting analysis \cite{CMS:2021sch} and ATLAS lepton jet search \cite{ATLAS:2022bll} (see footnote \ref{footnote:leptonjet}), as well as projected limits for ILC \cite{Jeanty:2022cwr} and CODEX-b \cite{Aielli:2022awh}. CMS and ILC have sensitivity over the whole c$\tau$ range, but currently only two lifetime points are available. MATHUSLA \cite{MATHUSLA:2022sze}, ANUBIS \cite{Bauer:2019vqk},
AL3X \cite{Gligorov:2018vkc} and FACET \cite{Cerci:2021nlb} likely also have sensitivity.
    \textbf{Right:} Assuming $S$ decays exclusively to hadrons, we show the existing CMS limits from the displaced jets \cite{CMS:2020iwv} and muon system \cite{CMS:2021juv} searches, together with projected limits for the ILC \cite{Jeanty:2022cwr}, FCCee/CEPC \cite{Alipour-Fard:2018lsf}, CODEX-b \cite{Aielli:2022awh}, ANUBIS \cite{Bauer:2019vqk} and MATHUSLA \cite{MATHUSLA:2022sze}.  The ILC has sensitivity over the whole c$\tau$ range, but currently only two lifetime points are available. Its shape should be similar to that of the FCC-ee/CEPC curve. For a number of curves, the original references provided different mass points than shown here. In those cases, a slight interpolation/extrapolation was performed by rescaling the $c\tau$ axis accordingly.
AL3X \cite{Gligorov:2018vkc} and FACET \cite{Cerci:2021nlb} likely also have sensitivity.
    }
    \label{fig:higgstoLLP}
\end{figure}

In the left hand panel of Figure \ref{fig:higgstoLLP}, we show a benchmark point for $m_S=0.5$ GeV with a 100\% branching ratio to muons. At this time, the two most relevant analysis strategies on this front are searches for displaced lepton jets with the muon systems \cite{ATLAS:2022bll} and trigger-level searches for low mass (displaced) dimuon pairs \cite{CMS:2021sch}. In both cases, innovative trigger strategies are crucial to unlock the full potential of the analysis: For instance, in \cite{ATLAS:2022bll}, ATLAS supplements their existing three muon strategy with a dedicated muon scan trigger, specifically designed to find pairs of collimated muons.\footnote{An earlier HL-LHC projection for this analysis \cite{ATL-PHYS-PUB-2019-002} is not included in  Figure \ref{fig:higgstoLLP}, as its projected sensitivity has already been superseded by \cite{ATLAS:2022bll} due to improvements in the trigger. At this time, no updated HL-LHC projection is available.\label{footnote:leptonjet}} The CMS trigger-level analysis \cite{CMS:2021sch} was covered in section \ref{sec:scouting}. Auxiliary LLP detectors located at low rapidity, such as MATHUSLA \cite{MATHUSLA:2022sze}, CODEX-b \cite{Aielli:2022awh} and ANUBIS \cite{Bauer:2019vqk} can have very good sensitivity, if the low invariant mass vertex can be reconstructed. Finally, a future lepton collider like the ILC, FCC-ee, and CEPC would enjoy much lower backgrounds than the LHC and would have very competitive sensitivity, especially at low $c\tau$~\cite{Alimena:2022hfr,Bernardi:2022hny}. We show the ILC projection from \cite{Jeanty:2022cwr}, and expect that FCC-ee and CEPC would have similar sensitivity, once rescaled for the available integrated luminosity.

The right hand panel of Fig.~\ref{fig:higgstoLLP} shows a benchmark point for $m_S=10$ GeV with a 100\% branching ratio to hadrons. This signature is targeted by a number of searches, though most are applicable to slightly higher values of $m_S$. Here we show the CMS displaced jet \cite{CMS:2020iwv} and muon system \cite{CMS:2021juv} searches. The former makes use of special displaced jet trigger, while the latter currently relies on a missing momentum trigger. A dedicated trigger on clusters of activity in the muon system is however being developed. Central, auxiliary LLP detectors such as MATHUSLA \cite{MATHUSLA:2022sze}, CODEX-b \cite{Aielli:2022awh}, ANUBIS \cite{Bauer:2019vqk} and AL3X \cite{Gligorov:2018vkc} are expected to have excellent sensitivity. Due to the higher number of Higgs bosons produced at the LHC they are expected to perform similarly or better than future lepton machines for $c\tau\gtrsim 1$ m. For  $c\tau\lesssim 1$ m the ILC, FCC-ee or CEPC would likely outperform the HL-LHC due the lower backgrounds and lesser trigger challenges associated with a lepton collider~\cite{Alimena:2022hfr,Bernardi:2022hny}. Finally, the potential of auxiliary detectors at future facilities has already been investigated in the context of both high energy lepton  \cite{Chrzaszcz:2020emg} and hadron colliders~\cite{Bhattacherjee:2021rml}.

Because of its forward acceptance, LHCb is less suited for detecting LLPs from exotic Higgs decays. Its excellent vertex resolution and Kaon identification capabilities however still give it an edge in scenarios where the LLP decays predominantly to Kaons~\cite{CidVidal:2019urm}.

While we have necessarily restricted ourselves to two example benchmarks, the qualitative features carry over to other analyses and models, such as the heavy neutral leptons and axion-like particles. In particular \emph{i)} in order to maximize the potential of the LHC for light LLPs, it is essential to invest in and develop innovated trigger strategies, as well as one or more low rapidity, auxiliary LLP detectors. \emph{ii)} A future lepton collider such as the ILC, CEPC or FCC-ee is expected to outperform the sensitivity of the LHC at low $c\tau$ in particular.

\subsubsection{High mass displaced vertices} \label{sub:heavyNeutralLLPs}

There are many compelling candidates for heavier LLPs, for which triggers and background suppression are less challenging. Supersymmetry in particular provides a number of such candidates, and large hierarchies of scales are well motivated in e.g. split supersymmetry, gauge mediation or models with R-parity violation. In split supersymmetry, the scalar superpartners are assumed to be extremely heavy and thus trivially avoid flavor bounds. The gauginos and higgsinos are assumed to be near the weak scale, to facilitate grand unification. This means that the decay width of the gluino in particular is suppressed by extremely off-shell squarks, making the gluino an LLP. In models of gauge mediation, the lightest supersymmetric particle (LSP) tend to be the gravitino. The next-to-lightest supersymmetric particle (NLSP) is typically a stau or neutralino, which decays to SM states plus a gravitino. This decay is suppressed by the scale of supersymmetry breaking, which is typically many orders of magnitude larger than the weak scale. Finally, in models where the gravitino is not the LSP, the introduction of R-parity violation renders the LSP unstable, whether it is a neutralino or a sfermion. Since at least some of the R-parity violating coupling are expected to be very small to avoid bounds on proton decay, a macroscopic lifetime for the LSP is well motivated. For all these signatures, one clearly benefits from the highest possible collider energies. Naturally, colored states (gluinos, squarks) are better probed with a high energy hadron machine, while lepton colliders are more sensitive to electroweak particles (electroweakinos, sleptons). 

\begin{figure}
    \centering
    \includegraphics[width=0.45\textwidth]{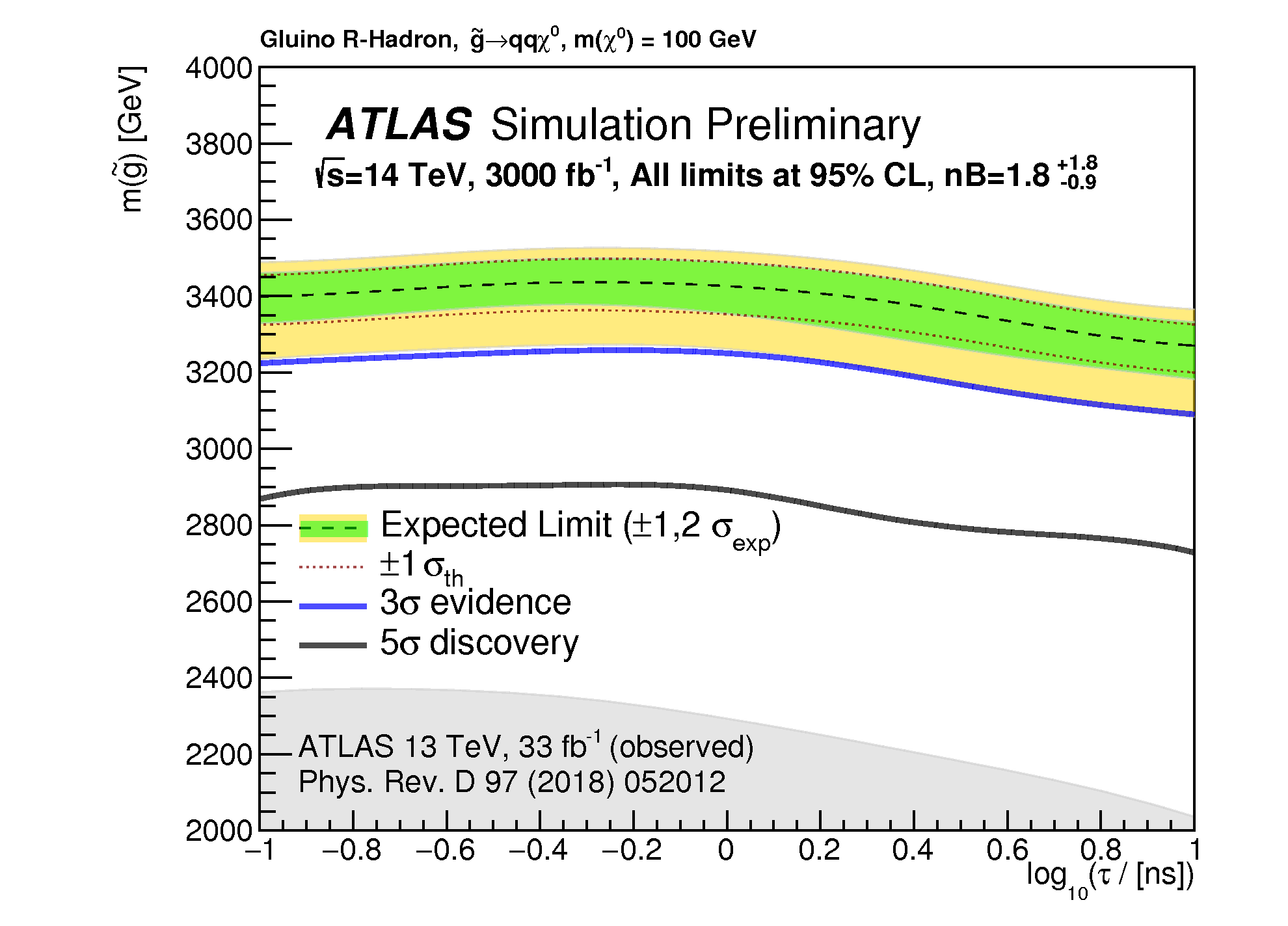}
    \includegraphics[width=0.45\textwidth]{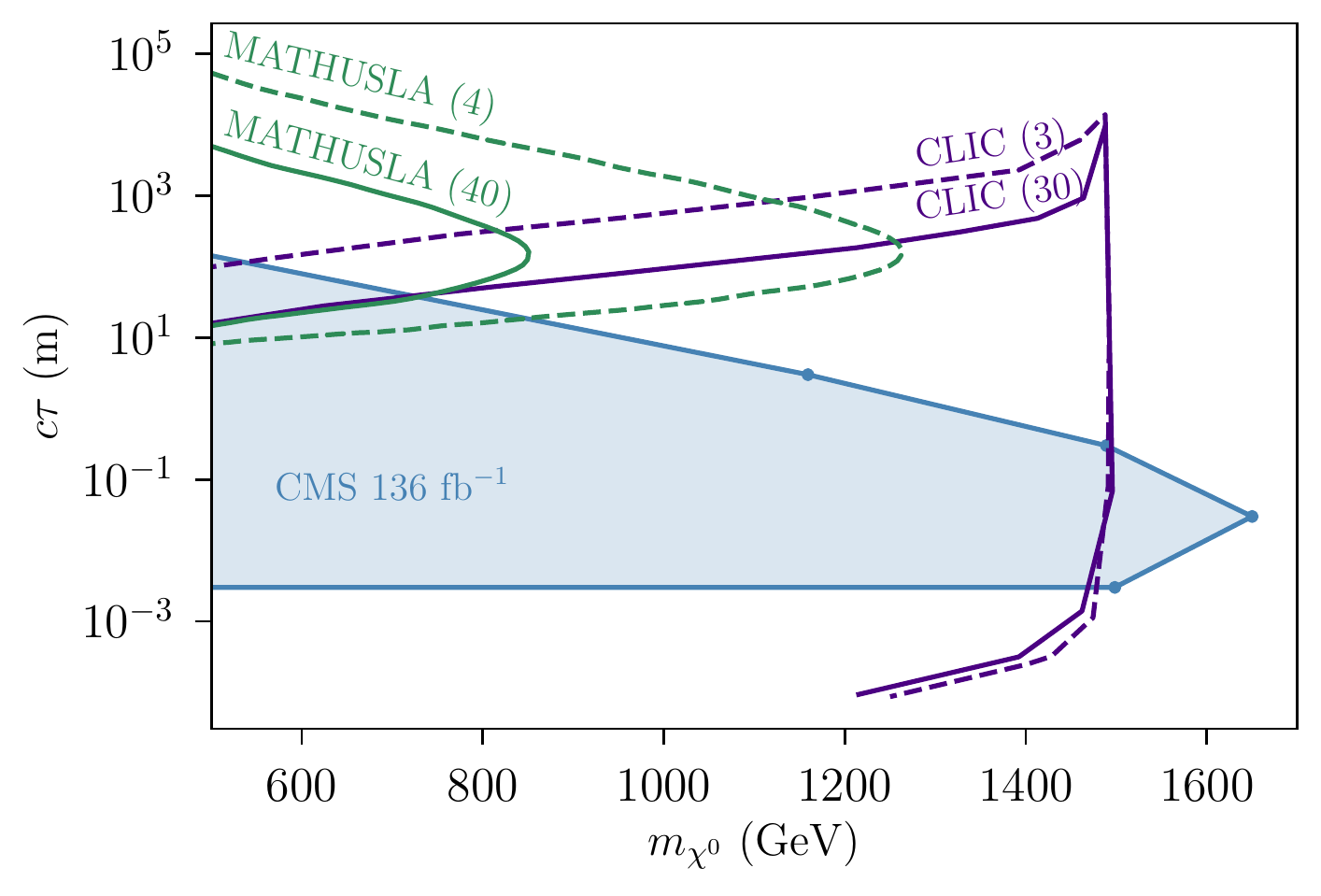}
    \caption{\textbf{left:} Existing limit \cite{ATLAS:2017tny} and HL-LHC projected sensitivity \cite{ATL-PHYS-PUB-2018-033} for long-lived gluinos. \textbf{Right:}  Existing CMS limit for Higgsino LLP, calculated from \cite{CMS:2020iwv}, along curves corresponding to 4 (dashed) and 40 (solid) event signal yield for MATHUSLA \cite{MATHUSLA:2022sze} and and curves corresponding to 3 (dashed) and 30 (solid) event signal yield for CLIC \cite{ILCInternationalDevelopmentTeam:2022izu}}
    \label{fig:heavyLLP}
\end{figure}

As a colored example, the right hand panel of Fig.~\ref{fig:heavyLLP} shows the existing limits \cite{ATLAS:2017tny} and HL-LHC sensitivity projections \cite{ATL-PHYS-PUB-2018-033} for the displaced decay $\tilde g \to q\bar q \tilde \chi^0$. We see that the HL-LHC can improve the limit with about 1 TeV and, more importantly, covers a non-excluded mass range of about 500 GeV in which a 5$\sigma$ discovery could be made. As an example of electroweak production, we take a long-lived Higgsino decaying to a displaced vertex, as occurs in models of R-partity violation and gauge mediation. Here ATLAS, CMS \cite{CMS:2020iwv} and MATHUSLA \cite{MATHUSLA:2022sze} have sensitivity, in addition to a high energy, linear $e^+e^-$ machine such as CLIC \cite{ILCInternationalDevelopmentTeam:2022izu}. The CMS limit shown in is a reinterpretation of the limit in \cite{CMS:2020iwv}, where the Higgsino cross sections from the SUSY cross section working group were used \cite{Fuks:2012qx,Fuks:2013vua,susyxsec}. At this time, no up to date HL-LHC projection is available.

\subsubsection{Dark showers}
\label{sub:DarkShowers}

\begin{figure}
    \centering
    \includegraphics[width=0.45\textwidth]{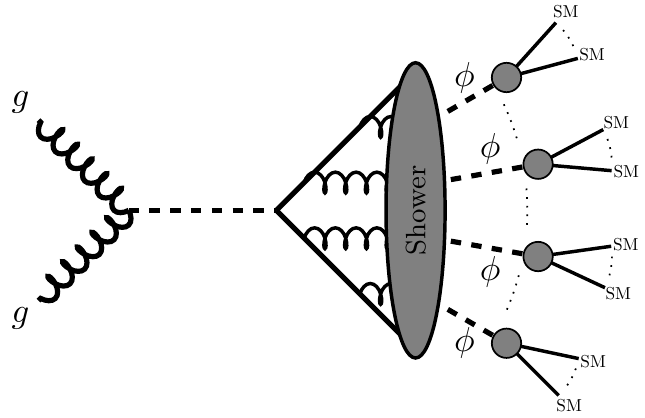}
    \caption{Schematic diagram of a dark shower event, image adapted from \cite{Knapen:2016hky}.  }
    \label{fig:showerdiagram}
\end{figure}

Dark showers are a particularly subtle manifestation of LLPs that can arise in the context of Hidden Valley models \cite{Strassler:2006im}. Hidden valley models feature one or more confining, secluded gauge groups with complicated dynamics. Figure \ref{fig:showerdiagram} schematically shows the topology of a dark shower event: First, a heavy particle, SM or otherwise, decays into a pair of dark sector ``quarks'' which undergo shower and hadronization. The shower and hadronization steps may resemble those in QCD, or could be radically different \cite{Strassler:2008bv}. Some of the dark sector ``hadrons'' can subsequently decay back to the standard model, while others could be stable or extremely long-lived. The particles that do decay visibly can have a wide range of lifetimes. A priori, it may seem that the enhanced multiplicity of the LLPs will make searching for them increasingly trivial. While there are scenarios for which this is the case, the decay products also tend to be softer and are frequently not isolated from one another. 

As a result, specialized strategies are needed, as reviewed in detail in \cite{Albouy:2022cin}. While major progress was made of the last few years, theoretical and experimental challenges remain: On the theory front, more order in the vast space of model building space is needed to arrive at a comprehensive set of recommendations. This may be achieved by imposing additional theory prior, developing simplified models or both. Further Monte Carlo development also remains needed, though a major step forward was made in \cite{Albouy:2022cin}. On the experimental side, one must continue to innovate with analysis strategies, both online and offline, and critically examine when isolation criteria and/or $p_T$ threshold could potentially be relaxed in exchange for a higher number of LLPs in the event. For example, the use of future timing layers has been shown to be beneficial \cite{Chekanov:2020xco}.

\clearpage


\section{Dark Matter}\label{Sec:DMFP}

For a general introduction and motivation of DM searches at collider desceribed in this Chapter, we refer back to the introductory section \ref{Sec:DM}. 

For the 2020 update of the European Strategy for Particle Physics, the Physics Briefing Book~\cite{Strategy:2019vxc} included Section 8.5 dedicated to the estimated sensitivities of searches for several key types of benchmark models of DM, including supersymmetric searches (Wino and Higgsino DM), comparisons of collider sensitivities to a subset of the LHC simplified models (mediators with axial-vector or scalar couplings to SM fermions and Dirac fermionic DM) and searches for “portal” sectors motivated by lighter, feebly coupled thermal relic DM. 
Section 9 of the Briefing Book compared projected collider constraints to direct detection and indirect detection experiments for a selected set of these models. 

In the present report, we discuss the status of collider WIMP searches focusing on work that has been submitted to the Snowmass process, including supersymmetric scenarios (Section \ref{sec:dm:wimp}), Higgs portal models (Section \ref{Sub:HiggsPortal}). updates of the projections for simplified collider DM models, along with the extrapolations to non-collider searches (Section \ref{Sec:DMst}), and new collider results and considerations on portal DM models and models of dark sectors including DM candidates (Section \ref{Sec:OtherDM}).
\subsection{Testing the simplest/minimal WIMP models (EW multiplets) and their extensions}
\label{sec:dm:wimp}
Among the WIMP scenarios, one particularly simple case is the dark matter particle being the lightest member
of an electroweak (EW) multiplet. Most familiar examples are the Higgsino (a Dirac fermion doublet) and the wino (a Majorana Fermion triplet) in the context of supersymmetry. At the same time, more general cases have also been considered \cite{Cirelli:2005uq,Cirelli:2009uv}. This is a very predictive scenario. In the simplest case, the interaction strengths are the SM gauging couplings.  The only free parameter, the mass of the dark matter particle, $m_{\chi}$,  is fixed by the by requiring thermal relic abundance matches the observation \cite{Planck:2018vyg}. These so called thermal target dark matter masses are typically in the TeV range \cite{Belotsky:2005dk,Hisano:2006nn,Cirelli:2007xd,An:2016gad,Mitridate:2017izz,DelNobile:2015bqo}. In particular, they are 1.1 TeV and 2.8 TeV for the Higgsino and the wino, respectively. Higher EW representations have higher thermal target masses. 

\begin{figure}[htp]

\includegraphics[width=\textwidth]{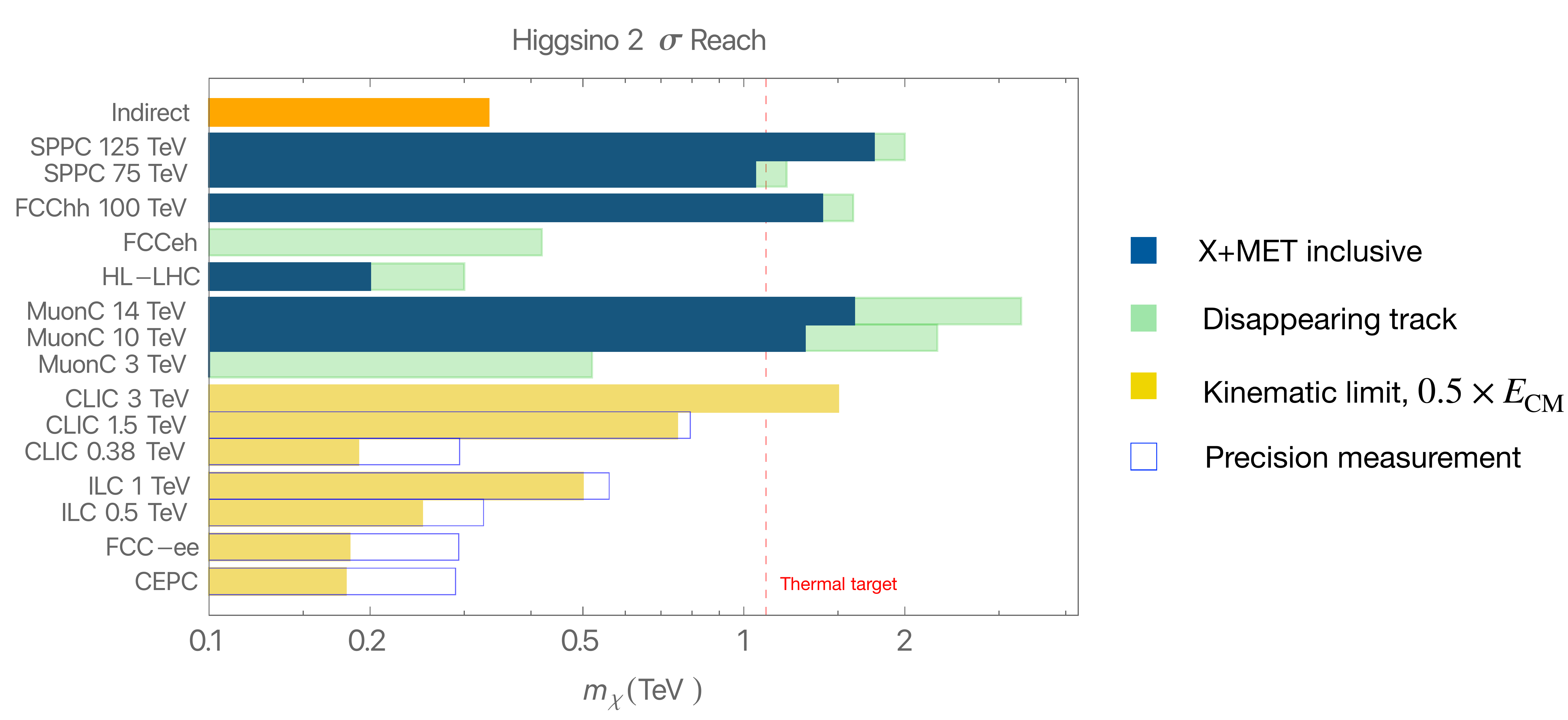}
\includegraphics[width=\textwidth]{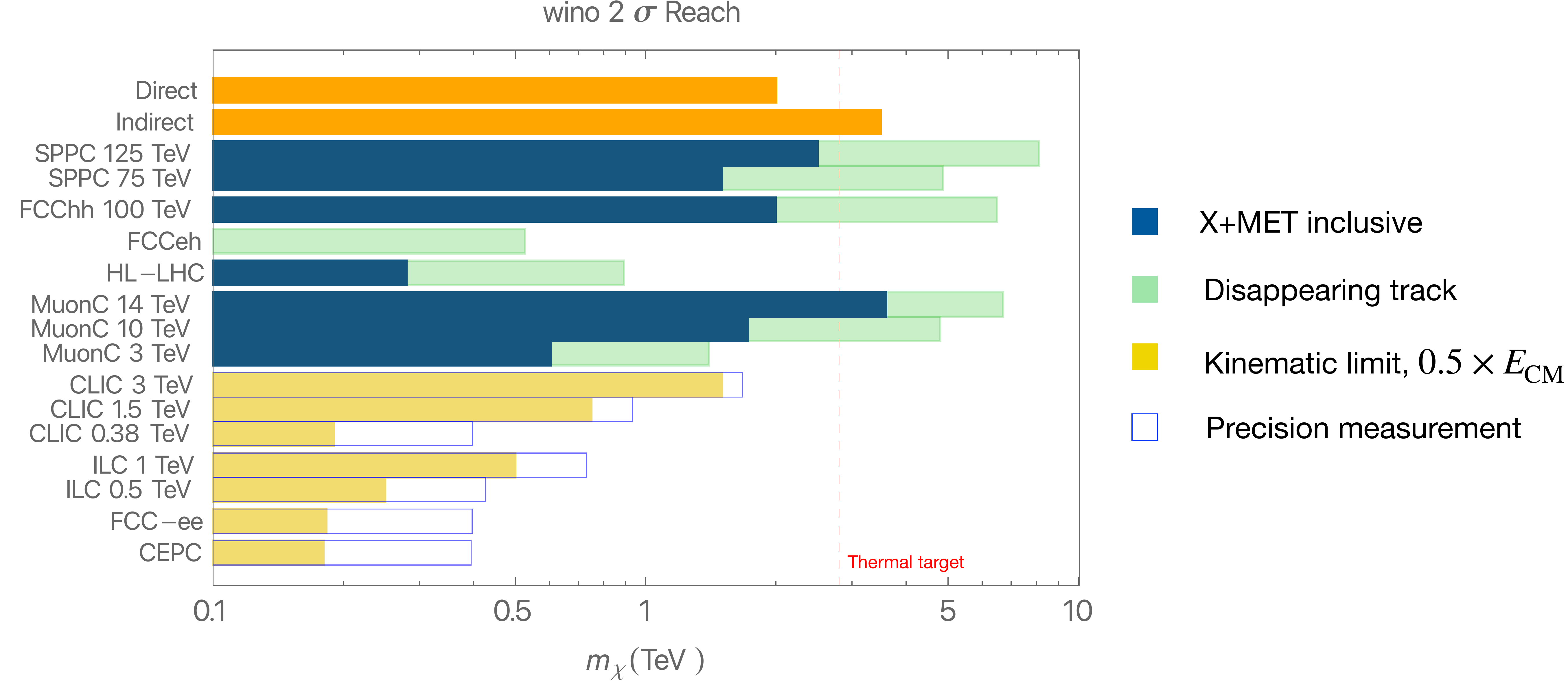}
    \caption{A summary of the reach of future colliders for simple WIMPs. For comparison, the reaches of the direct and indirect detections are also included.}
    \label{fig:WIMPSummary}
\end{figure}

Covering these cases is among the main physics drivers for future high energy colliders. A summary of the $2 \sigma$ reaches of the Higgsino and wino at future colliders  is shown in \autoref{fig:WIMPSummary}. An earlier summary can be found in the Physics Briefing Book  for the European Strategy for Particle Physics Update 2020 \cite{Strategy:2019vxc}. In the last couple of years, there have been new studies on the reach of a high energy muon collider \cite{Han:2020uak,Han:2022ubw,Capdevilla:2021fmj,Bottaro:2021srh,Bottaro:2021snn,Bottaro:2022one,Black:2022qlg}. These results are also included here.

With only gauge interactions,
the production and decay for the EW multiplets are highly predictable. Since the mass splittings  between the charged and neutral states are expected to be small, typically of the order of a few hundred MeV from EW loops in the minimal scenario, the decay products will be very soft, most likely escaping the detection. In this case, the main signal at high energy colliders is large missing energy-momenta recoiling against energetic SM particles.  At hadron colliders, the dominant channel is jets$+$MET \cite{Low:2014cba,Han:2018wus}. At high energy leptons, there are a number of channels \cite{Han:2020uak,Bottaro:2021snn,Han:2022ubw,Bottaro:2022one,Black:2022qlg}, with SM EW gauge bosons and leptons in the final state. Among the various possibilities, the importance of mono-muon at Muon collider is emphasized by \cite{Han:2020uak}, and the importance of the mono-W channel is emphasized by \cite{Bottaro:2021snn,Bottaro:2022one}. It is worth emphasizing that this class of signals is relatively insensitive to the mass splitting between the members of the EW multiplet. Hence, they are more robust against variations beyond the minimal scenario. 

Besides the conventional missing mass search, the loop-induced mass splitting among the component states of the EW multiplet also results in a disappearing track signature which can be used to enhance the reach. This signal, while powerful, is very sensitive to the mass splitting. Hence, the results presented here only apply to the minimal models. The reach of this channel also depends on the details of the detector design, which  still have not been finalized in many cases. Nevertheless, preliminary estimates of the reach have been made, and they serves as rough guides to the capability of future colliders. The reach at hadron colliders are taken from \cite{Saito:2019rtg,CidVidal:2018eel} with simple rescaling in cases not covered by those studies. The estimates for high energy muon colliders have been made in \cite{Han:2020uak,Capdevilla:2021fmj,Han:2022ubw}, with \cite{Capdevilla:2021fmj} modeled the effect of the beam induced background by using simulation carried out for $E_{\rm CM}=1.5$ TeV. 

In \autoref{fig:WIMPSummary}, for lepton colliders without a dedicated study, we indicate the kinematic limit. This is obviously only an upper bound. From the studies for muon colliders, we see that reaching the kinematic limit is non-trivial. While it is possible in the wino case with the use of disappearing track, it is likely a significant overestimate for the Higgsino case.

Beyond the direct production of the DM particles, the virtual effect of the EW multiplet can offer another probe for this scenario. This is particularly interesting for lepton collider with relative low $E_{\rm CM}$, where electroweak precision measurement (such as the $W$ and $Y$ parameters) can extend the reach of such colliders beyond its kinematic limit, with $m_{\chi}|_{\rm kin. \ limit}= 0.5 \times E_{\rm CM}$ \cite{DiLuzio:2018jwd}. The result presented here is taken from \cite{Strategy:2019vxc}. 

\begin{figure}[htp]
\begin{center}
\includegraphics[width=0.75\textwidth]{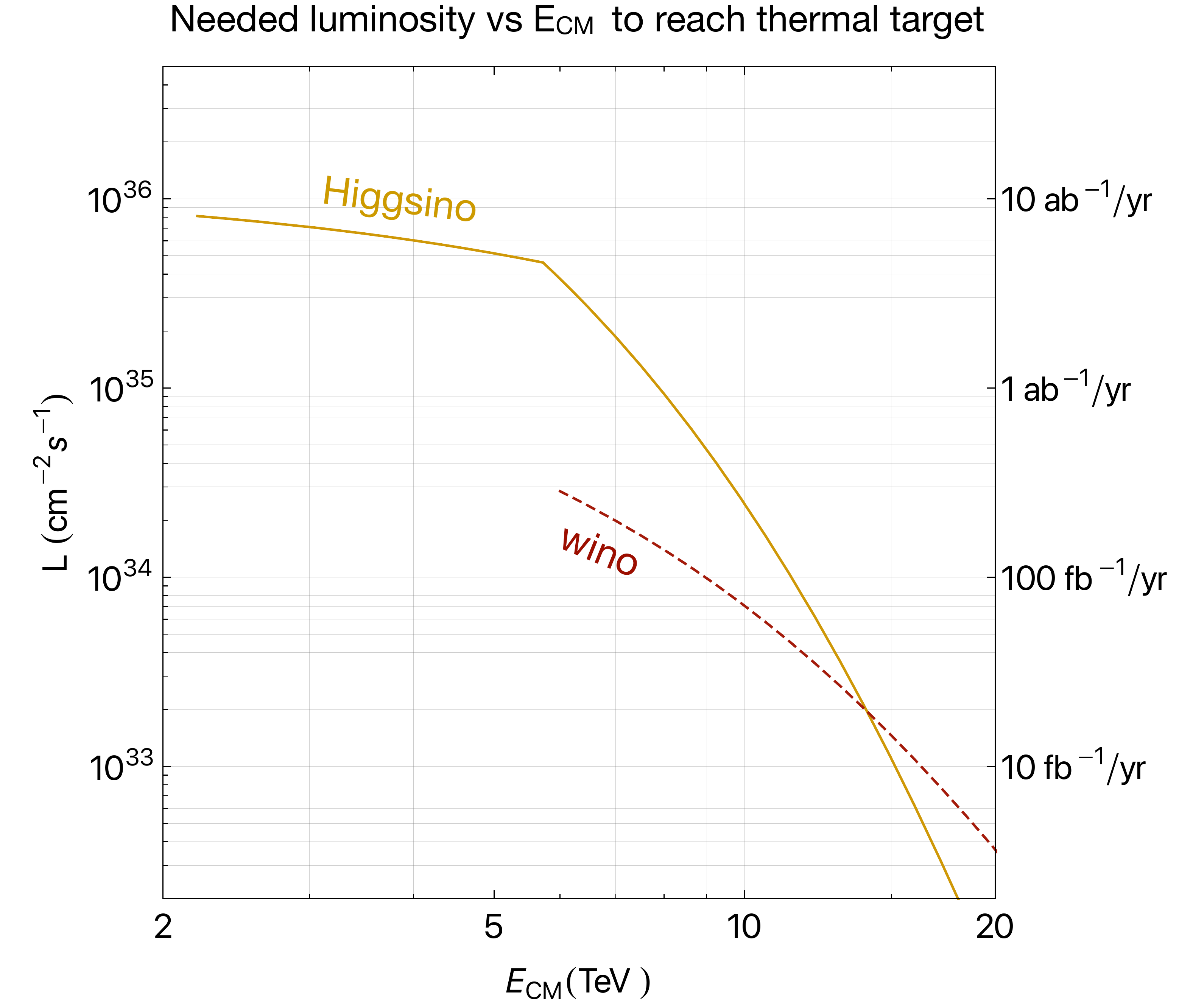}
\end{center}
    \caption{The luminosity needed, assuming a 10 year run plan, to reach thermal target as a function of $E_{\rm CM}$ at high energy lepton colliders. This is built on the available studies for muon colliders. The figure is taken from \cite{Liu:2022rua}.}
    \label{fig:LumiWIMP}
\end{figure}

In addition to collider probes, both direct detection and indirect detection can probe the minimal WIMP scenario. The direct detection cross section for wino is above (by less than one order of magnitude) the neutrino floor \cite{Chen:2018uqz}, and DARWIN can probe this scenario with $m_\chi = 2^{+3}_{-1}$ TeV. The indirect detection limit from H.E.S.S.\cite{Hryczuk:2019nql} also covered the thermal mass target for the wino scenario, although this is subject to uncertainties in modeling the DM halo. For Higgsino, future indirect detection from CTA could be sensitive to the thermal mass target. At the same time, with the assumption of a small Majorana mass splitting,  its direct detection cross section is below the neutrino floor. For completeness, we have included the results (taken from \cite{Strategy:2019vxc}) of direct and indirect detection in \autoref{fig:WIMPSummary}.

In \autoref{fig:LumiWIMP},  we show the required luminosity as a function of the lepton collider center-of-mass energy for Higgsino and Wino in solid and dashed purple lines, respectively. For concreteness, a 10 year run plan is assumed. The result for different run plans can be scaled from the figure in a straightforward manner. At low center-of-mass energy, the search sensitivity is driven by the inclusive missing energy searches, as the boost of the pair-produced charged Higgsino is low. Hence, the lifetime is not long enough to support a high efficiency of the ``disappearing track" signatures. Beyond 6~TeV center-of-mass energy, the search sensitivities are driven by the disappearing track searches. Different lepton colliders will have different sources of background for such a signature, and future studies will be of particular importance with concrete beam and detector designs. We can see from the figure that a 10~TeV lepton collider could achieve the Higgsino and Wino dark matter goal robustly.

Extension of the minimal scenario can include nearby states and more coannihilation channels. In this case, the dark matter do not need to be in a EW multiplet. The interaction with the SM is more model dependent. If the mass splittings are sufficiently This inclusive signal remains the best channel in this case. 


\subsection{Testing DM with the Higgs boson}
\label{Sub:HiggsPortal}

Other simple scenarios of WIMP dark matter involve a SM particle, such as the Higgs or the Z bosons, which interacts with the DM particle through a gauge invariant operator. These are known as the Higgs and Z portals 
\cite{Englert:2011yb, Djouadi:2011aa, Arcadi:2014lta}. 
In these scenarios, the Higgs or the Z boson can decay into a pair of invisible particles with WIMP DM properties and a mass smaller than half the mass of the mediator particle.  DM pair production can also be mediated from off-shell Higgs and Z, but the reach of such channels are typically not very strong. 
The Z portal is heavily constrained by precision measurements at LEP \cite{Arcadi:2014lta, Escudero:2016gzx}, so we focus on the Higgs portal as a benchmark for collider searches at future colliders. 
Since it is not a higher dimensional EFT operator, the Higgs portal is also likely to a sizable interaction between the SM and the dark sector. 

Since the SM branching ratio of the Higgs into invisible particles (neutrinos) via a pair of Z bosons is very small (at the per-mille level, 0.11\% of the total decays), observing an excess of invisible Higgs decays 
would signal the presence of new physics that could be connected to WIMP DM. 
Additional invisible Higgs decays can also be inferred via precision measurements of Higgs properties, as their presence would also modify the SM Higgs couplings. 

\begin{figure}[!htp]
 \centering
 \includegraphics[width=\textwidth]{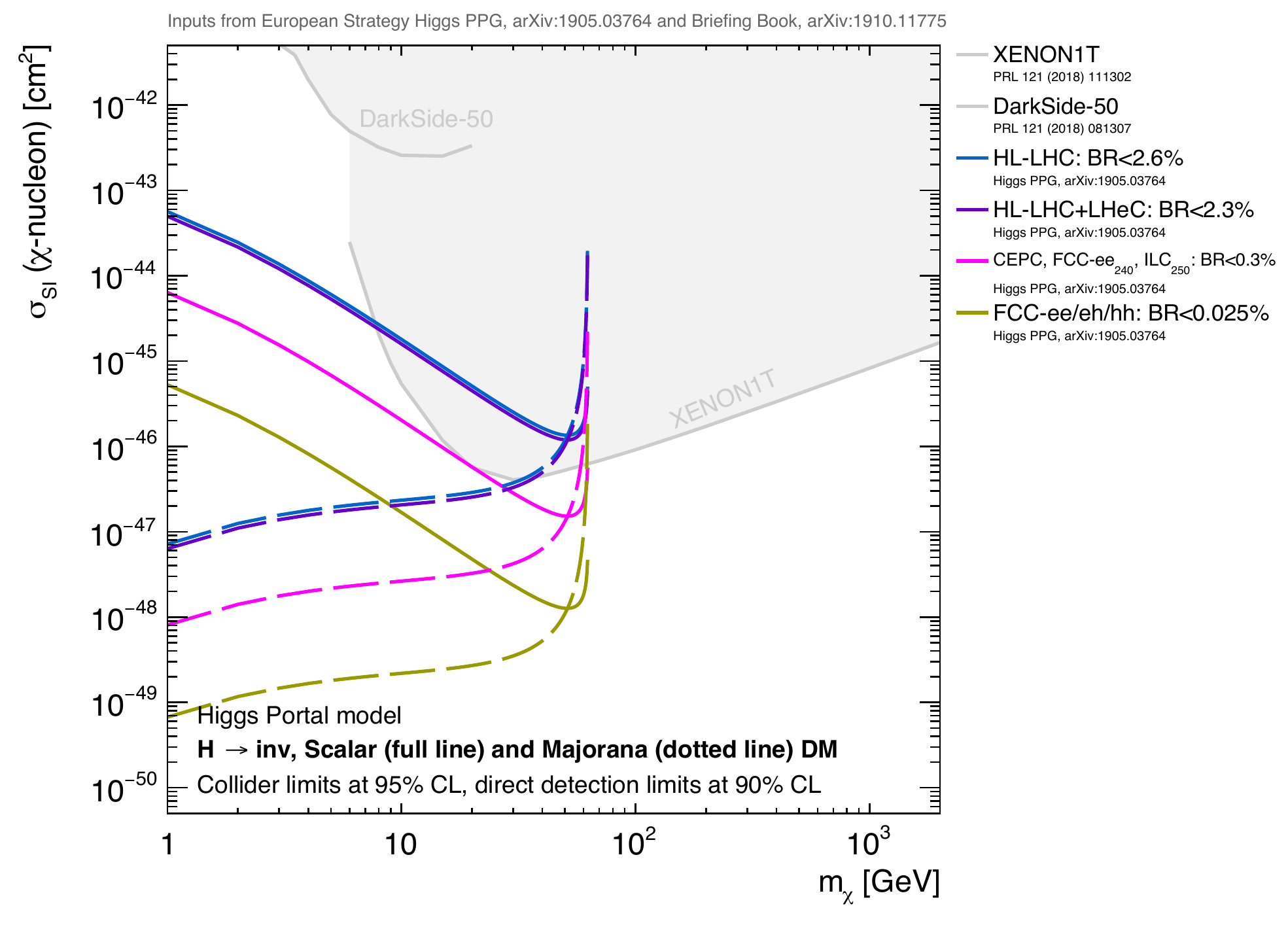}
  \caption{Projected limits from future colliders (direct searches for invisible decays of the Higgs boson) on the spin-independent WIMP-nucleon scattering cross-section versus DM mass plane, for a simplified model with the Higgs boson decaying to scalar or Majorana invisible (DM) particles. Collider limits are shown at 95\% CL and current direct detection limits at 90\% CL. Collider searches and current DD experiments exclude the areas above the curves.}
  \label{fig:HiggsInvisible}
\end{figure}

Projected upper limits on the Higgs to invisible branching fraction from future colliders have been obtained in the context of the Briefing Book of the Update of the European Strategy of Particle Physics \cite{deBlas:2019rxi,Strategy:2019vxc}.
As detailed in Section \ref{Sec:DiscVsLimit}, these constraints can be used as a proxy for the sensitivity to Higgs decays into DM particles.
They can be translated to the nucleon-DM scattering plane using an Effective Field Theory approach \cite{Fox:2011pm, Hoferichter:2017olk} and compared to current constraints from Direct Detection experiments. 
This is shown in Fig. \ref{fig:HiggsInvisible}, adapted from \cite{deBlas:2019rxi,Strategy:2019vxc}.

From Fig. of \cite{Strategy:2019vxc}
one can see that a future hadron collider is the only collider that would be able to probe the SM Higgs to invisible branching ratio, and would have the best reach for this type of model. 

The DM candidates considered in this plot are a scalar or a Majorana fermion particle. The contributed whitepaper \cite{Zaazoua:2021xls} also studies the case in which DM is a vector particle, where the validity of the EFT is limited \cite{Baek:2014jga}, and shows that the strength of the constraints depends on the  nature and parameters of the theory used to complete the EFT at higher energies. 
The authors of Ref. \cite{Zaazoua:2021xls}, based on discussion with the authors of \cite{Hoferichter:2017olk} also propose to extend the results of Fig. \ref{fig:HiggsInvisible} to DM masses below the GeV since the translation between collider parameters of the theory (upper limit on the Higgs branching ratios) and DM-nucleon cross-section are still valid in all regimes of momentum transfer.

\subsubsection{Models with additional scalars}

The Higgs portal model can also be extended by introducing additional Higgs-like scalars that couple to DM particles (see also Sec. \ref{Sec:OtherDM} for other models with Higgs mixing and an additional singlet extension). 

As examples of such models, we present two benchmarks from the submissions to the Topical Group EF10 in Refs. \cite{Kalinowski:2022fot,ATL-PHYS-PUB-2022-018, CMS-PAS-FTR-22-005,CMS-PAS-FTR-18-007}: a two-Higgs doublet model that includes a pseudoscalar mediator to dark matter (2HDM+a) (see Ref. \cite{LHCDarkMatterWorkingGroup:2018ufk} and references therein) and the Inert Doublet Model (IDM) (see Ref. \cite{Gustafsson:2010zz} and references therein for a review, and \cite{Kalinowski:2020rmb} for future collider prospects).

The 2HDM+a extends the two Higgs doublet model by adding a pseudo scalar that effectively serves as a portal to dark matter, and mixes with the pseudo scalar within the 2HDM, see e.g.  \cite{LHCDarkMatterWorkingGroup:2018ufk} and references therein, as well as \cite{Cabrera:2020lmg} for more specific considerations on the available parameter space for kind of models. 
Given the rich phenomenology and interplay between different searches, this model is used as benchmark for current LHC and HL-LHC searches. 
Recent projections of searches for the particles predicted by these models for HL-LHC are found in Refs. \cite{ATL-PHYS-PUB-2022-018, CMS-PAS-FTR-22-005}. 
The most recent projection from CMS using boosted Higgs bosons produced in association with the pseudoscalar DM mediator $a$ gives an indication of the parameter space that can be probed with the HL-LHC dataset as a function of the masses of the two pseudo scalars in the model: signals with pseudoscalars masses of 250 GeV for the DM mediator $a$ and 1 to 1.6 TeV for the heavy pseudoscalar $A$ could be discovered with the full HL-LHC dataset. 
Additionally, given that the current constraints set a lower mass scale for the additional heavy scalars, these models can be best tested at colliders with a sufficient centre-of-mass energy (1.5 TeV case of electron-positron colliders) \cite{Kalinowski:2022fot}. 

The Inert Doublet Model includes an additional doublet in addition to the SM Higgs doublet. 
A discrete $Z_2$ symmetry, introduced to avoid tree-level flavor-changing neutral currents, ensures that the lightest particle within the new doublet is stable. 
This mechanism provides a viable DM candidate. 
Ref. \cite{Kalinowski:2022fot} shows that the couplings of this models are strongly constrained from above by Higgs to invisible branching ratios, and imposing the relic density constraint bounds these couplings from below. 
This means that at future colliders we should be able to rule out the simplest scenario where the inert scalar constitutes all the dark matter, still keeping in mind that the relic constraint can be circumvented if additional DM annihilation channels are present. 




\subsection{Dark Matter: Simplified models}\label{Sec:DMst}

A common feature of the interactions described by the SM is the presence of a particle that mediates the interaction, such as the photon/Z/W for the electroweak interactions or the gluon for the strong interaction. 
This can be mirrored and extended in models that include new interactions between the SM and a dark sector that contains DM candidates, as in Ref.~\cite{Abercrombie:2015wmb} where the interaction between dark matter and dark sector is mediated by new (BSM) particles. 
This section discusses future collider sensitivities to a selection of these simplified models.

The simplified models in Ref.~\cite{Abercrombie:2015wmb} are classified by the type of coupling between the mediator particle and the SM/DM particles.
The mediator can be a vector, an axial vector, a scalar and a pseudo scalar particle. 
Minimum Flavour Violation is satisfied in these models, to evade flavour constraints. 
This leads to a Yukawa coupling structure for scalar and pseudoscalar mediators, where the couplings to fermions are proportional to the SM Higgs couplings. 
The particle spectrum of these models is kept as simple as possible, adding only the mediator and a single kind of DM particle to the SM, where the DM particle a Dirac fermion. 
This choice does not influence collider studies significantly, and it is based on the majority of studies using this choice in literature at the time of the Dark Matter Forum report. 
In particular, the collider phenomenology does not change if the particle is a Dirac fermion, a pseudo-fermion or a Majorana fermion. 
The mediator can decay into both the DM particle (invisible decays) and into SM quarks and leptons (visible decays), where the coupling strength can be considered a free parameter of the model. 
These simplified models by no means exhaust the possibilities for DM benchmarks, even limiting to models with a limited number of new particles; for example, Ref.~\cite{Ghosh:2022zef} studies how complementary collider, direct-, and indirect-detection searches can probe a more complete model in which singlet Majorana DM is coupled to SM fermions through scalar fermion partners.

In this report and in the accompanying Ref.~\cite{Boveia:2022jox}, we follow the recommendations of the LHC Dark Matter Working Group~\cite{10.1016/j.dark.2019.100365.Boveia.2020,10.1016/j.dark.2019.100377.Albert.2019} and use the models involving Dirac fermion DM discussed in Ref.~\cite{Abercrombie:2015wmb} to understand how future collider experiments would extend current LHC results. 
We also adopt the recommendations~\cite{10.1016/j.dark.2019.100365.Boveia.2020,10.1016/j.dark.2019.100377.Albert.2019} for comparing collider sensitivity to direct- and indirect-detection experiments, used in the European Strategy Update Briefing Book Ref.~\cite{Strategy:2019vxc}. 

\subsubsection{Inputs}

The majority of the collider projections have been provided to other Snowmass topical groups and reinterpreted in terms of these simplified models, as described in Ref. \cite{Boveia:2022jox}.

The inputs in terms of dijet decays of vector mediator models for HL-LHC and FCC-hh are described in more detail in Sec.\ref{Sec:NB} and references therein (\cite{Harris:2022kls}). 
New HL-LHC dilepton projections were derived for Snowmass based on \cite{ATLAS:2019mfr}.
Future collider sensitivity for invisible mediator decays are derived from jet+MET searches at the LHC, HL-LHC and FCC-hh. 

Projected sensitivities to the simplified models were prepared by Energy Frontier topical groups for the most relevant collider searches for the mediator, including both its invisible decays (the ``mono-jet'' search) and visible decays (dijet and dilepton resonance searches). Estimates for both HL-LHC (\cite{hllhc-monojet},\cite{Harris:2022kls},\cite{ATL-PHYS-PUB-2018-044}) and FCC-hh (\cite{Harris:2015kda},\cite{Harris:2022kls}) were made available.

\subsubsection{Fixed-coupling results}

This section focuses on future hadron colliders, as the projections for lepton colliders have only been discussed recently and, if time allows, will be included in the next version of this report. 
Lepton colliders such as ILC should also powerfully constrain the models when the mediator coupling to leptons is non-zero~\cite{Kalinowski:2021tyr}. 
As an illustration, Refs.~\cite{Boveia:2022jox,Black:2022qlg} examine the prospects for a mono-photon search at a future muon collider in the context of a dark matter candidate produced directly through electroweak interactions~\cite{Han:2020uak}.

\begin{figure}[htb!]
\centering
\subfloat[New scenario with varied SM couplings: $g_q=0.1$, $g_{\chi}=1.0$, $g_l=0.1$]{\label{subfig:vector-hl-lhc-massmass}\includegraphics[width=0.49\textwidth]{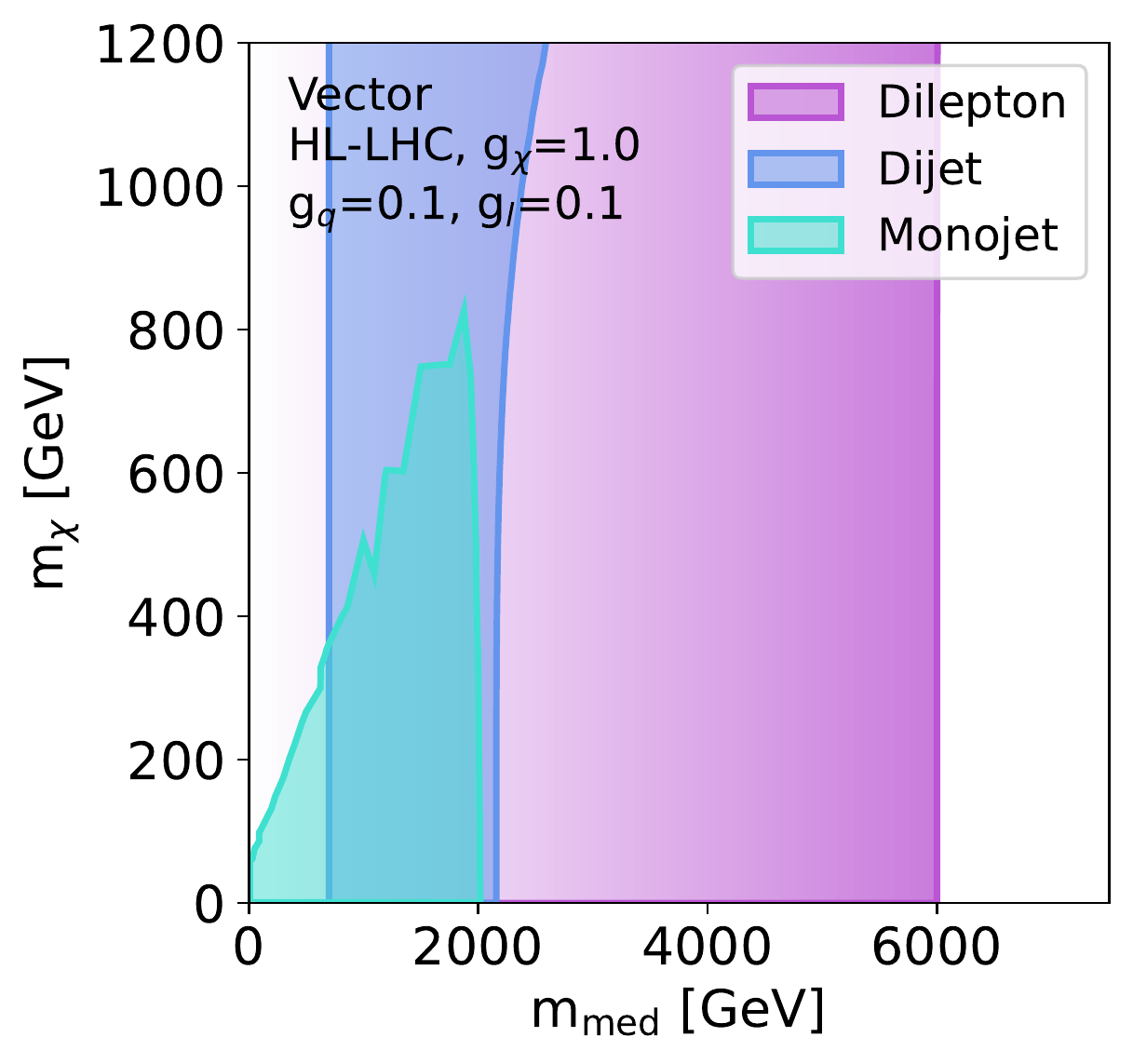}}
\subfloat[$g_q=0.1$, $g_{\chi}=1.0$, $g_l=0.0$]{\label{subfig:vector-fcc-massmass}
\includegraphics[width=0.49\textwidth]{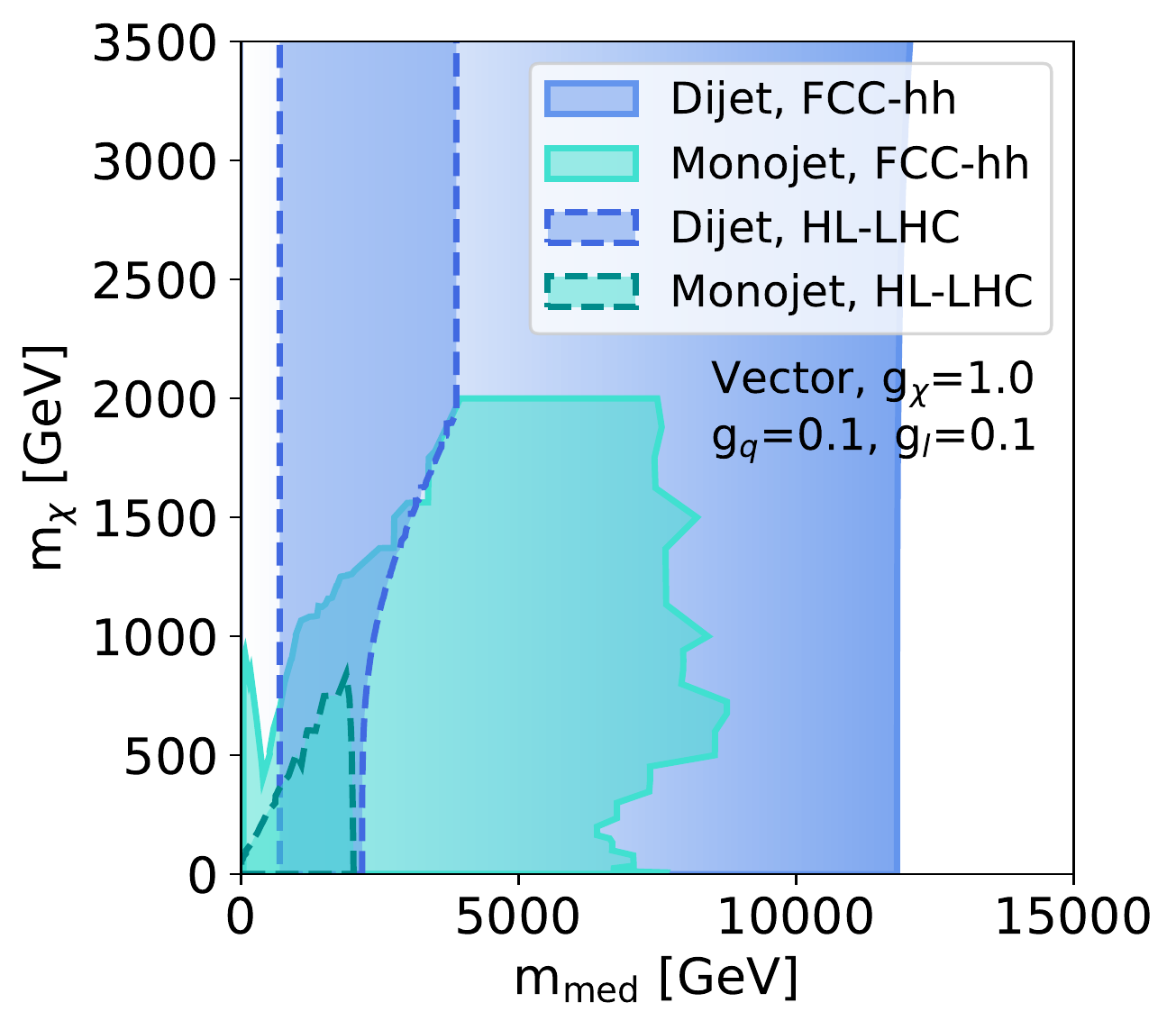}}
\caption{HL-LHC and FCC-hh projected limits for individual analyses in the vector model and with coupling values couplings: $g_q=0.1$, $g_{\chi}=1.0$ for \protect\subref{subfig:vector-hl-lhc-massmass} HL-LHC including lepton couplings of $g_l=0.1$ and \protect\subref{subfig:vector-fcc-massmass} HL-LHC shown together with FCC-hh without lepton couplings. The shading for FCC-hh represents the fact that the lower bound in mediator mass for dijet searches can go lower than what the studies in Ref.\cite{Boveia:2022jox} show, see the text for discussion.}
\label{fig:hl-lhc-fcc-massmass}
\end{figure}

Figure~\ref{fig:hl-lhc-fcc-massmass}, adapted from Ref.\cite{Boveia:2022jox}, shows the expected sensitivities of HL-LHC and FCC-hh xperiments to a heavy $Z^\prime$ mediator with vector couplings to quarks, leptons, and fermionic DM. 
The contours indicate the combinations of DM mass and mediator mass excluded by the mono-jet, dijet, or dilepton searches, assuming the lifetime of the HL-LHC and one possible set of fixed values of the couplings of the mediator to quarks, leptons, and DM. 

The right-hand side of this figure also shows how these same constraints grow stronger with the greater kinematic reach and integrated luminosity of the FCC-hh. 
It is worth noting that the lower bounds for both HL-LHC and FCC-hh for mediator (resonance) searches shown in the figure are very likely to be much higher than what can be reached with non-standard data taking techniques such as data scouting / trigger-level analysis described in Section \ref{sec:scouting} and in Refs. \cite{CMS:2016ltu,ATLAS:2018qto}. 
Such techniques will be required to ensure optimal coverage of all phase space for resonance searches around the electroweak scale with the increased datasets of future colliders. 

As illustrated in Ref.~\cite{Abercrombie:2015wmb}, the production and decay of a mediator with a different spin structure of its couplings (e.g., pure axial-vector rather than pure vector couplings to SM and DM fermions) would give similar results; see the appendix of Ref.~\cite{Boveia:2022jox} for a quantitative example. 
The results for a scalar model would be quantitatively different, due to the Yukawa coupling structure and to the much smaller cross-sections -- projections for a scalar model can be found in the European Strategy Briefing Book \cite{Strategy:2019vxc} with inputs from Refs. \cite{ATL-PHYS-PUB-2018-036,Harris:2015kda}. 
In the scalar case, it should be noted that while colliders can provide unique sensitivity to scalar-mediator models, Ref.~\cite{liu:2022vkm} discusses possible sensitivity in this region from gravitational wave observatories.

\FloatBarrier

\subsubsection{Extrapolation to lower coupling values}
\label{sub:extrapolationDMst}

The above results assume ad-hoc values of the mediator couplings in order to depict how sensitivities vary with DM and mediator masses. Sampling  the five dimensional parameter space of the model with full Monte Carlo simulation was beyond the scope and resources of the study. 
For this reason, Ref.~\cite{Strategy:2019vxc} did not study this coupling dependence. 
To show how the picture changes with other coupling values, here we use methods developed in Ref~\cite{Albert:2022xla} to analytically extrapolate the collider limits to lower values of the mediator coupling to quarks, leptons or dark matter, and to show the impact of changing the mass ratio between mediator and dark matter particles.

\begin{figure}[htb!]
\centering
\subfloat[]{\label{subfig:gqscan-dmLight}\includegraphics[width=0.49\textwidth]{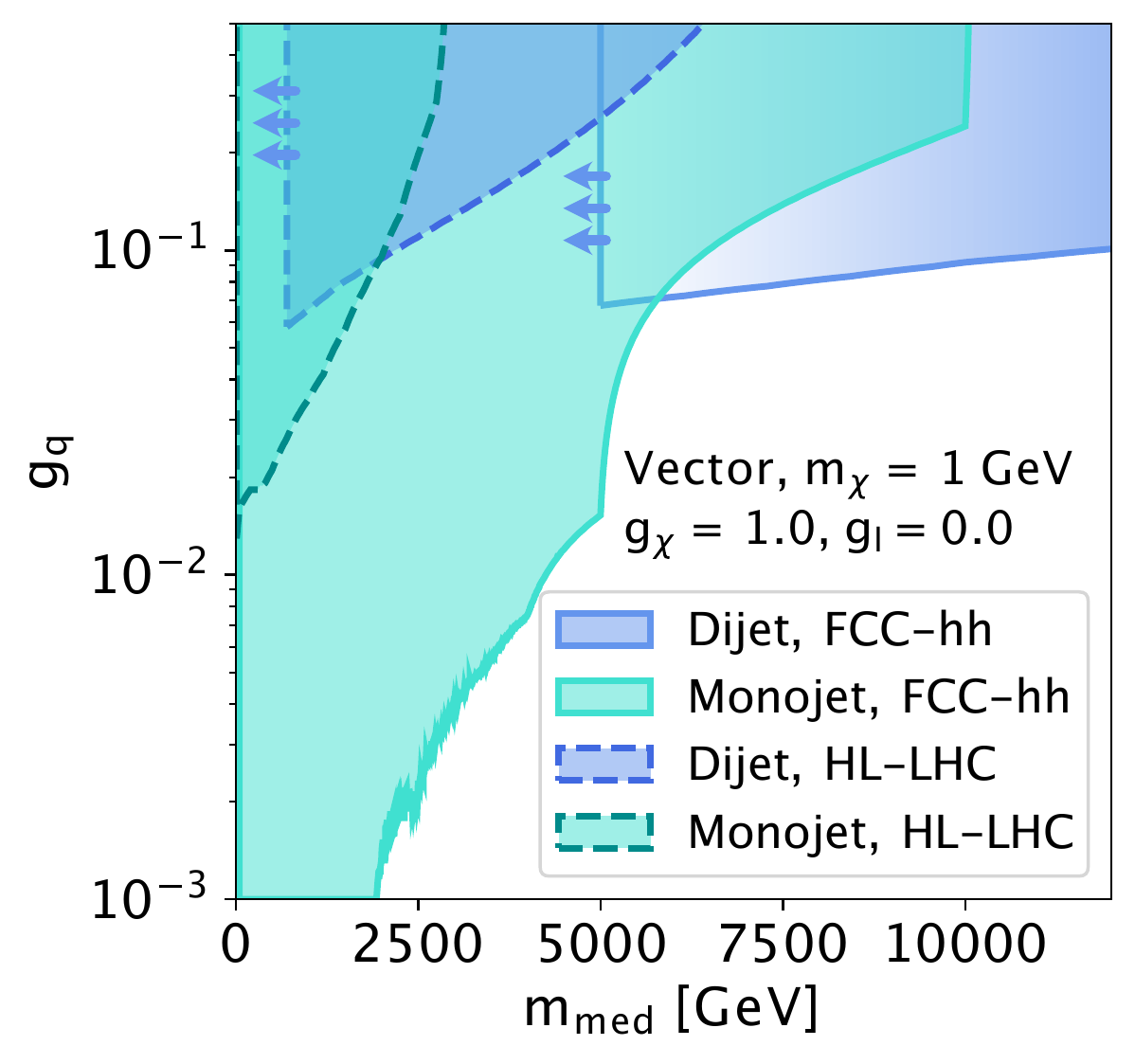}}
\subfloat[]{\label{subfig:gchiscan-dmLight}\includegraphics[width=0.49\textwidth]{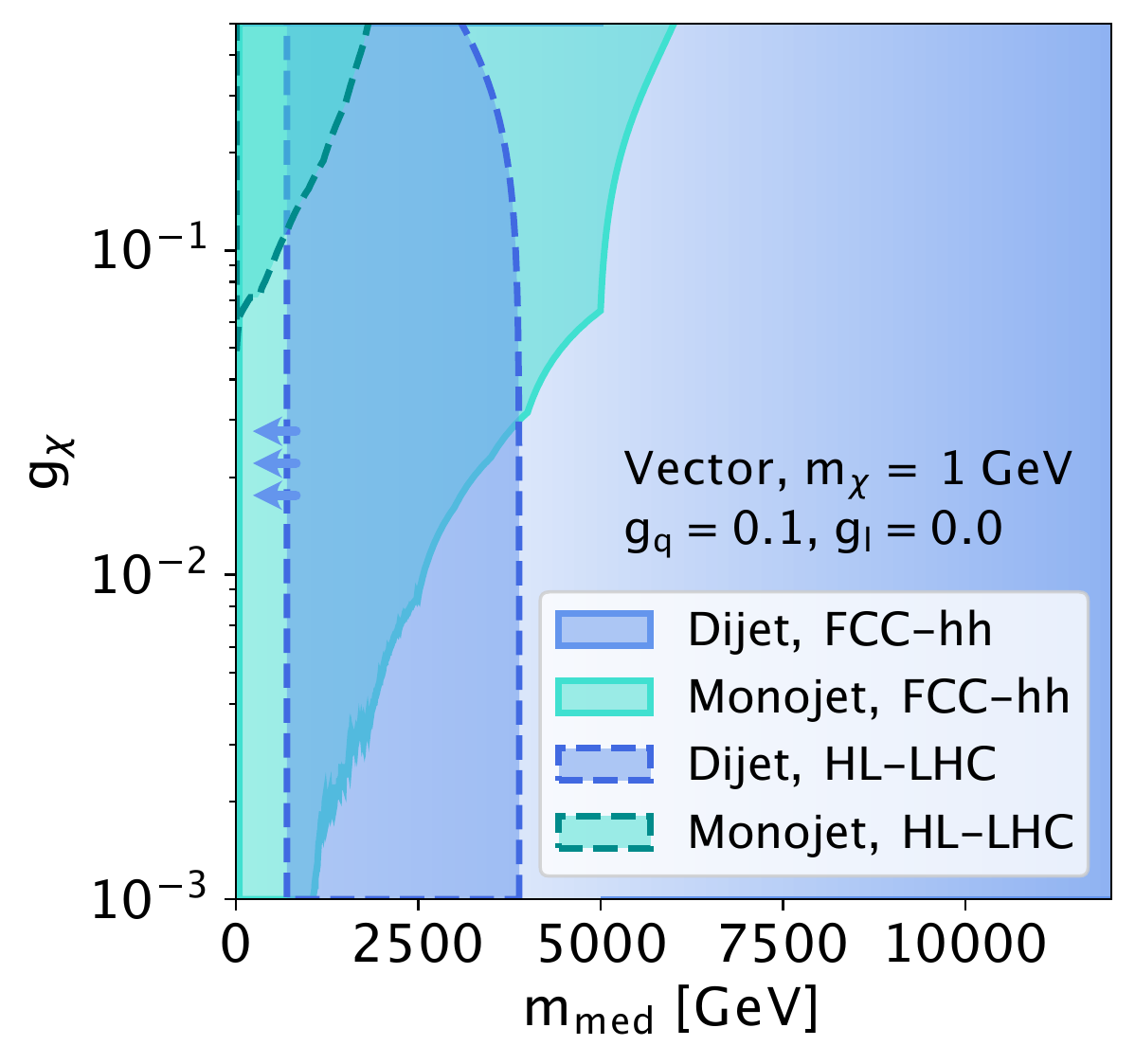}}
\\
\subfloat[]{\label{subfig:glscan-dmLight}\includegraphics[width=0.49\textwidth]{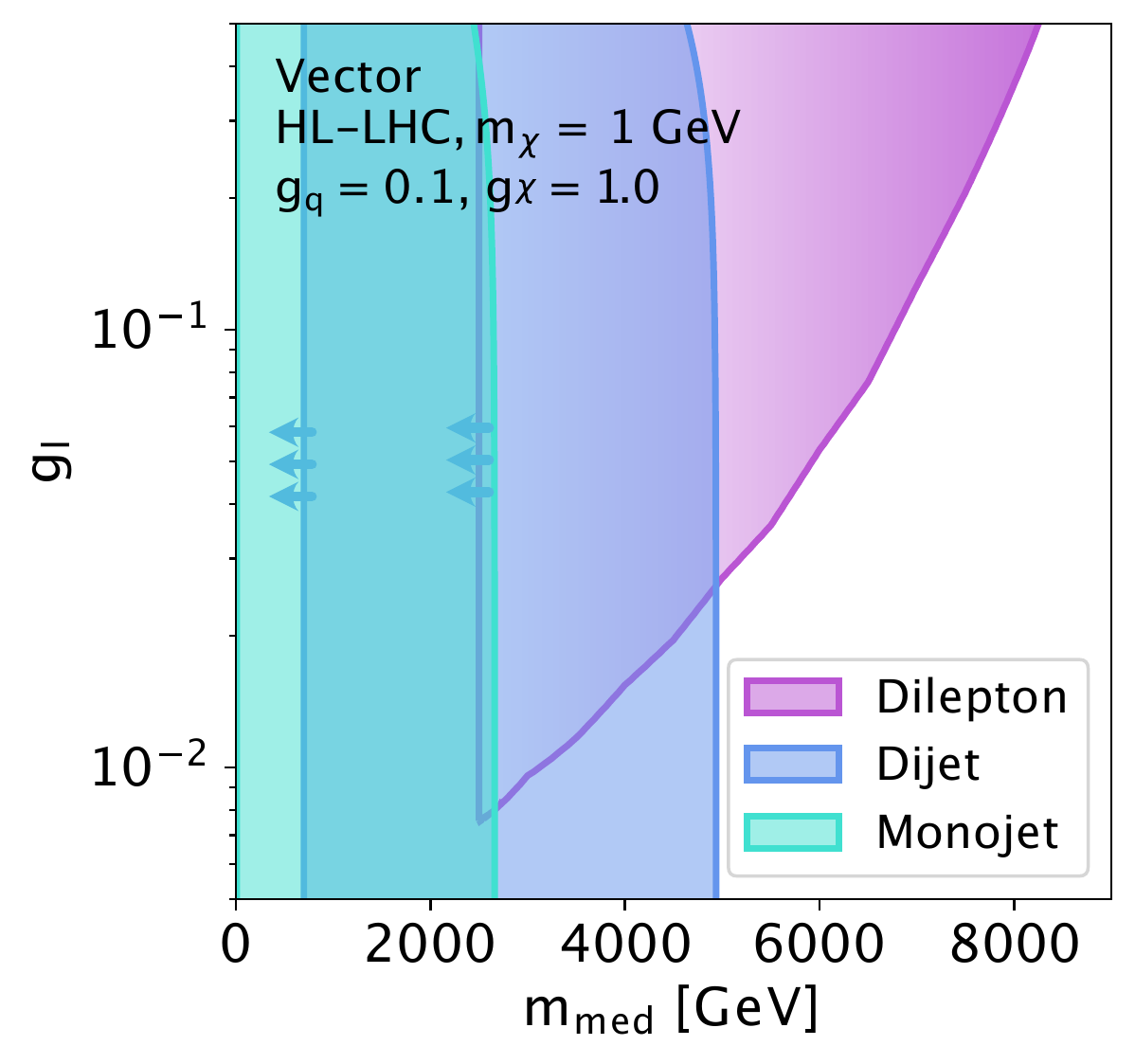}}
\caption{Projected exclusion limits on the couplings $g_q$ (\protect\subref{subfig:gqscan-dmLight}), $g_\chi$ (\protect\subref{subfig:gchiscan-dmLight}), and $g_l$ (\protect\subref{subfig:glscan-dmLight}) for a vector mediator at the HL-LHC and FCC-hh (for quark and DM couplings only). The result is shown as a function of the mediator mass $m_{med}$; the mass of the DM candidate is fixed to 1 GeV in all cases. The coupling on the $y$ axis is varied while the other two couplings are fixed: in~(\protect\subref{subfig:gqscan-dmLight}), $g_\chi$=1.0 and $g_l$=0.0; in~(\protect\subref{subfig:gchiscan-dmLight}), $g_q$=0.1 and $g_l$=0.0; and in~\protect\subref{subfig:glscan-dmLight}, $g_q$=0.25 and $g_\chi$=1.0. Similar considerations as Figure~\ref{fig:hl-lhc-fcc-massmass} apply in terms of the lower mediator mass bounds for resonance searches. 
The arrows in the lower edge of the contours indicate that other searches for lower mass mediators that are normally performed at colliders could be sensitive to these models, but are not shown because the inputs received focused on the highest mediator masses only.}
\label{fig:couplinglimits-hl-lhc-allanalyses}
\end{figure}

The results in Figure~\ref{fig:couplinglimits-hl-lhc-allanalyses} show, as expected, that future hadron colliders are powerful probes of the simplified mediator models when the quark coupling is significant. 
On the other hand, lepton colliders are especially important when this is not the case, especially if the lepton couplings are the dominant way the mediator connects to the SM, or when there is a large amount of mixing to Z/Higgs/photon. 
When considering spin-1 production in a polarized lepton collider, it is possible to obtain a substantial enhancement in the production cross section when polarized left handed electrons collide with right handed positrons. This substantial enhancement allows for a large sensitivity to searches for invisible decays in the missing energy plus photon final state. Furthermore, the knowledge of the missing energy allows for the reconstruction of a missing mass peak that can further enhance an invisible search. The combination of these effects enable an enhanced sensitivity for vector dark matter mediators that couple the leptons. This sensitivity is comparable to the projected sensitivity for resonant mediator searches in the dilepton final state. In fact, for a mediator mass roughly equal to that of the ILC ultimate energy with its total luminosity, we can exceed lepton-coupled projected mediator sensitivity expected at the HL-LHC upgrade\cite{Kalinowski:2021vgb}.

Figure~\ref{fig:couplinglimits-hl-lhc-allanalyses} shows that both HL-LHC and FCC-hh can cover a wide range of mediator masses and couplings for scenarios where the dark matter particle is relatively light (1 GeV), with FCC-hh dijet searches as expected dominating at higher mediator masses, while monojet searches are most powerful at lower mediator masses. 

Figure~\ref{fig:couplinglimits-hl-lhc-allanalyses} also highlights that searches for visible and invisible decays of the mediator are complementary where their sensitivity overlaps. 
In case of a simultaneous discovery of a resonance in the dijet or dilepton invariant mass spectrum and of an excess of invisible transverse momentum, interpreting the results in terms of couplings for different searches can shed light on the nature of the DM-SM interactions.  

The power to discover or exclude a massive vector mediator at the HL-LHC and FCC-hh decreases as the coupling $g_q$ decreases, as this also decreases the production cross-section of the mediator in $pp$ collisions. 

\begin{figure}[htb!]
\centering
\subfloat[]{\label{subfig:gdm_monojet_1GeV}\includegraphics[width=0.49\textwidth]{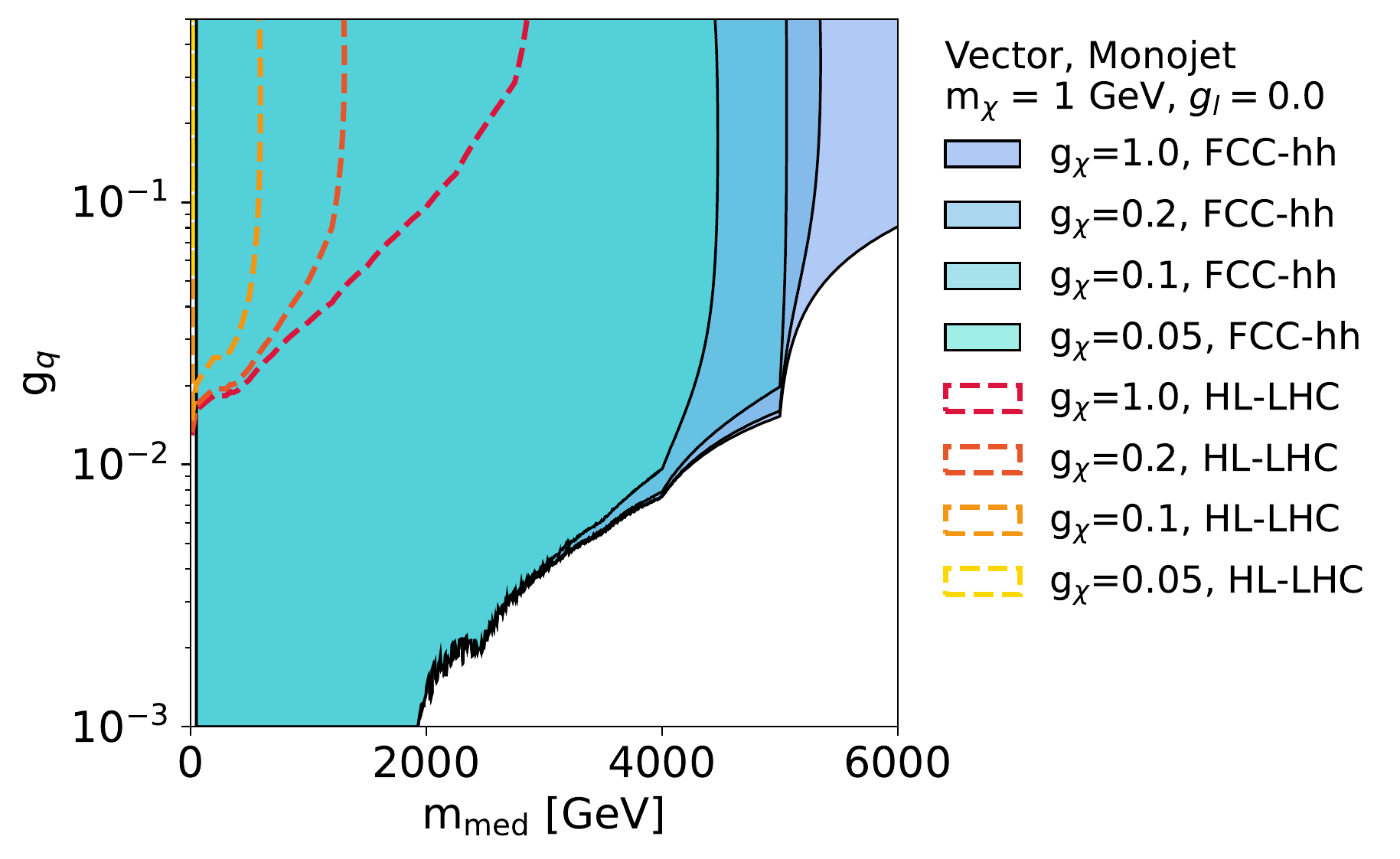}}
\subfloat[]{\label{subfig:gdm_dijet_1GeV}\includegraphics[width=0.49\textwidth]{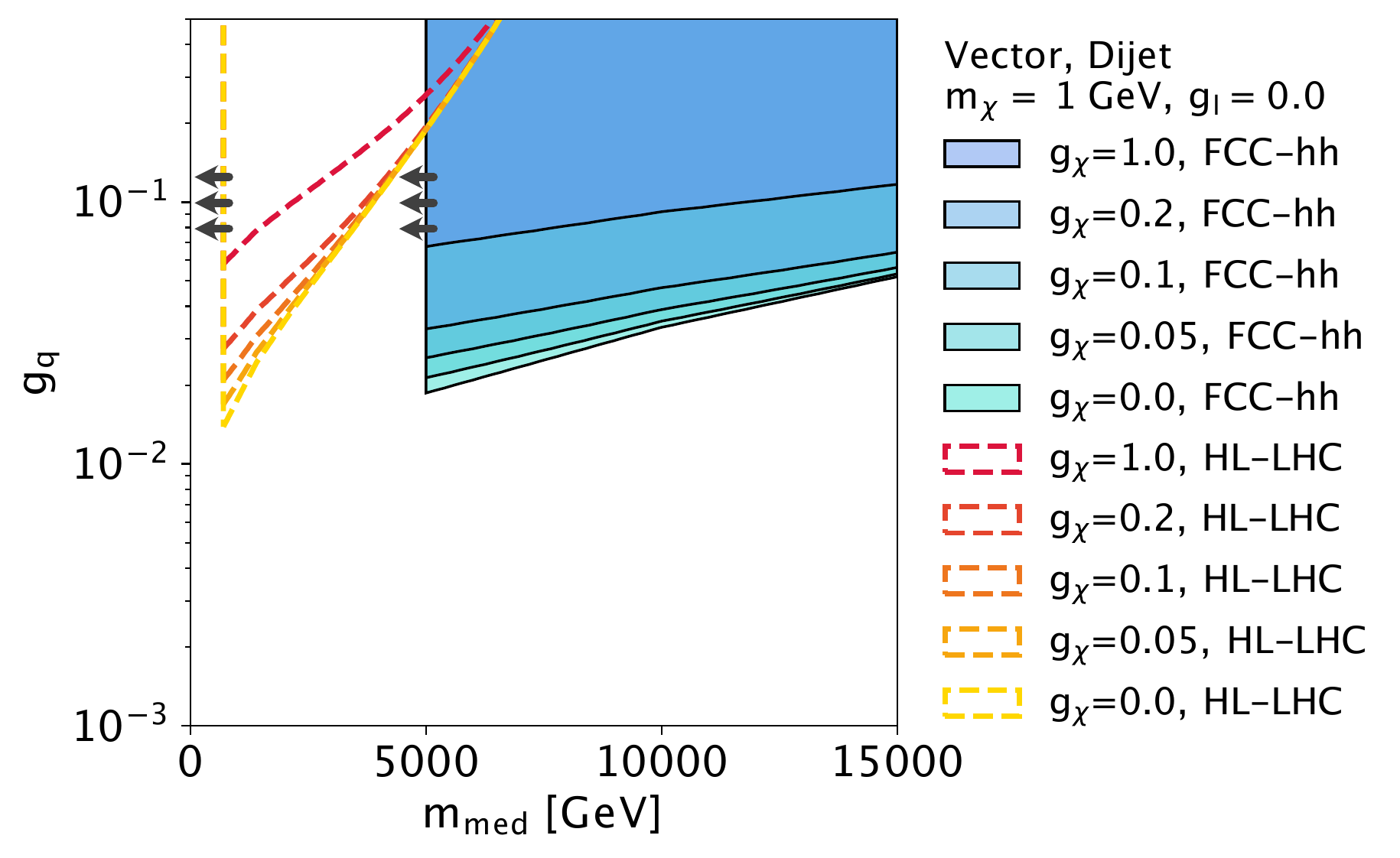}}
\\
\subfloat[]{\label{subfig:gdm_monojet_1GeV}\includegraphics[width=0.49\textwidth]{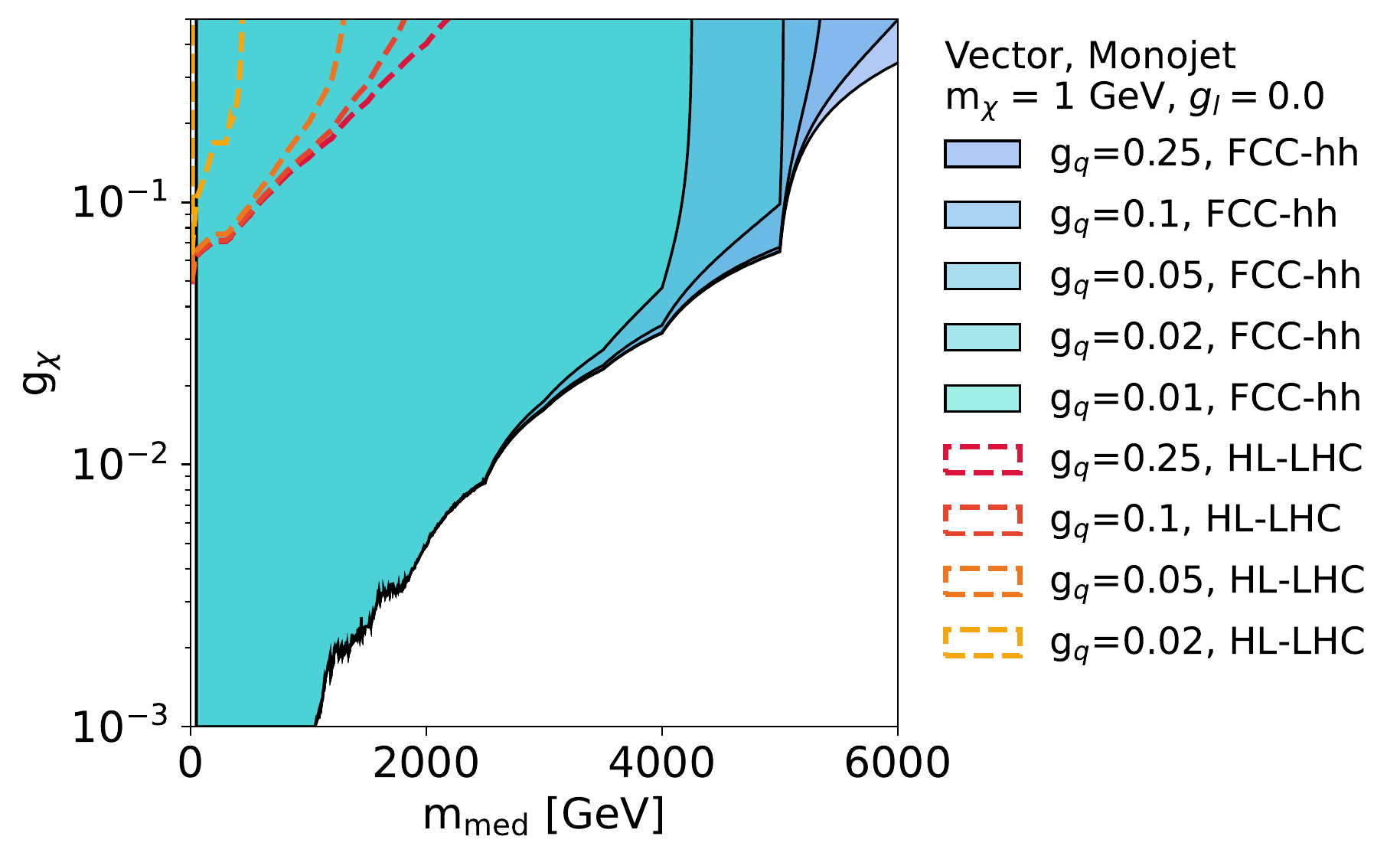}}
\subfloat[]{\label{subfig:gdm_dijet_1GeV}\includegraphics[width=0.49\textwidth]{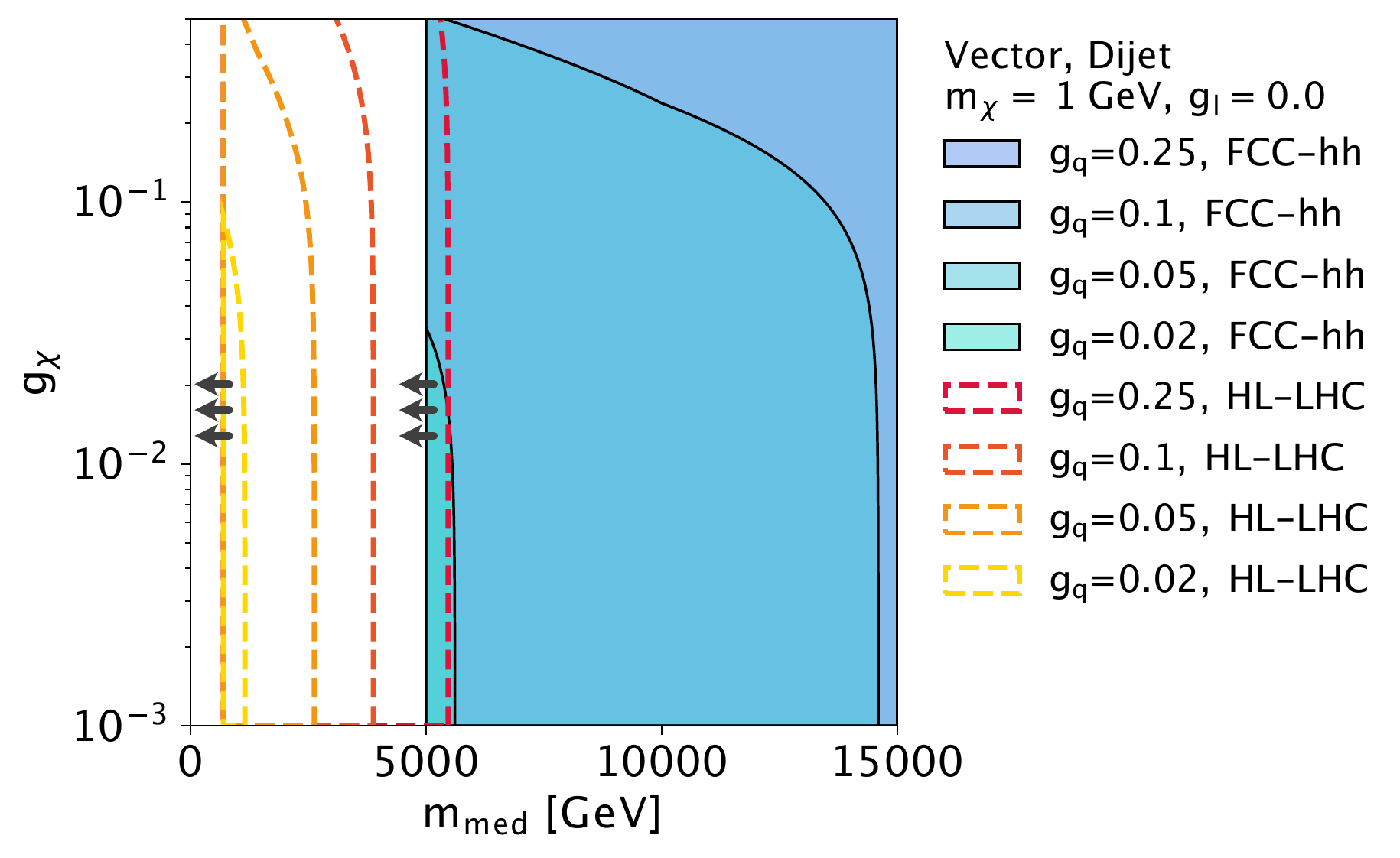}}

\caption{Projected exclusion limits on the couplings $g_q$ (top row) and $g_\chi$ (bottom row) for a vector mediator at the HL-LHC and FCC-hh. The result is shown as a function of the mediator mass $m_{med}$; the mass of the DM candidate is fixed to 1 GeV in all cases. The coupling on the $y$ axis is plotted as the other coupling is scanned as shown in the legend. Similar considerations as Figure~\ref{fig:hl-lhc-fcc-massmass} apply in terms of the lower mediator mass bounds for resonance searches. The arrows in the lower edge of the contours indicate that other searches for lower mass mediators that are normally performed at colliders could be sensitive to these models, but are not shown because the inputs received focused on the highest mediator masses only.}
\label{fig:couplinglimits-scans}
\end{figure}

The plots in Fig. \ref{fig:couplinglimits-scans} include a scan on the coupling not shown on the y axis as well, and show further that the dependence of the HL-LHC and FCC-hh experiments sensitivity to the value of the mediator’s couplings to SM fermions (which control the production and visible decays of the mediator) and the value of the DM coupling (which controls the relative proportions of visible and invisible decays). The dijet projections in Fig. \ref{fig:couplinglimits-scans} show how visible searches can also strongly constrain the $g_\chi$ coupling in these simplified models. 

Across the scenarios considered, collider bounds are at their strongest in cases of TeV-scale mediator masses for monojet, while dijet can reach higher mediator masses. This result is largely independent of the other model parameters.

\FloatBarrier

\subsubsection{Comparison with DM searches at other frontiers}

Besides the comparisons between searches at a given future collider, and comparisons that can be made between future collider facilities, one can also use the simplified models framework of Ref.~\cite{Abercrombie:2015wmb} to make comparisons of search coverage across proposed collider, direct-detection and indirect-detection searches, in the manner recommended by the LHC Dark Matter Working Group and performed for the recent European Strategy Briefing Book~\cite{Strategy:2019vxc}.

\begin{figure}[!htp]
\centering
\subfloat[Monojet searches]{\label{subfig:couplingscan-dd-monojet-hllhc}\includegraphics[width=0.49\textwidth]{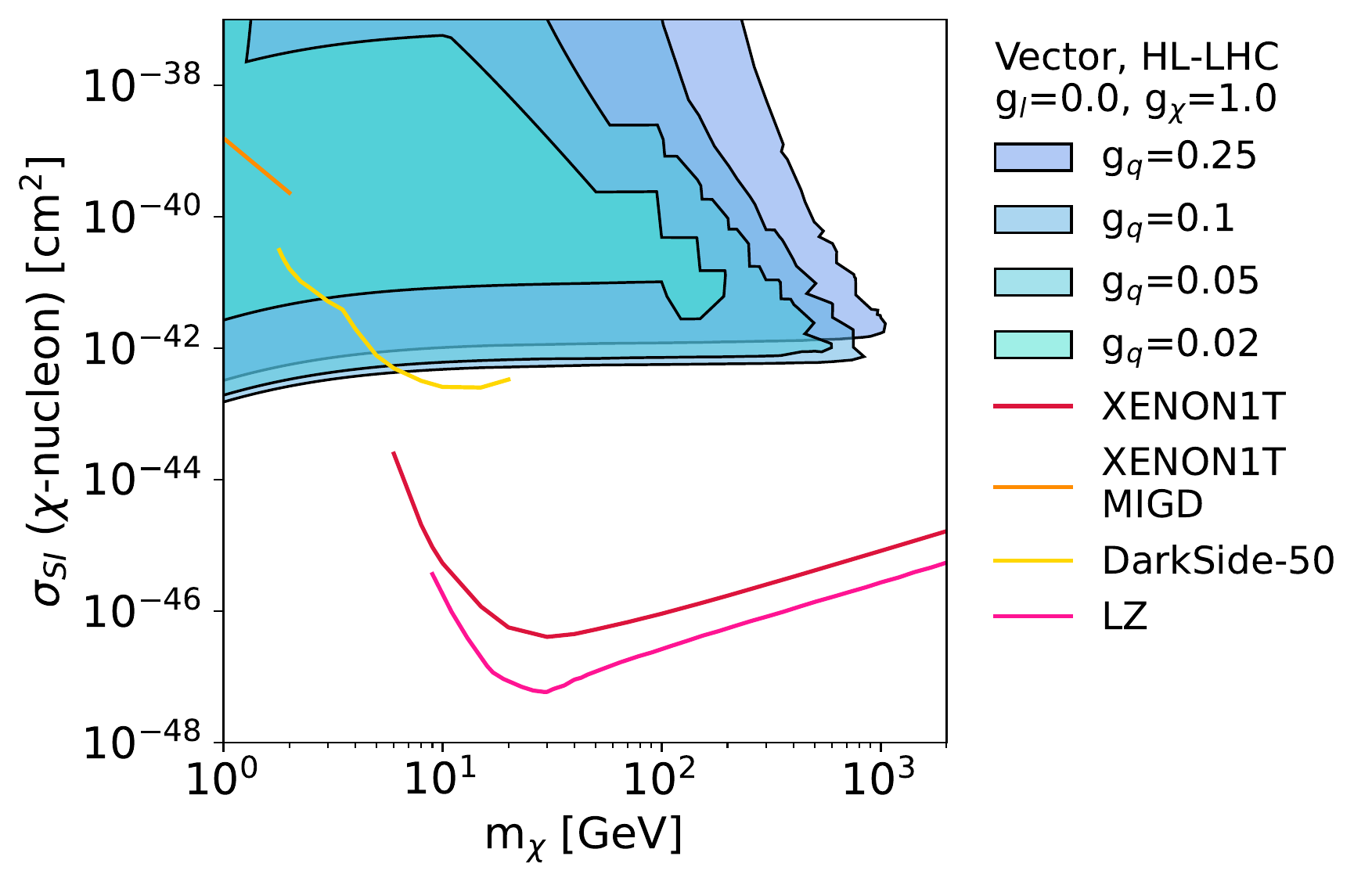}}
\subfloat[Dijet searches]{\label{subfig:couplingscan-dd-dijet-hllhc}\includegraphics[width=0.49\textwidth]{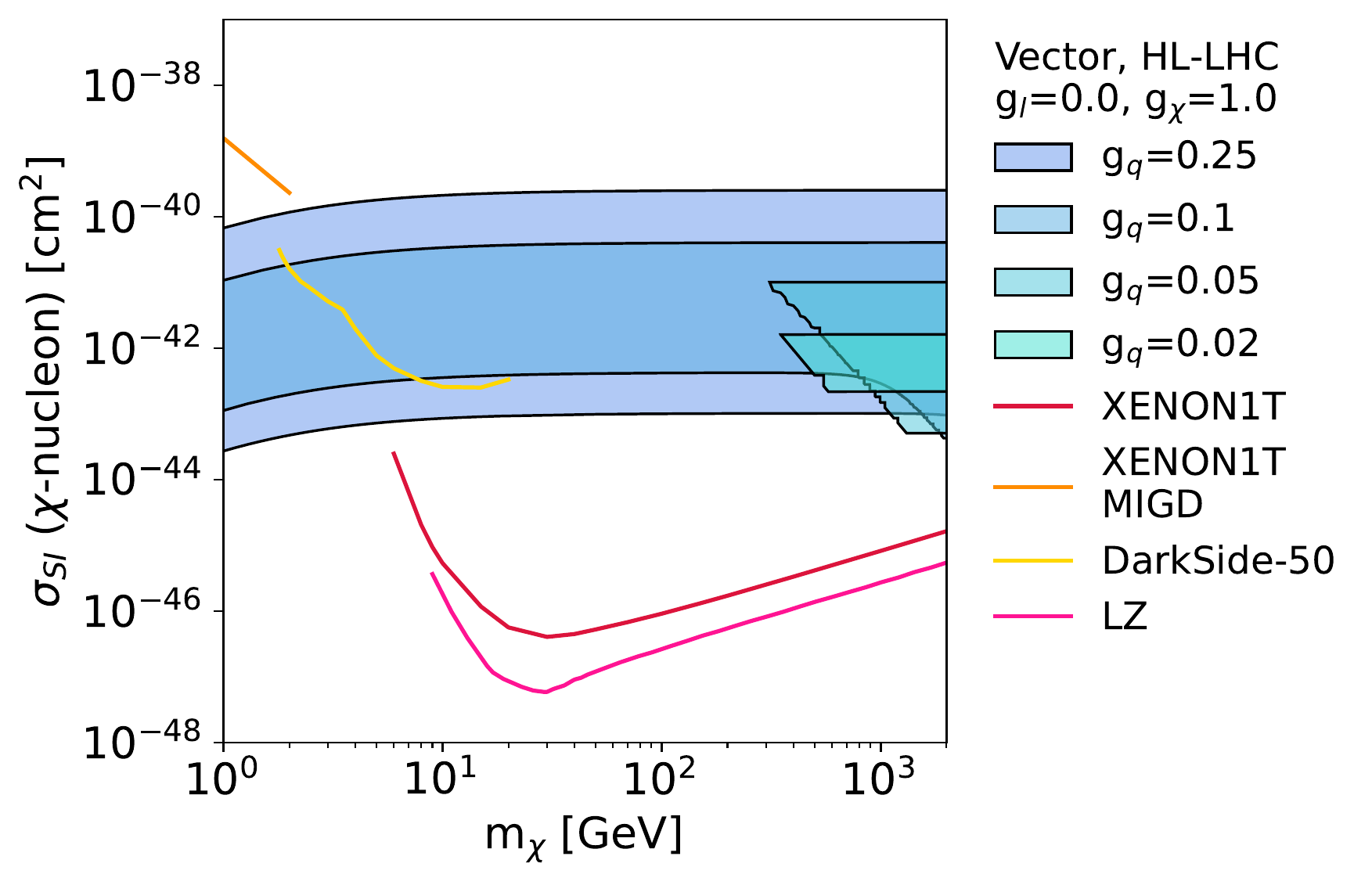}}\\
\subfloat[Monojet searches]{\label{subfig:couplingscan-dd-monojet-fcchh}\includegraphics[width=0.49\textwidth]{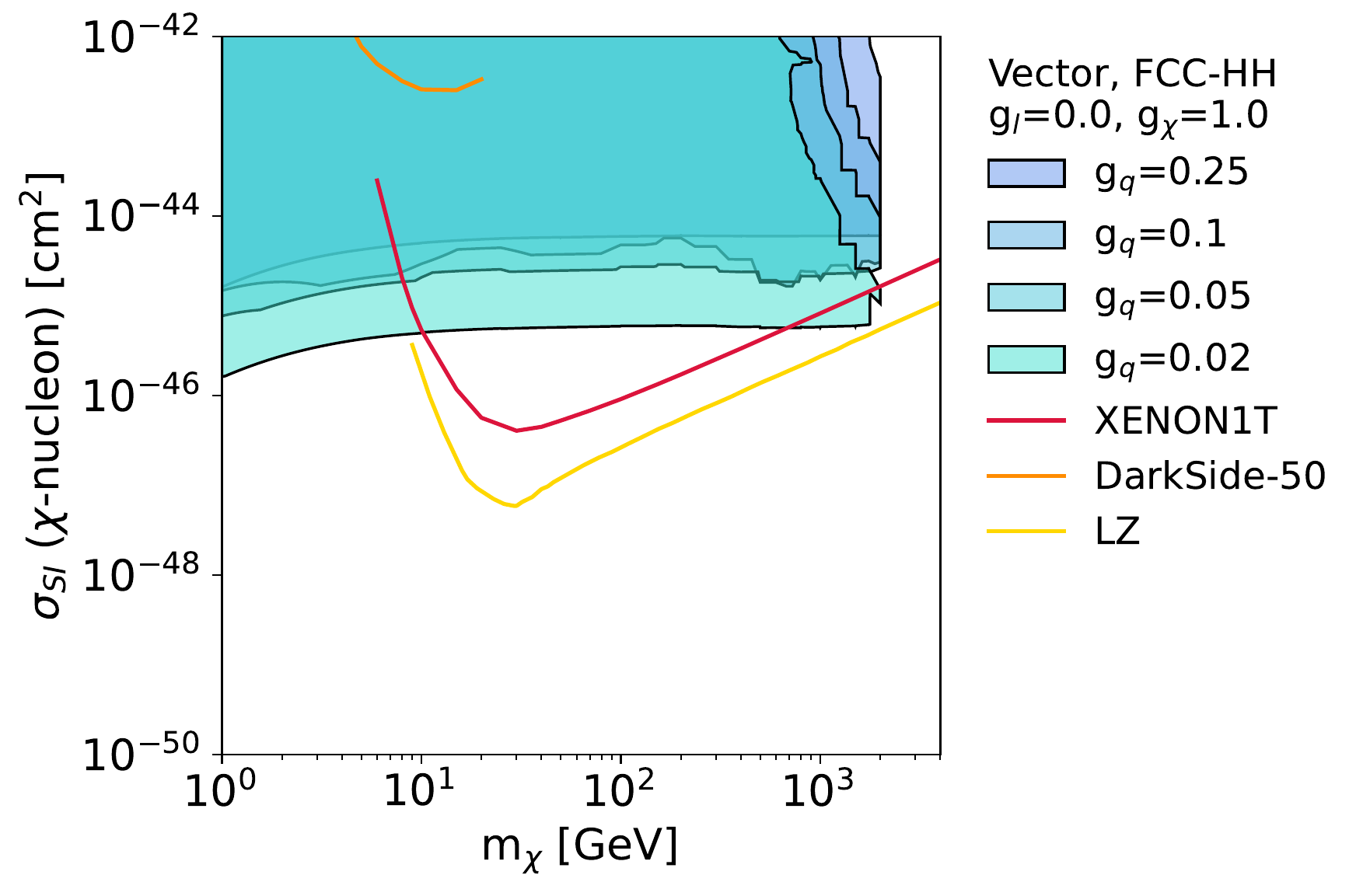}}
\subfloat[Dijet searches]{\label{subfig:couplingscan-dd-dijet-fcchh}\includegraphics[width=0.49\textwidth]{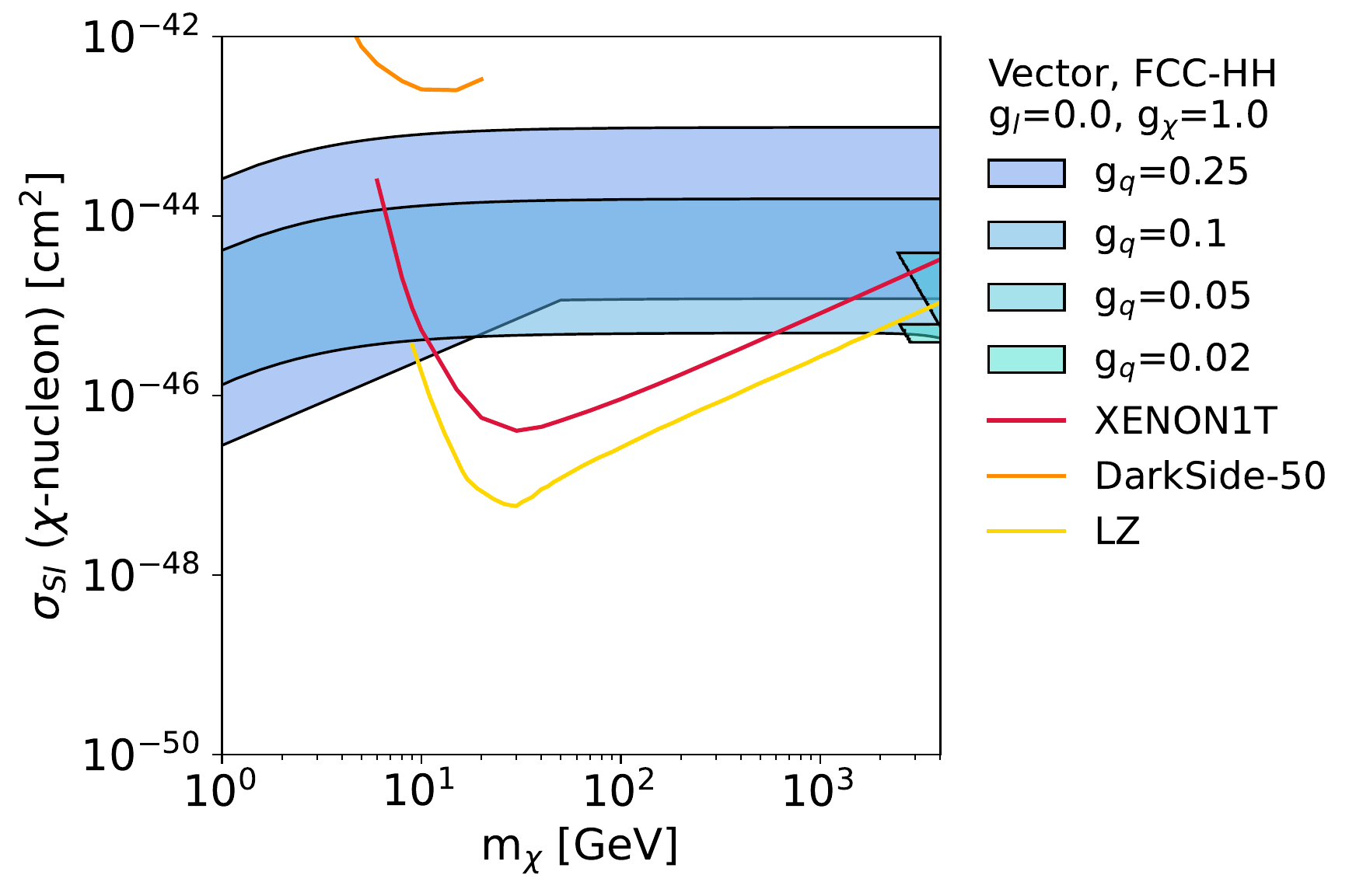}}
\caption{Effects on exclusions of $\sigma_{\mathrm{SI}}$ when varying the $g_q$ coupling for the monojet (\protect\subref{subfig:couplingscan-dd-monojet-hllhc}) and dijet (\protect\subref{subfig:couplingscan-dd-dijet-hllhc}) signatures at the HL-LHC, and for the monojet (\protect\subref{subfig:couplingscan-dd-monojet-fcchh}) and dijet (\protect\subref{subfig:couplingscan-dd-dijet-fcchh}) signatures at FCC-hh. The dark matter coupling is held fixed to $g_{\mathrm{DM}}=1$; there is no coupling to leptons. Existing limits from direct detection experiments~\cite{DarkSide:2018bpj,XENON:2018voc,XENON:2019zpr,LZ:2022ufs} are shown for context. A re-analysis of the DarkSide-50 data~\cite{DarkSide-50:2022qzh} appeared after this figure was finalized.}
\label{fig:hl-lhc-dd-separate-si}
\end{figure}

The conversion of collider results on the spin-independent DM-nucleon cross section vs DM mass plane is performed in Fig. \ref{fig:hl-lhc-dd-separate-si} using procedures from Refs.~\cite{10.1016/j.dark.2019.100365.Boveia.2020,10.1016/j.dark.2019.100377.Albert.2019}). 
A scan of the quark and DM couplings shows that as future collider searches reach smaller $g_q$, their coverage in the DM-nucleon cross-section plane improves. 
When the coupling sensitivity limit approaches (e.g. in the case of $g_q$=0.02 for HL-LHC), collider projections gradually disappear. This is especially evident for dijet results, where the quark coupling enters in both mediator production and decay. 
Similar results are obtained when scanning $g_{\chi}$. 

\begin{figure}[!htp]
 \centering
 \includegraphics[width=0.75\textwidth]{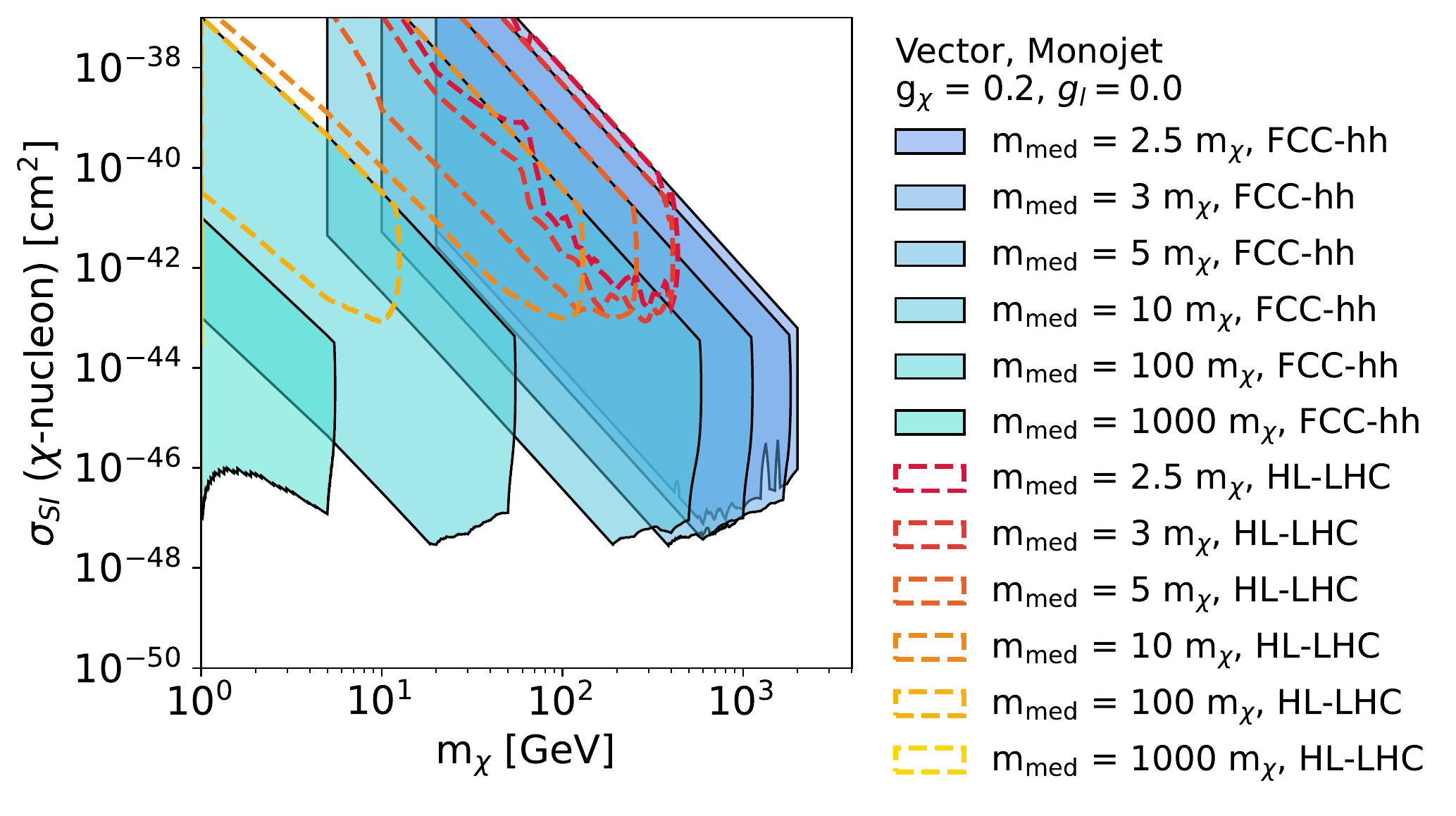}
  \caption{Projected limits from HL-LHC and FCC on the DM-nucleon cross-section vs DM mass plane, when varying the ratio between mediator and DM mass.}
  \label{fig:massRatio}
\end{figure}

One can also scan the ratio between the mediator mass and the dark matter mass (which are also free parameters in this kind of simplified models) while fixing one of the two couplings. 
This kind of scan shows a different projection of collider results with respect to the plot in Fig. \ref{fig:hl-lhc-dd-separate-si}, as it singles out individual mass hypotheses in the simplified model. 
An example of this scan for the monojet searches is shown in Fig. \ref{fig:massRatio}. 
One can see that imposing a specific requirement for the ratio of the masses instead of a specific requirement on the coupling changes the collider sensitivity and the excluded region depends on what that ratio is \footnote{The shape of the exclusions in this plane can be understood because the DM nucleon cross-section scales like the inverse of the fourth power of the DM mass when the mass ratio is fixed.}.


The main message from these plots remains the same as in the European Strategy Briefing Book~\cite{Strategy:2019vxc}. 
In a scenario where particle dark matter is discovered at a direct- or indirect-detection search, Figure~\ref{fig:hl-lhc-dd-separate-si} illustrates---in a necessarily model-dependent fashion for the specific simplified model considered---the parameter space of the model in which collider searches for invisible particles would also be sensitive to production of the mediator. 
In roughly these regions, both types of searches would have complementary discovery potential, as discovery at a direct-detection experiment would be combined with further study of the type and properties of the interaction between the DM and the Standard Model at a hadron collider.

The figure also indicates where collider searches for invisible particles would supplement the search coverage of the other DM experiments with unique sensitivity. 
Nevertheless, even in these regions, it would be essential to confirm that the collider discovery is indeed associated with galactic dark matter when considering models where the DM particle is stable\footnote{Collider searches for WIMPs have unique sensitivity to models of decaying DM. This is realized for example via R-parity violating higher-dimensional operators that induce LSP decays on less than cosmological time scales, but that keep the LSP stable on collider time scales.}. 
Due to the compact scale of collider experiments, a collider discovery in these scenarios could be confirmed as DM with the help of additional experiments beyond the Energy Frontier (i.e. targeted direct and/or indirect DM searches).

Further effects such as coannihilation can also alter DM production in the early universe and detection today.  Ref.~\cite{Baker:2018uox}, for example, presents a simplified $t$-channel model with a multi-TeV dark matter candidate which exhibits interesting complementarity between collider and direct detection searches.  Collider searches are relatively insensitive to whether DM is Dirac or Majorana, while this has a significant impact on direct detection prospects.


\FloatBarrier

\subsection{Beyond WIMPs: Dark Matter portals and other models}\label{Sec:OtherDM}

Models featuring portal or mediator particles that connect SM particle to dark matter and dark sectors have been widely described in literature (see e.g. the first few references of Ref. \cite{Rizzo:2022qan} as a starting point). 
The models of BSM mediation already described in Section \ref{Sec:DMst} are motivated by the presence of a thermal relic candidate with a mass in the GeV-TeV scale and with an interaction strength compatible with that of the weak force (a WIMP). 
Thermal relic DM with much lighter masses (below the GeV) can also be produced by much feebler interactions between the portal particle and the SM. 
This is a strong motivation for the recent experimental and theoretical efforts in extending the WIMP paradigm to lower DM masses and even smaller couplings \cite{Bertone:2018krk,sm21:rf-report}. 
The signatures generated by feebly coupled models often include particles with long lifetimes, also described in Sec. \ref{Sec:LLP}. 
Optimal coverage for this kind of signatures requires a dedicated experimental toolkit for experiments at high energy collider, as well as dedicated experiments.   

The wide and relatively unexplored range of possible parameters of portal models covering a wide range of DM masses can be used to highlight complementary avenues to search for DM with high- and low-transverse-momentum (forward) experiments \cite{Feng:2022inv} at particle colliders, with experiments at accelerators \cite{sm21:rf-report}, as with low-threshold Cosmic Frontier experiments \cite{Essig:2022dfa}. 

\subsubsection{Portals considered for DM at collider}

In this section, we briefly outline a limited number of types of portal/mediator models that have been used as search targets by the collider and accelerator community, and already featured in the European Strategy Update Briefing Book \cite{Strategy:2019vxc}. 
We focus on the models where relevant work has been submitted to the Snowmass process. 
A more detailed description of portals used as benchmarks for the Snowmass accelerator community can be found in the Snowmass Rare Processed and Precision Frontier Report~\cite{sm21:rf-report}.

\paragraph{Simplified models for DM at colliders and portal models for DM at accelerators}

The simplified models collected and developed by the Dark Matter Forum and Working Group for early LHC searches \cite{Abercrombie:2015wmb} include a new mediator/portal particle between the DM and the SM, and are described in Section \ref{Sec:DMst}. 
The simplified models that are relevant for this Section are those where the mediator particle is a vector, a scalar or a pseudo-scalar.  

The Physics Beyond Colliders Study Group report \cite{Beacham:2019nyx} features a list of minimal portal models, focusing on the lowest dimensional 
portals between the SM and a dark sector. 
In this Section, we will focus on these minimal implementations of the dark photon (vector), dark Higgs (scalar) and axion (pseudoscalar) portals, where the only additions to the SM are the portal particle and the DM particle. 

The purposes of these two groups of models are different: the mediator models in \cite{Abercrombie:2015wmb} were designed to be as simple as possible and span a broad parameter space to point out uncovered areas without a specific emphasis on theoretical or cosmological consistency, while the portal models in \cite{Beacham:2019nyx} are more theoretically self-consistent and have been used to compare the sensitivity of different experiments to thermal light dark matter. 

As an example of the qualitative differences between these two group of models, one can compare the vector model from \cite{Abercrombie:2015wmb} and the dark photon model in \cite{Beacham:2019nyx}. 
In the vector model, the couplings to DM and SM particles are treated as independent variables and interference with the Z can be neglected, since it is used as a benchmark for mediator masses much higher than the Z. In the dark photon model, the kinetic mixing leads to specific relationships between couplings of dark photon and the SM photon and Z. 
This means that in the dark photon model there are fewer degrees of freedom for the generation of different collider signatures, but on the other hand the theoretical consistency and the mixing with SM gauge bosons makes the dark photon model a more realistic search target for light mediator masses. 

In any case, the two groups of models are closely related to each other, and it is possible to recast results interpreted using mediator models into portal models and vice versa, as seen in the following section \ref{sec:FeaturesOfPortals}. 

\paragraph{Portal models for dark sector searches}

To expand on the portals mentioned in the previous paragraph, one can make the dark sector more complex. This class of models can be considered part of the larger family of Hidden Valley models \cite{Strassler:2006im}. 

In Hidden Valley models, the interaction between the SM and the dark particle spectrum can be realized through a mediator particle, e.g. a dark photon or a dark Higgs. 
The dark sector is “hidden" by the rarity of the interactions, and it may contain dark matter candidates.

More specifically, Ref. \cite{Knapen:2021eip} (and references therein) consider a set of heavy mediators that decays to lighter dark sector states, and one or more of these states may also serve as dark matter candidates. 
These portals have been taken as the starting point for building benchmarks leading to dark showers signatures in the contributed whitepaper in Ref.\cite{Albouy:2022cin} and summarized in previous Section \ref{sub:DarkShowers}. 

In the following section \ref{sec:FeaturesOfPortals}, we will focus on qualitative and conceptual aspects of including a dark matter candidate in HV models and dark showers, as discussed in the \href{https://indico.cern.ch/category/12893/}{Dark Showers project meeting in May 2022}. 

\subsubsection{Features of portals for DM searches at colliders}
\label{sec:FeaturesOfPortals}

\paragraph{Complementarity of visible and invisible searches}

As already mentioned in Section \ref{Sec:DMst}, when portal particles are produced via interactions with ordinary matter, they will also decay back into ordinary matter as well as into dark matter particles. 
This leads to the possibility of obtaining further information on the nature of the portal and of the dark matter candidate from both visible and invisible searches at colliders.   

For the dark photon (vector) portal with kinetic mixing, decays are predominantly into visible or invisible particles, depending on the dark photon - DM mass ratio and on their coupling. 
The full exploration of the parameter space for this kind of benchmarks requires a number of different experiments: an observation of a new vector particle decaying invisibly or into specific kinds of SM particles can be mapped to further observations (or lack thereof) in other channels in order to elucidate the full set of properties for the new vector particle. 

In a similar vein, the branching ratios of s-channel portal models to more complex dark sectors (e.g. semi-visible jets from dark showers) can also be constrained using the decays of the portal particle to the same SM particles that produced it.  
In order to focus on the parameter space that we should target for these models with future colliders, it would be useful to provide tools that allow for easy reinterpretation of resonance searches such as those in \cite{Harris:2022kls}. 

\paragraph{Collider inputs to the complementarity between low and high energy experiments}

Owing to the similarities between mediator and portal models, it is possible to reinterpret collider searches for invisible particles in terms of lighter (beyond WIMP) dark matter scenarios. 
This is a new development during the Snowmass process, starting from the results presented in the European Strategy Update. 

In Ref. \cite{Boveia:2022jox}, the CMS search for invisible particles in the jet+MET final state \cite{CMS:2021far} is interpreted in terms of the dark photon, dark Higgs and axion portal models. 


In the \textbf{dark photon portal model}, the portal particle has very small vector couplings with SM particles, induced by kinetic mixing with the SM photon. 
One can consider this model as a 
specialization of the vector mediator model from Section \ref{Sec:DMst}, (also called \texttt{LHC DM} in the following). 
The dark photon model also includes mixing between the dark mediator and the Z boson \cite{Curtin_2014,Curtin_2015}, with a mixing angle $\theta_a$, which does not feature in the \texttt{LHC DM} vector mediator model. 

Ref. \cite{Boveia:2022jox} uses the \textit{HAHM} implementation from  Refs.\cite{Curtin_2014,Curtin_2015} to extend and reinterpret the results from the most recent CMS search for invisible particles produced in association with an energetic jet \cite{CMS:2021far} in terms of the dark photon parameters. 
The study also projects these results to the full HL-LHC dataset. 
In this study, the hidden Higgs boson coupling is turned off by decoupling the dark Higgs, to isolate the effect of the Z mixing and to effectively reduce the HAHM model to the "minimal" dark photon model. 

\begin{figure}[!hptb]
     \centering
     \includegraphics[width=0.9\textwidth]{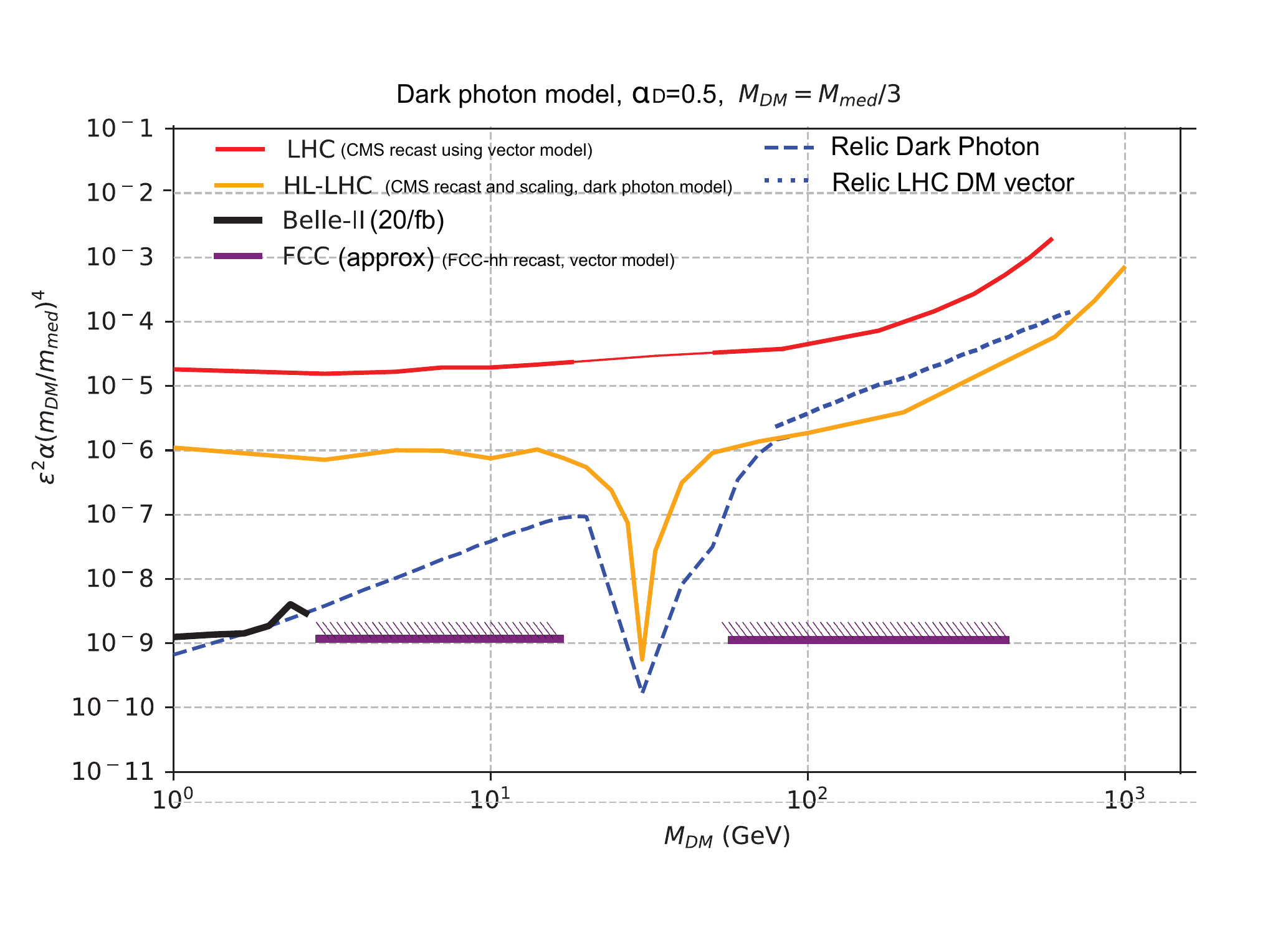}
     \caption{Comparison of the expected constraints from reinterpreting the results of \cite{CMS:2021far} in terms of the simplified vector mediator model (\textit{LHC DM}) and dark photon model. 
    The blue "relic" lines represent the minimum parameter combinations that would reproduce the observed thermal relic density for each model, with the expected deviation for the dark photon model around the Z resonance. Orange lines forecast the increased sensitivity of this search for the dark photon model with HL-LHC (via cross-section scaling of the CMS result),  FCC-hh (reinterpreting the results in Section \ref{sub:extrapolationDMst}) and from Belle-II 20/fb \cite{ Belle-II:2022cgf,Belle-II:2018jsg}.}
     \label{fig:dark_photon_1}
\end{figure}

The results of this reinterpretation are shown in Figure \ref{fig:dark_photon_1} for a DM mass range 3 GeV - 1 TeV, plotted on the plane  $y$ - dark photon mass, with $y = \epsilon^2 \alpha_D (m_{DM}/m_{med})^4$ where $\epsilon$ is the dark photon - SM photon mixing parameter, 
is the coupling constant between the dark photon and DM set as  $\alpha_D=\frac{g_{DM}^2}{4\pi}$ set to 0.5, 
and $m_{DM}$ and $m_{med}$ are the DM mass and the dark photon mass respectively. The dark photon mass is fixed to 3 times the DM mass, following the choice in the Rare Processes and Precision Frontier for invisible DM benchmarks, to allow the mediator to decay to DM on-shell and to have a predictive model for the relic density. 
We note in any case that scenarios where the mediator cannot decay directly to DM can be considered equally plausible, and can be favorable for collider searches since they privilege visible decays of the mediators that could be discovered in resonance searches (see Section \ref{Sec:NB}). 
Figure \ref{fig:dark_photon_1} also highlights the differences between the two models, due to the Z-dark photon mixing that is neglected in the \texttt{LHC DM} vector model due to its predominant use for higher mediator/DM mass scenarios, while in the region away from $m_{DM} = m_Z/3$ the two models are equivalent. 

These results can be used to advocate that lower energy experiments at accelerators and high energy colliders can cover together a number of well-motivated scenarios where the model can produce the entirety of the relic dark matter in the universe (also called \textit{thermal target} scenarios in the Rare Processes and Precision Frontier), using the dark photon as a portal for dark matter. 
It is also worth mentioning that additional processes and particles beyond those foreseen in the simple portal models (e.g. additional particles at higher energy scales, or multi-component DM) can also deplete or increase the early universe dark matter abundance, so that the relic constraints are satisfied outside the thermal target parameter values.  

Furthermore, we see that the full HL-LHC dataset is needed to reach a sensitivity to the thermal relic milestone for DM masses above 100 GeV, while FCC-hh is needed to cover the full range of DM masses. While below 3 GeV B-factory experiments and other accelerator experiments covered in the Rare Processes and Precision Frontier dominate the sensitivity, the region above 3 GeV is only partially covered and requires high energy collider experiments to be fully explored. 
We note that we also expect lepton colliders to play an important role in discovering this kind of models as discussed in \ref{sub:extrapolationDMst} but we did not have any inputs available. 

The \textbf{dark Higgs/scalar portal} features a new scalar $S$ with characteristics similar to the SM Higgs boson, that can decay into DM and SM particles. The dark Higgs and the SM Higgs mix, with an angle $\theta$. 
This benchmark model can be discovered or constrained using both searches for invisible decays of the Higgs boson and precision measurements of the Higgs couplings. 

\begin{figure}[htp]
     \centering
         \centering
         \includegraphics[width=0.45\textwidth]{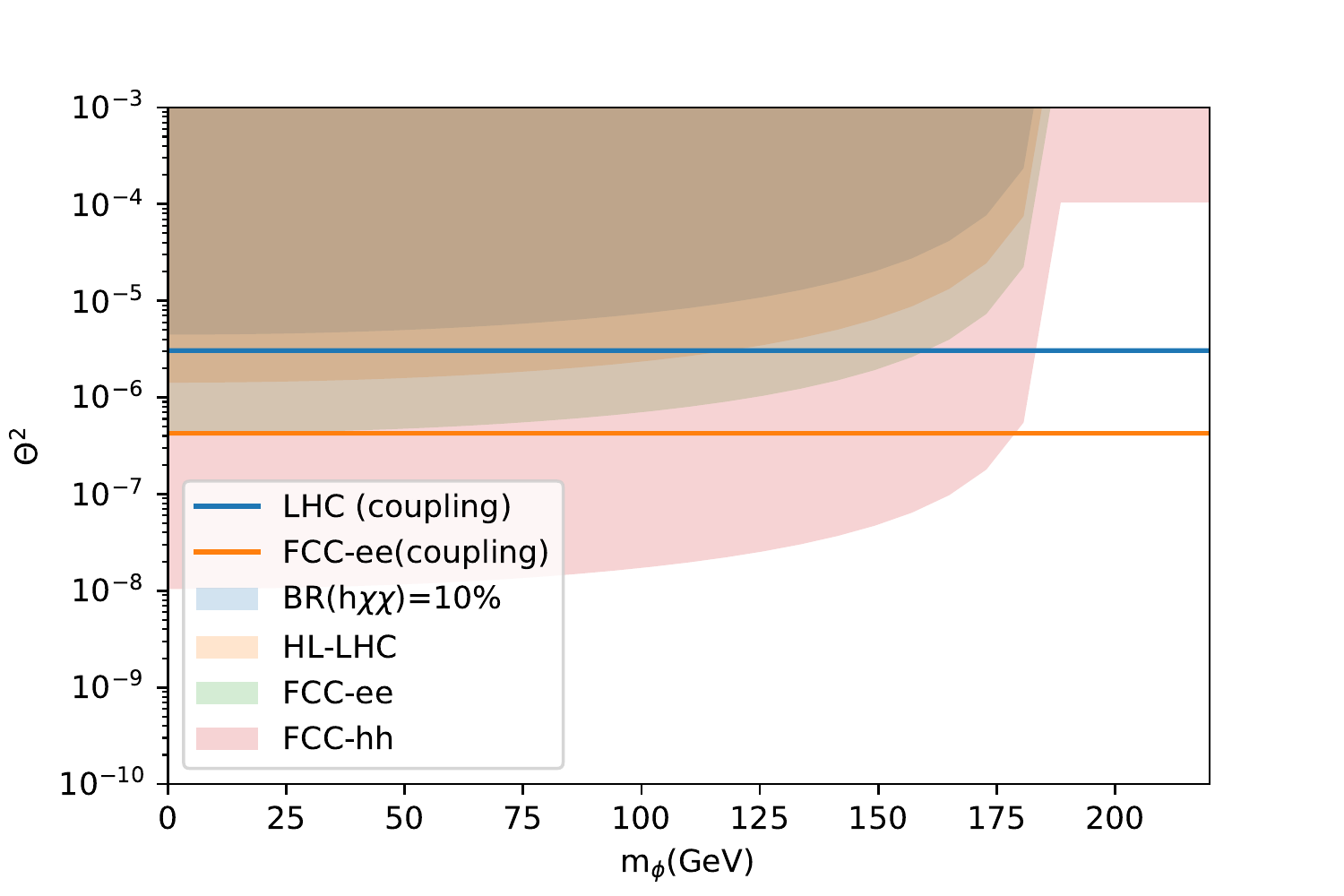}
         \includegraphics[width=0.45\textwidth]{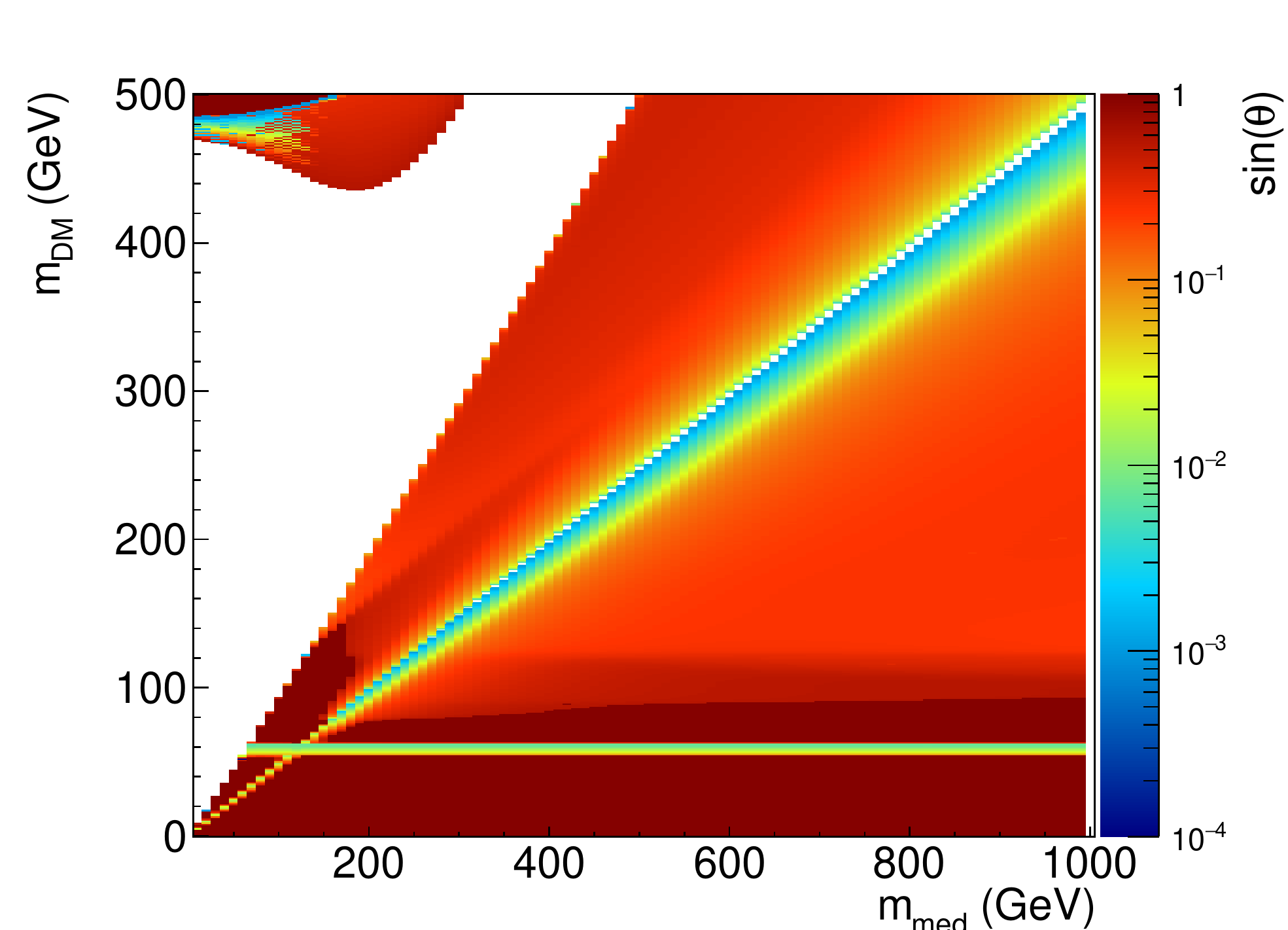}
         \caption{(Left) minimium mixing angle for the Higgs to invisible search when directly applying this search to the singlet mixing model. The solid lines indicate the constraints coming from indirect bounds on the Higgs couplings.  (Right) minimum allowed mixing angle for a model containing a Dark Higgs that mixes with the standard model Higgs boson.}
         \label{fig:darkhiggs}
\end{figure}

Figure~\ref{fig:darkhiggs} presents a reinterpretation of the constraints on the the mixing angle $\theta^{2}$ from the Higgs to invisible decays as a function of the scalar mass, assuming a dark matter mass to mediator ratio of $m_{\chi}/m_{\phi}=\frac{1}{3}$.
The values of $sin(\theta)$ required to fulfil the measured relic density are also shown in the right-hand side of Fig.\ref{fig:darkhiggs}. 
This plot confirms and extends the one in the RF06 contributed whitepapers “Dark Matter Production at Intensity-Frontier Experiments” \cite{Krnjaic:2022ozp}. 

This plot shows that for light dark Higgs and DM masses the model would overproduce DM (where the z axis saturates at $sin(\theta)$=1), but for larger masses there are still values of $sin(\theta)$ that lead to large mixing angles that could be in reach for HL-LHC searches. 

Collider searches for missing transverse energy can also be easily interpreted in terms of the gluon couplings $c_{g}$ of an axion-like particle that also decays into DM, in terms of the \textbf{axion portal} model.
Ref. \cite{Curtin:2018mvb} uses published constraints of the CMS search for a pseudoscalar mediator for this reinterpretation. 
LHC results constrain the gluon coupling to be roughly below $10^{-3}$, and future colliders will be able to extend this by an order of magnitude. 

\paragraph{Complementarity between low and high mass particle searches}

Most portal models require the presence of both low and high mass particles to be self-consistent, even though it is not immediately obvious from the formulation of minimal benchmarks. 
Two works submitted to the Snowmass Study come to this conclusion for the vector/dark photon and dark Higgs portal \cite{Rizzo:2022qan,Batell:2018fqo, Batell:2021xsi}.

In \cite{Rizzo:2022qan}, new particles at the TeV scale at loop-level are needed to generate the kinetic mixing interaction that characterizes dark photon models. 
This leads to new signatures discoverable by colliders, with a rich phenomenology given that these new particles carry additional dark charges on top of the SM ones. 
There are a variety of phenomenological implications depending on the specific nature of these new particles (called “portal matter”). 
While indirect signs of such particles can be seen at electron-positron colliders, their production mechanisms and masses require a high energy collider such as HL-LHC or FCC-hh to be discovered and studied directly.

\begin{figure}[h!]
     \centering
         \centering
         \includegraphics[width=0.7\textwidth]{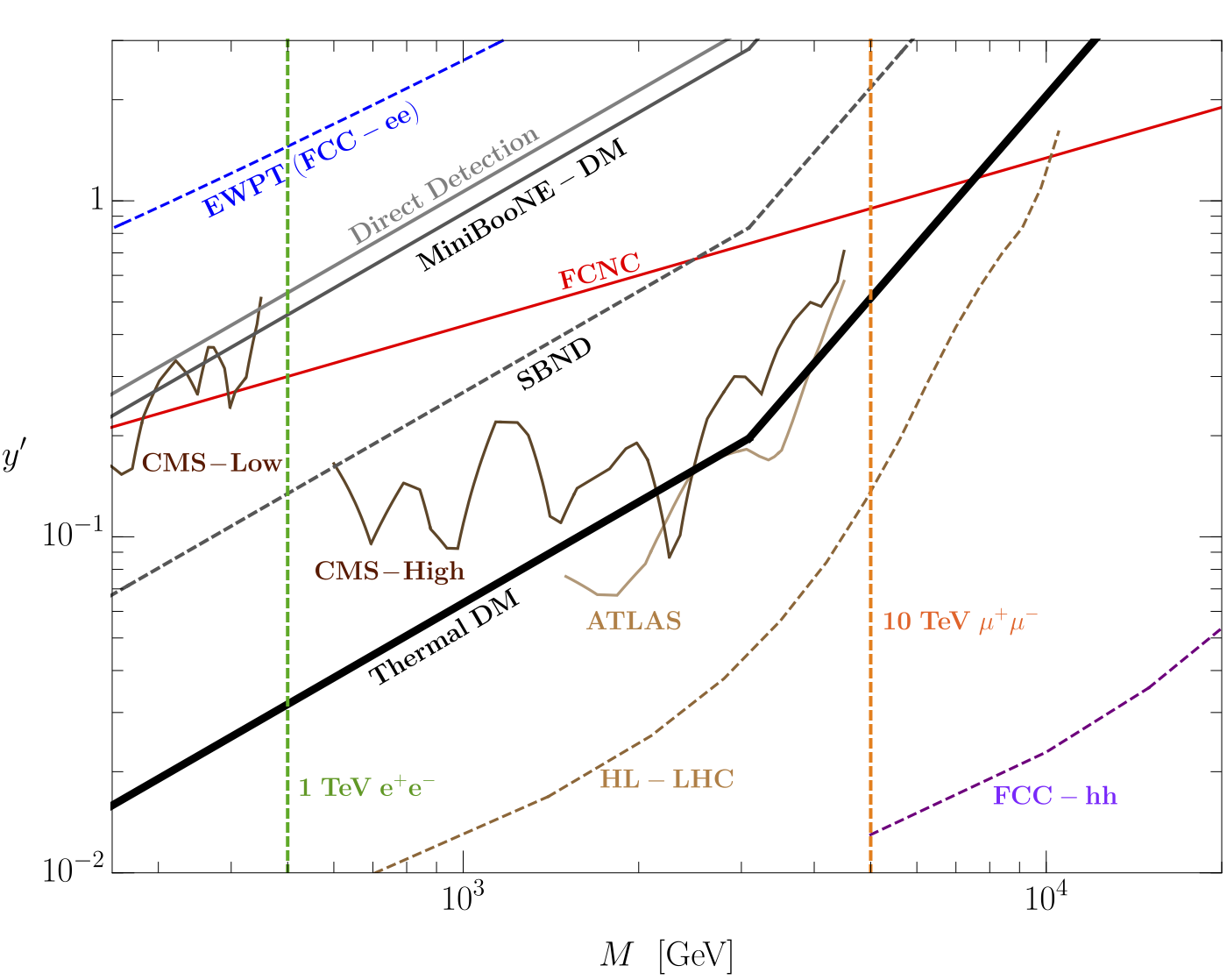}
         \caption{Existing constraints and future prospects for the up-specific scalar mediator $S$ in the renormalizable completion described in \cite{Batell:2018fqo, Batell:2021xsi}.
         together with parameter choices leading to the correct thermal dark matter abundance. 
         }
         \label{fig:flavorfulScalarResonances}
\end{figure}

The work in \cite{Batell:2018fqo, Batell:2021xsi,Egana-Ugrinovic:2019wzj,Egana-Ugrinovic:2019dqu} shows that if a light scalar singlet (the dark Higgs) decaying to light dark matter couples predominantly to one SM fermion family, then new heavy states are required to keep the model self-consistent at higher collider energies. 
In the example given, the Higgs boson mass and the dark matter mass are both of the order of a GeV, while the new particles needed to complete this model have masses of the order of a TeV. 
This is shown in Fig. \ref{fig:flavorfulScalarResonances}. 

These considerations are generically applicable to different types of portal models, and encourage the use of complementary probes for the low-energy phenomenology of this model and higher-energy particles required for the completion. 
Light portal particles (e.g. light dark Higgs, dark photon) with feeble couplings can be discovered at Rare and Precision Frontier experiment, while the higher-energy particles can be produced and discovered at higher energy colliders such as the Future Hadron Collider or the Muon Collider. 
A corresponding discovery in low-threshold Cosmic Frontier experiments will determine  the cosmological nature of the dark matter particle. 

Together, information from the Energy, Rare Processes and Precision and Cosmic Frontiers can shed light on the nature of the dark matter particle as well as on the nature of its interactions.  

\paragraph{Considerations on dark matter in dark sectors from Hidden Valley models}

The presence of a DM candidate within the dark sector was one of the original motivations of hidden valley models. 
When focusing on confining dark sectors, one can make an analogy between QCD and dark QCD \cite{Bai:2013xga}. Similarly to how QCD leads to stable massive particles (e.g. the proton), a QCD-like dark interaction gives rise to stable dark hadrons that can be used as DM candidates.

Dark matter was however not the main focus for hidden valley or dark shower models, since it is nearly always possible to engineer a viable DM candidate among the non-charged states of the dark sector, including a thermal relic (see e.g. \cite{Bernreuther:2019pfb,Cline:2016nab,Mies:2020mzw} or via new mechanisms for the relic density see e.g. \cite{Asadi:2021yml}. 
As for other collider searches, the phenomenology of dark shower models is not sensitive to the presence of a DM candidate specifically, but rather to the presence of stable states that escape detection at experiments at colliders. 

Following a common theme of this Section, it is worth emphasising the complementarity between collider searches for dark showers that can probe the stable or long-lived dark sector content, and experiments that can connect dark sector particles to cosmological dark matter \footnote{Additional information on the dark sector spectrum can come from interactions with the lattice community, see e.g. \cite{Kulkarni:2022bvh}}.

Even though the connection between colliders and cosmology is model-dependent, the Snowmass dark showers community encourages future efforts to show cosmological and collider information on the same plot. 
These plots can also be used (in the same way as in other domains, e.g. BRN targets) to motivate parameter scans where a selection of benchmarks satisfies cosmological constraints, and to help connect future discoveries from one Frontier to another.

\subsubsection{Facilities for dark matter searches alongside high energy colliders}


Portal models beyond the WIMP are also a dedicated search target for facilities or experiments that are designed to profit from the presence of a high energy beam or from complementary information recorded at collider experiments. 

More experiments of this kind are discussed in detail in Section \ref{Sub:ForwardDetectors} and in the RF06 Topical Group report in the Rare and Precision Frontier. 
In this Section we will focus on the Forward Physics Facility proposed to be installed in the long shutdown between the LHC Run-3 and the start of the HL-LHC \cite{Feng:2022inv}, which can house a number of such experiments. 

Collisions of high energy beams produce a large flux of secondary particles in the forward region. These can be a source of new, feebly interacting particles (via interaction with targets, bremsstrahlung or decays), potentially creating a “dark matter beam” that can be investigated by dedicated experiments placed downstream from the standard detectors. 

Facilities such as the FPF are relevant for the future of DM at Colliders, as their experiments an be optimised to cover a number of beyond-WIMP scenarios that complement what can be discovered at collider, while exploiting the same colliding beams. 
In particular, FLArE at the FPF \cite{Batell:2021blf} would be sensitive to thermal relic DM, and other experiments such as FASER2 \cite{FASER:2018eoc} and FORMOSA \cite{Foroughi-Abari:2020qar} would be sensitive to a number of dark matter-related scenarios covered earlier in this chapter, for example light invisible particles from decays of the vector / dark photon portal, inelastic DM, asymmetric DM, strongly-interacting millicharged DM and mediator particles such as dark photons and flavor-philic dark Higgs bosons. 

For a comprehensive description of the models and parameter space covered by the FPF experiments, we refer to the RF06 Topical Group report. 
Here, we highlight that the opportunities for dark matter discoveries (as well as of neutrino measurements relevant to cosmic DM experiments) offered by such a facility would increase the potential of the HL-LHC program even further, in a region that is optimal for searches for light thermal relic DM candidates, as well as for dark sectors and other BSM physics.  
The development and building of such a facility can be performed in parallel to current experiments and at a relatively contained cost. 

\subsubsection{Other models of DM at colliders}

The dark sector could be much richer than what is described above, which can lead to qualitatively different signals. 
An example is the dynamical dark matter (DDM) scenario summarized in \cite{Dienes:2022zbh}, in which the number of states in the dark sector is large and perhaps even infinite with different masses and lifetimes. 
The DM in the universe is comprised of an ensemble of dark states. 
An individual state in this ensemble can decay into others plus SM states. 
As a result, the signal of such a dark sector at high energy colliders are qualitatively different. 
In particular, they produce smooth kinematical distributions and long decay chains which are different from many of the standard DM and dark sector signals which rely on the assumption of a few relevant new particles. 
The work is still on going to fully understand the best way of searching for such a dark sector and the potential of the future colliders to cover such scenarios. 

\FloatBarrier

\clearpage

\section{Other signatures}\label{Sec:Gen}

The incredible variety of physics accessible at energy-frontier colliders makes it impractical to try to be comprehensive in the review and discussion of the reach of future facilities. The examples showed so far represent a selected set of signatures and models that have representative significance in the fundamental questions this frontier is aiming to answer.

Many of the investigations pursued at high energy colliders also interleave with other frontiers. Several examples have been shown already and an additional important one is presented in Section~\ref{sec:clfv}, where the interplay between the intensity and energy frontier on charged-lepton flavor violating processes is explored. 

At the same time, maintaining a wide-open approach to unexpected sources of physics beyond the SM is as important and necessary to fully exploiting the discovery power of high-energy colliders. Section~\ref{sec:anomaly-exp} touches on the emerging techniques in need of further development in this direction.


\subsection{Charged-lepton flavor violation}
\label{sec:clfv}
Processes where the flavor of charged leptons is not conserved are so suppressed in the Standard Model that, if observed, they would provide indisputable evidence for physics beyond the SM. Charged lepton flavor violation (CLFV) is therefore a powerful tool to search for BSM physics and a diverse program of experimental searches is currently underway in the US, Europe and Asia probing a number of processes, spanning many energy scales. Moreover, next-generation experiments and facilities will continue to extend the parameter space and, in many cases, enable searches with unprecedented sensitivity. Searches focusing on the tau sector and that are relevant for this report will be conducted at Belle2 and LHC (Run 3 and HL-LHC) in the near-term and at the Super $\tau$ Charm Facility (STCF), the Electron-Ion Collider (EIC) and FCC-ee on a longer timescale. These experiments will be sensitive to CLFV predicted in many BSM models ranging from Supersymmetry and 2HDM to models with heavy neutrinos, leptoquarks, compositeness and new heavy bosons. Searches at these experiments will enable complementary probes of CLFV at different energy scales and will be crucial for identifying the underlying sources of LFV and the underlying mediation mechanism. Ref.~\cite{CLFV} provides an overview of the experimental status of current searches for LFV in the $\tau$ sector and also reviews future prospects. A brief summary is presented here.

In the near term, Belle2 and LHCb will be major contributors and will achieve one to two order of magnitude increase in sensitivity with respect to BABAR and Belle, probing branching ratios down to $10^{-10}$ – $10^{-9}$ over a wide range of final states. Expected exclusion limits at 90\% CL on $B(\tau^{-} \rightarrow \mu^{-}\mu^{+}\mu^{-})$  for ATLAS and CMS for the HL-LHC dataset corresponding to an integrated luminosity of 3000~fb$^{-1}$ will be comparable and range from $1.0 \times 10^{-9}$ (ATLAS) to $3.7 \times 10^{-9}$ (CMS). It is also expected that ATLAS and CMS will search for LFV in Higgs and Z boson decays using the HL-LHC dataset. These searches will build upon those performed using the full Run 2 dataset in the $H \rightarrow e\tau$, $H \rightarrow \mu\tau$,  $Z \rightarrow e\tau$ and $Z \rightarrow \mu\tau$ final states.

Longer term, the EIC will have the ability to select the polarization direction for both electrons and protons in the source and will open the door for precision studies that can significantly test the SM in particular in the area of  electron to $\tau$ transitions. Studies evaluating the reach on the $e \rightarrow \tau$ transition mediated by a 1.9~GeV leptoquark indicate potential to improve current sensitivity by up to an order of magnitude.

The STCF detector will study the benchmark CLFV processes, $B(\tau^{-} \rightarrow \mu^{-}\mu^{+}\mu^{-})$ and $\tau^{-} \rightarrow \mu^{-}\gamma$ . The sensitivity of $B(\tau^{-} \rightarrow \mu^{-}\mu^{+}\mu^{-})$  using one year’s worth of data is estimated to be of $1.4 \times 10^{-9}$ at 90\% CL while the sensitivity of $\tau^{-} \rightarrow \mu^{-}\gamma$ is found to be consistently within the range $(1.2 - 1.8) \times 10^{-8}$ at 90\% CL. In addition, STCF can study the CLFV decay via the process $J/\psi \rightarrow l\tau$ at an expected sensitivity of $4.0 \times 10^{-9}$ at 90\% CL or better. 

The FCC-ee program will open the door to a very rich tau physics program, including the ability to probe the same set of LFV tau decays measured by the B-factories, with sensitivities in the $10^{-10}$ – $10^{-9}$ range.

Overall, we are entering an interesting era wherein current and future experiments will explore uncharted territory in the search for LFV. Searches performed at the next-generation experiments and facilities will probe new physics and improve current limits by few orders of magnitude in the next decades.






\subsection{Anomaly detection}
\label{sec:anomaly-exp}

Theoretical priors are a very important lamppost to guide searches for BSM physics. 
On the other hand, a robust program of exploration sensitive to unexpected BSM phenomena is of uttermost importance to harvest at the best the physics potential of present and future colliders. 

Historically, broad programs of searches for unexpected BSM physics in collider data has been carried out using automatic tools that compare expectations from details physics simulations to data in a large variety of final states~\cite{CDF:2008voc,D0:2000dnz,ATLAS:2018zdn,CMS:2020ohc}. This strategy is very powerful in analyzing a large variety of final states but relies on accurate simulations of a variety of SM processes. 
Recently, advances in artificial intelligence (AI) and machine learning (ML) have allowed new strategies to start developing, broadly referred to as \textit{anomaly detection}\cite{Karagiorgi:2021ngt}. Most of these strategies rely less on accurate simulations of all expected backgrounds and instead make use of data itself to build the SM expectation.  

One area where anomaly detection techniques are being developed is the familiar "bump-hunt" search strategy, searching for new resonances showing in the invariant mass of reconstructed objects as leptons, photons and hadronic jets. 
To exemplify, Ref.~\cite{Chekanov:2021pus} proposes an approach that combines the information of objects reconstructed in the event to define a "feature" space and uses autoencoders to analyze this feature space and identify possible "outlier" regions that SM physics in unlikely to populate. The strategy is exemplified in the search for dijet resonances on LHC simulated data, showing that good separation between SM backgrounds and a variety of BSM benchmarks can be obtained without prior knowledge of such BSM models.

Anomaly detection is certainly going to be integral part of the physics program at future colliders. Ref~\cite{Gonski:2021jek} explores the use of anomaly detection in searches for new resonances in $e^+e^-$ colliders below the nominal center-of-mass energy by analyzing collisions with initial-state radiation (radiative return). The paper shows that data in side-band regions around the target mass search region can be used to inform a weakly-supervised learning of the background features from data itself and the approach is successful in finding new resonances in simulated $e^+e^-$ collision data using a Deep Sets model with Particle Flow Networks.

As the sophistication of these techniques grows, it is equally important to ensure a stimulating environment where new ideas can flourish and be subsequently applied in future colliders. Ref~\cite{Kasieczka:2021xcg} reports on an open challenge where new techniques were submitted to discover a variety of new physics processes that were simulated and distributed as a blinded dataset. The paper describes the various approaches that were received and underlines how such exercises are fundamental in the development of new ideas. The importance of cultivating a vibrant research environment in this area cannot be overstated.

\section{Conclusion (same as Executive Summary)}\label{Conc}

\FloatBarrier

\section{Acknowledgments}
The conveners would like to thank the following people for their work in writing parts of this report, producing some of the summary plots and help us in implementing comments received from the community: Kaustubh Agashe, Juliette Alimena, Sebastian Baum, Mohamed Berkat, Kevin Black, Marco Drewes, Gwen Gardner, Tony Gherghetta, Josh Greaves, Maxx Haehn, Phil C. Harris, Robert Harris, Julie Hogan, Suneth Jayawardana, Abraham Kahn, Jan Kalinowski, Juraj Klaric, Simon Knapen, Ian M. Lewis, Katherine Pachal, Matthew Reece, Tania Robens, Carlos E.M. Wagner, Riley Xu, Felix Yu, Aleksander Filip \.Zarnecki.

The work of S.B. supported by NSF grant No. PHYS-2014215, DoE HEP QuantISED award No. 100495, and the Gordon and Betty Moore Foundation Grant No. GBMF7946.
Work by A.B. is supported by the US Department of Energy (Grant DE-SC0011726).
The work of the work of G.G. is supported by DOE grant DE-SC0007901.
The work of T.G. is supported in part by DOE grant DE-SC0011842 at the University of Minnesota.
The work of R.H. was supported by the Fermi National Accelerator Laboratory, managed and operated by Fermi Research Alliance, LLC under Contract No. DE-AC02-07CH11359 with the U.S. Department of Energy.
The work of J.H. is supported by NSF award 2110972.
The work of A.K. is supported by DOE grant DE-SC0007901.
S.K. was supported by the Office of High Energy Physics of the U.S. Department of Energy under contract DE-AC02-05CH11231.
The work of I.M.L. is supported by DOE grant DE-SC0017988. 
Z.L. has been supported in part by the DOE grant DE-SC0022345 at the University of Minnesota.
K.P.'s research is supported by TRIUMF, which receives federal funding via a contribution agreement with the National Research Council (NRC) of Canada.
C.W. has been partially supported by the U.S. Department of Energy under contracts No. DEAC02-06CH11357 at Argonne National Laboratory. The work of C.W. at the University of Chicago has been supported by the DOE grant DE-SC0013642. 
The work of S.~Pagan~Griso is supported by the Office of High Energy Physics of the U.S. Department of Energy under contract DE-AC02-05CH11231.
The work of T.~Bose is supported in part by DOE grant DE-SC0017647 at the University of Wisconsin-Madison.
N.~R.~S. is supported by the U.S. Department of Energy under contract DE-SC0007983. 
The work of the work of R.X. is supported by DOE grant DE-SC0007901.
The work of F.Y. is supported by the Cluster of Excellence PRISMA$^+$, ``Precision Physics, Fundamental Interactions and Structure of Matter" (EXC 2118/1) within the German Excellence Strategy (project ID 39083149).

\clearpage

\bibliographystyle{JHEP}
\bibliography{main}

\appendix
\section{Tables for LHC/ATLAS Sources}\label{Ap:SUSY}

\begin{table}[h!]
    \caption{Sources for Light Squark Limits}
    \label{tab:ef08_lightsquark}
    \centering
\begin{tabular}{|l|l|l|}
\hline
 Collider & Method & Reference  \\
  \hline\hline
 LHC Run-2 & ATLAS data analysis &  \cite{ATLAS:2020syg} \\
           & CMS   data analysis &  \cite{CMS:2019zmd} \\
\hline
 HL-LHC    & ATLAS Collider Reach &  Run-2 \cite{ATLAS:2020syg} re-scaled \\
           & CMS Collider Reach &    Run-2 \cite{CMS:2019zmd} re-scaled \\
\hline
HE-LHC (27 TeV)     & ATLAS Collider Reach & Run-2 \cite{ATLAS:2020syg} re-scaled \\
                    & CMS Collider Reach &  Run-2 \cite{CMS:2019zmd} re-scaled\\
\hline 
FCC-hh (100 TeV) & ATLAS Collider Reach &  Run-2 \cite{ATLAS:2020syg} re-scaled \\
                 & CMS Collider Reach &    Run-2 \cite{CMS:2019zmd} re-scaled \\
                 & Dedicated Study    & \cite{Golling:2016gvc}   \\
\hline
ILC/C$^3$ (1 TeV)  &  $\sqrt{s}/2$ &  \\
CLIC/Muon (3 TeV)  &  $\sqrt{s}/2$ & \cite{Aime:2022flm} \\
Muon (10 TeV)  &  $\sqrt{s}/2$ & \cite{Aime:2022flm} \\
Muon (30 TeV)  &  $\sqrt{s}/2$ & \cite{Aime:2022flm} \\
\hline
\end{tabular}
\end{table}

\begin{table}[h!]
    \caption{Sources for Gluino Limits}
    \label{tab:ef08_gluino}
    \centering
\begin{tabular}{l|l|l}
 Collider & Method & Reference  \\
  \hline
 LHC Run-2 & ATLAS data analysis & \cite{ATLAS:2020syg} \\
           & CMS   data analysis & \cite{CMS:2019zmd} \\
\hline
 HL-LHC    & ATLAS Collider Reach &  Run-2 \cite{ATLAS:2020syg} re-scaled \\
           & CMS Collider Reach &    Run-2 \cite{CMS:2019zmd} re-scaled \\
           & Dedicated Study    & \cite{CidVidal:2018eel} \\
\hline
HE-LHC (27 TeV)     & ATLAS Collider Reach &  Run-2 \cite{ATLAS:2020syg} re-scaled \\
                    & CMS Collider Reach &    Run-2 \cite{CMS:2019zmd} re-scaled \\
                    & Dedicated Study & \cite{CidVidal:2018eel} \\
\hline 
FCC-hh (100 TeV) & ATLAS Collider Reach & Run-2 \cite{ATLAS:2020syg} re-scaled \\
                 & CMS Collider Reach &   Run-2 \cite{CMS:2019zmd} re-scaled \\
                 & Dedicated Study    & \cite{Golling:2016gvc} \\   
\hline
\end{tabular}
\end{table}

\begin{table}[h!]
\caption{Sources for Stop Squark Limits}
\label{tab:ef08_stop_sources}
\centering
\begin{tabular}{l|l|l}
 Collider & Method & Reference  \\
  \hline
 LHC Run-2 & ATLAS data analysis & \cite{ATL-PHYS-PUB-2022-013}  \\
           & CMS   data analysis & \cite{CMS:2021beq}  \\
\hline
 HL-LHC    & ATLAS Study & \cite{ATL-PHYS-PUB-2018-021}  \\
           & ATLAS Collider Reach & Run-2 \cite{ATL-PHYS-PUB-2022-013} re-scaled \\
           & CMS Collider Reach & Run-2 \cite{CMS:2021beq} re-scaled \\
\hline
HE-LHC (27 TeV)     & ATLAS Collider Reach & Run-2 \cite{ATL-PHYS-PUB-2022-013} re-scaled \\
                    & CMS Collider Reach & Run-2 \cite{CMS:2021beq} re-scaled \\
\hline 
FCC-hh (100 TeV) & ATLAS Collider Reach & Run-2 \cite{ATL-PHYS-PUB-2022-013} re-scaled \\
                 & CMS Collider Reach & Run-2 \cite{CMS:2021beq} re-scaled \\
                 & Dedicated Study    & \cite{Gouskos:2642475} \\   
\hline
ILC/C$^3$ (1 TeV)  &  $\sqrt{s}/2$ & n/a \\
CLIC/Muon (3 TeV)  &  $\sqrt{s}/2$ & n/a \\
Muon (10 TeV)  &  $\sqrt{s}/2$ & n/a \\
Muon (30 TeV)  &  $\sqrt{s}/2$ & n/a \\
\hline
FCC-ee Precision & Analysis of precision measurement predictions & \cite{Essig_2017} \\
CEPC Precision   & Analysis of precision measurement predictions & \cite{Essig_2017} \\
\hline
\end{tabular}
\end{table}

\begin{table}[h!]
    \caption{Sources for Wino-Chargino Limits}
    \label{tab:ef08winochargino}
    \centering
\begin{tabular}{l|l|l}
 Collider & Method & Reference  \\
  \hline
 LHC Run-2 & ATLAS data analysis & \cite{ATLAS:2021yqv} \\
           & CMS   data analysis & \cite{CMS:2018xqw} \\
\hline
 HL-LHC    & ATLAS Collider Reach & \cite{ATLAS:2021yqv} \\
           & CMS Collider Reach & \cite{CidVidal:2018eel} \\
           & CMS Dedicated (hadronic)       &  \cite{ATL-PHYS-PUB-2022-018} \\
\hline
HE-LHC (27 TeV)     & ATLAS Collider Reach & \cite{ATLAS:2021yqv} \\
                    & CMS Collider Reach & \cite{CidVidal:2018eel} \\
\hline 
FCC-hh (100 TeV) & ATLAS Collider Reach &  \cite{ATLAS:2021yqv} re-scaling\\   
                 & CMS Collider Reach & n/a \\
                 & Dedicated Study    &  \cite{Golling:2016gvc}\\   
\hline
ILC/C$^3$ (1 TeV)  &  $\sqrt{s}/2$ & \cite{Antusch_2018} \\
CLIC/Muon (3 TeV)  &  $\sqrt{s}/2$ & \cite{CLIC:2018fvx}\\
Muon (10 TeV)  &  $\sqrt{s}/2$ & n/a\\
Muon (30 TeV)  &  $\sqrt{s}/2$ & n/a\\
\hline
\end{tabular}
\end{table}

\begin{table}[h!]
    \caption{Sources for Higgsino Limits}
    \label{tab:ef08higgsino}
    \centering
\begin{tabular}{l|l|l}
 Collider & Method & Reference  \\
  \hline
 LHC Run-2 & ATLAS data analysis &  \cite{ATLAS:2021moa}\\
           & CMS   data analysis &  \cite{CMS:2021edw}\\
\hline
 HL-LHC    & ATLAS Collider Reach & Run-2 \cite{ATLAS:2021moa} re-scaled \\
           & CMS Collider Reach & Run-2 \cite{CMS:2021edw} re-scaled \\
           & ATLAS Dedicated Study    & \cite{ATLAS:2018jjf} \\
           & CMS Dedicated Study    & \cite{CMS:2018qsc} \\   
 \hline
HE-LHC (27 TeV)     & ATLAS Collider Reach & Run-2 \cite{ATLAS:2021moa} re-scaled  \\
                    & CMS Collider Reach & Run-2 \cite{CMS:2021edw} re-scaled \\
\hline 
FCC-hh (100 TeV) & ATLAS Collider Reach & Run-2 \cite{ATLAS:2021moa} re-scaled\\
                 & CMS Collider Reach & Run-2 \cite{CMS:2021edw} re-scaled \\
\hline
ILC/C$^3$ (1 TeV)  &  $\sqrt{s}/2$ &  \\
CLIC/Muon (3 TeV)  &  $\sqrt{s}/2$ & \\
Muon (10 TeV)  &  $\sqrt{s}/2$ & \\
Muon (30 TeV)  &  $\sqrt{s}/2$ & \\
\hline
\end{tabular}
\end{table}

\begin{table}[h!]
    \caption{Sources for Smuon Limits}
    \label{tab:ef08_smuon}
    \centering
\begin{tabular}{l|l|l}
 Collider & Method & Reference  \\
 \hline
 LHC Run-2         & ATLAS data analysis & \cite{ATLAS:2019lff} \\
 (prompt)          & CMS   data analysis & \cite{CMS:2018eqb} \\
 \hline
 LHC Run-2         & ATLAS data analysis & \cite{ATLAS:2020wjh} \\
 (long-lived)      & CMS   data analysis & \\
\hline
 HL-LHC    & ATLAS Collider Reach & Run-2 \cite{ATLAS:2019lff, ATLAS:2020wjh} re-scaled \\
           & CMS Collider Reach & Run-2 \cite{CMS:2018eqb} re-scaled \\
           & Dedicated Study (long-lived, not included in plot) & \cite{CidVidal:2018eel}  \\
\hline
HE-LHC (27 TeV)     & ATLAS Collider Reach & Run-2 \cite{ATLAS:2019lff, ATLAS:2020wjh} re-scaled \\
                    & CMS Collider Reach & Run-2 \cite{CMS:2018eqb} re-scaled \\
\hline 
FCC-hh (100 TeV) & ATLAS Collider Reach & Run-2 \cite{ATLAS:2019lff, ATLAS:2020wjh} re-scaled \\
                 & CMS Collider Reach & Run-2 \cite{CMS:2018eqb} re-scaled \\
\hline
\end{tabular}
\end{table}

\begin{table}[h!]
    \caption{Sources for Stau Limits}
    \label{tab:ef08_stau}
    \centering
\begin{tabular}{l|l|l}
 Collider & Method & Reference  \\
  \hline
 LHC Run-2 & ATLAS data analysis & \cite{stau:atlaslhc} \\
           & CMS   data analysis &  \\
\hline
 HL-LHC    & ATLAS Collider Reach & Run-2 \cite{stau:atlaslhc} re-scaled \\
 & CMS Collider Reach &  \cite{CMS:2018imu} \\
 & ATLAS dedicated study & \cite{ATLAS:2018diz} \\
\hline
HE-LHC (27 TeV)     & ATLAS Collider Reach & Run-2 \cite{stau:atlaslhc} re-scaled \\
                    & CMS Collider Reach &  \\
\hline 
FCC-hh (100 TeV) & ATLAS Collider Reach & Run-2 \cite{stau:atlaslhc} re-scaled \\
                 & CMS Collider Reach &  \\
                 & Dedicated Study    & \cite{Gouskos:2642475} \\   
\hline
ILC/C$^3$ (1 TeV) & Dedicated Study &  \cite{NunezPardodeVera:2022izz}  \\
CLIC (3 TeV)  &  Private Communication & \cite{Strategy:2019vxc}\\
\hline
\end{tabular}
\end{table}

\FloatBarrier

\end{document}